\documentclass[12pt]{report}
\usepackage{usfthesis}  				
\usepackage[numbers]{natbib}
\usepackage{amsfonts}
\usepackage{amsmath}
\usepackage{braket}
\renewcommand{\headrulewidth}{0pt}
\begin{document}

\thispagestyle{empty}

\begin{center}

\vspace*{0.6in}

\begin{doublespace}
{
Field Quantization for Radiative Decay of Plasmons in Finite and Infinite Geometries
}
\end{doublespace}

\vspace*{0.4in}
by
\vspace*{0.57in}

Maryam Bagherian

\vspace*{0.8in}

A dissertation submitted in partial fulfillment\\
of the requirements for the degree of\\
Doctor of Philosophy\\
Department of Mathematics \& Statistics\\
College of Arts and Sciences\\
University of South Florida\\

\vspace*{0.4in}

Major Professor: Sherwin Kouchekian, Ph.D.\\
Brian Curtin, Ph.D. \\
Ali Passian, Ph.D.\\
Ivan Rothstein, Ph.D.\\
Boris Shekhtman, Ph.D.

\vspace*{0.4in}
Date of Approval \\
March 8, 2019

\vspace*{0.3in}

\vspace{\baselineskip}

Copyright \copyright\, 2019, Maryam Bagherian

\end{center}

\newpage


\thispagestyle{empty}

\renewcommand\baselinestretch{2.0}\selectfont

\begin{center}
\vspace*{0.55in}
{\normalsize\bf DEDICATION}\\
\vspace{0.35in}
To Farshid; brother and example. 
\end{center}

\newpage

\thispagestyle{empty}

\begin{center}
\vspace*{0.55in}
{\normalsize\bf ACKNOWLEDGMENTS}
\end{center}

 I would like to sincerely appreciate the great encouragement, support and guidance I received from my advisor Dr.~Kouchekian throughout this research. It is not easy to overestimate the help and guidance I received from Dr.~Passian to whom I would like to manifest my gratitude. They chose to give me a chance and devote a large amount of their time and energy for my betterment, while they had no reason to believe in me or help me in anyway.  

 I would also like to thank  Dr.~Rothstein for his contribution to this research as well as being a member of my dissertation committee. Thanks are due to  Dr.~Curtin and Dr.~Shekhtman for agreeing to serve on my dissertation committee, which is greatly appreciated.

My deepest gratitude goes to my beloved family, each and every single one of them, for their lifelong love and support. This dissertation would not have been completed without their constant encouragement and kindness.

 Lastly, I offer my regards to all of those who supported me in any respect during the completion of this work. 

\setlength{\parindent}{0.5in}

\setlength{\parindent}{0in}



{ 

\renewcommand{\thepage}{\roman{page}}

{
\renewcommand\baselinestretch{0.8}\selectfont
\setcounter{page}{0} 
{\long\def\footnotemark[#1]{}
\tableofcontents}
\newpage
}


\titlespacing{\chapter}{0pt}{45pt}{2.5ex}


\singlespace
\begingroup
\renewcommand\addvspace[1]{}
\setlength\parskip{\baselineskip}
\endgroup
\newpage

\addcontentsline{toc}{chapter}{\listfigurename}
\begingroup
\renewcommand\addvspace[1]{}
\setlength\parskip{\baselineskip}
\listoffigures
\endgroup
\newpage

}  

{\renewcommand{\thepage}{\roman{page}}	
\setlength{\parindent}{0.5in}
\addcontentsline{toc}{chapter}{\normalsize\textnormal{{\MakeUppercase\abstractname}}}

\begin{center}
\vspace*{0.65in}
{\normalsize\bf ABSTRACT}
\end{center}


We investigate field quantization in high-curvature geometries. The models and calculations can help with understanding the elastic and inelastic scattering of photons and electrons in  nanostructures and probe-like metallic domains. The results find important applications in high-resolution photonic and electronic modalities of scanning probe microscopy, nano-optics,  plasmonics, and quantum sensing. 
Quasistatic formulation, leading to nonretarded quantities, is employed and justified on the basis of the nanoscale, here subwavelength, dimensions of the considered domains of interest.
Within the quasistatic framework, we represent the nanostructure material domains with  frequency-dependent dielectric functions. Quantities associated with the normal modes of the electronic systems, the nonretarded plasmon dispersion relations, eigenmodes, and  fields are then calculated for several geometric entities of use in nanoscience and nanotechnology. 
From the classical energy of the charge density oscillations in the modeled nanoparticle, we then derive the Hamiltonian of the system, which is used for quantization. 
The quantized plasmon field is obtained and, employing an interaction Hamiltonian derived from the first-order perturbation theory within the hydrodynamic model of an electron gas, we obtain an analytical expression for the radiative decay rate of the plasmons. 
The established treatment is applied to multiple geometries  to investigate the quantized charge density oscillations on their bounding surfaces. Specifically, using one sheet of a two-sheeted hyperboloid of revolution, paraboloid of revolution, and cylindrical domains, all with one infinite dimension, and the finite spheroidal and toroidal domains are treated. 
In addition to a comparison of  the paraboloidal and hyperboloidal results,  interesting similarities are observed for the paraboloidal domains with respect to the surface modes and radiation patterns of a prolate spheroid, a finite geometric domain highly suitable for modeling of nanoparticles such as quantum dots. The prolate and oblate spheroidal calculations are validated by comparison to the spherical case, which is obtained as a special case of a spheroid. 
In addition to calculating the potential and field distributions, and dispersion relations, we study the angular intensity and the relation between the emission angle with the rate of radiative decay. 
 The various morphologies are compared for their plasmon dispersion properties, field distributions, and radiative decay rates, which are shown to be consistent. 
 For the specific case of a nanoring, modeled in the toroidal geometry, significant complexity arises due to an inherent coupling among the various modes. Within reasonable approximations to decouple the modes, we study the radiative decay channel for a vacuum bounded single solid nanoring by quantizing the fields associated with charge density oscillations on the nanoring surface. Further suggestions are made for future studies. The obtained results are relevant to other material domains that model a nanostructure such as a probe tip, quantum dot, or  nanoantenna.

\newpage}
\titlespacing{\chapter}{0pt}{37pt}{2ex}
\renewcommand{\thepage}{\arabic{page}}	
\setcounter{page}{1}					
\setlength{\parindent}{0.5in}
\chapter{INTRODUCTION}\label{1}
\label{Introduction}

\pagenumbering{arabic} \pagestyle{plain} 

\section{Introduction}\label{into}

Interaction of photos with surface plasmons has been studied extensively during the past five decades, particularly in relation to their application in scanning probe microscopy (SPM), they present a high potential for emerging applications in fields such as quantum sensing~\cite{garapati,garapati2, ben,qafm,tlf:thin}. It is known as a branch of microscopy that forms images of surfaces using a physical probe. Throughout these studies, various geometries have been investigated. In particular, Ritchie together with Crowell, Little, Ashley and Ferell \cite{Crowell, Little_Paper, B, Rit, Ritchie} have studied the interaction of general and special cases of surface of a metallic sphere and oblate spheroid with photons and derives the very first relations describing the field quntization of surface plasmons. Inspired by these results, quantization relations for a special case of a long string-like cylinder has been studied by Burmistrova \cite{BURMISTROVA}. In this work, we wish to investigate the interaction in more detail as well as to extend the obtained results to both new finite and infinite geometries such as cylinder, paraboloid, hyperboloid, prolate spheroid and  the non-simply connected case of a torus. It should be pointed out that the \emph{surface plasmon-photon interaction} (SPP)  can be analyzed for various cases such as absorption, emission, Thomson and Rayleigh scattering, and  scattering cross-sections. Our main focus in this study, however, is the emission case and its application to the decay rate.  As we shall see later in this chapter, there is a close relation between both cases of emission and absorption in terms of the interaction matrix element. As a result, our study also addresses the absorption phenomena \cite{B,GENZEL}.

Throughout this dissertation, we consider a metallic particle confined in vacuum whose shape is described by one of the above mentioned geometries. The conduction electrons and ion centers in a metal together to form a plasma. The authors in \cite{pines}, while studying the collective oscillations in the metallic plasma, showed that through the plasma system exhibits resonances along the directions that charge density waves propagate, with no effect of damping. Ritchie \cite{Ritchie} in 1957 showed that the electron density can be supported at the surface of a bounded plane free-electron gas. In \cite{Ferrell1, SternFerrell}, it has been discussed both the property of plasmons and their interaction with charged particles. The properties of plasmons and details on how they interact with light or charges particles, such as photons, are also discussed and reviewed in \cite{FIS,MASRI, BARKER, RIT34,raether}. To describe the properties of a metal, mostly from a dynamical point of view, the complex-valued \emph{dielectric function} $\varepsilon(\omega)$ was used. Generally speaking, the dielectric function specifies the relation between the frequency and the wave-vector in a certain metal. It is also known as \emph{permitivity}. The main assumption here is that in the incident of large wave-vectors, the dielectric function would be no longer wave-vector-dependent \cite{pines}. 

Throughout this chapter, we adopt the conventions of Ritchie mostly presented in \cite{Little_Paper}, unless otherwise specified. 
The \emph{dielectric function}, $\varepsilon(\omega)$, is mostly used to express certain properties of a metal. It is a complex and frequency-dependent function which in general also depends on wave vector of the plasmon field, $k_p \equiv 2\pi/ \lambda_p$ where $\lambda_p$ denotes the plasmon wavelength. However, under certain suitable assumptions, the wave vector dependency could be ignored \cite{pines}. Under the assumption that the wavelength is large enough, the dielectric function can be treated as independent from the wave vector. This requirement is met for all cases which have been investigated in this work.\\

\noindent \textbf{Free-electron-gas dispersion relation}

We start by considering the case of a spring oscillator, with no damping forces. The equation of motion, using \emph{Newton's second law}, can be written as \cite{may}:
\begin{equation}\label{may1}
F(\mathbf r_i)=m\, \frac{d^2 \mathbf r_i}{dt^2} , 
\end{equation}
where $m$ denotes the mass and $\mathbf r_i$ is the displacement vector of $i$-th electron with respect to the equilibrium position. \emph{Hooke's} law, on the other hand, gives:
\begin{equation}\label{may2}
F(\mathbf r_i)=-k\, \mathbf r , 
\end{equation}
where $k$ denotes the spring constant. From Eqs.~\eqref{may1} and \eqref{may2}, one could write the second order differential equation: 
\begin{equation}\label{may3}
m\, \frac{d^2 \mathbf r_i}{dt^2} =-k\, \mathbf r_i, 
\end{equation}
whose solutions are given by:
\begin{equation}\label{may4}
 \mathbf r_i(t)= A\cos(\omega_0t)+ B\sin(\omega_0t), 
\end{equation}
for some constants $A$ and $B$, where $\omega_0 = \sqrt{\frac{k}{m}}$ is the \textit{resonant frequency}. The equation of motion for the $i$-th electron, including both the driving forces and the \emph{damping} $\gamma$ is written as \cite{fox}:
\begin{equation}\label{may5}
\frac{d^2 \mathbf r_i}{dt^2}+ \gamma \frac{d \mathbf r_i}{dt}+ \omega^2_0 \mathbf r_i =- \frac{e}{m_e}\, \vec E, 
\end{equation}
where $e$, $m_e$ denote the electric charge and the mass of an electron, respectively, and $\vec E$ represents the \emph{electric field}. The terms $\omega^2_0 \mathbf r_i$ and $- \frac{e}{m_0}\, \vec E$ represent the \emph{restoring} and electric forces, respectively. The time dependent electric field of a light wave inducing oscillations is given by: 
\begin{equation}\label{may6}
\vec E(t)= \cos(\omega t+ \theta), 
\end{equation}
where $\omega$ is the frequency of the system and $\theta$ is denotes the phase of the wave. Eq.~\eqref{may6} could be written as:
\begin{equation}\label{may7}
\vec E(t)=\vec E_0 \, \text {Re}\, ({e^{-(i\omega t+\theta)}}), 
\end{equation}
where $\vec E_0$ denotes the amplitude. We are mostly interested in those solutions of Eq.~\eqref{may5} which have the form similar to Eq.~\eqref{may7}, namely, 
\begin{equation}\label{may8}
\mathbf r_i(t)=\vec R_0 \,  \text{Re} ({e^{-(i\omega t+\tilde \theta)}}),  
\end{equation}
where $ \vec R_0$ and $ \tilde \theta$ represent the amplitude and phase of the oscillations, respectively. Substituting Eqs. ~\eqref{may7} and \eqref{may8} in Eq.~\eqref{may5}, one may write:
\begin{equation}\label{may9}
\left( -\omega^2 \,  -i \gamma \, \omega+\omega^2_0\right)  \vec R_0 = - \frac{e}{m_0}\, \vec E_0, 
\end{equation}
and hence: 
\begin{equation}\label{may10}
 \vec R_0=  \frac{-e\, \vec E_0}{m_e \left( -\omega^2 \,  -i \gamma \, \omega+\omega^2_0\right)}\, , 
\end{equation}
where $\omega$ is the frequency of the electric field. The \emph{electric polarization vector} $\vec P$ is given by \cite{jackson}:
\begin{equation}\label{P2}
\vec P= n_0\, p(t), 
\end{equation}
where $p(t)$ denotes the \emph{dipole moment} per unit volume and $n_0$ is the number of effective electrons per unit volume. One can write:
\begin{equation}\label{P3}
 n_0 p(t)= -n_0 \, q\, \mathbf r_i =  \frac{n_0 e^2 \, \vec E_0}{m_e \left[ \omega^2_0-\left( \omega^2 \,  +i \gamma \, \omega\right) \right]}. 
\end{equation}
The \emph{displacement} vector $\vec D$ throughout the volume of the gas is written as \cite{garrity}:
\begin{equation}\label{D0}
\vec D= \vec E+4\pi \, \vec P. 
\end{equation}
Assuming the substance is isotropic, the relation between polarization vector and electric field is given by:
\begin{equation}\label{0-E}
 \vec P= -\frac{\chi_e(\omega)}{4\pi }\, \vec E, 
\end{equation}
where $\chi_e(\omega)$ denotes the \emph{electrical susceptibility} of a dielectric material. Using Eq.~\eqref{0-E}, one may write Eq.~\eqref{D0} as:
\begin{equation}\label{D2}
\vec D= \frac{1}{4\pi}\, \varepsilon(\omega)\, \vec E, 
\end{equation}
where we put:
\begin{equation}\label{D3}
\varepsilon(\omega)=1+ \chi_e(\omega), 
\end{equation}
denoting \emph{dielectric function} for a materiel. Substituting Eqs.~\eqref{P3} and \eqref{D3} in Eq.~\eqref{D2}, one finds the dielectric function for a material as:
\begin{equation}\label{1-28}
\varepsilon (\omega) = 1+ \frac{\omega_p^2}{\omega_0^2 - \omega^2 - i\gamma \omega},
\end{equation}
where $\omega_p$ denotes the \emph{bulk plasma frequency} also known as \emph{plasmon} frequency, and is given by:
\begin{equation}\label{1-29}
\omega_p^2= \frac{4\pi\, n_0\, e^2}{m_e}, 
\end{equation}
 in \emph{Gaussian's} unit.  The \emph{Drude model} for the dielectric function of metals is obtained in the assumption that the free electrons in metals are not bound to any atoms and are not subject to restoring forces. This implies the spring constant $k$ to be zero and therefore $\omega_0=0$ \cite{Bo,fox}. Using this in Eq.~\eqref{1-28} leads to:
\begin{equation}\label{1-30}
\varepsilon (\omega) =1- \frac{\omega_p^2}{\omega(\omega +i\gamma)}.
\end{equation}
In the absence of damping forces, the free-electron-gas dielectric function reduces to the well-known expression:
\begin{equation}\label{0-eps1}
\varepsilon(\omega)= 1- \frac{\omega^2_p}{\omega^2}.
\end{equation}
In the generalized Drude model one may assume $\gamma$ is frequency-independent and that it is complex valued for $\varepsilon$ to undergo \emph{Kramers Kroing} dispersion relation, in which the real part stays constant with respect to plasma energy, while the imaginary part could be neglected. Thus the simple Drude model stays valid throughout all the steps up  to plasma energy. 
If one ignores the damping force, then the real dielectric function of a free electron gas could be expressed by Eq.~\eqref{0-eps1}. In order to determine the value of dielectric function for surface plasmon oscillations, we impose the requirement of continuity of electric potential and normal component of the displacement vector across the surface of the metal. This is where the geometric effects of the surface  plays a role in finding the values of dielectric function $\varepsilon(\omega)$ and hence the value of frequency $\omega$ itself. We will discuss these geometric effects in the coming chapters thoroughly. 

Throughout the whole discussion, we took on the assumption that all the interactions occur without a delay, i.e. they are instantaneous. In other words, the system is considered in a way that electrostatic solutions are applicable. This gives us a relatively fair approximation as long as our assumption of having large wavevectors for plasmons (comparing to that of light) is valid. This large-wavevector region is known in the literature the as nonretarted region. This allows us to consider \emph{Poisson} and \emph{Laplace}'s equations instead of wave equation to determine the allowed frequencies. What one might loose under this assumption is thoroughly discussed by Ritchie in \cite{RIT34}.  We devote this chapter to describing our approach regarding the  radiative decay of the surface plasmons of an excited metallic surface which mostly follows the one given in \cite{B}. Furthermore, the geometric effect of the curvatures for such metallic surface has been investigated thoroughly in \cite{PassianCurve}. 

\section{Preliminaries}
The electrical size is determined by comparing physical dimension of an electronic structure and the signal wavelength. Let us consider a field which varies at low enough frequency when compared to its wavelength. Therefore, considering a small lapse of time, one could assume that the system remains in an internal equilibrium. In other words, the time delay due to wave propagation from one point to any other point can be ignored if the circuit geometry is relatively small compared to the signal wavelength (which prevents the field distribution to vary significantly). Under these circumstances, quantities like potential and surface current (which will be defined later in this chapter) will not be dependent on position and time \cite{sui}. We start this section with elaborating the mathematical formulation in greater detail.

Let us recall \emph{Maxwell's equations} in their most standard formulations as: 
\begin{align}
\vec \nabla \cdot \vec E &= 4\pi\, \rho,\label{Max2} \\
\vec \nabla \times \vec E&= -\frac{\partial \vec B}{\partial t}, \label{0-Max2} \\
\vec \nabla \cdot \vec B&=0, \label{0-Max3}\\
c^2 \, \vec \nabla \times \vec B&= \mathbf J+ \frac{\partial \vec E}{\partial t}, \label{0-Max4}
\end{align}
where $\vec E$, $\vec B$, $\mathbf J$, and $\rho$ denote \emph{electric} field, \emph{magnetic} field, \emph{current} and \emph{charge density}, respectively. As customary, $c$ indicates the speed of light. 
In the absence of any magnetic field $\vec B$, one could set:
\begin{eqnarray}\label{M1}
&&\vec E= - \left[\frac{1}{c}\, \frac{\partial \mathbf A}{\partial t}+	\vec \nabla \Phi(\mathbf r,t)\right], \label{Max1}
\end{eqnarray}
where $\Phi(\mathbf r,t)$ and $\mathbf A$ are \emph{scalar potential} and \emph{vector potential} field, respectively, and $\mathbf r$ denotes the position vector (see any classical reference on this topic, also \cite{garrity,harris,jackson,SET,Sakurai,SIRQFT}). Furthermore, if the wave speed is relatively small compared to $c$, one may ignore the first term in the right-hand side of Eq.~\eqref{Max1}. Under this assumption, we can write Eq.~\eqref{Max1} as: 
\begin{eqnarray}\label{M2}
&&\vec E= - \vec \nabla \Phi(\mathbf r,t).
\end{eqnarray}
By means of Eqs.~\eqref{Max2} and \eqref{M2}, we may assume that the scalar potential $\Phi$ satisfies Poisson's equation:
\begin{equation}\label{1-1P2}
\vec \nabla^2 \Phi(\mathbf r,t)= -4\pi \rho. 
\end{equation}
In a charge-free system, Poisson equation is known as Laplace's equation:
\begin{equation}\label{1-1L}
\vec \nabla^2 \Phi(\mathbf r,t)= 0.
\end{equation}
It is of fundamental importance that under proper boundary conditions the solution to either Poisson or Laplace's equations is unique \cite{med}. In terms of boundary conditions, they are discussed in the next section, for the convenience of the reader and the sake of self consistency.

\section{General Approach} 

It is well-known that the retarded potentials are due to time-varying charge densities which are associated with surface plasmons \cite{Bohm}. 
In the non-retarded limit, the instantaneous potentials are calculated assuming the source is a static charge density. 
Retardation for the fields can be ignored near the surface of a particle if the dimensions of the particle are much less than the wavelength of light at frequency $\omega$. 
\par
\begin{figure}[H]
	\centering
	\includegraphics[width=4.5in]{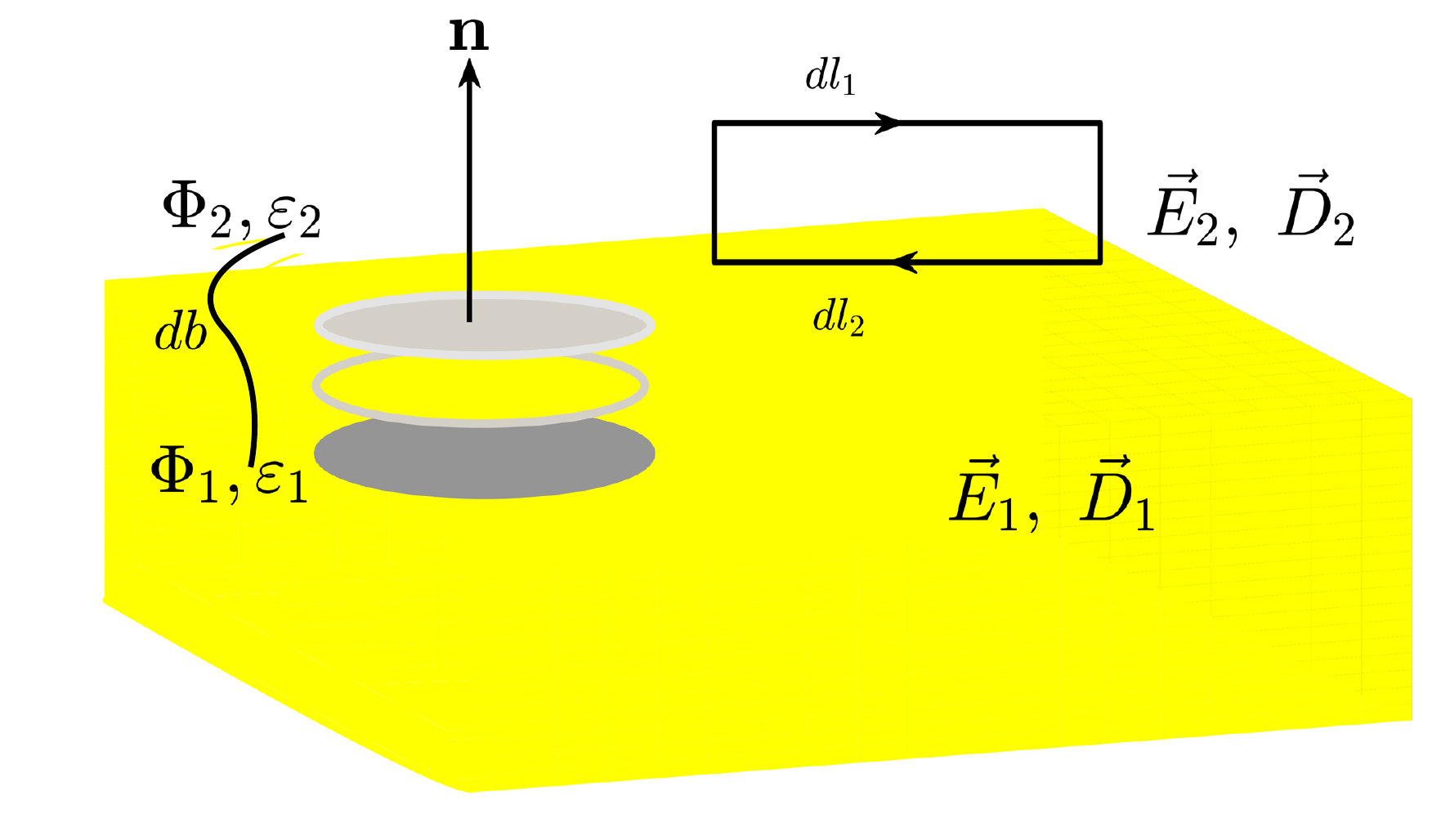}\\	
	\caption[A tiny pillbox of volume $\mathcal V$ at the interface between two]{A tiny pillbox of volume $\mathcal V$ at the interface between two media with dielectric functions $\varepsilon_1$ and  $\varepsilon_2$. Vector $\mathbf n$ is the normal vector to the surface. }
	\label{FBD}
\end{figure}

\begin{figure}[H]
	\centering
	\includegraphics[width=5in]{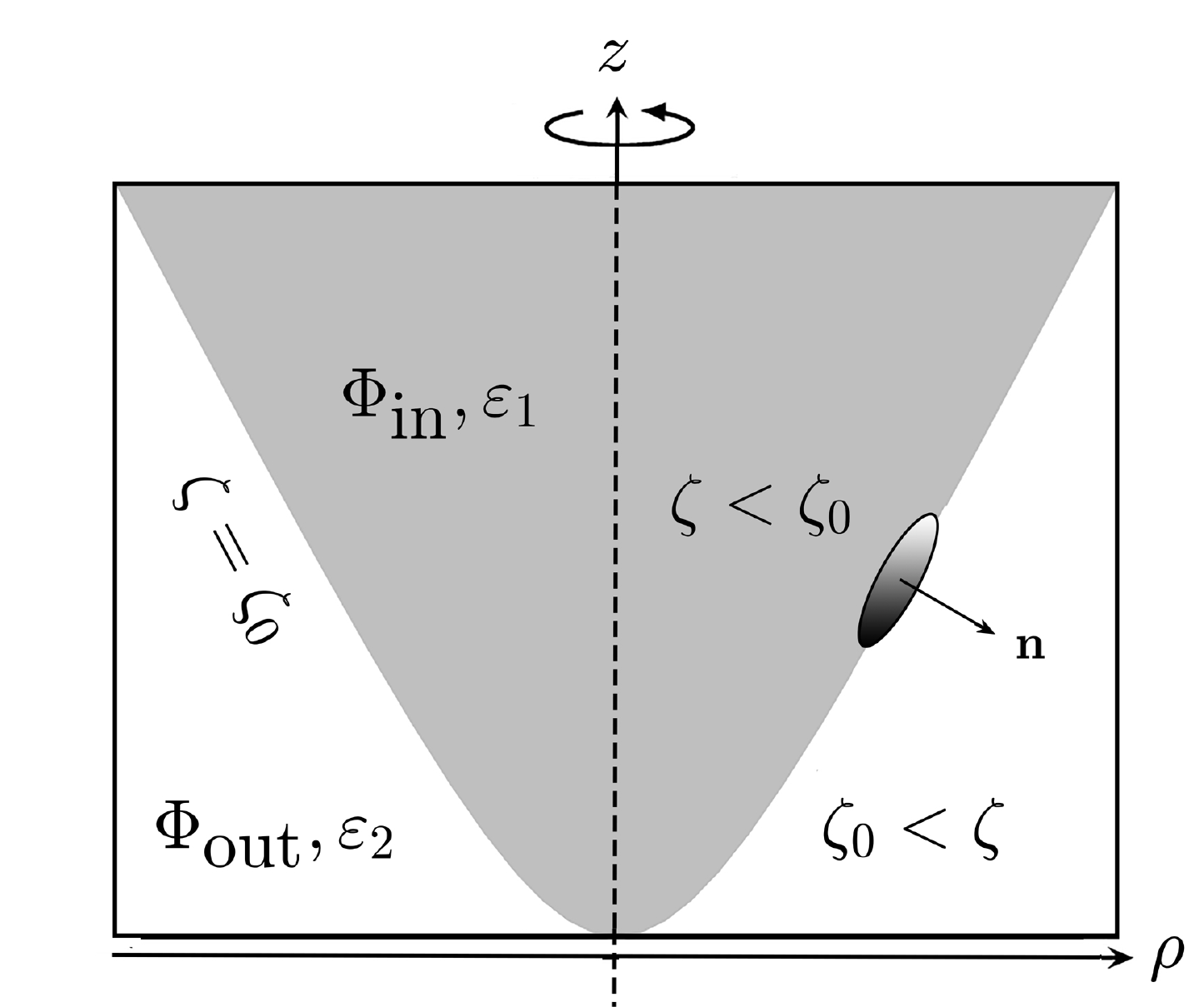}\\	
	\caption[The solid surface of revolution region defined by fixing $\zeta=\zeta_0$, ]{The solid surface of revolution region defined by fixing $\zeta=\zeta_0$, along with inside and outside scalar potentials $\Phi_{\text{in}}$ and $\Phi_{\text{out}}$, respectively. The $\varepsilon_1$ and $\varepsilon_2$ denote the dielectric constant for the inside and outside regions. $\mathbf n$ denotes the normal unit vector. } 
	\label{B}
\end{figure}

 We start by considering a pillbox with volume $\mathcal V$ at the interface of two media with dielectric function $\varepsilon_1$ and $\varepsilon_2$, such that the areas filled with each material are the same (see Fig.~\ref{FBD}). Maxwell's equations imply the continuity of the tangential component of the electric field $\vec E$ and the normal component of displacement vector $\vec D$. We determine these boundary conditions using \emph{Gauss} and \emph{Stoke's} laws \cite{gauss}. 
 In view of  Eq.~\eqref{D2}, for any external sources in vacuum with $\varepsilon(\omega)=1$, using Gauss's law for a continuous charge density $\rho$, we can write: 
\begin{equation}\label{Pi1}
\oint_{\partial \Pi} \vec D \cdot \mathbf n \, d\mathcal A= \int_{\Pi} \rho \,d\mathcal V.
\end{equation}
Utilizing the identity:
\begin{equation}\label{Pi3}
\oint_{\partial \Pi} \vec D \cdot \mathbf n \, d\mathcal A=(\vec D_2-\vec D_1)\cdot \mathbf n\,  \Delta \mathcal A, 
\end{equation}
where $\vec D_1$ and $\vec D_2$ correspond to displacement vectors of the inside and outside regions and $\mathbf { n}$ is the surface unit normal vector. Since the polarization charge is only confined to the surface, we may write:
\begin{equation}\label{Pi2}
\int_{\Pi} \rho \,d\mathcal V= \int_{\partial \Pi} \sigma d\mathcal A= \sigma\, \Delta \mathcal A.
\end{equation}
It follows from Eqs.~\eqref{Pi3} and \eqref{Pi2}:
\begin{equation}\label{B1}
(\vec D_2-\vec D_1)\cdot \mathbf n=\sigma, 
\end{equation}
where $\sigma$ notes the {surface charge} density on the surface boundary $\partial \Pi$. Eq.~\eqref{B1} presents no polarization charge and therefore indicates that the normal component of displacement vector $\vec D$ has a shift equals to surface charge density $\sigma$. This is the first boundary condition. Using Eq.~\eqref{D2}, Eq.~\eqref{B1} may be also written as:
\begin{equation}\label{B2}
(\varepsilon_2 \, \vec E_2-\varepsilon_1\, \vec E_1)\cdot \mathbf n=4\pi\, \sigma. 
\end{equation}
Making use of Eq.~\eqref{M2}, we may also write:
\begin{equation}\label{B3}
\varepsilon_1\nabla \Phi_1 \cdot \mathbf{ n}\,  \Big|_{\partial \Pi}-
\varepsilon_2\nabla \Phi_2 \cdot \mathbf {n}\,  \Big|_{\partial \Pi}= 4\pi\, \sigma, 
\end{equation}
the second boundary condition is widely used to determine the dielectric function and, as a consequence, normal mode frequencies for the given geometry. Using the definition $\oint_C\vec E\cdot dl= 0$ on a closed path $C$, it follows that:
\begin{equation}\label{B4}
\int_{A}^{B} \vec E\cdot dl= - (\Phi_{B}  - \Phi_{A}). 
\end{equation}
In virtue of {Stoke's} theorem, we have:
\begin{equation}\label{B5}
\int_{\partial \Pi} (\vec \nabla\times\vec E) \cdot \mathbf n\,  d\mathcal A= 0.
\end{equation}
Under electrostatic condition, we derive the identity $\vec \nabla \times E=0$. Considering a closed loop as shown in Fig.~\ref{FBD} and applying the line-integral along the loop, we obtain:
\begin{equation}\label{B6}
\oint_C \vec E \cdot dl= \int (\vec E_2- \vec E_1) \cdot dl = 0, 
\end{equation}
having used the fact that $dl_1= -dl_2$. In vector notation, since Eq.~\eqref{B6} holds for any component $dl$, it makes the tangential component of $\vec E$ to be continuous along the boundary. Therefore, 
\begin{equation}\label{B7}
(\vec E_2- \vec E_1)\times \mathbf n=0. 
\end{equation}
By means of Eq.~\eqref{D2}, it follows from Eq.~\eqref{B7} that:
\begin{equation}\label{B8}
\frac{\vec D_1}{\varepsilon_1}= \frac{\vec D_2}{\varepsilon_2}. 
\end{equation}
Moreover, using Eq.~\eqref{M2}, we can obtain:
\begin{equation}\label{B9}
\Phi_{2} -\Phi_{1} = \int (\vec E_2- \vec E_1)\cdot db, 
\end{equation}
where $db$ is shown in Fig.~\ref{FBD}. Once $db\to 0$, the right hand side of Eq.~\eqref{B9} vanishes. Hence at the boundary we have:
\begin{equation}\label{B10}
\Phi_{1} =\Phi_{2}. 
\end{equation}
Given a coordinate system $(\zeta, \xi, \varphi)$  with the scaling factors :
\begin{equation}
h_\zeta= \left| \frac{\partial \mathbf {r}}{\partial \zeta}\right |, \quad
h_\xi= \left| \frac{\partial \mathbf { r}}{\partial \xi}\right |, \quad
h_\varphi= \left| \frac{\partial \mathbf{  r}}{\partial \varphi}\right |, 
\end{equation}
we consider a body $\Pi$ with dielectric function $\varepsilon_1$ immersed in a medium with dielectric function $\varepsilon_2$. Boundary surface $\partial \Pi$, finite or infinite, is described via revolution through an angle $\varphi$ about the $z$--axis resulting in the usual azimuthal symmetry and is defined by fixing one of the coordinates, say $\zeta=\zeta_0$ \cite{Ency}. Furthermore, we denote the inside and outside regions of $\Pi$ by $\zeta<\zeta_0$ and $\zeta>\zeta_0$ as well as their respective potentials as $\Phi_{\text{i}}$ and $\Phi_{\text{o}} $, respectively, noting that: 
\begin{equation}\label{in-out}
\begin{cases}
\nabla^2 \Phi_{\text{i}} = 0, \quad &\zeta<\zeta_0, \\
\nabla^2 \Phi_{\text{o}} = 0, \quad &\zeta_0<\zeta,
\end{cases}
\end{equation}
while at the boundary $\partial \Pi$, 
\begin{equation}\label{BD1}
\Phi_{\text{i}}	\, \Big|_{\zeta=\zeta_0}= \Phi_{\text{o}}	\, \Big|_{\zeta=\zeta_0},
\end{equation}
\begin{equation}\label{BD22}
\varepsilon_1 \nabla \Phi_{\text{i}} \cdot \mathbf{ n}\,  \Big|_{\zeta=\zeta_0}=
\varepsilon_2	\nabla \Phi_{\text{o}} \cdot \mathbf {n}\,  \Big|_{\zeta=\zeta_0}.
\end{equation}
Considering the quotient $\varepsilon_1/\varepsilon_2$, we may assume with no loss of generality, $\varepsilon_2=1$ which corresponds the dielectric constant of vacuum, and $\varepsilon_1=\varepsilon$. Thus Neumann boundary condition given in Eq.~\eqref{BD22} can be rewritten as:
\begin{equation}\label{BD2}
\varepsilon \nabla \Phi_{\text{i}} \cdot \mathbf{ n}\,  \Big|_{\zeta=\zeta_0}=
\nabla \Phi_{\text{o}} \cdot \mathbf {n}\,  \Big|_{\zeta=\zeta_0}, 
\end{equation}
An arbitrary surface of revolution along with inside and outside potentials and dielectric constants are illustrated in Fig.~\ref{B}. \\ 
\clearpage
\noindent \textbf{Laplace equation}

Since the non-retarded potential $\Phi$'s inside and outside parts satisfy the Laplace equation (see Eq.~\eqref{in-out}), we consider general form of the Laplacian in a coordinate system $(\zeta, \xi, \varphi)$ :
\begin{equation}\label{0-L}
\vec \nabla^2 \Phi= \frac{1}{h_\zeta\, h_\xi\, h_\varphi}\, \Bigg\{\frac{\partial}{\partial \zeta} \left[ \frac{h_\xi\, h_\varphi}{h_\zeta}\, \frac{\partial \Phi}{\partial \zeta}\right] + \frac{\partial}{\partial \xi} \left[ \frac{h_\zeta\,  h_\varphi}{h_\xi}\, \frac{\partial \Phi}{\partial \xi}\right]+\frac{\partial}{\partial \varphi} \left[ \frac{h_\zeta\, h_\xi}{h_\varphi}\, \frac{\partial \Phi}{\partial \varphi}\right]
\Bigg \}.
\end{equation}
A coordinate system is called separable if it obeys \emph{Robertson} condition (see \cite{morse}, p.~655). Except the case of a nanoring in which a quasi-separation of variables holds, all considered coordinate systems in this manuscript satisfy Robertson condition and are separable. The case of a nanoring is treated separately in Chapter \ref{ring}. 
Inspired by the azimuthal symmetry assumption and hence the $2\pi$-periodicity of the potential $\Phi$, an ansatz of the form:
\begin{equation}\label{0-p}
\Phi_m= Z(\zeta)\, E (\xi)\, e^{im\varphi}, 
\end{equation}
where $m=0, \pm 1, \pm 2, \dots$ is used. As a consequence, we obtain two \emph{Sturm-Liouville} problems in $\zeta$ and $\xi$ (see \cite{morse} p.~656 and Appendix A, Eqs.~\eqref{0-SL}--\eqref{0-SL2}). Denoting by $F_{mn}(\zeta)$, $\tilde F_{mn}(\zeta)$, and $G_{mn}(\xi)$, $\tilde G_{mn}(\xi)$ the linearly independent pair of solutions to the corresponding obtained Sturm-Liouville problems in $Z(\zeta)$ and $E (\xi)$, where $n$ stands for the second separation constant, 
the general form of eigenfunctions for the Laplacian may be written as:
\begin{equation}
\Phi_{mn}(\zeta, \xi, \varphi)= \begin{cases} F_{mn}(\zeta)\\ \tilde F_{mn}(\zeta) \end{cases}
\times \begin{cases} G_{mn}(\xi)\\ \tilde G_{mn}(\xi) \end{cases} \times e^{im\varphi}.
\end{equation}
For simplicity, we may assume that $n$ takes discrete values, i.e. $n=0, \pm 1, \pm 2, \dots$. However, as we shall see in Chapter \ref{3}, we will also deal with real continuous spectrum with respect to $n$. Utilizing the \emph{superposition} principle, we may write the general solution as:
\begin{equation}
\Phi(\zeta, \xi, \varphi, t)= \sum\limits_{m,n\in \mathbb{Z}} \, \mathcal D_{mn}(t)\,\begin{cases} F_{mn}(\zeta)\\ \tilde F_{mn}(\zeta) \end{cases}
\times \begin{cases} G_{mn}(\xi)\\ \tilde G_{mn}(\xi) \end{cases} \times e^{im\varphi}, 
\end{equation}
where $\mathcal D_{mn}(t)$ is a general Fourier component with respect to time $t$ and they are considered as the non-retarded, time-dependence amplitudes at time $t$. Consequently, the Inside and outside potentials, $\Phi_{\text{i}}$ and $\Phi_{\text{o}}$, could be defined by considering the asymptotic behavior of the eigenfunctions $F_{mn}(\zeta)$, $\tilde F_{mn}(\zeta)$  and $G_{mn}(\xi)$, $ \tilde G_{mn}(\xi)$ as the potential has to satisfy the following conditions:
\begin{eqnarray}\label{NBC}
\begin{cases}
&	\Phi(\mathbf r,t)~\text{is finite at every point of space and for all $t$, }\\
&\lim \limits	_	{r\to\infty}		\Phi(\mathbf r,t)=0, \quad \text{where} \quad r=|\mathbf {r}|.
\end{cases}
\end{eqnarray}
Without loss of generality, we may assume the sets $ F_{mn}(\zeta)\, G_{mn}(\xi)$ and $ \tilde F_{mn}(\zeta)\, G_{mn}(\xi)$ determine finite potentials for the  inside and the outside of the geometry, respectively. Using Eq.~\eqref{NBC} together with boundary condition given in Eq.~\eqref{BD1}, we may write:
\begin{eqnarray}
	\Phi_{\text{i}}(\mathbf r,t)&=&\sum\limits_{m,n\in \mathbb{Z}} \, \mathcal D_{mn}(t)\, F_{mn}(\zeta) \tilde F_{mn}(\zeta_0) 
G_{mn}(\xi)\, e^{im\varphi}, \qquad \zeta\le \zeta_0, \label{in}\\
	\Phi_{\text{o}}(\mathbf r,t)&=&\sum\limits_{m,n\in \mathbb{Z}} \, \mathcal D_{mn}(t)\,   F_{mn}(\zeta_0) \tilde F_{mn}(\zeta) 
G_{mn}(\xi)\, e^{im\varphi}, \qquad \zeta_0\le \zeta. \label{out}
\end{eqnarray}

\section{Harmonic Oscillator Model}\label{HA}

In this section, we will show that amplitudes $\mathcal D_{mn}(t)$ undergo harmonic oscillator motion. We start by deriving two relations between surface charge density and quasi-static potential using the \emph{total charge} for the whole space. Equating the two relations gives us an equation in amplitudes $\mathcal D_{mn}(t)$ and their second time-derivative $\ddot {\mathcal D}_{mn}(t)$ known as the harmonic oscillator equation of motion, see Eq.~\eqref{0-HA}. The total charge $Q$ is defined by:
\begin{equation}\label{Q1}
Q=\int_{\text{volume}} \rho \, d\mathcal V,
\end{equation} 
where $\rho$ denotes the volume-charge density. Since the polarization charge is only confined to the surface, Eq.~\eqref{Q1} may be written as:
\begin{equation}\label{Q2}
Q= \int_{\text{surface}} \sigma \, d\mathcal A, 
\end{equation} 
where $\sigma$ denotes the surface charge density and $d\mathcal A$ is the surface element. Equating Eq.~\eqref{Q1} with Eq.~\eqref{Q2}, we have:
\begin{equation}\label{Q3}
\int_{\zeta<\zeta_0}^{\zeta>\zeta_0} \int \int \rho \, h_\zeta h_\xi h_\varphi \, d\zeta d\xi d\varphi= 
\int \int_{\zeta =\zeta_0}\sigma  \,  h_\xi h_\varphi \, d\xi d\varphi . 
\end{equation} 
It follows now from Eq.~\eqref{Q3} that
 \begin{equation}\label{0-25}
\rho= \frac{\delta (\zeta-\zeta_0)}{h_\zeta} \sigma, 
\end{equation}
where $\delta$ denotes the \emph{Dirac delta} function. In view of Poisson's equation given in Eq.~\eqref{1-1P2}, one finds: 
\begin{equation}\label{0-26}
\vec \nabla^2 \Phi = -\frac{4\pi\, \delta(\zeta-\zeta_0)}{h_\zeta}  \sigma.
\end{equation}
On the other hand, using the electric polarization vector given by \cite{jackson}:
\begin{equation}\label{P1}
\vec P= n_0 (-e) \, \vec u,
\end{equation}
where $\vec u$ denotes the \emph{charge displacement} of the system. Assuming there is no free-charge density on surface, then using Eq.~\eqref{D0}, we can write:
\begin{equation}\label{D1}
\nabla \cdot  \vec D= \nabla \cdot (\vec E + 4\pi \vec P)= 0. 
\end{equation}
It follows from  Eq.~\eqref{Max2}  that $\vec \nabla \cdot  \vec P= -\rho$, which in view of Eq.~\eqref{Q1} and  \emph{Divergence Theorem} gives:
\begin{equation}
Q= -\int _{\text{volume}}  \vec \nabla \cdot \vec P\, d\mathcal V
= -\oint _{\text{surface}}  \vec P\cdot \mathbf{ n} \, d\mathcal A= \int _{\text{surface}} \vec P\cdot \mathbf{\vec e}_{\zeta}\big|_{\zeta=\zeta_0} \, d\mathcal A,
\end{equation} 
where $\mathbf{ n}=-\mathbf{\vec e}_{\zeta}$. 
Comparing the above equation with Eq.~\eqref{Q2} and using Eq.~\eqref{P1}, one can find:
\begin{equation}\label{sigma0}
\sigma= \vec P \cdot \mathbf{\vec e}_{\zeta}\big| _{\zeta=\zeta_0}= -n_0 e \, \vec u\cdot\,  \mathbf{\vec e}_{\zeta}\big| _{\zeta=\zeta_0}.
\end{equation}
The surface charge due to polarization
is also derived as the motion of free electrons in the metal surface due to a time-harmonic external electric field, following standard descriptions of anomalous dispersion \cite{SET}. We can thus also describe this surface charge as $
\sigma=-n_0\, e\, \vec u. $
The electrostatic force on the electron current per unit volume is $
F=-n_0e\nabla\Phi_{\text {i} }.$
Putting these together through Newton's force law, since  the acceleration is the force per unit mass, we have:
\begin{equation}\label{0-ddotu}
\ddot {\vec  u} =\frac{e}{m_e} \nabla\Phi_{\text {i}}. 
\end{equation}
Differentiating both sides of Eq.~\eqref{sigma0} twice with respect to time gives:
\begin{equation}\label{sigmaddot}
\ddot{\sigma}= -n_0 \, e\,  \ddot{\vec{u}}\cdot \mathbf{\vec e}_{\zeta}\big| _{\zeta=\zeta_0}, 
\end{equation}
which together with Eq.~\eqref{0-ddotu} implies:
\begin{eqnarray}\label{0B34}
\ddot{\sigma}=-\frac{\omega_p^2}{4\pi }\, \mathbf{\vec{e}}_\zeta\, \cdot\vec \nabla\Phi_{\text {i}}|_{\zeta=\zeta_0}.
\end{eqnarray}
Utilizing the identity:
\begin{eqnarray}\label{0-iden}
\mathbf{\vec{e}}_\zeta\, \cdot\vec \nabla= \frac{1}{h_\zeta}\, \frac{\partial }{\partial \zeta}, 
\end{eqnarray}
one may write:
\begin{equation}\label{0-sddot1}
\ddot{\sigma}=-\frac{\omega_p^2}{4\pi } \bigg(
\frac{1}{h_{\zeta}}\, \frac{\partial\Phi_{\text {i}}}{\partial\zeta}
\bigg)\Bigg|_{\zeta=\zeta_0}.
\end{equation}
This equation is also provided in \cite{B} without a detailed proof. Differentiating Eq.~\eqref{0-26} twice with respect to time $t$ gives:
\begin{equation}\label{0-sddot}
\vec \nabla^2 \ddot{\Phi }= -\frac{4\pi\, \delta(\zeta-\zeta_0)}{h_\zeta}  \ddot{\sigma}.
\end{equation} 
By equating Eq.~\eqref{0-sddot1}, in terms of ${\mathcal D}_{mn}(t)$, and Eq.~\eqref{0-sddot}, in terms of second time-derivatives of amplitudes $\ddot 	{\mathcal D}_{mn}(t)$, we obtain the harmonic oscillator equation with the frequencies $\omega_{mn}$ of the form:
\begin{equation}\label{0-HA}
\ddot 	{\mathcal D}_{mn}(t) + \omega_{mn}^2 \, \mathcal D_{mn}(t)=0.
\end{equation}
The importance of Eq.~\eqref{0-HA} is the fact that the  (allowed) resonant  values of the dielectric function $\varepsilon$ can also 
be independently calculated from this transcendental equation obtained by imposing the quasistatic boundary conditions at the bounding surfaces of the domain within which the scalar electric field satisfies the Laplace equation \cite{zw}.\\

\noindent \textbf{Classical energy}

Once the scalar potential and charge density are known, we shall be able to calculate the \textit{total energy} of the system. In classical mechanics, the total energy $E$, also known as \textit{total mechanical energy}, of a position dependent force, in our case surface plasmons, is given by:
\begin{equation}\label{0-39}
E=V+T, 
\end{equation}
where $V$ and $T$ denote \textit{potential} and \textit{kinetic} energy, respectively. This is nothing more than an expression of the conservation of mechanical energy principle \cite{hasbun,buhler}. We shall start by calculating potential energy. \\
The potential energy is given by the volume integral \cite{B}:
\begin{equation}\label{0-35}
V= \frac12 \int_{\zeta\le \zeta_0} \rho\,  \Phi_{\text{i}}\, d\mathcal V, 
\end{equation}
where $\mathcal V$ is the volume of the geometry of interest. Using Eq.~\eqref{0-25}, one could write:
\begin{equation}\label{0-36}
V= \frac12 \int_{\zeta \le \zeta_0} \frac{\delta (\zeta-\zeta_0)}{h_\zeta}\,  \sigma \, \Phi _{\text{i}}\, d\mathcal V.
\end{equation}
Since inside and outside scalar potential $\Phi_{\text{i}}$ and $\Phi_{\text{o}}$ coincide on the surface, namely the first boundary condition given in Eq.~\eqref{BD1}, it is a direct computation using any of the relations of inside or outside scalar potentials. Utilizing the  $\delta$ function, we can write:
\begin{equation}\label{0-37}
V =\frac12 \int_{\text{surface}}
\sigma\,   {\Phi_{\text{i}}}\bigg|_{\zeta=\zeta_0}  \, 
d\mathcal A.
\end{equation}
Using the identity $d\mathcal A= h_{\xi }h_{\varphi} \, d\xi  d\varphi$, we find:
\begin{equation}\label{0-V}
V =\frac12 \int\int  
\sigma  {\Phi_{\text{i}}}\bigg|_{\zeta=\zeta_0}  
h_{\xi }h_{\varphi} \, d\xi d\varphi. 
\end{equation}
The kinetic energy $T$ of the oscillations is given by the volume integral \cite{B}:
\begin{equation}\label{0-T}
T= \frac{m_e n_0}{2}
\int_{ \textrm{volume}}
\dot{\vec{u}}  \cdot{\dot{\vec{u}}}\, d\mathcal V.
\end{equation}
Depending on whether time-dependent amplitudes obey pure harmonic oscillator or not, we may take different steps in order to find kinetic energy. 
In all the cases investigated here, amplitudes satisfy the relation given in Eq.~\eqref{0-HA} but the case for nano-ring. We proceed by assuming the general case and we defer the discussion for nano-ring to Chapter \ref{ring}. 
We may use the relation given in Eq.~\eqref{0-ddotu} to write: 
\begin{equation}\label{0-udd1}
\ddot {\vec  u} =\frac{e}{m_e} \, \vec \nabla\Phi_{\text {i}}= \frac{e}{m_e}\, \vec \nabla\ \sum\limits_{mn} \mathcal D_{mn}(t) \Psi_{mn}, 
\end{equation}
where we put: 
\begin{equation}\label{0-psi}
\Phi_{\text {i}}= \sum\limits_{mn}\Phi_{mn}= \sum\limits_{mn} \mathcal D_{mn}(t)\Psi_{mn}. 
\end{equation}
Now, using Eq.~\eqref{0-HA}, we may write:
\begin{equation}
\mathcal D_{mn}(t)= -\frac{\ddot {\mathcal D}_{mn}(t)}{\omega_{mn}^2}. 
\end{equation}
Putting this in Eq.~\eqref{0-psi}, we could write:
\begin{equation}\label{0-psi1}
\Phi_{\text {i}}= \sum\limits_{mn}\ddot {\Phi}_{mn}=-\sum\limits_{mn} \frac{\ddot {\mathcal D}_{mn}(t)}{\omega_{mn}^2}\, \Psi_{mn}. 
\end{equation}
Replacing this is Eq.~\eqref{0-udd1}, we obtain:
\begin{equation}
\ddot {\vec  u} =- \frac{e}{m_e}\, \vec \nabla\sum\limits_{mn} \frac{\ddot {\mathcal D}_{mn}(t)}{\omega_{mn}^2}\, \Psi_{mn}. 
\end{equation}
Integrating both sides of the above equation with respect to time $t$ gives:
\begin{equation}\label{0-udot}
\dot{\vec  u}=- \frac{e}{m_e}\, \vec \nabla\sum\limits_{mn} \frac{\dot{\mathcal D}_{mn}(t)}{\omega_{mn}^2}\, \Psi_{mn}= - \frac{e}{m_e}\, \vec \nabla\sum\limits_{mn} \frac{	\dot{\Phi}_{mn}}{\omega_{mn}^2}. 
\end{equation}
Using the latter expression for $\dot{\vec u}$ in Eq.~\eqref{0-T}, we may write: 
\begin{equation}\label{0-T1}
T= \frac{ e^2\, n_0}{2\, m_e}
\int_{ \textrm{volume}}
\left( \vec \nabla\sum\limits_{mn} \frac{	\dot{\Phi}_{mn}}{\omega_{mn}^2}\right) 
\cdot  \left( \vec \nabla\sum\limits_{m'n'} \frac{	\overline{\dot{\Phi}_{m'n'}}}{\omega_{m'n'}^2}\right) \, d\mathcal V, 
\end{equation}
where $\overline{\dot{\Phi}_{mn}}$ denotes the conjugate of ${\dot{\Phi}}_{mn}$. The fact that scalar potential is real-valued allows us to replace it with its complex conjugate. Here, \emph{Fubini's Theorem} \cite{conway} allows us to change the order of the summation and integration, assuming the integral is bounded. Using above argument as well as plasmon-bulk frequency relation given in Eq.~\eqref{1-29}, Eq.~\eqref{0-T1} could be rewritten as:
\begin{equation}\label{0-T2}
T= \frac{\omega_p^2}{8\pi}\, \sum\limits_{mn}\, \sum\limits_{m'n'}\frac{1}{\omega_{mn}^2\, \omega_{m'n'}^2}
\int_{ \textrm{volume}}
\left( \vec \nabla 	\dot{\Phi}_{mn}\right) 
\cdot \left( \vec \nabla 	\overline{\dot{\Phi}_{m'n'}}\right) \, d\mathcal V. 
\end{equation}
Since Laplace equation is independent from time $t$, it is easy to see that if $\Phi_{\text {i}}$ satisfies the Laplace equation, so does its time-derivative  $\dot\Phi_{\text{i}}$. We can use the following general identity: 
\begin{equation}\label{1-49}
\vec \nabla
\dot\Phi \cdot
\vec \nabla \dot\Phi=
\vec \nabla\cdot
\left ( \dot\Phi
\vec \nabla
\dot\Phi \right ),
\end{equation}
to write the kinetic energy as: 
\begin{equation}\label{1-50}
T= \frac{\omega_p^2}{8\pi}\, \sum\limits_{mn}\, \sum\limits_{m'n'}\frac{1}{\omega_{mn}^2\, \omega_{m'n'}^2}
\int_{ \textrm{volume}}
\vec \nabla\cdot \, 
\left( 
\dot\Phi_{mn}\, 
\vec \nabla
\overline{\dot\Phi_{m'n'}}
\right) 
\, d\mathcal V. 
\end{equation}
The volume integral in Eq.~\eqref{1-50}, using Divergence Theorem could be transformed to a surface integral through the following relation:
\begin{eqnarray}\label{1-51}
T= \frac{\omega_p^2}{8\pi}\, \sum\limits_{mn}\, \sum\limits_{m'n'}
\frac{1}{\omega_{mn}^2\, \omega_{m'n'}^2}
\int_{ \textrm{surface}}
\left( 
\dot\Phi_{mn}\, 
\vec \nabla \overline{\dot\Phi_{m'n}}
\right) 
\bigg |_{\zeta=\zeta_0}\, 
\cdot \, 
\mathbf{\vec e}_{\zeta}
\, d\mathcal A.
\end{eqnarray}
Using the identity given in Eq.~\eqref{0-iden}, we find the final expression for kinetic energy as: 
\begin{eqnarray}\label{Sp-K}
T=\frac{\omega_p^2}{8\pi }   
\sum\limits_{mn} 
\sum\limits_{m'n'}
\frac{1}{\omega_{mn}^2\, \omega_{m'n'}^2}
\int\, \int 
\dot \Phi_{mn} \, \frac{\partial\overline{\dot \Phi_{m'n'}}}{\partial \zeta}\, 
\frac{h_\xi\, h_\varphi}{h_\zeta}\, 
d\xi d\varphi .	
\end{eqnarray}
From Eqs.~\eqref{0-V} and \eqref{Sp-K}, the classical energy is obtained by using Eq.~\eqref{0-39}. Depending on amplitudes being real or complex valued, they could be written in the following form:
\begin{equation}\label{0-D}
\mathcal D_{mn}(t)= \frac{\mathbf{\alpha_{mn}}}{2\omega_{mn}} \big (d_{mn}\pm d^*_{mn}\big ) , 
\end{equation}
with time-dependent operators $d_{mn}$ directly proportional to $e^{i\omega_{mn}t}$, where $d^*_{mn}$ is its respective complex conjugate. The purely imaginary case could be assumed by multiplying the real case of Eq.~\eqref{0-D}  by $i$. For the simplicity of the discussion, we may consider the case for which amplitudes are real-valued. The goal is now to determine the values of $\mathbf{\alpha_{mn}}$. First find the time-derivative of amplitudes $\mathcal D_{mn}(t)$ using Eq.~\eqref{0-D}, i.e.: 
\begin{equation}\label{0-Dd}
\dot{	\mathcal D}_{mn}(t)=i\, \frac{\mathbf{\alpha_{mn}}}{2} \big (d_{mn}- d^*_{mn}\big ). 
\end{equation}
Using Eqs.~\eqref{0-D} and \eqref{0-Dd} to write the total energy $E$ in terms of $d_{mn}$ and $d^*_{mn}$ and comparing the new expression with \emph{Hamiltonian} operator for the quantum mechanical harmonic oscillator given as \cite{pines}:
\begin{equation}\label{0-H}
H\equiv \sum\limits_{mn} \frac{\hbar \omega_{mn}}{2}\, \left({\hat d}^\dagger_{mn}\, \hat d_{mn}+ \hat d_{mn}\, \hat d^\dagger _{mn}\right), 
\end{equation}
where $\hbar$ is the \emph{Planck's} constant . The following substitution allows us to find the values for $\mathbf {\delta_{mn}}$:
\begin{equation}\label{0-S}
\left( d_{mn}\, , \, d^*_{mn}\right)  \rightarrow \left( {\hat d}_{mn}\, , \, \hat d^\dagger _{mn}\right). 
\end{equation}
The operators $\left( {\hat d}_{mn}\, , \, \hat d^\dagger _{mn}\right)$ are interpreted as the non-commuting \emph{boson} creation and annihilation operators, respectively. 
Calculations of the interactions between surface plasmons and other excitations are often time complicated, yet have been simplified by quantizing the fields (second quantization) \cite{Ritchie}. The total classic energy was calculated to help us finding coefficients in Hamiltonian operator. In the next section, we will use the Hamiltonian operator to calculate matrix element and probability amplitudes for interactions.

\section{Quantum Field \& Second Quantization }\label{QF}

\par
Among the many quantum numbers that distinguish particles, we can count their spin \cite{zeid, maggiore, Folland}. There are two main families of particles: \emph{bosons}, with integer spin and \emph{fermions}, with half-integer spin. Photons, gluons and vector bosons are examples of bosonic particles and electrons, neutrinos, and quarks are classified as fermionic particles.

There have been two main approaches to quantum field theory and on how to study the interaction between particles and surfaces. One approach is due to Feynman: His celebrated path-integral approach based on \emph{Feynman's diagram}. Two detailed accounts can be found in \cite{zeid, todorov}. This approach does not rely on operators and the theory behind the operator in \emph{Hilbert} spaces. Another approach, by means of operator theory, can be traced back to Heisenberg, Born, Jordan, Dirac, Pauli, and von Neumann. This method, more formal than the previous one, is more difficult to apply to practical computations, due to the nonlinearities of the interactions. In \cite{zeid}, the crucial aspect of a quantum field is that it can be treated as a system with infinitely many quantum particles that can be created and annihilated. The main purpose of second quantization is to describe quantum particles, either bosons or fermions, in terms of operators. The number of particles is assumed to be infinite. We first set our full system as the combined surface plasmon and photon field. Then we consider the ground state or zero-population for both plasmon and photon as:
\begin{equation}
|0\rangle.
\end{equation}
A state consisting of precisely $N$ plasmon particles is shown by the notation:
\begin{equation}
|\text{Plasmon} \rangle\equiv | \nu _{m_1n_1} \, \cdots \, \nu _{m_Nn_N}\, \cdots \rangle, 
\end{equation}
where $\nu _{mn}$'s are the number of plamons in the state $(mn)$. Similarly, the state consisting of exactly $\tilde N$ photons can be represented by:
\begin{equation}
|\text{Photon} \rangle\equiv | \nu _{s_1 q_1} \, \cdots \, \nu _{s_{\tilde N} q_{\tilde N}}\, \cdots \rangle, 
\end{equation}
where $\nu _{ s q}$'s show the number of photons with wavevector $s$ and polarization $q=1,2$. Polarization vectors are chosen based on the fact that they are perpendicular to each other and to the direction of travel. It is customary to represent them by s-polarization and p-polarization vectors. We shall discuss it in more details in Chapter \ref{3}.
Our main purpose  concerns the investigation of interacting quantum fields in rigorous mathematical terms. We shall use the fact that the operators describing photons only act on and hence change the photon state. The same situation holds for the plasmon operators while acting over the full system. The states being orthonormal results in the Kronecker delta function as:
\begin{equation}
\langle \cdots \nu_{n_k} \, \cdots \, \nu_{n_1} \big| \nu_{n'_1}\, \cdots \, \nu_{n'_k}	\cdots  \rangle= \delta_{n_1n'_1}\, \cdots \, \delta_{n_kn'_k}\, \cdots, 
\end{equation}
for some arbitrary states $1, \cdots, k$. Considering these facts, we are now able to write a general state for the combined plasmon and photon field in the form:
\begin{equation}
|\text{Photon} \rangle\, |\text{Plasmon} \rangle\equiv\, | \nu _{s_1 q_1} \, \cdots \, \nu _{s_{\tilde N} q_{\tilde N}}\rangle\,  | \nu _{m_1n_1} \, \cdots \, \nu _{m_Nn_N}\, \cdots\, \rangle. 
\end{equation}
Next, we shall discuss how creation and annihilation operators, $\hat d_{mn}$ and $\hat d^\dagger_{mn}$ acting on an arbitrary state, including ground state of the field. A prototype of commutation relations for these operators when act on bosonic particles is given in \cite{zeid, maggiore}:
\begin{equation}\label{0-r1}
\langle \, 0\, |\, \hat d _{mn}\, \hat d^\dagger _{m'n'}\, |\, 0\rangle=\delta_{mm'}\, \delta_{nn'}, 
\end{equation}
for all plasmon states $mn$ and $m'n'$ and 
\begin{equation}\label{0-r2}
\langle 0\, |\, \hat a_{s'q'}\, \hat a^*_{sq}\, |\, 0\rangle=\delta(s-s')\, \delta_{qq'}.
\end{equation}
with $\hat a$ and $\hat a^*$, creation and annihilation operators for photon and for all photon states $sq$ and $s'q'$.

\section{Quantization of the Electromagnetic field }\label{1-f}

In view of Maxwell's equations (recall Eqs.~\eqref{Max2}--\eqref{0-Max4}), both electric and magnetic fields can be expressed in terms of vector potential $\mathbf A$ as:
\begin{align}
\vec E&= - 	\vec \nabla \Phi(\mathbf r,t)-\frac{\partial \mathbf A}{\partial t}, \\
\vec B&= \vec \nabla \times \mathbf A. 
\end{align}
 In case of absence of any sources of the field, it is possible to choose the \emph{Coulomb gauge} condition as:
\begin{equation}\label{1-1A}
\vec \nabla\cdot  \mathbf A = 0.
\end{equation}
This condition is also known as \emph{transversality} condition since the vector potential under this condition is purely transverse and could be expressed as \cite{harris,hatfield}:
\begin{equation}\label{A}
{\mathbf A}=\sum\limits_{q=1,2}
\int \frac{1}{(2\pi)^3}
\sqrt{\frac{\hbar c^2 }{\omega_s}}\,
\mathbf {\hat e}_q\, 
\left[a_{sq}(t)\, e^{i\,  s\cdot \mathbf{r}}+  
a^*_{sq}(t)\, e^{-i\, s\cdot \mathbf{r}}\, \right] d^3s\,,
\end{equation}
where $\hbar\omega_s$ is the photon energy, the polarization vector $\mathbf {\hat e}_q$ is perpendicular to $s$ for both values of polarization on index $q=1,2$ and $ a_{sq}(t)$ and its conjugate $ a^*_{sq}(t)$ are the photon operators, such that:
\begin{equation}\label{A1}
a_{sq}(t)= a_{sq}(0)\, e^{-i\omega_st}, 
\end{equation}
therefore:
\begin{equation}\label{A2}
\dot a_{sq}(t)=-i\omega_s\, a_{sq}(t), 
\end{equation}
which is considered as the equations for motion of the field for all $s$. To write the photon field as a sum with discrete momentum eigenstates as opposed to the continuous representation, we have to make some specific choices. We consider our electromagnetic field to be confined to a volume $\mathcal{V},$ which is normally taken to be represented by a cube over which we impose periodic boundary conditions. 
Since the electromagnetic energy confined to this volume is independent of the shape of the volume \cite{Bennett, Belinsky, Folland}, we take as our quantization volume as a box $\mathcal B$, with side $L$. 
The transition to the discrete sum then follows the \emph{lattice strategy} in quantum electrodynamics~\cite{zeid} which we summarize as follows:
\begin{enumerate}
	\item Periodic free field: Classical free field is represented in terms of finite Fourier coefficients. The first step is to confine the photons and electrons within the above-mentioned box $\mathcal B$.
	\item We impose periodic boundary conditions by assuming that fields have the period $L$ with respect to the position of particles. Observe that this new condition on boundary by no means contradicts the boundary conditions we first defined in Eqs.~\eqref{BD1} and \eqref{BD2}.
	\item We introduce a momentum space which yields finite Fourier series. 
	\item The Fourier coefficients of the finite Fourier series are replaced by creation and annihilation operators we introduced in this section. This is the bridge from classical mechanics to quantum mechanics. 
	\item Using finite number of creation and annihilation operators, the matrix elements could be formed. 
	\item Finally, we let the finite box $\mathcal B$ tend to infinity, in other words, we undertake the limit of $L\to \infty$. We let  the period $L$ become infinite.  
\end{enumerate} 
In practice, we perform the transition:
\begin{eqnarray}\label{0- FV}
\frac{1}{(2\pi)^{3/2}}	\int d^3 s\rightarrow\frac{1}{\sqrt{\mathcal{V}}}\sum\limits_{ s},
\end{eqnarray} 
to write the discretized vector potential as:
\begin{eqnarray}\label{0-A3}
\mathbf A=
\sum\limits_{s}
\sum\limits_{q=1,2}
\sqrt{\frac{\hbar c^2 }{\mathcal V \omega_s}}
\mathbf{\hat e}_q
\big (	\hat a_{sq}	\,e^{i\mathbf{s\cdot r}}	+\hat a^{\dagger}_{sq}   \,e^{-i\mathbf{s\cdot r}}	\big ). 
\end{eqnarray}
where the photon operators have been replaced by $\hat a_{sq}\, , \hat a^\dagger_{sq}$, annihilation and creation operators for photon. Following Eq.~\eqref{0-r2}, we replace Dirac delta function $\delta(s-s')$ by the discrete \emph{Kronecker delta} function $\delta_{ss'}$ and obtain new  the commutation relations as:
\begin{equation}\label{1-cm}
[\hat a_{sq},\hat a^{\dagger}_{sq}]=\delta_{qq'}\delta_{ss'}.
\end{equation} 
Given the explicit form of vector potential $\mathbf A$, the goal is to calculate the \emph{probability amplitude} and \emph{interaction Hamiltonian}.\\

\noindent \textbf{Probability amplitude and matrix elements}

The probability amplitude that a surface plasmon in a given state, defined by $m',n'$ will interact with a photon, described by $s',q'$ is expressed as:
\begin{equation}
P(m'n'\to s'q')=\Big |\langle 0|\nu_{s'q'}\, H_{\text{int}}\, \nu^*_{m'n'}|0\rangle\Big |^2, 
\end{equation}
where $H_{\text{int}}$ denotes the interaction Hamiltonian. We shall call: 
\begin{equation}
\mathbf{\mathcal M}=\langle 0|\nu_{s'q'}\, H_{\text{int}}\, \nu^*_{m'n'}|0\rangle\, , 
\end{equation}
as the interaction \emph{matrix element}. We could define various matrix elements of the interaction Hamiltonian based on the type of interaction.  A collection of matrix elements are given in \cite{B} and we list them here:
\begin{enumerate}
	\item Direct scattering (elastic, including the Thomson limit):
	\begin{equation}\label{Mds}
	\mathbf{\mathcal M}_{ds}=
	\left\langle 0\left| \hat a_{q_{f}}(s_{f})\, 
	H^{(0)}_{RP}~\hat a^*_{q_i}(s_{i})\right| 0\right\rangle,
	\end{equation}
	where index $f$ shows final state while index $i$ indicates initial state and $H^{(0)}_{RP}$ denotes the direct scattering interaction Hamiltonian, given by: 
	\begin{equation}\label{Hds}
	H^{(0)}_{RP}=
	\frac{n_0e^2}{2mc^2}
	\int_{\zeta\le \zeta_0}\mathbf A\cdot\mathbf A\, d\mathcal V.
	\end{equation}
	where $\zeta_0$ is the shape parameter. This Hamiltonian is also known as the zeroth-term of Hamiltonian. 

	\item Emission (radiative decay of surface plasmons):
	\begin{equation}\label{Mem}
	\mathbf{\mathcal M}_{em}= 
	\left\langle 0\left| \hat a_{q_{f}}(s_{f})~
	H^{(1)}_{RP}~\hat d^{\dagger}_{m_in_i}\, \right| 0\right\rangle,
	\end{equation}
	where
	\begin{equation}\label{Hem}
	H^{(1)}_{RP}=\frac{1}{c}\int \mathbf J\cdot\mathbf A\, d\mathcal V.
	\end{equation}
	The term $H^{(1)}_{RP}$ represents the interaction of one photon and one plasmon, which can be used to predicct the creation of a plasmon by a photon or the decay of a plasmon into a photon.

	\item Absorption:
	\begin{equation}\label{Mabs}
	\mathbf{\mathcal  M}_{abs}= \left\langle 0\left| \hat d_{m_fn_f}\, H^{(1)}_{RP}\, \hat a^*_{q_{i}}(s_{i})\right| 0\right\rangle, 
	\end{equation}
	which is the Hermitian dual to Eq.~\eqref{Mem}, since it represents the inverse process. 
	\item Total elastic scattering:
	Here $\kappa$ is the plasmon damping factor and the matrix element is obtained as shown below:
	\begin{equation}\label{Mtes}
	\mathbf{\mathcal M}_{tes}= 	\mathbf{\mathcal M}_{ds}+\sum_{mn}\frac{1}{\hbar}\left[\frac{\mathbf{\mathcal M}_{em}	\mathbf{\mathcal M}_{abs}}{\omega_{s_{i}}-\omega_{km}+(i\kappa/2)}-\frac{\mathbf{\mathcal M}^*_{em}	\mathbf{\mathcal M}^*_{abs}}{\omega_{s_{f}}+\omega_{km}-(i\kappa/2)} \right] .
	\end{equation}
	\item Inelastic scattering:
	\begin{equation}\label{Mc}
	\mathbf{\mathcal M}_{c}= \left\langle 0\left| \hat a_{q_{f}}(s_{f}) \hat d_{m_fn_f}\, H^{(2)}_{RP}\, \hat a^*_{q_{i}}(s_{i})\right| 0\right\rangle,
	\end{equation}
	where 
	\begin{equation}\label{Hc}
	H^{(2)}_{RP}=\frac{e^2}{2mc^2}\int \mathbf{\hat n}\, \mathbf A\cdot\mathbf A\, d\mathcal V.
	\end{equation}
	This term involves two photons and one plasmon and can describe, for example, the inelastic scattering of a photon in creating a plasmon. 
\end{enumerate}

\noindent \textbf{ Interaction Hamiltonian}

Here, to describe the plasmon-photon  interaction, we resort to the hydrodynamical formulation of the electron gas by Crowell and Ritchie~\cite{Crowell} as:
\begin{eqnarray}\label{0-FPC}
&&\vec\nabla \frac{\partial}{\partial t} \Psi ({\bf r}, t)= -\frac em \vec\nabla \Phi({\bf r}, t)+ \frac{\xi^2}{n_0}\vec\nabla n({\bf r}, t), \\
&&\vec\nabla^2 \Phi({\bf r}, t)=4\pi e \, n({\bf r}, t), \\
&&\vec\nabla^2\Psi({\bf r}, t) =\frac{\partial}{\partial t} n({\bf r}, t)/n_0,  
\end{eqnarray}
where $\Phi(\bf r, t)$, $\Psi(\bf r, t)$ and $n(\bf r, t)$ are the electric potential, velocity potential, and electronic density, respectively, in the electron gas, while $n_0$ is the electronic density in the undisturbed state of the electron gas and $\xi$ is the propagation speed of the disturbance through electron gas. By linearizing these equations, employing perturbation theory, 
Ritchie obtained the first order interaction Hamiltonian given as:
\begin{equation}\label{0-Int}
{H}_{\text{int}}=
\frac{1}{c}
\int\mathbf{J\cdot A}\, d\Omega, 
\end{equation}
where $\mathbf J$ denotes the \emph{current density}. This interaction has been used previously to describe the emission of  photons via plasmon decay on finite surfaces of an oblate spheroid modeling  silver nanoparticles evaporated on a dielectric substrate~\cite{Little_Paper}. Here, we will apply this Hamiltonian to modeling the creation of a plasmon on the surface  by a photon or the decay of a plasmon and emission of a photon~\cite{B}. This application requires the explicit determination of the current density operator. Let us first consider the current given by:
\begin{equation}\label{0-J}
\mathbf{J}=-n_e e \, \dot{\vec u}.
\end{equation}
We define:
\begin{equation}\label{0-psidot}
\dot \Psi =- \frac{e }{m_e}\, \sum\limits_{mn} \frac{	\dot{\Phi}_{mn}}{\omega_{mn}^2}, 
\end{equation}
in which we make the replacement of $\dot{D}_{mn}(t)$ with its equivalent given in Eq.~\eqref{0-Dd}, and use Eq. ~\eqref{0-udot} to write:
\begin{equation}\label{0-J1}
\mathbf{J}= n_e \, e \, \vec \nabla\dot \Psi. 
\end{equation}
Thus one can write: 
\begin{equation}
H_{\text{int}}= -\frac{n_0e}{c}\int(\vec \nabla\dot{\Psi})\cdot \mathbf{A} \, d\mathcal V.
\end{equation}
Using the vector identity: 
\begin{equation}
(\vec \nabla\dot{\Psi})\cdot \mathbf A=\vec \nabla\cdot
(\dot{\Psi}	\mathbf A)-\Psi(\vec \nabla\cdot \mathbf A).
\end{equation}
Imposing the Coulomb gauge condition given in Eq.~\eqref{1-1A}, and the divergence theorem, recalling that the current only exists on the surface of the body, we can find: 
\begin{equation}
H_{\text{int}} =  -\frac{n_0e}{c}\int\vec \nabla\cdot
\left(\dot\Psi \mathbf A \right)\, d \mathcal V=-\frac{n_0e^2}{c}
\int_{\zeta=\zeta_0}
\left(\dot \Psi  \mathbf A \right)
\, d\mathcal A,  
\end{equation}
which gives the interaction Hamiltonian as: 
\begin{eqnarray}\label{0-IntH}
H_{\text{int}}  =   -\frac{n_0\, e}{c}
\int\, \int 
\left(	\dot\Psi \mathbf A	\cdot	\mathbf{\hat e}_{\zeta}	\right)\, 
h_\xi	h_\varphi \,	d\xi 	d\varphi. 
\end{eqnarray}
Once $H_{\text{em}}  $ is known, can calculate the emission matrix element given in Eq.~\eqref{Mem}. 
Using Eqs.~\eqref{0-A3} and \eqref{0-psidot}, we write:
\begin{multline}\label{0-IntH1}
H_{\text{int}}  =   -\frac{n_0\, e}{c}
\int\, \int 
\Bigg\{\bigg[	- \frac{e }{m_e}\, \sum\limits_{mn} \frac{	\dot{\Phi}_{mn}}{\omega_{mn}^2} \bigg]\\
\times \bigg[\sum\limits_{s}
\sum\limits_{q=1,2}
\sqrt{\frac{\hbar c^2 }{\mathcal V \omega_s}}
\left (	\hat a_{sq}	e^{i\mathbf{s\cdot r}}	+\hat a^{\dagger}_{sq}   e^{-i\mathbf{s\cdot r}}	\right )
\times \left( 	\mathbf{\hat e}_q\cdot	\mathbf{\hat e}_{\zeta}	\right) \bigg] \Bigg\}\, 
h_\xi	h_\varphi \,	d\xi 	d\varphi. 
\end{multline}
Using Fubini's Theorem in Eq.~\eqref{0-IntH1}, we obtain:
\begin{multline}\label{0-IntH2}
H_{\text{int}}  =   \frac{n_0\, e^2}{m_e} \sum\limits_{mn}\, \sum\limits_{s}
\sum\limits_{q=1,2} \sqrt{\frac{\hbar  }{\mathcal V \omega_s}}
\int\, \int 
\Bigg\{\bigg(	  \frac{	\dot{\Phi}_{mn}}{\omega_{mn}^2} \bigg)\\
\times \bigg[		
\left (	\hat a_{sq}	\, e^{i\mathbf{s\cdot r}}	+\hat a^{\dagger}_{sq}   \, e^{-i\mathbf{s\cdot r}}	\right )
\times \left( 	\mathbf{\hat e}_q\cdot	\mathbf{\hat e}_{\zeta}	\right) \bigg] \Bigg\}\, 
h_\xi	h_\varphi \,	d\xi 	d\varphi. 
\end{multline}
We use this equation in the relevant transition matrix element for photon emission, given by Eq.~\eqref{Mem} where $\hat a_{q_fs_f}$ and $\hat d^\dagger_{m_i n_i}$ describe  the final photon state and the initial plasmon state, respectively. It is noteworthy to emphasize again that photon operators act only on the photon state and plasmon operators act only on the plasmon state \cite{Folland, JRTPE, zeid, maggiore, dick}. The question of which particular operators, either creation or annihilation, give rise to a non vanishing matrix element depends entirely on initial and final states. The following  commutation relation is our main reference to calculate the probability amplitude and therefore, the matrix element: 
\begin{equation}\label{0-R}
\left\langle 
0\left| a_{q_f s_f}\, \Big[ \hat a^\dagger _{qs}\, \hat d_{mn}\, \Big] \hat d^\dagger_{m_i n_i} \right|  	0
\right\rangle = \delta _{ s s_f}\, \delta_{q q_f}\, \delta_{mm_i}\, \delta_{nn_i}. 
\end{equation}
Using all these commutation relations, we could discard the summations in Eq.~\eqref{0-IntH2} and can proceed to calculate the matrix element. Having calculated the emission matrix element, we have all the ingredients to discuss and calculate the radiative decay in the following section. 

\section{Radiative Decay Rate}\label{rad}

In the study of the phenomenon of radiation in the quantum domain, the starting point is to talk about vector potential $\mathbf A$ \cite{Sakurai}. Imposing guage condition given in Eq.~\eqref{1-1A}, the fields, either electric or magnetic, which are derived from vector potentials are called \emph{radiation} fields. This term is often used to describe, improperly, the vector potential itself.  \emph{Fermi's} formalism based on emission interaction Hamiltonian, derived in the previous section, is called the radiation or Coulomb gauge method. A theory once quantized, provides transparent details about the process of bosonic particles, here photons, being emitted, absorbed and scattered. Let us now specify our discussion with respect to emission process. We consider a solid angle element $d\Omega$ through which a photon is emitted. The number of allowed states in an energy interval can be written as: 
\begin{equation}\label{0-o1}
\rho_{d\Omega}= \frac{\mathcal V\omega_{s}^2}{(2\pi)^3}\frac{d\Omega}{\hbar c^3},
\end{equation}
with $s$ pointing into the solid angle $d\Omega$, This equation also is known as the the density of a single photon \cite{Sakurai}. Fermi's Golden rule \cite{jackson} gives the transition probability in the form:
\begin{eqnarray}\label{0-g1}
\gamma_{fi}=\frac{2\pi}{\hbar}\, |\mathcal M_{em}|^2 \rho_f,
\end{eqnarray}
where $ \mathcal M_{em}$ represents the matrix element given in Eq.~\eqref{Mem} and $\rho_f$ is the density of final states. The total radiative decay rate, $\gamma_{mn}$, is obtained by summing over final photon states as:
\begin{eqnarray}\label{0-g2}
\gamma_{mn}=\sum_{s}\sum_{q=1,2}\gamma_{fi}.
\end{eqnarray}
If we let the quantization volume, $\mathcal V$, become large enough so that the sum over $ s$ changes into an integral, then we could write:
\begin{eqnarray}\label{0-g3}
\gamma_{mn}=\sum_{q=1,2} \int \gamma_{fi}\, d\mathcal V.
\end{eqnarray}
We could now write, using Eq.~\eqref{0-o1}:
\begin{equation}
\frac{\partial\gamma_{mn}}{\partial \Omega}=\sum_{q=1,2}\frac{\mathcal V}{(2\pi)^3} 
\int\gamma_{fi}\frac{\omega_{\vec s}^2}{\hbar c^3}\,  d\omega_{s}. 
\end{equation}
Lastly, using Eq.~\eqref{0-g1}, we find:
\begin{equation}\label{0-dr}
\frac{d\gamma_{mn}}{d \Omega}=
\sum_{q=1,2}\frac{\mathcal V}{(2\pi)^3} 
\bigg [\frac{2\pi}{\hbar^2}~|\mathcal M^q_{em}|^2 \bigg ]_{\omega_s=\omega_{mn}}\frac{\omega_{mn}^2}{ c^3}, 
\end{equation}
which gives the radiative decay rate per solid angle for two polarization vectors $q=1,2$.
Eq. ~\eqref{0-dr}, as one can easily see, becomes independent from the quantization volume $\mathcal V$, also known as normalized volume. In calculating Eq.~\eqref{0-dr}, we also require to expand the plane wave $
e^{\pm i\, s\mathbf{\cdot r}},$
using the transition expressions from the given coordinate system, here $(\zeta, \xi, \varphi)$, to Cartesian coordinates. Often time, the resulting wave functions are difficult to calculate. Depending on the case, we either avoid direct calculations of wave functions using rotating and scaling the wave vector $s$ with algebraic substitutions, or we used numerical calculations to compute integrals due to the absence of analytical solutions.

In the next chapter, we apply the whole procedure given in the current chapter to two infinite geometries as solid paraboloid and hyperboloid and one finite geometry, prolate spheroid. The results for these two cases have been provided in \cite{Bag2018, Bagh20191}.

\setlength{\parindent}{0.5in}
\chapter[RADIATIVE DECAY ON PARABOLOIDAL, HYPERBOLOIDAL,\hspace{0.5in}\\
\null \hspace{0.25in} \& PROLATE SPHEROIDAL SURFACES]
{RADIATIVE DECAY ON PARABOLOIDAL, HYPERBOLOIDAL, \& PROLATE SPHEROIDAL SURFACES}\label{3}
\footnotetext[1]{Portions of this chapter were reprinted from: Bagherian, M. and Kouchekian, S. and Rothstein, I. and Passian, A., Quantization of surface charge density on hyperboloidal and paraboloidal domains with application to plasmon decay rate on nanoprobes,  125413, vol. 98, Sep 2018, with permission from \emph{Phys. Rev. B} \\
Permission is included in Appendix C.}
\label{ch:3}


\section{\label{sec:level1}Introduction}
Materials confined to microscopic elongated "probe-like" domains, in addition to having been  tremendously enabling in various forms of scanning probe microscopy (SPM), hold great potential for emerging applications in fields such as quantum sensing~\cite{garapati,garapati2, ben,qafm}. Recent interest in the properties of such tip-shaped material domains is reflected in the demonstration of laser pulse induced electron emission from a gold solid tip under grating coupled plasmon excitation~\cite{gulde,muller0,muller}.
These applications make use of the excitation and resonant properties of surface modes  on bounding surfaces and interfaces of metallic, dielectric, and metallo-dielectric domains that take the form of a tip~\cite{PassianCurve}. Examples of systems that use a probe tip include scanning tunneling microscope, photon scanning tunneling microscope, apertured and apertureless nearfield scanning optical microscope, nanoantennas, and processes such as tip-enhanced spectroscopy and lithography. The intriguing excitations, typically studied near the tip apex, are expected to receive contributions not only from stationary modes, such as those occurring at the surfaces of finite nanoparticles, but also from non-stationary modes  propagating at the infinite interfaces.
The theoretical and modeling tools for investigating the response of these  systems and their dependence specifically upon the geometric characteristics have been indispensable in the  development of these applications. In particular, analytical techniques that lend themselves to provide complete or partial information on the system are often regarded as necessary not only for obtaining the system response (e.g., energy distribution in the nearfield of the nanoparticles), but also for elucidating  the inner working of the systems (e.g., the contributing eigenstates and eigenvalues). Calculation of geometric and material dependencies of surface mode excitation, decay and scattering on the bounding surfaces of nanoscale domains are both instructive and necessary for better design and fabrication.    

\begin{figure}
	\begin{center}
		\includegraphics[width=4.5in]{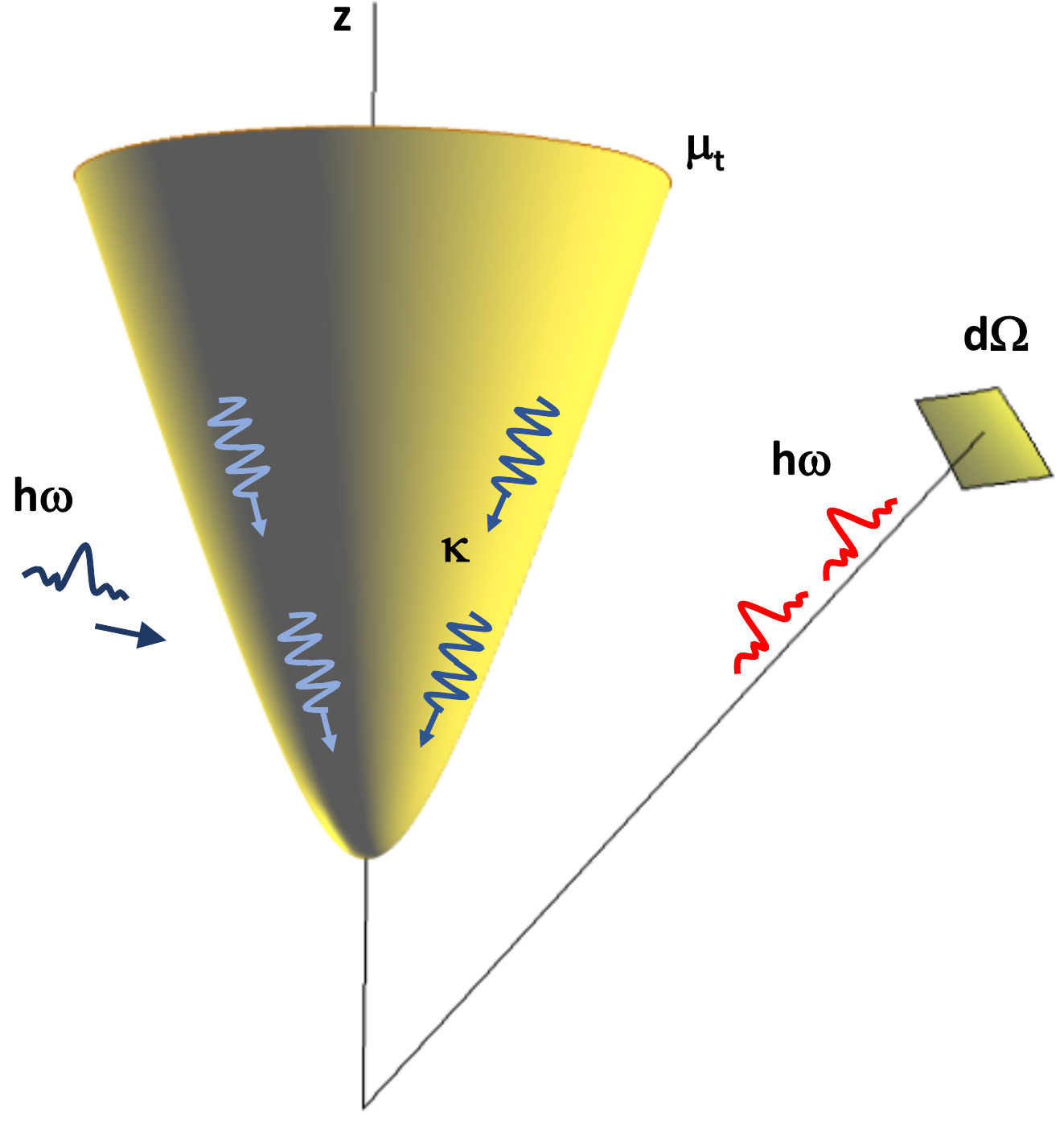}\\
		\caption[Modeling systems: One sheet of a two-sheeted hyperboloid]{Modeling systems: One sheet of a two-sheeted hyperboloid  of revolution modeling a nanotip or a nanostructure with local curvature. Surface modes of momentum $\kappa$, e.g., excited by incoming photons $h\omega$, decay radiatively into a solid angle $d\Omega$. The curvature of the tip apex is set by the $\mu_0$ defining the hyperboloidal surface. Here, $\mu_0 = \cos\theta_0$, where $\theta_0$ is the angle between the $z$ axis and an asymptote to the hyperboloidal surface such that small $\theta_0$ yields a sharp probe while $\theta_0\to \pi/2$ corresponds to $xy$ plane. The apex point   $z_{\text{min}} = z_0 \mu_0,$  near the focal point of the hyperboloid, is set by the scale factor $z_0$, as in Eq.~\eqref{5-AHC1}.}
		\label{system}
	\end{center}
	
\end{figure}

Here, we investigate the radiative decay rate of plasmons by quantizing the surface modes engendered on the surface of a metallic probe modeled as one sheet of a two-sheeted hyperboloid of revolution, shown in Fig.~\ref{system}. This geometry offers an elegant adaptability not only for the description of the local curvature of a fabricated probe but also for the modeling  of nearly planar interfaces~\cite{PassianCurve}. In addition, it has the property that the hyperboloidal domain translates along its symmetry axis when changing the opening angle $\theta_0$, that is,  smaller $\mu_0 = \cos \theta_0,$ yields smaller gap $z_{\text{min}}$, the apex distance to the origin $o$ in Fig.~\ref{system}. To provide a basis for comparison, we quantize the surface charge density oscillations on the useful system of a paraboloid of revolution, which offers
a similar apex morphology but a different asymptotic behavior away from the apex.  
Importantly, the apex and off-apex curvature of a paraboloid of revolution presents a more natural topology for comparison of its spectral and scattering properties with that of a finite body of similar curvature, e.g., a spheroidal domain. Therefore, for the sake of validation, we extend our investigations to study the radiative decay of plasmons  excited on a prolate spheroid, which owing to its finite volume presents a more  tangible system.

Following the steps given in Chapter 1, we organize this chapter as follows: In section~\ref{para}, we treat the paraboloidal plasmons. Here, within the quasi-static framework, representing the material domain with a frequency-dependent dielectric function, we derive the nonretarded plasmon dispersion relations, eigenmodes, and fields. From the classical energy of paraboloidal charge density oscillations, we then derive the Hamiltonian of the system. 
We then proceed to quantize the plasmon field and, employing an interaction Hamiltonian derived from the first order perturbation theory within the hydrodynamic model of an electron gas~\cite{Ritchie,Crowell}, obtain an analytical expression for the radiative decay rate of the plasmons. 
Having established the full treatment of the paraboloidal system, in section~\ref{hyper}, we proceed to investigate the quantized charge density oscillations on the surface of one sheet of a two-sheeted hyperboloidal of revolution. In both these sections the use of non-retarded potentials and dispersion relations is justified due to the sub-wavelength dimension of the tip.   
We would also discuss our findings and compare both the paraboloidal and hyperboloidal results. An interesting comparison of the paraboloidal domain can be made with respect to the surface modes and radiation patterns of a prolate spheroid, a finite geometric domain highly suitable for modeling of nanoparticles such as a quantum dot. In specific cases, we further validate the results using computational techniques to obtain the lower energy eigenmodes and farfield radiation patterns.  Quantum calculations that take into account the geometric effects of the bounding surfaces of the material domains are important  to corroborate experimental observations in nanophysics.
Here, our goal is to calculate the probability amplitude that a surface plasmon  in a given initial state, engendered near the apex region of a probe-shaped material domain, 
will emit a photon into a given final state.
To model a tip-like domain, we will employ three specific cases of surfaces of revolution: a single sheet of a two-sheeted hyperboloid, a paraboloid, and a prolate spheroid. Fig.~\ref{system} shows an example of a hyperboloidal domain and its relation with solid angle $d\Omega$ (see Eq.~\eqref{0-dr}).  
\clearpage
\clearpage

	%

\section{Paraboloidal Surfaces}\label{para}

In the following, we will first treat a paraboloid, for which we begin by seeking pertinent classical quantities.
The paraboloidal domain allows the parametric study of the various scattering processes as functions of the local curvature without displacement of the domain. The quasi-static solution of the electric scalar potential  near a paraboloidal domain has been reported in \cite{petrin}. \\
The parabaloidal coordinates $(\xi,\eta,\varphi)$,  
are related to the rectangular coordinates by
	\begin{align}\label{5-Pc}
			x&=a~\xi~\eta\cos\varphi, \\
			y&=a~\xi~\eta\sin\varphi, \\
			z&=\frac{a}{2}~(\xi^2-\eta^2),
	\end{align}
with the corresponding  scale factors
	\begin{equation}\label{5-sf}
			h_\xi=h_\eta=a\sqrt{\xi^2 + \eta^2},\qquad
			h_\varphi=a\,\xi  \,\eta,
	\end{equation}
where $0\leq\varphi < 2\pi$ denotes the usual azimuthal angle, $a>0$ is a dimensionless constant to be determined later, and the two coordinates $\eta,\xi\geq 0$ are such that the  surfaces of constant $\eta>0$ and  $\xi>0$ describe upward and downward paraboloids of revolution about the $z$-axis, respectively.

\subsection{Nonretarded Potential \& Dispersion Relations}

\begin{figure}
	\begin{center}
		\includegraphics[width=6in]{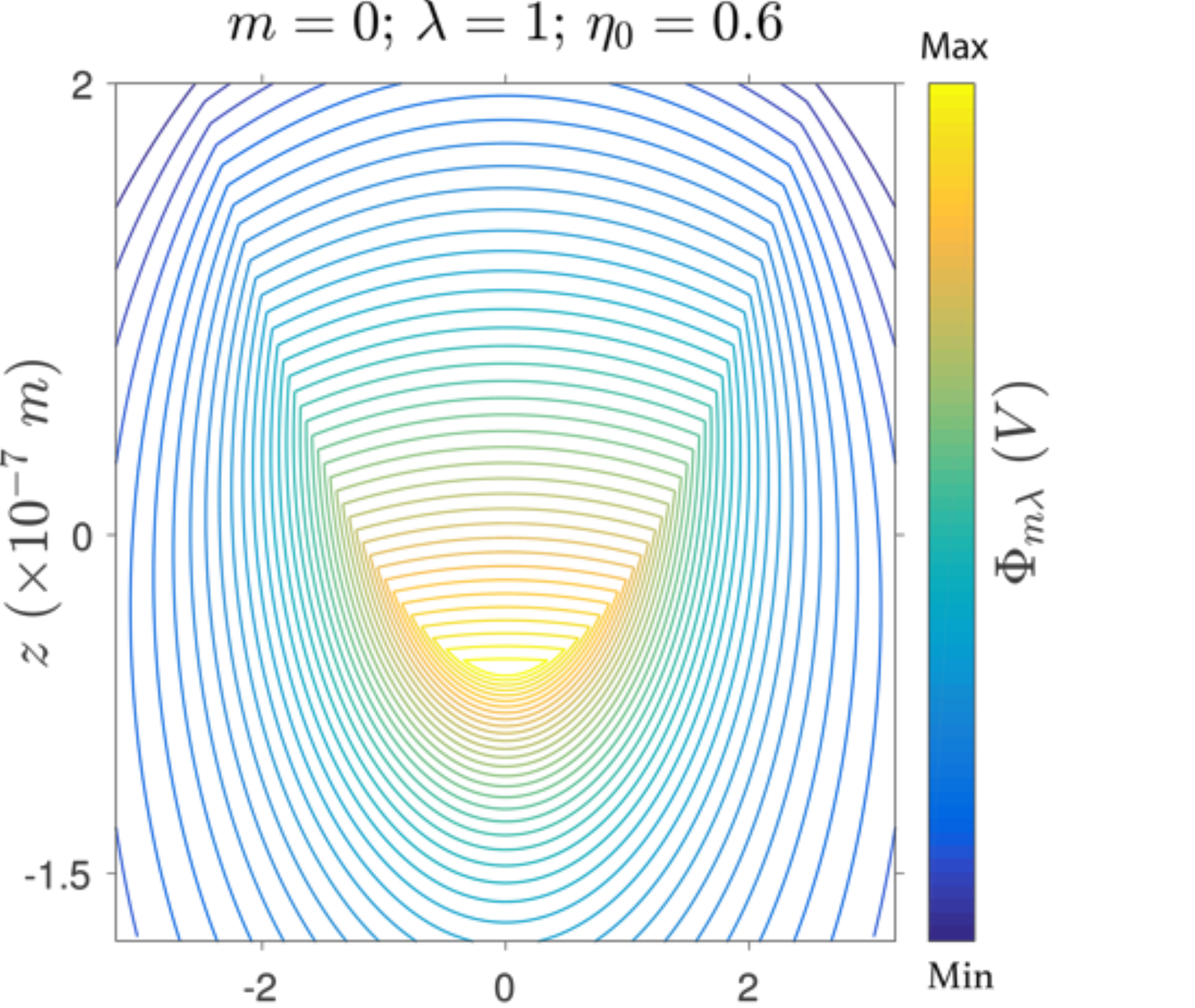}\\
		\caption[Modeling systems for the spatial distribution of the lowest ]{Modeling systems for the spatial distribution of the lowest lying eigenmodes of the quasi-static electric potential for the paraboloidal domain. For the same mode index $m$, optimizing the apex curvature overlap within the same spatial $zx$ domains, and analysing the potential distribution, leads to the determination of the corresponding continuous eigenvalues $\lambda$ of the paraboloid. The geometric parameter $\eta_0=60~$(nm) determines the form of the considered domains. }
		\label{5-pot-p}
	\end{center}
\end{figure}

We consider a vacuum-bounded solid paraboloid of revolution defined by $\eta=\eta_0$, via the coordinates $(\xi,\eta,\varphi)$ given in Eqs.~\eqref{5-Pc}--\eqref{5-sf}. 
Denoting the frequency $\omega$ dependent dielectric function of the paraboloidal material domain with $\varepsilon(\omega)$, and we consider  the outside medium to be vacuum with dielectric constant $1$ (see explanation after Eq.~\eqref{BD2}). 
Following the procedure outlined through Eqs.~\eqref{0-p}--\eqref{out}, assuming the ansatz $\Phi(\xi,\eta,\varphi)=F(\lambda\xi)G(\lambda\eta)e^{im\varphi}$ results in two Sturm-Liouville problems, where $F$ satisfies the Bessel equation: 
\begin{eqnarray}\label{ASLV1}
\begin{aligned}
\left[
\frac{d^2}{d\xi^2}+\frac{1}{\xi}\frac{d}{d\xi} +\left(1- \frac{m^2}{\xi^2}\right)
\right]
F_m(\lambda\xi)=0, 
\end{aligned}
\end{eqnarray}
and $G$ satisfies the modified Bessel equation:
\begin{eqnarray}\label{ASLV2}
\begin{aligned}
\left[
\frac{d^2}{d\eta^2}+\frac{1}{\eta}\frac{d}{d\eta}-
\left(	1+\frac{m^2}{\eta^2}	\right)
\right]
G_m(\lambda\eta)=0.
\end{aligned}
\end{eqnarray}
The solutions to Eq.~\eqref{ASLV1} are the Bessel functions of the first and second kind $J_m(\lambda\xi)$ and $Y_m(\lambda\xi)$, whereas  solutions to Eq.~\eqref{ASLV2} are the modified Bessel Functions $I_m(\lambda\eta)$ and $K_m(\lambda\eta)$. These eigenfunctions form a basis for the solution of corresponding equation (see Appendix B) \cite{LSSFA,bessel:exp0,bessel:exp}.
Considering 
the two resulting  Sturm-Liouville problems Eqs.~\eqref{ASLV1} and \eqref{ASLV2}, 
with unbounded domain $\eta,\xi\in[0,\infty)$, lead to 
a continuous spectrum of real eigenvalues and eigenfunctions~\cite{PK} in terms 
of Bessel and modified Bessel functions given by Eqs.~(10.3.64) and (10.3.65) \cite{morse}, the reader is also encouraged to read \cite{MSFTH, BIBF}. 
Using the fact that the potential 
is bounded on the $z$-axis and vanishes as $r \to\infty$ (see Eq.~\eqref{NBC}), 
together with the asymptotic behavior of the Bessel functions given in Appendix B, Eq.~\eqref{B-Bessel}, we denote the potentials with $\Phi_{\text{i}}$ and $\Phi_{\text{o}}$, for the interior
and the exterior domains, respectively, and utilize the Heaviside function $\Theta$, given in Eq.~\eqref{1-15}, with the half-maximum convention  $\Theta(0)=\frac12$ to write the total potential as: 
\begin{equation}
\Phi({\mathbf r},t)=\Theta(\eta_0-\eta)
\Phi_{\text{i}}({\mathbf r},t)+
\Theta(\eta-\eta_0)
\Phi_{\text{o}}({\mathbf r},t),
\end{equation}
or explicitly: 
\begin{multline}\label{pin-pout}
\Phi({\mathbf r},t)=
\sum_{m,p}		\ S_m^p(\varphi)	
\int_0^\infty 		A_{m\lambda p}(t)		J_m(\lambda\xi) 
\bigg [ 	\Theta(\eta_0-\eta)  
I_m(\lambda\eta) K_m(\lambda\eta_0) 	\\
+	\Theta(\eta-\eta_0)
I_m(\lambda\eta_0)	
K_m(\lambda\eta)\bigg ] 
\, d\lambda,	
\end{multline}
where $m=0, 1, 2, \cdots,$ and $p=0,1,$ while $A_{m\lambda p}(t)$ are the time $t$ dependent amplitudes to be determined by the boundary conditions, and $\{S^p_m(\varphi)\}$ indicate the azimuthal symmetry of the eigenmodes, explicitly: 
\begin{equation}
S_m^p(\varphi)  =(2-\delta_{0m})\delta_{0p}\cos{m\varphi}+\delta_{1p}\sin{m\varphi},
\end{equation}
satisfying the orthogonality relation given: 
\begin{equation}\label{5-ASmpO}
\int_0^{2\pi}  
S_m^p(\varphi) 
S_{m'}^{p'}(\varphi) \, d\varphi = 
\pi  \hat \delta_{mp}, \delta_{mm'} \delta_{pp'},
\end{equation}
where 
\begin{equation}\label{5-delta1}
\hat \delta_{mp}=  4\delta_{0p} +\delta_{1p}- 2~\delta_{m0} \delta_{0p}.
\end{equation}
Fig.~\ref{5-pot-p} shows the spatial distribution of the lowest lying eigenmodes of the quasi-static electric potential for the paraboloid.\\ \\

\clearpage
\clearpage
\noindent The Laplacian of $\Phi$ in paraboloidal coordinates is given by (see Eq.~\eqref{0-L}):
\begin{equation}\label{ALp}
\vec\nabla^2=\frac{1}{a^2(\xi^2+\eta^2)}
\bigg [
\frac{\partial^2}{\partial\xi^2}
+\frac{1}{\xi}\frac{\partial}{\partial\xi}+
\frac{\partial^2}{\partial\eta^2}
+\frac{1}{\eta}\frac{\partial}{\partial\eta} 
+	\left(	\frac{1}{\xi^2}+\frac{1}{\eta^2}	\right)
\frac{\partial^2}{\partial\varphi^2}
\bigg ].
\end{equation}
Since the Heaviside function depends on the coordinate $\eta,$ we  only need  to consider $\frac{\partial \Phi}{\partial\eta}$ and $\frac{\partial^2 \Phi}{\partial\eta^2}$; The partial derivatives of $\Phi$ with respect to the  other coordinates, $\xi$ and  $\varphi$, can be trivially expressed in terms the corresponding partials of $\Phi_{\text{i}}$ and $\Phi_{\text{o}},$ respectively. We can write: 
\begin{equation}\label{eta0}
\frac{\partial \Phi}{\partial\eta}=
\Theta(\eta_0-\eta)
\frac{\partial \Phi_{\text{i}}}{\partial\eta}+
\Theta(\eta-\eta_0)
\frac{\partial \Phi_{\text{o}}}{\partial\eta}
-	\delta(\eta_0-\eta)\Phi_{\text{i}}+
\delta(\eta-\eta_0)\Phi_{\text{o}},
\end{equation}
where the last two terms of the above expression cancel out due to the fact that $\Phi_{\text{i}} =\Phi_{\text{o}}$ at the boundary $\eta=\eta_0.$ Thus,
\begin{equation}\label{eta1}
\frac{\partial \Phi}{\partial\eta}= 
\Theta(\eta_0-\eta)
\frac{\partial \Phi_{\text{i}}}{\partial\eta}+
\Theta(\eta-\eta_0)
\frac{\partial \Phi_{\text{o}}}{\partial\eta},
\end{equation} 
and as a result:
\begin{equation}\label{eta2}
\frac{\partial^2 \Phi}{\partial\eta^2}=	
\Theta(\eta_0-\eta)
\frac{\partial^2 \Phi_{\text{i}}}{\partial\eta^2}+ 
\Theta(\eta-\eta_0)
\frac{\partial^2\Phi_{\text{o}}}{\partial\eta^2} 
- \delta(\eta_0-\eta)
\frac{\partial \Phi_{\text{i}}}{\partial\eta}+
\delta(\eta-\eta_0)
\frac{\partial \Phi_{\text{o}}}{\partial\eta}.
\end{equation}
Substituting  Eqs.~\eqref{eta1} and \eqref{eta2}, together with partials derivatives of $\Phi$ with respect to $\xi$ and $\varphi,$ into the  Laplacian Eq.~\eqref{ALp} gives
\begin{equation}\label{Lp1}
\vec\nabla^2\Phi =  \frac{1}{a^2(\xi^2+\eta^2)}
\bigg [	\delta(\eta-\eta_0)
\left( 		\frac{\partial \Phi_{\text{o}}}{\partial\eta}-		\frac{\partial \Phi_{\text{i}}}{\partial\eta}
\right) 
+	\Theta(\eta_0-\eta) \vec \nabla^2\Phi_{\text{i}}+
\Theta(\eta-\eta_0)	\vec	\nabla^2\Phi_{\text{o}} 	\bigg ].
\end{equation}
Treating the above expression in the sense of a distribution in $\eta$ only  and noting that  $\vec \nabla^2\Phi_{\text{i}}$ and  $\vec \nabla^2\Phi_{\text{o}}$ vanish for $\eta < \eta_0$ and $\eta > \eta_0,$ respectively, we obtain the following identity for the Laplacian of $\Phi$:
\begin{equation}\label{5-LPhi}
\vec\nabla^2\Phi= \frac{\delta(\eta-\eta_0)}{a^2(\xi^2+\eta^2)} 
\left( 
\frac{\partial \Phi_{\text{o}}}{\partial\eta}-
\frac{\partial \Phi_{\text{i}}}{\partial\eta}
\right) 
=-	\frac{4 \pi}{h_\eta} \sigma  \delta(\eta-\eta_0),
\end{equation} 
where $h_\eta$, given by Eq.~\eqref{5-sf}, is a scale factor of the paraboloidal system.   \\

\noindent \textbf {Surface charge density}

\begin{figure}
	\begin{center}
		\includegraphics[width=6in]{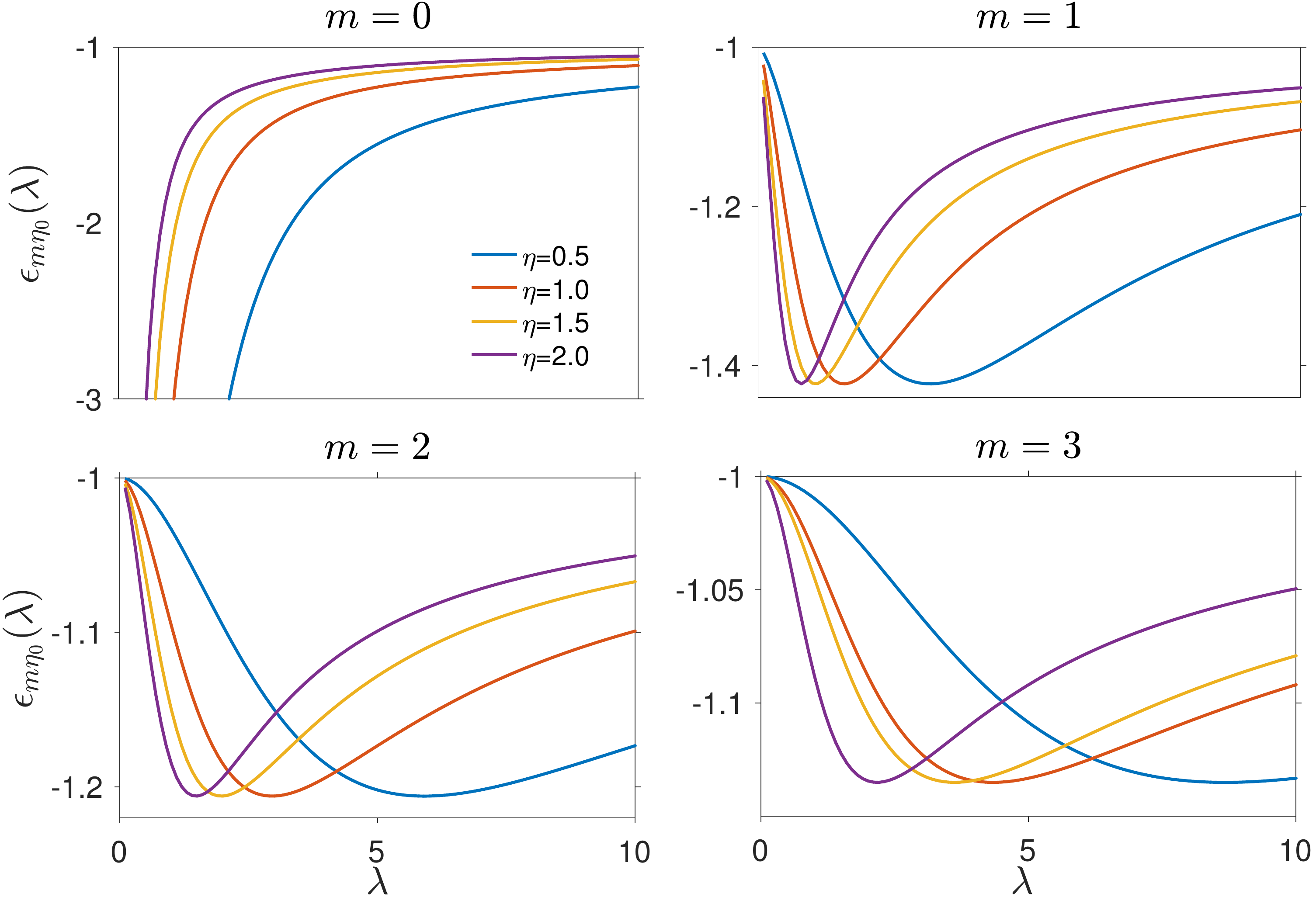}\\ 		
		\caption[Paraboloidal nonretarded surface plasmon  dispersion relations]{
			Paraboloidal nonretarded surface plasmon  dispersion relations. The resonance values of the dielectric function $\varepsilon$ are shown for low lying modes as a function of  the continuous eigenvalue $\lambda$ for a paraboloid .The surfaces of the paraboloid is set by the parameter $\eta_0$. The discrete modes are denoted by $m$ for the azimuthal oscillations.        
		}
		\label{5-eps-p}
	\end{center}
\end{figure}

From  Eq.~\eqref{5-LPhi} and using the relation given in Eq.~\eqref{0-sddot}, one can solve for the charge density to obtain:
\begin{equation}\label{ASigma1}
\sigma =- \frac{1}{ 4\pi a \sqrt{\xi^2+\eta^2}} 
\eval{
	\left(
	\frac{\partial \Phi_{\text{o}}}{\partial\eta}-
	\frac{\partial \Phi_{\text{i}}}{\partial\eta}
	\right)
}_{\eta=\eta_0}.
\end{equation}
The expression on the right-hand side of Eq.~\eqref{ASigma1} can be easily calculated from Eq.~\eqref{pin-pout} as:
\begin{equation}
\eval{\left( \frac{\partial \Phi_{\text{o}}}{\partial\eta}-
	\frac{\partial \Phi_{\text{i}}}{\partial\eta}\right) }_{\eta=\eta_0}=
\sum_{m,p}S^p_m(\varphi)
\int_0^\infty 			\lambda
A_{m\lambda p}(t)  			
J_m(\lambda\xi)		
\mathcal W_m (\lambda \eta_0)\, d\lambda,
\end{equation}
where $\mathcal W_m(\cdot)$ denotes the Wronskian given by:
\begin{equation}\label{AW}
\mathcal{W}_m(z)=
I_m(z)
K'_m(z)-
I'_m(z) K_m(z).
\end{equation}
In view of the identity $\mathcal W_m(z)=-\frac1z$ ($z\ne 0$) for the modified Bessel functions \cite{LSSFA}, and from 
Eqs.~\eqref{ASigma1} and \eqref{AW} we obtain:
\begin{equation} \label{5-sigma}
\sigma=\dfrac{1}{4\pi a \eta_0\sqrt{\xi^2+\eta_0^2}} 
\sum_{m,p}
S^p_m(\varphi)		
\int_0^{\infty}
A_{m\lambda p}(t) 
J_m(\lambda\xi)  d\lambda. 
\end{equation}	
From the definition of the polarization,  the surface charge $\sigma$ 
can be written as relation provided in Eq.~\eqref{0B34} and it follows that:
\begin{equation}\label{5-sigmaP_2}
\ddot \sigma  =  -\frac{\omega_{p}^2}{4\pi a\sqrt{\xi^2+\eta_0^2}}
\sum_{m,p}
S^p_m(\varphi)
\int_0^\infty  \lambda A_{m\lambda p}(t)	J_m(\lambda\xi)		
I'_m(\lambda\eta_0)
K_m(\lambda\eta_0)						
\, d\lambda.
\end{equation}
Differentiating the charge density in Eq.~\eqref{5-sigma} twice with respect to time $t$ and equating it with
Eq.~\eqref{5-sigmaP_2}, it follows from the orthogonality of system $\{S_m^p\}_{m,p}$ given in Eq.~\eqref{5-ASmpO}, that for each fixed $m$ and $p$:
\begin{equation}\label{5-Sigma_3}
\int_0^{\infty}
J_m(\lambda\xi) 
\big [ \ddot A_{m\lambda p}(t)  
+ \omega^2_{m\lambda} 
A_{m\lambda p}(t)  \big ] \, d\lambda =0,
\end{equation}	
where 
\begin{equation}\label{5-wml}
\omega^2_{m\lambda}=\omega_{p}^2 \, \lambda \eta_0 \,
I'_m(\lambda\eta_0) \,
K_m(\lambda\eta_0).
\end{equation}
\noindent Utilizing the orthogonality relation for Bessel functions given by (see Eq.~(11.59) \cite{AWMMP}):
\begin{eqnarray}\label{5-ABesselO}
\int_{0}^{\infty} 	\xi 
J_{m}(\lambda\xi)		J_{m}(\lambda'\xi)		d\xi=
\frac{\delta(\lambda-\lambda')}{\lambda} ,
\end{eqnarray}
for $m\geq-1,$ and $\lambda, \lambda'>0,$ it follows from Eq.~\eqref{5-Sigma_3} that the amplitudes $A_{m\lambda p}(t)$ undergo harmonic oscillations
at continuous frequencies $\omega_{m\lambda}$ given by Eq.~\eqref{5-wml}, that is:
\begin{equation}\label{5-Bml}
\ddot A_{m\lambda p}(t)  + 
\omega^2_{m\lambda}A_{m\lambda p}(t)=0. 
\end{equation}	
The harmonic behavior of the field amplitudes will play an essential role in the possibility of analytically calculating the energy of the paraboloidal charge system prior to quantization.\\
Moreover, as it has already been discussed in Section \ref{HA}, the obtained $\omega_{m\lambda}$ in Eq.~\eqref{5-wml} satisfies the Drude model given in Eq.~\eqref{0-eps1}, namely:
\begin{equation}\label{epsp}
1-\varepsilon_{m\lambda}=	\omega^2_{p}\,	\omega_{m\lambda}^{-2}.
\end{equation}
The resonance values of the dielectric function $\varepsilon$ given in Eq.~\eqref{epsp} for a few modes are shown in Fig.~\ref{5-eps-p}.
\clearpage

\clearpage

\subsection{Classical Energy}

Having obtained closed form expressions for the potential and induced surface charge density, we are now in the position to calculate the potential energy $V$ due to the polarization charge distribution $\rho$ given by Eq.~\eqref{0-35}. 
Since $\rho$  is only confined to the paraboloidal surface $\eta=\eta_0$ and vanishes elsewhere in the space, one may integrate over an infinitesimal thin cover across the boundary: $
\eta_0 - \epsilon \leq \eta \leq \eta_0+\epsilon.$
Letting $\epsilon \to 0^+$ and  making the observation  $\rho h_{\eta}=\delta(\eta-\eta_0)\sigma$,
the potential energy can be expressed as a surface integral given in Eq.~\eqref{0-V}.
From Eq.~\eqref{pin-pout}, the potential $ \Phi_{\text{i}}\bigg|_{\eta=\eta_0} $  is given by:
\begin{equation}\label{5-pin1}
\Phi_{\text{i}}\bigg|_{\eta=\eta_0}= \sum_{m,p} S_m^p(\varphi)   
\int_0^\infty  A_{m\lambda p}(t)  I_m(\lambda\eta_0)  K_m(\lambda\eta_0) J_m(\lambda\xi) \, d\lambda.
\end{equation}
Substituting Eq. ~\eqref{5-pin1}  and the expressions for $\sigma,$  $h_{\xi},$ and $h_{\varphi}$ from Eqs.~\eqref{5-Sigma_3} and \eqref{5-sf} into Eq.~\eqref{0-V} gives	
\begin{multline}\label{5-V1}
V	=\frac{a}{8\pi}   \sum_{m',p'}\sum_{m,p}   \int_0^{\infty}  \int_0^{2\pi}  
S^p_m(\varphi)   S^{p'}_{m'}(\varphi)   \\
\times \Bigg  [	\int_0^{\infty}  A_{m'\lambda' p'}(t)  I_{m'}(\lambda'\eta_0)  K_{m'}(\lambda'\eta_0)  J_{m'}(\lambda'\xi) \, d\lambda'\\
\times  	 \int_0^{\infty}  A_{m\lambda p}(t)  J_m(\lambda\xi) \, d\lambda 
\Bigg ] \xi \, d\varphi d\xi.
\end{multline}
From the orthogonality relations for the trigonometric system $\{S^p_m(\varphi) \}$ given in Eq.~\eqref{5-ASmpO}, 
we can write Eq.~\eqref{5-V1} as:
\begin{multline}\label{V2}
V=\frac{a}{8} \sum_{m,p}   \hat \delta_{mp}
 \int_0^{\infty} \Bigg [
\int_0^{\infty}  A_{m\lambda' p}(t)  I_m(\lambda' \eta_0)  K_m(\lambda' \eta_0)
J_m(\lambda'\xi) \, d\lambda' \\
\times  \int_0^{\infty}  A_{m\lambda p}(t)  J_m(\lambda\xi) \, d\lambda\Bigg ]
\xi \, d\xi.
\end{multline}
Finally, in view of the orthogonality relation for Bessel functions given by Eq.~\eqref{5-ABesselO}, we obtain we obtain the potential:
\begin{equation}\label{5-V}
V =  \frac{a}{8} 
\sum_{m,p}   \hat\delta_{mp} 
\int_{0}^\infty 	\frac{I_{m}(\lambda\eta_0)  K_{m}(\lambda\eta_0)}{\lambda}  
\big [ A_{m\lambda p}(t)  \big ]^2 
\, d\lambda,
\end{equation}
where $\hat\delta_{mp}$ is given in Eq.~\eqref{5-delta1}.

Note that in the above calculations, we have changed the order of integration to simplify the obtained expressions. This is done in view of Fubini's theorem based on the absolute integrability of the above expressions. Our argument is based on the asymptotic behavior of Bessel functions (see Appendix B, Eqs.~\eqref{7-B-Asym}--\eqref{7-B-Asym2}) and the fact that the potential, and thus Eq.~\eqref{pin-pout}, is finite in the entire space. The same argument is also used in the other cases using the asymptotic behavior of the corresponding eigenfunctions.

We will now seek the   kinetic energy $T$ of the paraboloidal charge system. Employing the charge displacement vector, 
kinetic energy is given in Eq.~\eqref{0-T}.
To obtain an expansion for $\dot{\vec u},$ we note that in the expression for the potential $\Phi_{\text{i}}$ given by Eq.~\eqref{pin-pout}, one may use the harmonic oscillator equation Eq.~\eqref{5-Bml} for the amplitudes to get:
\begin{equation}
A_{m\lambda p}(t)=-\frac{\ddot A_{m\lambda p}(t)}{\omega^2_{m\lambda}}.
\end{equation}
As a result, integrating the charge displacement vector with respect to time, we obtain:
$m_e	\dot{\vec u}=	-e		\vec\nabla \dot\Psi,$
where
\begin{equation}\label{5-pdot}
\Psi({\bf r,t})=
\sum\limits_{m,p} S^p_m(\varphi) 
\int_0^{\infty}
\frac{ A_{m\lambda p}(t)}{\omega_{m\lambda}^2} 
J_{m}(\lambda\xi)
I_{m}(\lambda\eta)	
K_{m}(\lambda\eta_0)  \, d\lambda.
\end{equation}
With  this expression in the integral for kinetic energy, making use of the Gauss theorem \cite{gauss} and orthogonality relations for $\{S^p_m \}_{m,p}$ and Bessel functions, given in Eqs.~\eqref{5-ASmpO} and \eqref{5-ABesselO}, we thus calculate Kinetic energy by rewriting Eq.~\eqref{Sp-K} for the case of paraboloid as:
\begin{eqnarray}\label{5-T2}
T=\frac{e^2 n_0}{2m_e}
\int_{0}^{2\pi}
\int_{0}^{\infty}
\eval{\Big ( \dot{{\Psi}}  \frac{1}{h_\eta}\frac{\partial{\dot{{\Psi}}	}}{\partial\eta} \Big )}_{\eta=\eta_0} 
h_{\xi} h_{\varphi} d\xi d\varphi.
\end{eqnarray}
Using the definition of $\Psi$ given by Eq.~\eqref{5-pdot} and the expressions for the scale factors Eq.~\eqref{5-sf}, the right-hand side of Eq.~\eqref{5-T2} can be written as:
	\begin{multline}\label{5-T3}
				T=a \eta_0 \frac{e^2  n_0}{2m_e}
				\sum\limits_{m,p}	\sum\limits_{m',p'} 	\int_{0}^{2\pi}
				S^p_m(\varphi)	S^{p'}_{m'}(\varphi) 	\, d\varphi \\
				\times	\int_{0}^{\infty}				\bigg [ \int_0^{\infty}
				\frac{\dot A_{mp\lambda}(t)}
				{\omega_{m\lambda}^2}
				J_{m}(\lambda\xi) 		~	I_{m}(\lambda\eta_0)	K_{m}(\lambda\eta_0)		\, d\lambda		\\
				\times	\int_0^{\infty}
				\lambda 		\frac{{\dot A _{m'p'\lambda'}}(t)}		{\omega_{m'\lambda'}^2}
				J_{m'}(\lambda'\xi)	I'_{m'}(\lambda'\eta_0)	~	K_{m'}(\lambda'\eta_0)
				\, d\lambda'\bigg ]		\xi   	\, d\xi.	
	\end{multline}
Invoking the orthogonality relations, given in Eqs.~\eqref{5-ASmpO} and \eqref{5-ABesselO}, to the Eq.~\eqref{5-T3} by using a similar argument as the one given for the case of the potential,  it follows that
	\begin{equation}\label{5-T4}
			T=\frac{a\eta_0 \omega_{p}^2}{8}
			\sum\limits_{m,p}
			\hat \delta_{mp} 
				\int_0^{\infty} \frac{1}{\omega_{m\lambda}^4}
			\left[ \dot A_{mp\lambda}(t)\right] ^2   I_{m}(\lambda\eta_0)	 I'_{m}(\lambda\eta_0)	
			\left[ K_{m}(\lambda\eta_0)\right] ^2 	 \, d\lambda.	
	\end{equation}
Now in view of Eq.~\eqref{5-wml}, we have :
$$\eta_0 \omega_{p}^2 / \omega_{m\lambda}^2 = \big (\lambda K_{m}(\lambda\eta_0) I'_{m}(\lambda\eta_0) \big )^{-1}.$$ This substitution in Eq.~\eqref{5-T4} gives the kinetic energy as:
	\begin{equation}\label{5-T}
			T=\frac{a}{8}
			\sum\limits_{m,p}
			\hat \delta_{mp}
			\int_0^{\infty} \frac{I_{m}(\lambda\eta_0)  K_{m}(\lambda\eta_0)}{ \lambda \, \omega_{m\lambda}^2}
			\left[ \dot A_{m\lambda p}(t) \right] ^2    d\lambda.
	\end{equation}

In view of Eqs.~\eqref{5-V} and \eqref{5-T}, we find that the  total classical energy $E$ of the paraboloidal system takes the following form:
\begin{equation}\label{5-E0}
E=  \frac{a}{8}
\sum\limits_{m,p}
\hat\delta_{mp} 
\int_{0}^\infty 
\frac{	I_{m}(\lambda\eta_0)	  K_{m}(\lambda\eta_0)}{\lambda \, \omega_{m\lambda}^2} 
\Big \{ \big [	\dot A_{m\lambda p}(t)	\big ]^2 +
\omega_{m\lambda}^2	\big [	 A_{m\lambda p}(t) 	\big ] ^2 \Big \} \, d\lambda,
\end{equation}
which is the total energy of the surface plasmon field, suitable for quantization. \\

\subsection{Field Quantization \& Interaction of Plasmons with Photons}\label{INT_P}

To obtain the quantized plasmon field, the expression for the classical field $E$ needs to be rewritten in a suitable form. We begin by noting that since the potential together with the paraboloidal harmonics $I_{m}(\lambda\eta),$ 
$J_{m}(\lambda\xi),$ $K_{m}(\lambda\eta),$  and $S^p_m(\varphi)$  are all real-valued, we have that the amplitudes $A_{m\lambda p}(t)$ are real-valued and satisfy the harmonic oscillator equation Eq.~\eqref{5-Bml}. 
In analogous to Eq.~\eqref{0-D}, to pursue the conversion of $E$  into the Hamiltonian operator for a scalar boson field (plasmons are spinless quasi-particles) and we write: 
\begin{equation}\label{5-A_m}
A_{m\lambda p}(t)=
\frac{ \alpha_{mp\lambda}}{2\omega_{m\lambda}} 
~[a_{m\lambda p}(t)+a^*_{m\lambda p}(t)],
\end{equation}
where $a_{m\lambda p}$ are  complex-valued time dependent functions  proportional to $e^{-i\omega_{m\lambda} t},$ and   
$\alpha_{mp\lambda}$ are some real-valued constants to be determined later. The time derivative of  Eq.~\eqref{5-A_m} can now be expressed as:
\begin{equation}\label{5-Adot}
\dot A_{m\lambda p}(t)=
\frac{ i\alpha_{mp\lambda}}{2} ~
[a^*_{m\lambda p}(t)-a_{m\lambda p}(t)].
\end{equation}
Substituting Eqs.~\eqref{5-Adot} and \eqref{5-A_m} in the expression for classical energy given in Eq.~\eqref{5-E0}, we may write:
\begin{equation}\label{5-E}
E=\cfrac{a}{8}
\sum\limits_{m,p}
\hat\delta_{mp}
\int_0^{\infty}
\frac{I_{m}(\lambda\eta_0)	K_{m}(\lambda\eta_0) }{\lambda\omega_{m\lambda}^2}  					
\, 	\frac{\alpha^2_{mp\lambda}}{2}
\bigg(	a^*_{mp\lambda}	a_{m\lambda p}		+	a_{m\lambda p}	a^*_{mp\lambda} \bigg)
d\lambda.
\end{equation}
Performing field quantization (see Section~\ref{1-f}), we replace the amplitudes $a_{m\lambda p}(t)$ and $a^*_{m\lambda p}(t)$ with operator valued distributions $\hat{a}_{m\lambda p}$ and its conjugate $\hat{a}^\dagger_{m\lambda p},$ and following Eq.~\eqref{0-R}, we note the commutation relations:
\begin{equation}
[\hat{a}_{m\lambda p},\hat{a}^\dagger_{m'\lambda' p'}]=\delta_{mm'}\delta_{pp'}\delta(\lambda-\lambda'),
\end{equation}
owing to the continuous spectrum of the eigenvalues $\lambda $ that originates from the infinite axial dimension of the paraboloid.
Taking the normal ordered expansion of the non-commuting boson creation $\hat{a}^\dagger_{m\lambda p}$ and annihilation  $\hat{a}_{m\lambda p} $ operators, a  comparison  with the normal ordered expression of the Hamiltonian operator for a scalar boson field~\cite{B}, yields the Hamiltonian:
\begin{eqnarray}\label{5-E2}
\mathcal{H}_{sp} = \sum\limits_{m,p}
\int_0^{\infty}
\hbar\omega_{m\lambda}
\hat{a}^{\dagger}_{m\lambda p}\hat{a}_{m\lambda p}d\lambda,
\end{eqnarray}
upon choosing:
\begin{equation}\label{5-alpha}
\alpha^2_{m\lambda p}= 
\frac{8\,\hbar	\,\lambda \,	\omega^3_{m\lambda} \,\hat\delta_{mp}^{-1} 	}
{a 	\, I_{m}(\lambda\eta_0)	\ K_{m}(\lambda\eta_0)} .
\end{equation}
For the interaction of the plasmon system with a photon field with Hamiltonian $\mathcal{H}_p$, 
we require the plasmon-photon interaction Hamiltonian $\mathcal{H}_i$, which is required to determine the radiative decay rate of surface plasmons excited on the paraboloidal surface. \\
Our full system, plasmon field + photon field is described by the tensor product of the Fock spaces for the two constituent fields.

This interaction has been used previously to describe the emission of  photons via plasmon decay on finite surfaces of an oblate spheroid modeling  silver nanoparticles vacuum evaporated on a dielectric substrate~\cite{Little_Paper}. Here, we will apply this Hamiltonian to modeling the creation of a plasmon on the surface of the paraboloid by a photon or the decay of a paraboloidal plasmon and emission of a photon~\cite{B}. This application requires the explicit determination of the current density operator, which in light of the displacement operator: 
$\dot{\vec u}=	-(e/m_e)		\vec\nabla \dot\Psi,$
can be written as $
\mathbf{J} =-(n_0e^2/m_e)\vec \nabla\dot{\Psi}
$ (see Eqs.~\eqref{0-J}--\eqref{0-J1}).

To write the photon field as a sum with discrete momentum eigenstates as opposed to the continuous representation, we consider our electromagnetic field to be confined in a volume $\mathcal{V},$ which is normally taken to be represented by a cube. Taking the electromagnetic energy confined to a volume to be independent of the shape of the volume \cite{Bennett,Belinsky}, we take as our quantization volume a paraboloidal structure given by
$\xi=\eta=L$ with $L\gg \eta_0$. The volume is then found to be $ \frac{\pi}{2}L^6.$
In the paraboloidal coordinates, $\xi$ and $\eta$ have dimensions of length$^{1/2}$ and so our volume has dimension $L^3$. 
Hence, the discretized vector potential given in Eq.~\eqref{0-A3} for paraboloidal case may be written as:
\begin{eqnarray}\label{5-A3}
\mathbf{A}=
\sum\limits_{s}
\sum\limits_{j=1,2}
\sqrt{\frac{\hbar c^2 }{\mathcal V \omega_s}}
\mathbf{\hat e}_j
\big (	\hat c_{\mathbf{s}j}	e^{i\mathbf{s\cdot r}}	+\hat c^{\dagger}_{\mathbf{s}j}   e^{-i\mathbf{s\cdot r}}	\big ),
\end{eqnarray}
with $\mathbf{s}$ being the three-dimensional photon wavevector $\omega_s=\mathbf s c$, $\mathbf{\hat e}_j$ the polarization vector being perpendicular to $\mathbf{s}$  for both values of $j$ and  $\hat c^{\dagger}_{\mathbf{s}j}$ and $\hat c_{\mathbf{s}j}$, 
the photon creation and annihilation time-dependent operators, following Eq.~\eqref{1-cm}, satisfying the commutation relations:
		\begin{equation} 
		[\hat c_{\mathbf{s}j},\hat c^{\dagger}_{\mathbf{s}j}]=\delta_{jj'}\delta(\mathbf{s-s'}).
		\end{equation} 
For the physical quantities of interest in our work, 
we find that the quantization volume $\mathcal{V}$ cancels out in the calculations. The photon field Hamiltonian corresponding to the vector potential above is 
	\begin{equation} 
	\mathcal{H}_p  = \sum_{\mathbf{s} j }     \hbar \omega_{\mathbf{s}}    \hat c^\dagger_{\mathbf{s}j}   \hat c_{\mathbf{s} j }    .
	\end{equation} 
Taking $\mathbf{A}$ to be in the radiation gauge with both $\Phi=0$ and $\vec\nabla\cdot\mathbf{A}=0$ (see Section~\ref{1-f}), 
we note
		\begin{equation} 
		(\vec\nabla\dot{\Psi})\cdot\mathbf{A}=
		\vec\nabla\cdot(\dot{\Psi}\mathbf{A}),
		\end{equation} 
since the current is confined to the surface of the paraboloid, we have
		\begin{eqnarray}\label{5-H2}
		\mathcal{H}_{i} =   \frac{n_0e^2}{c m_e}
		\int_0^{2\pi}~\int_0^\infty
		\left(	\dot{\Psi}\mathbf{A}	\cdot	\mathbf{\hat e}_{\eta} 			\right)
		h_\xi 	h_\varphi \,	d\xi 	d\varphi.
		\end{eqnarray}
Differentiating $\Psi$ given by Eq.~\eqref{5-pdot} with respect to time and replacing  $\dot A_{m\lambda p}(t)$ using Eq.~\eqref{5-Adot}, we can now write the interaction Hamiltonian as:
		\begin{multline}\label{5-H3}
		\mathcal{H}_{i}  =  \frac{n_0e^2}{2i\,m_e}
		\sum\limits_s	\sum\limits_{j=1,2}		\sqrt{\frac{\hbar }{\mathcal V\omega_s}}
		(\mathbf{\hat{e}}_\eta\cdot\mathbf{\hat{e}}_j)
		\big (\hat c_{\mathbf{s}j}		e^{i\mathbf{s\cdot r}}	+
		\hat c^{\dagger}_{\mathbf{s}j}		e^{-i\mathbf{s\cdot r}}		\big )   \\
		\times \sum\limits_{m,p}
		\int_{0}^{2\pi}
		\int_{0}^{L}\bigg [
		\int_{0}^{\infty}
		S^m_p(\varphi)
		J_{m}(\lambda\xi)
		I_{m}(\lambda\eta_0)	K_{m}(\lambda\eta_0) \\
		\times	\frac{\alpha_{m\lambda p}(t)}{\omega_{m\lambda}^2}
		(\hat a^{\dagger}_{m\lambda p}-\hat a_{m\lambda p})d\lambda	\bigg ]
		h_\xi		h_\varphi \,		d\xi 	d\varphi , 
		\end{multline}
where we have taken the integral in $\xi$ to be from 0 to $L$, since the space is bounded by $\xi=L$.
\subsection{Emission of Light \& Radiative Decay }
Having obtained the explicit form of the interaction Hamiltonian, we now aim to calculate the probability amplitude that a surface plasmon in a given initial state, 
defined by $m,\lambda,p$ will emit a $j$-polarized photon in $\mathbf{s}$ state with momentum $\hbar\mathbf{s}$. 
Following Section~\ref{rad}, the total radiative decay rate for a given initial state, $\gamma_{m\lambda p}$, is obtained by summing over final photon states, 
$ \gamma_{m\lambda p}= \sum_{ s}		\sum_{j=1,2}		\gamma^{(j\mathbf{s})}_{m\lambda p},$
where by the Fermi Golden rule, the transition probabilities are:
		\begin{eqnarray}\label{Transition}
		\gamma^{(j\mathbf{s})}_{m\lambda p} =
		\frac{2\pi}{\hbar}
		~\left |\mathcal M^{(j\mathbf{s})}_{m\lambda p}\right |^2
		\delta(\omega_\mathbf{s}-\omega_{m\lambda}) ,
		\end{eqnarray}
with 
$ \mathcal{M}^{(j\mathbf{s})}_{m\lambda p}=  |\bra{0}  \hat c_{j_f\textbf{s}_f} \mathcal{H}_{i}~ \hat a^{\dagger}_{m_i\lambda_i p_i} \ket{0}| $ 
denoting  the emission matrix elements,
$f$ the final state, and $i$ the initial state, by means of the commutation properties:
		\begin{eqnarray}\label{5-AProp}
		\langle 	
		0|~\hat a_{m\lambda p}~\hat a^{\dagger}_{m'\lambda' p'}~|0  \rangle    && = 
		\delta_{mm'} 		\delta(\lambda-\lambda')
		\delta_{pp'},   \\
		\langle
		0|~\hat c_{\mathbf{s}'q'}~		\hat c^{\dagger }_{\mathbf{s}q}~|		0
		\rangle   &&	 = 		\delta(\mathbf s-\mathbf s') 		\delta_{qq'},
		\end{eqnarray}
the non-vanishing terms yield:
		\begin{equation}\label{5-M1}
		\mathcal{M}^{(j\mathbf{s})}_{m\lambda p}= 
		\frac{n_0e^2}	{2i\,m_e}
		\sqrt{	\frac	{\hbar}	{\mathcal V\omega_s}}
		\frac{\alpha_{m\lambda p}}{\omega^2_{m\lambda}}
		I_{m}(\lambda\eta_0)	K_{m}(\lambda\eta_0)\mathcal{I}^{(j)}_{m\lambda} , 
		\end{equation} 
where we have omitted $i$ and $f$ for convenience and we have set:
		\begin{equation} \label{5-I}
		\mathcal I^{(j)}_{m\lambda}=   \int_{0}^{2\pi}
		\int_{0}^{L} 		S^p_m(\varphi)
		(\mathbf{\hat{e}}_\eta		\cdot		\mathbf{\hat{e}}_j)
		J_{m}(\lambda\xi) 	e^{-i\mathbf{s\cdot r}}
		h_\xi h_\varphi     	d\xi d\varphi .
		\end{equation} 
For an arbitrary photon wavevector of the form:
		\begin{equation}
		\mathbf{s}=  \frac{\omega_s}{c}		(\cos\psi,	0	,	\sin\psi), 	
		\end{equation}
we may consider the polarization vectors $\hat{\mathbf{e}}_{j}$ for s-polarization and p-polarization, being perpendicular and parallel to $\hat{z}\mathbf s$-plane, respectively, noting that $\mathbf s \cdot \hat{\mathbf{e}}_{j} =0$.
Thus, we consider the polarization vectors $\hat{\mathbf{e}}_{1}=(0,1,0)$ and 
$\hat{\mathbf{e}}_{2}=(\sin\psi,0,-\cos\psi).$ 
Under these polarization conditions, Eq.~\eqref{5-I} leads  to two different integrals:
		\begin{equation} \label{5-I1}
		\mathcal I^{(1)}_{m\lambda}=a\eta_0	
		\int_{0}^{2\pi} 	\int_{0}^{L}			\xi^2		\sin\varphi  
		E_{m\lambda}(\eta,\xi)
		J_{m}(\lambda\xi) 
		S^p_m(\varphi)
		\,	d\xi d\varphi,
		\end{equation} 
and 
		\begin{equation} \label{5-I2}
		\mathcal I^{(2)}_{m\lambda}=a\eta_0
		\int_{0}^{2\pi}
		\int_{0}^{L}
		\Big ( 				\xi^2			\sin\psi			\cos\varphi+		\xi		\eta		\cos\psi		\Big )
		E_{m\lambda}(\eta,\xi)		J_{m}(\lambda\xi)  		S^p_m(\varphi)		d\xi d\varphi,
		\end{equation} 
where $	E_{m\lambda}(\eta,\xi)=e^{-i\mathcal A}$ with $\mathcal{A}$ given by:
		\begin{eqnarray}
		\mathcal A= -\frac{a\omega_s}{c}
		\left[		\xi		\eta_0		\cos\psi			\cos\varphi+
		\frac{\sin\psi(\xi^2-\eta_0^2)}{2}		 			\right]. \notag
		\end{eqnarray}
Thus, depending on the choice of the polarization vector  $\hat{\mathbf{e}}_{j}$, each integral contributes to a polarization state. In other words, $\mathcal I^{(1)}_{m\lambda}$ represents the s-polarization and $\mathcal I^{(2)}_{m\lambda}$ corresponds to the p-polarization.
For a photon emitted pointing to the solid angle $d\Omega$ as depicted  in Fig.~\ref{system}, final photon states result in a continuous energy spectrum. Hence, we let the quantization volume $\mathcal V$ go to infinity so that we recover a continuum set of $\mathbf s$-states, that is:
		\begin{eqnarray}
		\sum\limits_{\mathbf s}\to \frac{\mathcal V}{(2\pi)^3} \int s^2ds \int d\Omega, 
		\end{eqnarray}
where $\mathcal V/(2\pi)^3 $ is the density of photon states per polarization. In view of transition probability given in Eq.~\eqref{Transition}, the radiative decay rate per unit solid angle may be written as: 
$$	\frac{d\gamma_{m\lambda p}}{d \Omega}=\sum\limits_{j=1,2}  \frac{\mathcal V}{(2\pi)^3}\int \gamma^{(j\mathbf{s})}_{m\lambda p} s^2ds, $$
in which using $\omega_s=sc$, 
we obtain the final expression:
\begin{equation}\label{5-R}
\frac{d\gamma_{m\lambda p}}{d \Omega}=
\sum_{j=1,2}
\frac	{\mathcal V}	{(2	\pi)^3} 
\frac{\omega_{m\lambda}^2}	{ c^3}\bigg [
\frac	{2\pi}	{\hbar^2}~
\left|		\mathcal M^{(j\mathbf{s})}_{m\lambda p}		\right| ^2 		\bigg ]_{\omega_{s}=		\omega_{m\lambda}}	,
\end{equation}
which is observed to be independent of the volume $\mathcal V$ as $|\mathcal M^{(j\mathbf{s})}_{m\lambda p}|^2$ is proportional to $\mathcal V^{-1}$ in . Thus, we can take our quantization volume to be infinite and convert the sum over the photon states to an integral.
Hence, the paraboloidal decay rate per solid angle is given by:
\begin{equation} \label{5-R1}
\frac{d\gamma_{m\lambda p}}{d\Omega}	=
\frac{\lambda \, 	\hat\delta_{mp}^{-1} 	}		{a	\pi 	c^3	}
\left(		\frac{n_0e^2}{m_e}		\right)^2  	I_m(\lambda\eta_0)		K_m(\lambda\eta_0) 	
\big [	\big ( 	\mathcal I^{(1)}_{m\lambda}	\big )^2+
\big (  \mathcal I^{(2)}_{m\lambda}		\big )^2		\big ],
\end{equation} 
where $ \mathcal I^{(1)}_{m\lambda}	$  and $\mathcal I^{(2)}_{m\lambda}$ are given by Eqs.~\eqref{5-I1} and \eqref{5-I2}, respectively, and must be computed numerically.    
\section{Hyperboloidal Surfaces}\label{hyper}
Similarly to the paraboloidal case, the hyperboloidal domain, shown in Fig.~\ref{system}, is highly relevant for modeling of electronic and photonic response  of  probe-like nanostructures~\cite{PPRB}. Under the same assumptions as the paraboloidal case, we proceed as follow:
An arbitrary point in the Cartesian space can be expressed in terms of the prolate spheroidal coordinates $(\eta,\mu,\varphi)$  \cite{LSSFA,LSUWPAM} as:
		\begin{align}
		\begin{split}
		x(\zeta,\theta,\varphi)&=z_0\, 	\sinh{\zeta}	\sin{\theta}	\cos{\varphi},\\
		y(\zeta,\theta,\varphi)&=z_0\, 	\sinh{\zeta}	\sin{\theta}	\sin{\varphi},\\
		z(\zeta,\theta,\varphi)&=z_0\, 	\cosh{\zeta}	\cos{\theta},
		\end{split}
		\end{align}
in the domain defined by :
		\begin{equation}
		0	\le	\zeta<\infty	,\qquad 	0\le	\theta	\le	\pi, 	\qquad 0\le\varphi\le2\pi,
		\end{equation}
or alternatively, 
		\begin{align}\label{5-AHC1}
		x&=z_0	\, 	\sqrt{({\eta}^2-1)(1-\mu^2)}	\cos{\varphi},\notag\\
		y&=z_0	\, 	\sqrt{({\eta}^2-1)(1-\mu^2)}	\sin{\varphi},\\
		z&=z_0	\, 	{\eta}\, 	{\mu},\notag
		\end{align}
with the corresponding scale factors:
		\begin{align}\label{5-AH_SF}
		&h_{\eta}=z_0\, 	\sqrt{	\frac{\eta^2-\mu^2}	{\eta^2-1}}, \notag
		\quad h_{\mu}=z_0\, 	\sqrt{	\frac{\eta^2-\mu^2}{1-\mu^2}},\\
		&h_{\varphi}=z_0\, 	\sqrt{	(1-\mu^2)(\eta^2-1)},
		\end{align}
where $z_0>0$ is an overall scale factor, $1\le\eta<\infty,$  $-1\le	\mu	\le 1,$  and $0\leq\varphi < 2\pi$ denote the usual azimuthal angle. The  surfaces of constant  $\mu$ define hyperboloids of revolution about the $z$-axis, while the surfaces of constant  $\eta$ correspond to prolate spheroids. It is customary  to use the substitutions $\sinh \zeta=\eta$ and $\sin \theta=\mu$ with 
$ 1\le\eta<\infty$,  and  $-1\le\mu\le1;$  however, in this presentation  the surfaces of constant coordinates are no longer  geometrically meaningful. 
%
%
%
\noindent The radius vector is given by:
		\begin{equation}\label{5-H_r}
		\mathbf{\vec r}=z_0\, 	\sqrt{({\eta}^2-1)	(1-\mu^2)}	(\cos{\varphi}\hat i+
		\sin{\varphi}\hat j)+
		z_0\, 	{\eta}{\mu}\hat k
		\end{equation}
where $(\hat i, \hat j, \hat k)$ are the Cartesian unit vectors. The unit vectors for the prolate spheroidal coordinates are:
		\begin{align}\label{5-H_uv}
		\begin{split}
		\hat e_{\eta}	&=		\frac{1}{h_{\eta}}		\frac{\partial \mathbf{\vec r}}{\partial \eta } 
		=\sqrt{\frac{\eta^2-1}{\eta^2-\mu^2}}
		\left[	\frac{\eta\sqrt{1-\mu^2}}{\eta^2-1}
		\left(	\cos{\varphi}\hat i	+	\sin{\varphi}\hat j \right) +		\mu\hat k \right], \\
		\hat e_{\mu}&	=	\frac{1}{h_{\mu}}		\frac{\partial \mathbf{\vec r}}{\partial \mu }=
		\sqrt{\frac{1- \mu^2}{\eta^2-\mu^2}}
		\left[\frac{-\mu	\sqrt{\eta^2-1}}{1-\mu^2}
		\left(	\cos{\varphi}\hat i+\sin{\varphi}\hat j \right) 		+\eta\hat k \right],\\
		\hat e_{\varphi}&	=	\frac{1}{h_{\varphi}}	\frac{\partial \mathbf{\vec r}}{\partial \varphi }=
		(-\sin{\varphi}		\hat i		+		\cos{\varphi}\hat j		).
		\end{split}
		\end{align}
\subsection{Nonretarded Potential \& Dispersion Relations}
\begin{figure}
		\begin{center}
			\includegraphics[width=5.5in]{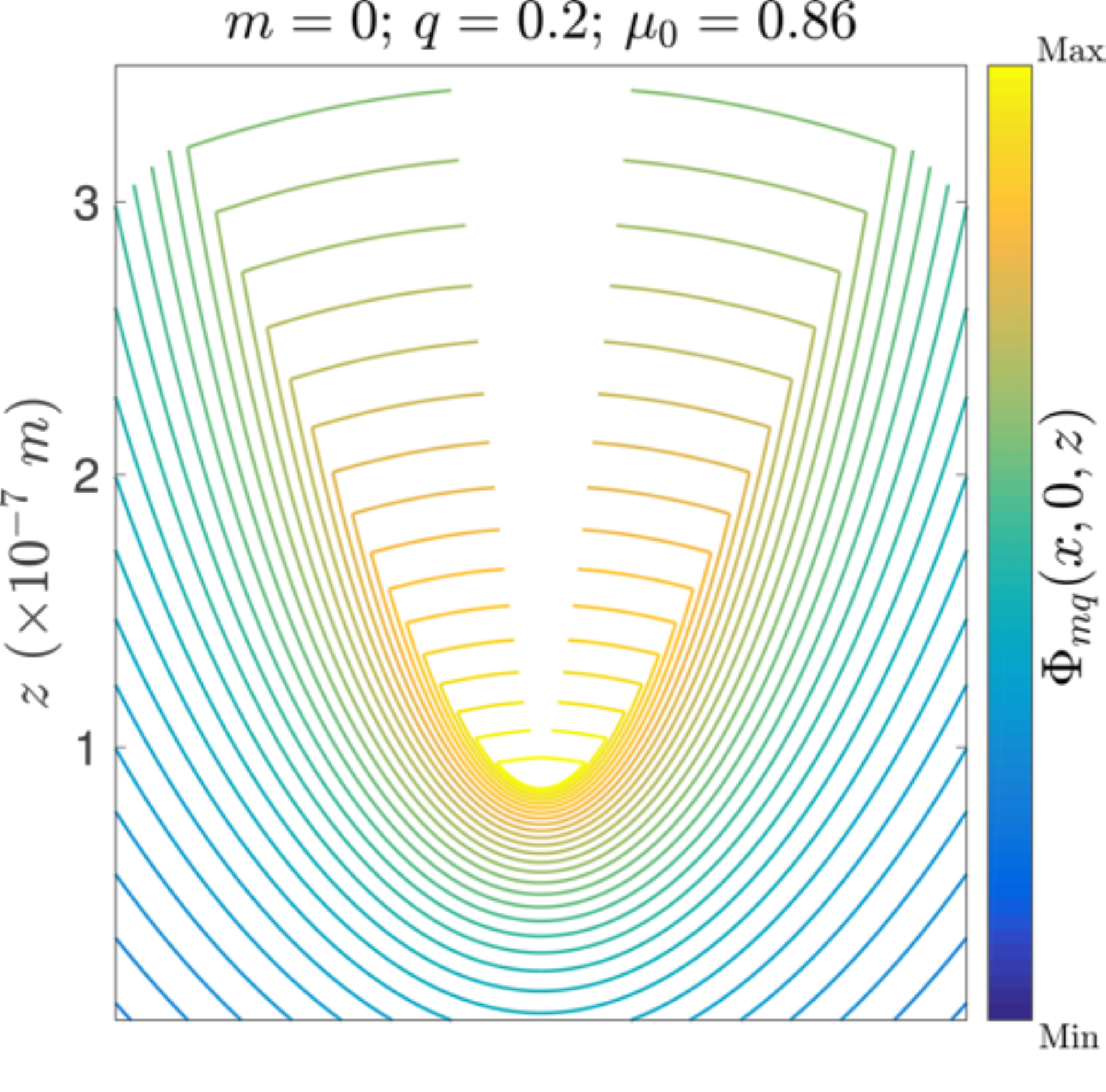}\\
			\caption[Modeling systems for the spatial distribution]{Modeling systems for the spatial distribution of the lowest lying eigenmodes of the quasi-static electric potential for the hyperboloidal domain. For the same mode index $m$, optimizing the apex curvature overlap within the same spatial $zx$ domains, and analysing the potential distribution, leads to the determination of the corresponding continuous eigenvalues $q$ of the hyperboloid The geometric parameter $\mu_0=86~$(nm) determines the form of the considered domains. }
			\label{5-pot-h}
		\end{center}
	\end{figure}

\begin{figure}
		\begin{center}
			\includegraphics[width=6in]{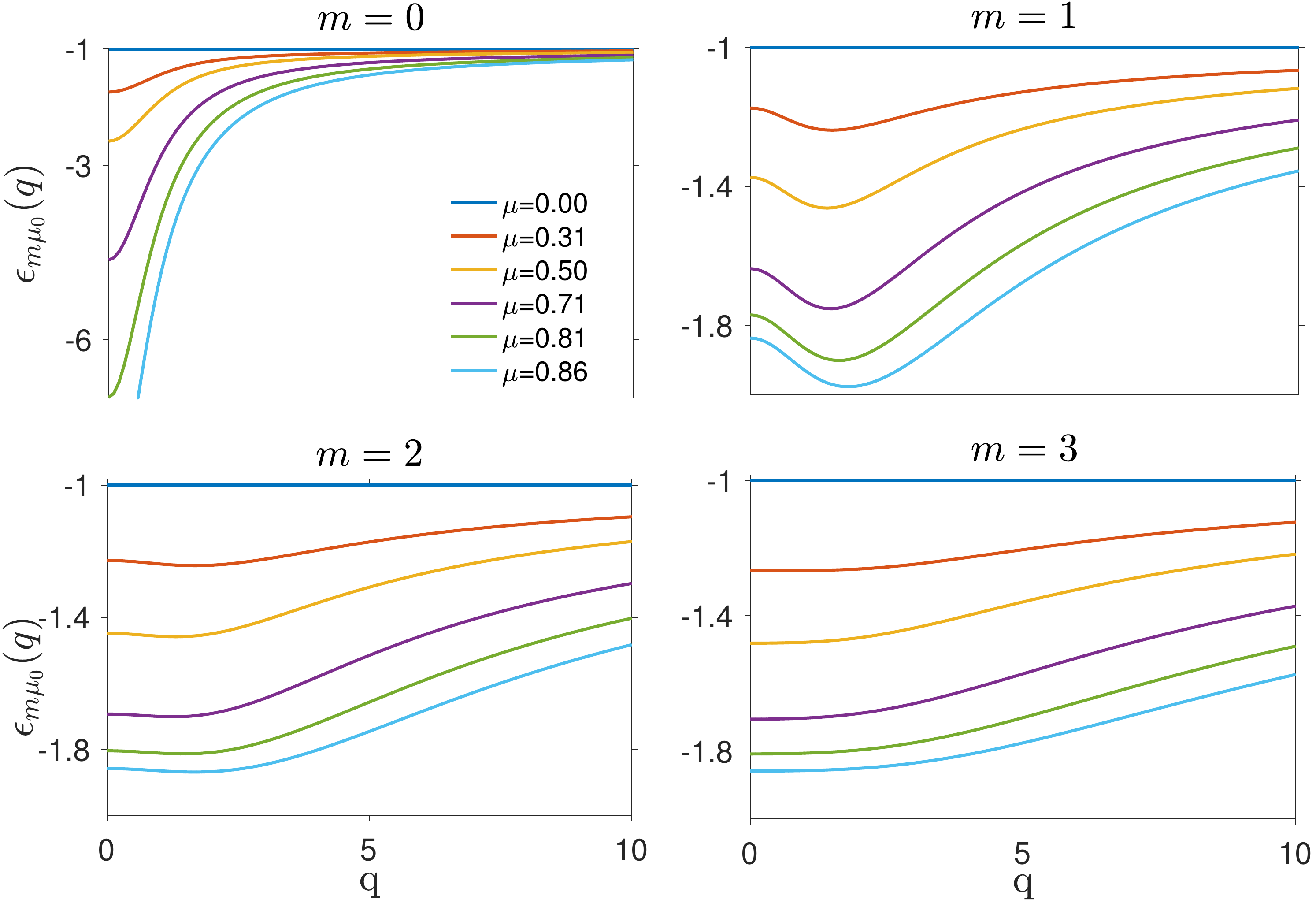}\\
			\caption[Hyperboloidal nonretarded surface plasmon  dispersion relations]{
				Hyperboloidal nonretarded surface plasmon  dispersion relations. The resonance values of the dielectric function $\varepsilon$ are shown for low lying modes as a function of  the continuous eigenvalue $q$ for a hyperboloid .The surfaces of the hyoerboloid is set by the parameter $\mu_0$. The discrete modes are denoted by $m$ for the azimuthal oscillations.        
					}
			\label{5-eps-h}
		\end{center}
	\end{figure}

Considering a solid hyperboloid of revolution  $\mu \geq \mu_0$ ($\mu_0 >0$) with dielectric constant $\varepsilon_1$
immersed in a medium  whose dielectric constant $\varepsilon_2$ is assumed to be $1,$ it follows that the potential of the electric $\Phi(\mu,\eta,\varphi)$ has to satisfy the Laplace equation $\vec \nabla^2\Phi=0$ everywhere except on the boundary surface $\mu=\mu_0,$ where the Laplacian is given by \cite{LSSFA}:
	\begin{multline}\label{AHLPhi}
			\vec\nabla^2 = \frac{1}{z_0^2(\eta^2-\mu^2)}	
			\Bigg\{ \frac{\partial}{\partial\eta}	
			\left[ (\eta^2-1) 	\frac{\partial}{\partial\eta}	\right] 
			+	\frac{\partial}{\partial\mu}		\left[ 		(1-\mu^2)
			\frac{\partial}{\partial\mu}		\right]  
			+		\left[ 	\frac{\eta^2-\mu^2}	{(\eta^2-1)	(1-\mu^2)}	\right] 
			\frac{\partial^2}{\partial\varphi^2}
			\Bigg\} .
	\end{multline}
Assuming the ansatz \cite{LSSFA} $\Phi(\mu,\eta,\varphi)=F(\eta)G(\mu)e^{im\varphi},$
one finds that $F$ and $G$ satisfy the  differential equations:
	\begin{align}\label{AHLpEq}
		\frac{d}{d\eta} \left[ (\eta^2 -1) \frac{dF_m}{d\eta}  \right] - \frac{m^2}{\eta^2 -1} F_m &= c \, F_m, \\
		\frac{d}{d\mu} \left[ (1- \mu^2) \frac{dG_m}{d\mu}  \right] - \frac{m^2}{1-\mu^2} G_m &= -c \, G_m.
	\end{align}
Letting $c=\nu (\nu+1)$ (see \cite{PK, Belova, LSSFA}), where $\nu=-\frac12+iq$ and $q\in [0,\infty),$ one can obtain a continuous spectrum of real eigenvalues and eigenfunctions  in terms of  the associated Legendre functions or conical functions $P^m_{-\frac12+iq}(z)$ (denoted by $P_{mq}(z)$) of complex lower index with 
$ -\infty < z < \infty$ satisfying the orthogonality relation \cite{PK,vn:ortho}:
	\begin{eqnarray}\label{AHor}
		\int_{1}^{\infty} 		P_{mq}(\eta)			P_{mq'}(\eta)d\eta=
		\frac{Z^m_q}{q	\tanh(\pi q)}		\delta(q-q'),
	\end{eqnarray}
with 
	\begin{align}\label{AHZ}
		Z_q^m=(-1)^m	{\displaystyle \prod_{k=1}^{m} } \left( q^2+\frac{(2k-1)^2}{4} \right).
	\end{align}
A bounded satisfactory real-valued solution to the first equation in Eq.~\eqref{AHLpEq} is given by  $P^m_{-\frac12+iq}(\eta)$  (see \cite{PK, Belova, LSSFA}), whereas a pair of satisfactory real-valued solutions to the second equation in Eq.~\eqref{AHLpEq} are given by $P^m_{-\frac12+iq}(\mu)$ and 
$P^m_{-\frac12+iq}(-\mu)$ 
\cite{LSSFA,z&k:table1,z&k:table2}. Avoiding the singularity at $\mu=-1$ \cite{LSSFA}, we shall choose the solution $P^m_{-\frac12+iq}(\mu)$ for the inside region $\mu_0 \leq \mu \leq 1$ and 
$P^m_{-\frac12+iq}(-\mu)$ for the outside region $-1 \leq \mu \leq \mu_0.$
Thus,
the inside and outside potentials $\Phi_{\text{i}}$  and $\Phi_{\text{o}}$  are expressed as sums over the Fourier harmonics 
	\begin{equation}\label{F1}	
		\int_0^{\infty} A_{mq}(t) P^m_{-\frac12+iq}(\eta) 	P^m_{-\frac12+iq}(\mu)\, dq \, e^{im\varphi},
	\end{equation}
and 
	\begin{equation}\label{F2}
		\int_0^{\infty} B_{mq}(t) P^m_{-\frac12+iq}(\eta) 	P^m_{-\frac12+iq}(-\mu)\, dq \, e^{im\varphi},
	\end{equation}
respectively, where $A_{mq}(t)$ and $B_{mq}(t)$ are complex valued  amplitudes and $m\in \mathbb{Z}$. Using the fact \cite{g&r:table} that 
	\begin{equation}
		\Gamma(\frac12+ m -iq) 		P^m_{-\frac12+iq}(z) =  
		\Gamma(\frac12- m -iq) P^{-m}_{-\frac12+iq}(z)
	\end{equation}
for all $z\in\mathbb{R}$ and $m \geq 0,$ it follows that one only needs to expand $\Phi_{\text{i}}$ and $\Phi_{\text{o}}$ in terms of 
$\cos m\varphi,$ for  $m=0,1, 2,  \dots.$ More precisely, we use the system  $H_m(\varphi)=(2-\delta_{0m})\cos\varphi$ satisfying the orthogonality relation:
	\begin{equation}\label{AHort}
		\int_0^{2\pi }		H_m(\varphi) 		H_{m'}(\varphi) \, d\varphi = \pi \hat\delta_{0m} \delta_{mm'}, 
	\end{equation}
where 
	\begin{equation}\label{delta2}
		\hat \delta _{0m}=1+\delta_{0m} .
	\end{equation}
Finally, imposing the continuity of the potential across the boundary $\mu=\mu_0,$  we find the inside and outside potentials,  which are expressed as:
	\begin{equation}
			\Phi=	\Theta(\mu-\mu_0)		\Phi_{\text{i}}(\mathbf r, t)	+	
			\Theta(\mu_0-\mu)		\Phi_{\text{o}}(\mathbf r, t),
	\end{equation}
or explicitly \cite{PPRB}:
	\begin{multline}\label{Hpin-Hpout}
			\Phi(\mathbf r,t)=	
			\sum\limits_{m=0}^\infty		H_{m} (\varphi) 
			\int_0^{\infty}		A_{mq}(t) 		P_{mq}(\eta)
			\Big [ \Theta(\mu-\mu_0)P_{mq}(-\mu_0)   \\
			\times
			P_{mq}(\mu) + \Theta(\mu_0-\mu)
			\, P_{mq}(\mu_0)
			P_{mq}(-\mu)\Big ] \, dq, 
	\end{multline}
where $A_{mq}(t)$ are real time dependent amplitudes and $	H_m(\varphi) = (2-\delta_{0m}) \cos m\varphi,$
and, for convenience, we use $P_{mq}(\cdot)$ to denote $P^m_{-\frac12+iq}(\cdot)$. Fig.~\ref{5-pot-h} shows the spatial distribution of the lowest lying eigenmodes of the quasi-static electric potential for the hyperboloid. The values of conical functions $P^m_{-\frac12+iq}(\cdot)$ at any given point were calculated through the algorithms provided in \cite{Gil,kolbig}.
\clearpage
\clearpage 
\noindent Applying the Laplacian in Eq.~\eqref{AHLPhi} to the potential Eq.~\eqref{Hpin-Hpout} and using a similar argument given in Eqs.~\eqref{eta0}--\eqref{Lp1}, we can write:
	\begin{align}
			\mathcal D_{\mu}&=-2\mu\frac{\partial }{\partial\mu}+(1-\mu^2)~\frac{\partial^2}{\partial\mu^2},\\
			\mathcal D_{\eta}&=~2\eta\frac{\partial }{\partial\eta}+(\eta^2-1)~\frac{\partial^2}{\partial\eta^2},\\
			\mathcal D_{\varphi}&=\frac{\eta^2-\mu^2}{(\eta^2-1)(1-\mu^2)}~\frac{\partial^2}{\partial\varphi^2},
	\end{align}
hence
	\begin{equation}
			\nabla^2\Phi=\mathcal D \Phi=
			\bigg [ \mathcal D_{\mu}+\mathcal D_{\eta}+\mathcal D_{\varphi}\bigg ] \Phi,
	\end{equation}
we have
	\begin{align}
			\frac{\partial \Phi}{\partial\mu}&=\Theta(\mu-\mu_0)\frac{\partial \Phi_{\text{i}}}{\partial\mu}+\Theta(\mu_0-\mu)\frac{\partial \Phi_{\text{o}}}{\partial\mu},\\
			\frac{\partial \Phi}{\partial\eta}&=\Theta(\mu-\mu_0)\frac{\partial \Phi_{\text{i}}}{\partial\eta}+
			\Theta(\mu_0-\mu)\frac{\partial \Phi_{\text{o}}}{\partial\eta},\\
			\frac{\partial \Phi}{\partial\varphi}&=\Theta(\mu-\mu_0)\frac{\partial \Phi_{\text{i}}}{\partial\varphi}+
			\Theta(\mu_0-\mu)\frac{\partial \Phi_{\text{o}}}{\partial\varphi},
	\end{align}
by imposing the first boundary condition $\mu=\mu_0$. Second derivatives could be calculated  as:
	\begin{align}
			\frac{\partial^2 \Phi}{\partial\mu^2}&=\delta(\mu-\mu_0)\frac{\partial \Phi_{\text{in}}}{\partial\mu}+\Theta(\mu-\mu_0)\frac{\partial^2 \Phi_{\text{in}}}{\partial\mu^2}-
			\delta(\mu_0-\mu)\frac{\partial \Phi_{\text{out}}}{\partial\mu}+\Theta(\mu_0-\mu)\frac{\partial^2\Phi_{\text{out}}}{\partial\mu^2}\\
			\frac{\partial^2 \Phi}{\partial\eta^2}&=\Theta(\mu-\mu_0)\frac{\partial^2 \Phi_{\text{in}}}{\partial\eta^2}+
			\Theta(\mu_0-\mu)\frac{\partial^2 \Phi_{\text{out}}}{\partial\eta^2},\\
			\frac{\partial ^2\Phi}{\partial\varphi^2}&=\Theta(\mu-\mu_0)\frac{\partial ^2\Phi_{\text{in}}}{\partial\varphi^2}+
			\Theta(\mu_0-\mu)\frac{\partial^2 \Phi_{\text{out}}}{\partial\varphi^2},
	\end{align}
hence the Laplace equation becomes
	\begin{eqnarray}
			\nabla^2\Phi&=&  (1-\mu^2) \delta(\mu-\mu_0)\left[  \frac{\partial \Phi_{\text{i}}}{\partial\mu}-\frac{\partial \Phi_{\text{o}}}{\partial\mu}\right]
			+\Theta(\mu-\mu_0)\mathcal D_{\mu}\Phi_{\text{i}}+\Theta(\mu_0-\mu)\mathcal D_{\mu}\Phi_{\text{o}} \notag \\
			&&+\Theta(\mu-\mu_0)\mathcal D_{\eta}\Phi_{\text{i}}+\Theta(\mu_0-\mu)\mathcal D_{\eta}\Phi_{\text{o}}
			+\Theta(\mu-\mu_0)\mathcal D_{\varphi}\Phi_{\text{i}}+\Theta(\mu_0-\mu)\mathcal D_{\varphi}\Phi_{\text{o}}	\notag \\
			&=&  (1-\mu^2) \delta(\mu-\mu_0)\left[  \frac{\partial \Phi_{\text{i}}}{\partial\mu}-\frac{\partial \Phi_{\text{o}}}{\partial\mu}\right]
			+\Theta(\mu-\mu_0)\mathcal D\Phi_{\text{i}}+\Theta(\mu_0-\mu)\mathcal D\Phi_{\text{o}}\notag \\
			&=&  (1-\mu^2) \delta(\mu-\mu_0)\left[  \frac{\partial \Phi_{\text{i}}}{\partial\mu}-\frac{\partial \Phi_{\text{o}}}{\partial\mu}\right]
			+\Theta(\mu-\mu_0)  \nabla^2\Phi_{\text{i}}+\Theta(\mu_0-\mu)  \nabla^2\Phi_{\text{o}},
	\end{eqnarray}
since Laplacian vanishes everywhere but on the surface of the hyperboloid, one finds
	\begin{equation}\label{AH_L3}
			\vec\nabla^2\Phi=
			- \frac{ 1-\mu^2 }{z_0^2(\eta^2-\mu^2)}
			\left(  \frac{\partial \Phi_{\text{o}}}{\partial\mu}	- 
			\frac{\partial \Phi_{\text{i}}}{\partial\mu}\right)
			\delta(\mu-\mu_0).
	\end{equation}
Using the expressions for  $\Phi_{\text{i}}$ and $\Phi_{\text{o}}$ in Eq.~\eqref{Hpin-Hpout}, we get
	\begin{equation}\label{AH_o-i}
			\eval{\left( \frac{\partial \Phi_{\text{o}}}{\partial\mu}-
				\frac{\partial \Phi_{\text{i}}}{\partial\mu}\right) }_{\mu=\mu_0}=\sum_{m}		H_m(\varphi) 
			\int_0^{\infty}	A_{mq}(t)\,		P_{mq}(\eta)  \mathcal W_{mq}(\mu_0) \, dq,
	\end{equation}
where the Wronskian for $\mathcal W_{mq}(\mu)$ for the conical functions is given by:
	\begin{equation}\label{AHW1}
			\mathcal W_{mq}(\mu)	=	P_{mq}(\mu)	\frac{d P_{mq}(-\mu)}{d\mu}-
			P_{mq}(-\mu)		\frac{d P_{mq}(\mu)}{d\mu}.
	\end{equation}
The exact value of Eq.~\eqref{AHW1} is given by \cite{PK}:
	\begin{equation}\label{AHW}
			\mathcal W_{mq}(\mu)=
			\frac{2		Z^m_q		\cosh(\pi q)	}	{	\pi		\sqrt{1-\mu^2}}	,
	\end{equation}
where $Z_q^m$ is given by Eq.~\eqref{AHZ}.\\
It follows from the orthogonality relation for the system $\{H_m(\varphi)\}$, given in Eq.~\eqref{AHort}, that for each fixed $m=0, 1, 2, \dots,$ we have:
	\begin{equation}\label{Hpq}
		\int_0^{\infty} 
		P_{mq}(\eta)	\bigg [ 			 \mathcal W_{mq}(\mu_0) \, 
		\ddot A_{mq}(t)  
		+ \omega_{p}^2		 		P_{mq}(-\mu_0)	
		P'_{mq}(\mu_0)		\, A_{mq}(t)\bigg ]  \, dq=0.
	\end{equation}
Applying the Van-Nostrand orthogonality relation for the conical functions given in  Eq.~\eqref{AHor} and the exact expression for the Wronskian given in Eq.~\eqref{AHW},
it follows from Eq.~\eqref{Hpq} that for each fixed $m$ and $q\geq 1$ the amplitudes $A_{mq}(t)$ satisfy the harmonic oscillator model :
	\begin{equation}
			\ddot A_{mq}(t) +	\omega^2_{mq}		A_{mq}(t)=0, 
	\end{equation}	
where the frequencies $\omega^2_{mq}$ are given by:
	\begin{equation}\label{Hwmq}
			\omega^2_{mq}=		\frac{\omega_{p}^2 \, \pi \sqrt{1-\mu_0^2}}{2Z^m_q \cosh(\pi q)}
			P_{mq}(-\mu_0) P'_{mq}(\mu_0).
	\end{equation}
Again, these resonant values  can also 
be independently calculated from the transcendental equation  generated by satisfying the boundary conditions. Dispersion relations $\varepsilon_{mq}$, using Drude model, are shown in Fig.~\ref{5-eps-h}. \\
Following Eqs.~\eqref{eta0}--\eqref{Lp1}, the surface charge density $\sigma$ on $\mu=\mu_0$ is found to be:
\begin{equation}\label{H_sig1}
\sigma=	 \frac{-1}{4\pi  z_0} 	
\sqrt{\frac{1-\mu^2}{\eta^2-\mu_0^2}}
\sum_{m}		H_m(\varphi)	 
\int_0^{\infty}			A_{mq}(t) 	
P_{mq}(\eta)  \mathcal W_{mq}(\mu_0) dq. 
\end{equation}
\clearpage
\clearpage

\subsection{Classical Energy \& Decay Rate}

To calculate the classical energy $E,$ we follow the exact same argument as in the case of a paraboloid outlined in previous section. We hence find potential and kinetic energy in hyperboloid as:
	\begin{equation}\label{AHV}
			V=\frac {z_0\, \omega_{p}^2}{8} (1-\mu_0^2)				\sum\limits_{m}		 \hat\delta_{0m}
			 \int_0^{\infty}			
			\frac{Z^m_q\, \big [ A_{mq}(t) \big ] ^2}	{q\tanh(\pi q)\, \omega^2_{mq}}  
			\big [	 P_{mq}(-\mu_0)		\big ] ^2   P_{mq}(\mu_0)			P'_{mq}(\mu) \, 		dq, 
	\end{equation}
and 
	\begin{equation}\label{AHT}
			T=\frac {z_0\,  \omega_{p}^2}{8} \, (1-\mu_0^2)
			\sum\limits_{m}		 \hat\delta_{0m}\,
			 \int_0^{\infty}			\frac{Z^m_q\, 	\big [ \dot A_{mq}(t) \big ] ^2 	}{q\tanh(\pi q)\, \omega_{mq}^4}
			\big [ 	P_{mq}(-\mu_0)\big ]^2 	
			P_{mq}(\mu_0)		\frac{\partial P_{mq}(\mu)}{\partial \mu} \, dq
	\end{equation}
Consequently, using  the orthogonality relations in the hyperboloidal case given by Eqs.~\eqref{AHort} and \eqref{AHor}, one finds the hyperboloidal energy:
	\begin{multline}\label{5-HE}
			E=\frac {z_0}{4\pi}	\sqrt{1-\mu_0^2} \sum\limits_{m}   
			\hat\delta_{0m}\\
			\times \int_0^{\infty}	
			\frac{\big [		Z^m_q 	\cosh(\pi q)	\big ]^2}{	q	\sinh(\pi q) \ \omega_{mq}^2} 
			P_{mq}(-\mu_0) 	\,
			P_{mq}(\mu_0)
			\,	\Big \{\big [ \dot A_{mq}(t)\big ] ^2	+
			\omega_{mq}^2	\big [ A_{mq}(t)\big ] ^2	\Big \} dq.
	\end{multline}
Using similar ansatz for $A_{mq}(t)$ and $\dot A_{mq}(t)$ as in Eqs.~\eqref{5-A_m} and \eqref{5-Adot},  the coefficients $\alpha^2_{mq}$  can be obtained by a comparison of  Eq.~\eqref{5-HE} with the Hamiltonian in Eq.~\eqref{5-E2} as:
	\begin{equation}\label{Halpha}
			\alpha^2_{mq}=  
			\frac{4\pi 	\hbar\, \hat\delta_{0m}^{-1}	 	}	{z_0\sqrt{1-\mu^2_0}}~
			\frac{q\, \omega^3_{mq}\, 		\sinh(\pi q) 				}{	\big [	Z^m_q	\cosh(\pi q)	\big ]^2	 P_{mq}(-\mu_0)\, P_{mq}(\mu_0)} ,		
	\end{equation}
where $\hat\delta_{0m}$ is given by Eq.~\eqref{delta2} and thus:
	\begin{multline}\label{5-HE1}
			E=\frac {z_0}{4\pi}	\sqrt{1-\mu_0^2} \sum\limits_{m}      \hat\delta_{0m}\\
				\times\int_0^{\infty}	
			\frac{\big [		Z^m_q 	\cosh(\pi q)	\big ]^2}{	q	\sinh(\pi q)	} 
			P_{mq}(-\mu_0) 	
		\, P_{mq}(\mu_0)
			\frac{\alpha^2_{mq}}{2\omega_{mq}^2}
			\big (	a^*_{mq}a_{mq}+a_{mq}a^*_{mq} \big )dq,
	\end{multline}

Utilizing Eq.~\eqref{5-A3} and the hyperboloidal analogue of Eq.~\eqref{5-H2}, setting up the plasmon current density, 
the interaction Hamiltonian for the photon emission can be expressed as:
\begin{multline}\label{5-HH}
\mathcal{H}_{i}  =  \frac{n_0e^2}{ 2i\, m_e}	\sum\limits_\mathbf{s}
\sum\limits_{j=1,2}
\sqrt{\frac{\hbar  }{	\mathcal V	\omega_s}}
(\mathbf{\hat{e}}_\mu			\cdot		\mathbf{\hat{e}}_j)
\big [ 	\hat c_{\mathbf{s}j}		e^{i\mathbf{s\cdot r}}		+		\hat c^{\dagger}_{\mathbf{s}j}		e^{-i\mathbf{s\cdot r}}\big ] \\
\times 		\sum\limits_{m}
\int_{0}^{2\pi}
\int_{1}^{L} \Bigg [
\int_0^{\infty}
H_m(\varphi)
P_{mq}(\eta)	P_{mq}(-\mu_0)		P_{mq}(\mu) \\
\times
\frac{	\alpha_{mq}	}{	 \omega^2_{mq}	}(	\hat a^{\dagger}_{mq}	-	\hat a_{mq})
dq \Bigg ]  h_\varphi h_\eta \, d\varphi		 d\eta . 
\end{multline}
Thus, we may calculate the hyperboloidal emission matrix element as:
\begin{equation}\label{5-HM}
\mathcal{M}^{(j\mathbf{s})}_{mq} =
\frac{n_0e^2}{2i\, m_e}
\sqrt{\frac{\hbar}{\mathcal V\omega_s}}
\frac{\alpha_{mq}}{\omega^2_{mq}}
P_{mq}(-\mu_0) 	P_{mq}(\mu_0)
\mathcal I^{(j)}_{mq},
\end{equation}
where for different polarization directions, we obtain:
\begin{equation}\label{5-HI1}
\mathcal I^{(1)}_{mq}=- z_0		\mu_0	\sqrt{1-\mu_0^2}	
\int_{0}^{2\pi}		
\int_{1}^{L}
\sqrt{\eta^2-1}		
P_{mq}(\eta)  
\sin\varphi		 
H_m(\varphi) E_{mq}(\eta, \varphi)
d\eta d\varphi,
\end{equation}
and 
\begin{multline}\label{5-HI2}
\mathcal I^{(2)}_{mq}= z_0 		 \sqrt{1-\mu_0^2}	
\int_{0}^{2\pi}		
\int_{1}^{L}
P_{mq}(\eta) 				 						
\bigg [		\mu_0		\sqrt{\eta^2-1}
\sin\psi		\cos\varphi		\\+	 	\sqrt{1-\mu_0^2}	 
\cos\psi		\eta
\bigg ]
H_m(\varphi) E_{mq}(\eta, \varphi) 	
d\eta d\varphi,
\end{multline}
where we have set $E_{mq}(\eta, \varphi)=	e^{	-i\mathcal B}$, with:
\begin{eqnarray}
\mathcal B=\frac{z_0\omega_s	}{c}	
\bigg [	 \sqrt{(\eta^2-1)	(1-\mu_0^2)}
\cos\psi		\cos\varphi			
+		\mu_0		\sin\psi		\eta 
\bigg ] . \notag
\end{eqnarray}
The wavevector $\mathbf{s}$ and the polarization vectors $\hat{\mathbf{e}}_1,~\hat{\mathbf{e}}_2$ remain  the same as before.
Moreover, similar to the case for paraboloid, the choice of the polarization vector  $\hat{\mathbf{e}}_{j}$ in each integral determines the polarization state. Hence, $\mathcal I^{(1)}_{mq}$ represents the s-polarization and $\mathcal I^{(2)}_{mq}$ corresponds to the p-polarization.
Using Eq.~\eqref{Halpha} in Eq.~\eqref{5-R}, and now the position vector in hyperboloidal coordinates given by Eq.~\eqref{5-AHC1}, 
we arrive at:
\begin{equation}\label{5-HR1}
\frac{	d\gamma_{mq}	}{	d \Omega}	=	
\frac{q	\sinh(\pi q) \hat\delta_{0m}^{-1}  }{4\pi z_0 c^3 
	\sqrt{1-\mu_0^2}}	
\left( 	\frac{n_0e^2}{m_e}	\right) ^2 
\frac{	P_{mq}(-\mu_0)		P_{mq}(\mu_0) \,  }{	\left[ 	Z^m_q	\cosh(\pi q)	\right] ^2} 
\,   \left[		\left(  \mathcal I^{(1)}_{mq}	\right)^2	+	\left(  \mathcal I^{(2)}_{mq}\right)^2	\right].
\end{equation}

\section{Prolate Spheroidal Surfaces}
In order to provide a geometric basis for comparison, we now treat the case of plasmon excitation on the surface of a prolate spheroidal domain. This structure presents an almost identical curvature to that of the paraboloid but encompasses a finite domain, making the interpretation of the surface modes and their associated radiation patterns more tangible. The closely related structure of an oblate spheroid has been employed in previous plasmon studies~\cite{Little_Paper}. In both prolate and oblate systems, a limiting case is that of a sphere~\cite{B,Crowell}, which can serve to validate the results. 
To obtain the quantized surface modes of the prolate spheroid, we closely follow the oblate case~\cite{Little_Paper}. 
A prolate spheroid is defined by fixing the coordinate $\eta=\eta_0$ in coordinate system $(\eta, \mu, \varphi)$ given in Eq.~\eqref{5-AHC1}. 
To avoid confusion, we use $\zeta$ in the case for prolate spheroid and recall the coordinate system $(\zeta, \mu, \varphi)$  as:
\begin{align}
\begin{split}
x(\zeta,\mu,\varphi)&=z_0\sqrt{({\zeta}^2-1)(1-\mu^2)}\cos{\varphi},\\
y(\zeta,\mu,\varphi)&=z_0\sqrt{({\zeta}^2-1)(1-\mu^2)}\sin{\varphi},\\
z(\zeta,\mu,\varphi)&=z_0{\zeta}{\mu},
\end{split}
\end{align}
in the domain defined by :
\begin{equation}
1\le\zeta<\infty,\qquad -1\le\mu\le 1, \qquad 0<\varphi\le 2\pi,
\end{equation}
Scale factors in this coordinate system are:
\begin{align}
\begin{split}
h_{\zeta}=z_0&\sqrt{\frac{\zeta^2-\mu^2}{\zeta^2-1}}		,\quad
h_{\mu}=z_0\sqrt{\frac{\zeta^2-\mu^2}{1-\mu^2}},\\
h_{\varphi}&=z_0\sqrt{(\zeta^2-1)(1-\mu^2)}.
\end{split}
\end{align}
The radius vector is 
\begin{equation}\label{PS-r}
\mathbf{\vec r}=z_0
\sqrt{({\zeta}^2-1)(1-\mu^2)}
(\cos{\varphi}\hat i+
\sin{\varphi}\hat j)+
z_0{\zeta}{\mu}\hat k.
\end{equation}
The unit vectors for the prolate spheroidal coordinates are
\begin{align}
\begin{split}
\hat e_{\zeta}&=\frac{1}{h_{\zeta}}
\frac{\partial \mathbf{\vec r}}{\partial \zeta} 	
=\sqrt{\frac{\zeta^2-1}{\zeta^2-\mu^2}}
\left[	\frac{\zeta\sqrt{1-\mu^2}}{\zeta^2-1}	
\left(\cos{\varphi}\hat i
+\sin{\varphi}\hat j \right) +\mu\hat k 	\right] ,\\
\hat e_{\mu}&=\frac{1}{h_{\mu}}
\frac{\partial \mathbf{\vec r}}{\partial \mu }=
\sqrt{\frac{1- \mu^2}{\zeta^2-\mu^2}}
\left[\frac{-\mu\sqrt{\zeta^2-1}}{1-\mu^2}
\left(\cos{\varphi}\hat i+
\sin{\varphi}\hat j \right) +
\zeta\hat k \right],\\
\hat e_{\varphi}&=\frac{1}{h_{\varphi}}
\frac{\partial \mathbf{\vec r}}{\partial \varphi }
=(-\sin{\varphi}\hat i
+\cos{\varphi}\hat j).
\end{split}
\end{align}
\subsection{Nonretarded Potential \& Dispersion Relations}
The quasi-static scalar potential for a spheroid defined by $\zeta=\zeta_0$ may then be written as \cite{speroidal:wf,z&k:spheric}:
\begin{equation}\label{4-Sp_P}
\Phi=	\Theta(\zeta_0-\zeta)		\Phi_{\text{i}}(\mathbf r, t)	+	
\Theta(\zeta-\zeta_0)		\Phi_{\text{o}}(\mathbf r, t),
\end{equation}
or explicitly:
\begin{multline}\label{PS-pin-pout}
\Phi({\bf r},t)=\sum\limits_{m,l,p} 
A_{mlp}(t)
Y^p_{lm}(\mu, \varphi)  
\bigg [ \Theta(\zeta_0-\zeta)P^m_{l}(\zeta)   
Q^m_{l}(\zeta_0)+
\Theta(\zeta-\zeta_0)P^m_{l}(\zeta_0) 	Q^m_{l}(\zeta)\bigg ],		
\end{multline}
for some real coefficients $A_{mlp}(t)$, with $l=1,2,3,\cdots $, $m=0, \cdots,l$, $p=0,1$
and 
$P^m_l(\cdot)$ and $Q^m_l(\cdot)$ being the associated Legendre polynomial of first and second kind, respectively, while $Y^p_{lm}(\mu, \varphi)$ are the real spherical harmonics given by \cite{g&r:table}: 
\begin{equation}\label{APS_Sh}
Y^p_{lm}(\mu, \varphi)= 
\sqrt{\frac{(2-\delta_{0m})	(2l+1)(l-m)!}{4\pi(l+m)!}} \,	 P^m_l(\mu)  
\left[ 	\delta^1_{p}\cos(m\varphi)+ 
\delta^{-1}_{p}\cos(m\varphi)\right],
\end{equation}
with associated Legendre functions $P^m_l(\cdot) $ and orthogonality relation \cite{Abra}:
\begin{equation}\label{APs_Sh1}
\int_0^{2\pi} 
\int_{-1}^{1}
Y^p_{lm}(\mu, \varphi)
Y^{p'}_{l'm'}(\mu, \varphi)
\, d\mu d\varphi= \delta_{ll'}\delta_{mm'}\delta_{pp'}, 
\end{equation}
and $\delta$ is the Kroneker delta symbol. Note that the first argument of the real spherical harmonic, here $\mu$, is also taken to be $\mu=\cos\theta$ for some $-\pi\le \theta\le \pi$ is certain references. Since we developed the subsequent in terms of $\mu$, we avoid using conventional angle $\theta$. 

\begin{figure}
	\begin{center}
		\includegraphics[width=5.5in]{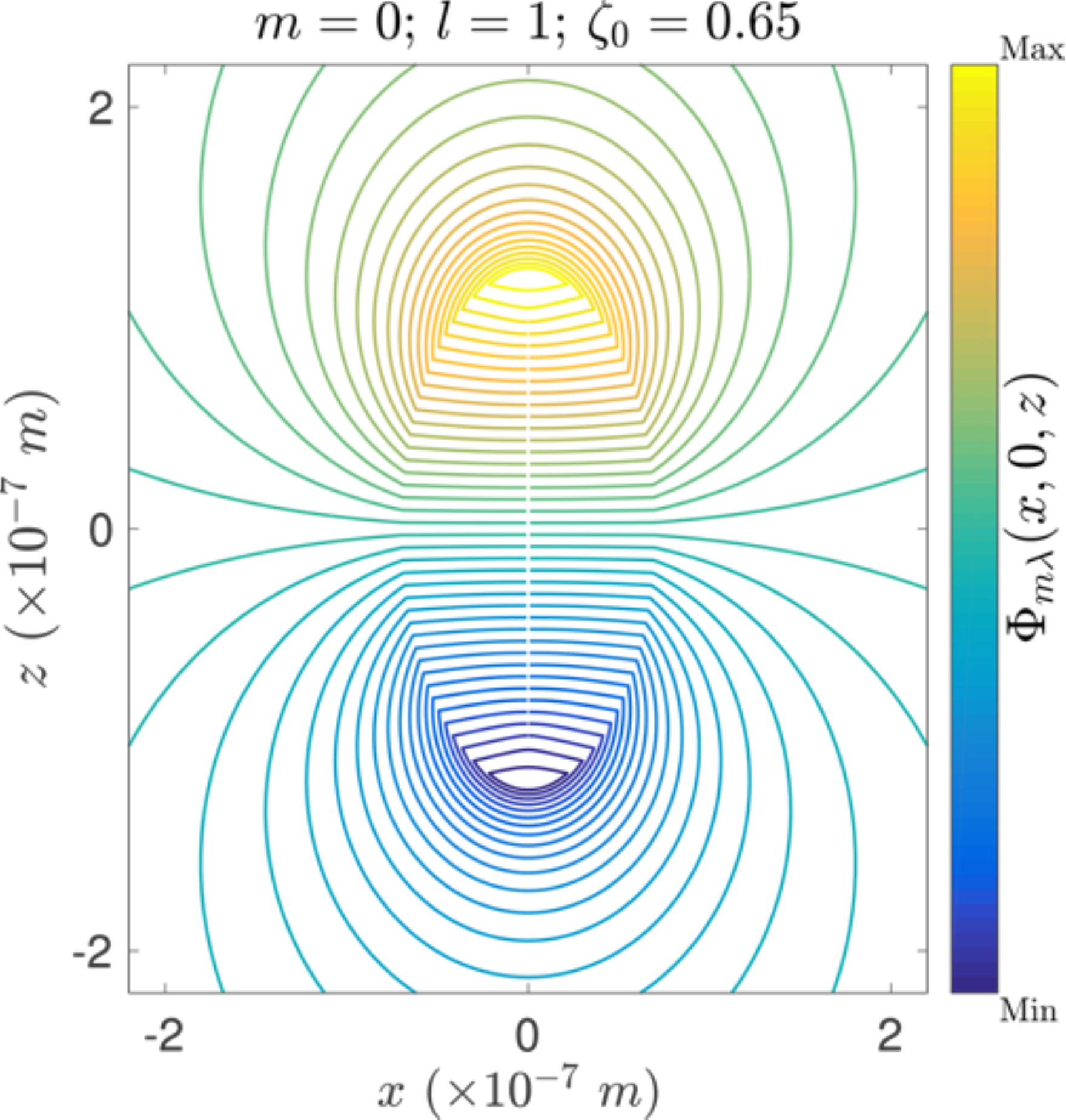}\\
		\caption[Figure shows the spatial distribution of the lowest ]{Figure shows the spatial distribution of the lowest lying eigenmodes of the quasi-static electric potential for spheroidal modeling domain investigated. For the same mode index $m$ as well as the discrete eigenvalue  $l$ of the prolate spheroid. The magnitude of the potential could be read by the color bar. Yellow color shows the highest intensity of the potential distribution while the dark blue shows the lowest. }
		\label{5-pot-sp}
	\end{center}
\end{figure}

\begin{figure}
	\begin{center}
		\includegraphics[width=6in]{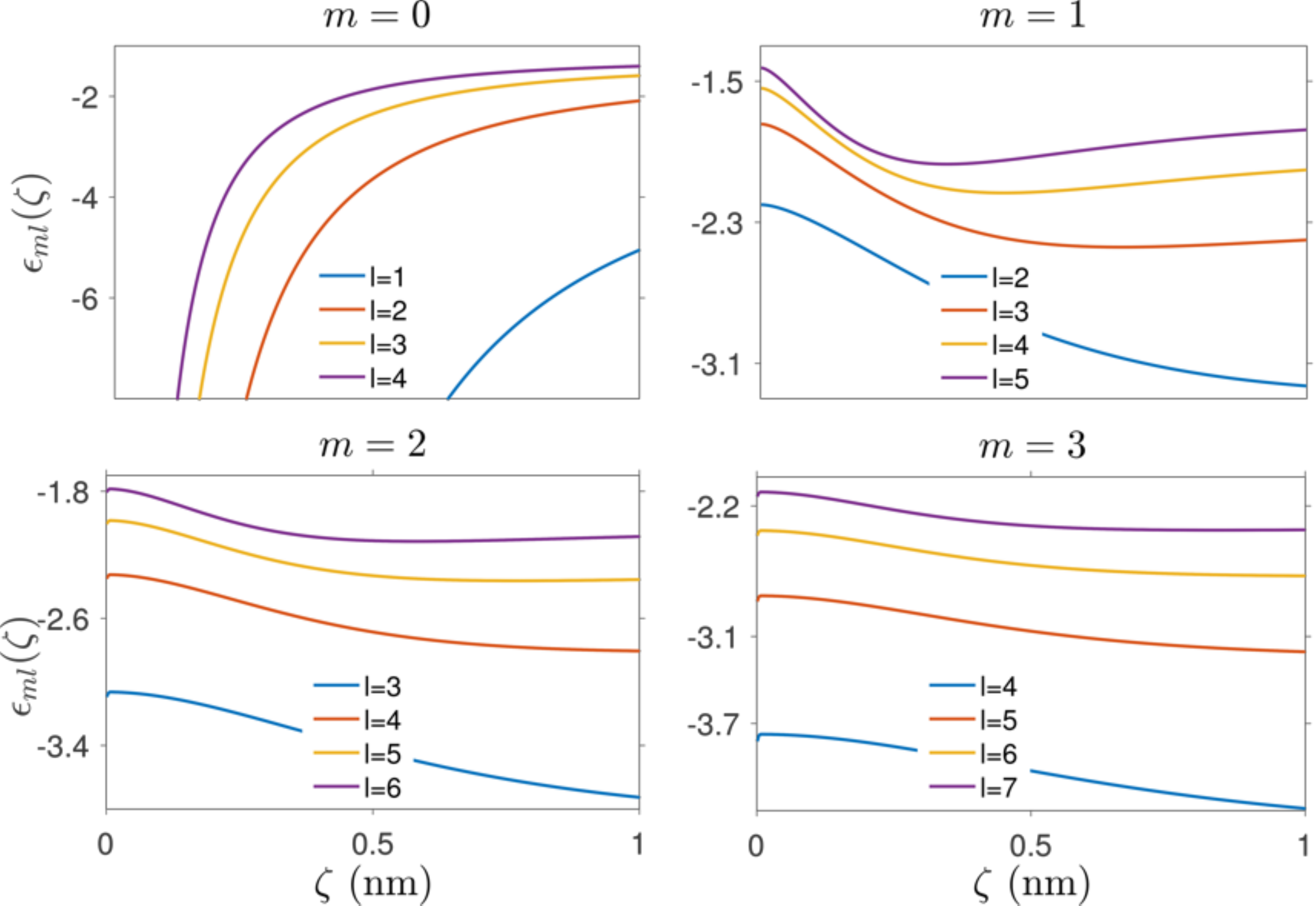}\\
		\caption[Prolate spheroidal nonretarded surface plasmon  dispersion relations]{
			Prolate spheroidal nonretarded surface plasmon  dispersion relations. The resonance values of the dielectric function $\varepsilon_{lm}$, given in Eq.~\eqref{Ps-eps}, are shown for low lying modes as a function of  the shape parameter $\zeta$ of a spheroid. Due to the relation between two discrete modes $l$ and $m$ for which $m< l$, second modes had to be specified as well. 
		}
		\label{5-eps-sp}
	\end{center}
\end{figure}

In Fig.~\ref{5-pot-sp}, the potential distribution for the fixed shape parameter $\zeta_0= 65~$(nm) and fixed lowest modes $(l,m)=(1,0)$ is shown. The magnitude of the potential could be read by the color bar. Yellow color shows the highest intensity of the potential distribution while the dark blue shows the lowest. As expected, the lowest mode shows the mono-pole approximation for the field distribution.  
\clearpage 
\clearpage 
With Heaviside function $\Theta$, the inside and outside potential could be written as given in Eq.~\eqref{4-Sp_P} as :
\begin{eqnarray}
\Phi=	\Theta(\zeta_0-\zeta)		\Phi_{\text{i}}(\mathbf r, t)	+	
\Theta(\zeta-\zeta_0)		\Phi_{\text{o}}(\mathbf r, t).
\end{eqnarray}
Laplace equation in this coordinate system is given by:
\begin{multline}\label{PS-L}
\vec \nabla^2\Phi=\frac{1}{z_0^2(\zeta^2-\mu^2)}
\bigg\{ \frac{\partial}{\partial\zeta}
\left[ (\zeta^2-1) 
\frac{\partial}{\partial\zeta}\right] + 
\frac{\partial}{\partial\mu}
\left[ (1-\mu^2) 
\frac{\partial}{\partial\mu}	\right] \\+
\left[ \frac{\zeta^2-\mu^2}{(\zeta^2-1)(1-\mu^2)}		\right] 
\frac{\partial^2}{\partial\varphi^2}
\bigg\} \Phi.
\end{multline}
We define the derivative operators $\mathcal D$ in each component $\zeta$, $\mu$ and $\varphi$ as:
\begin{align}
\mathcal D_{\zeta}&=~2\zeta\, \frac{\partial }{\partial\zeta}+(\zeta^2-1)~\frac{\partial^2}{\partial\zeta^2},\\
\mathcal D_{\mu}&=-2\mu\frac{\partial }{\partial\mu}+(1-\mu^2)~\frac{\partial^2}{\partial\mu^2},\\
\mathcal D_{\varphi}&=\frac{\zeta^2-\mu^2}{(\zeta^2-1)(1-\mu^2)}~\frac{\partial^2}{\partial\varphi^2},
\end{align}
we can put:
\begin{equation}
\vec \nabla^2\Phi=\mathcal D \Phi=\bigg [ \mathcal D_{\zeta}+\mathcal D_{\mu}+\mathcal D_{\varphi}\bigg ] \Phi,
\end{equation}
we write:
\begin{eqnarray}\label{PS_L2}
\vec \nabla^2\Phi=  \frac{\zeta^2-1}{z_0^2(\zeta^2-\mu^2)} 
\delta(\zeta-\zeta_0)
\left(  \frac{\partial \Phi_{\text{o}}}{\partial\zeta}-
\frac{\partial \Phi_{\text{i}}}{\partial\zeta}\right)+
\Theta(\zeta_0-\zeta) 
\nabla^2\Phi_{\text{i}}+
\Theta(\zeta-\zeta_0)
\nabla^2\Phi_{\text{o}}.
\end{eqnarray}
Since the Laplace equation vanishes inside and outside but on the surface, then we have 
\begin{eqnarray}\label{PS_L3}
\vec\nabla^2\Phi= 
\frac{\zeta^2-1}{z_0^2(\zeta^2-\mu^2)}
\delta(\zeta-\zeta_0)
\left(  \frac{\partial \Phi_{\text{o}}}{\partial\zeta}
-\frac{\partial \Phi_{\text{i}}}{\partial\zeta}
\right).
\end{eqnarray}
Using Eq.~\eqref{PS-pin-pout} along with Maxwell's equations give us the charge density as  $\rho=-\frac{\nabla^2\Phi}{4\pi}$, using the Laplacian we just obtained, we have 
\begin{eqnarray}\label{PS-rho}
\rho=- \frac{\zeta^2-1}{4\pi \, z_0^2(\zeta^2-\mu^2)} 
\delta(\zeta-\zeta_0)
\sum\limits_{l,m,p} 
A_{lmp}(t)
Y^p_{lm}(\mu, \varphi)
\mathcal W^m_l(\zeta),
\end{eqnarray}
where we have used the Wronskian identity as 
\begin{eqnarray}\label{PS-W}
\mathcal W^m_l(\zeta)=P^m_{l}(\zeta_0)      Q'^m_{l}(\zeta)-
Q^m_{l}(\zeta_0)	P'^m_{l}(\zeta)=
\frac{(-1)^{m}(l+m)!}{(\zeta^2-1)(l-m)!}.
\end{eqnarray}	
\noindent Using Eq.~\eqref{0-25} one can find 
\begin{eqnarray}\label{PS-s1}
\sigma= \frac{1}{4\pi} 
\sum\limits_{l,m,p} 
e_\zeta\cdot\left( \nabla\Phi_{\text{i}}-\nabla\Phi_{\text{o}}\right),
\end{eqnarray}
where $\sigma$ is the surface charge. We have 
\begin{eqnarray}\label{PS-s2}
\ddot \sigma=-\frac{1}{4\pi z_0}
\sqrt \frac{\zeta^2-1}{\zeta^2-\mu^2} 
\sum\limits_{l,m,p} 
\ddot{A}_{lmp}(t)
Y^p_{lm}(\mu, \varphi)
\frac{(-1)^m(l+m)!}{(\zeta^2-1)(l-m)!}.
\end{eqnarray}
\noindent On the other hand, Eq.~\eqref{0-sddot1} gives:
\begin{eqnarray}\label{PS-s3}
\ddot \sigma= -\frac{\omega_p^2}{4\pi z_0 }
\sqrt{\frac{\zeta^2-1}{\zeta^2-\mu^2}}
\sum\limits_{l,m,p} 
A_{lmp}(t)
Y^p_{lm}(\mu, \varphi)
P'^m_{l}(\zeta_0)  Q^m_l(\zeta_0).
\end{eqnarray}
Putting \eqref{PS-s2} and \eqref{PS-s3} equal, one finds frequency as 
\begin{eqnarray}\label{Ps-w1}
\omega^2_{lm}=
\frac{(-1)^{m+1} \, \omega^2_p(\zeta^2-1)(l-m)! }{(l+m)!}P'^m_{l}(\zeta_0) Q^m_{l}(\zeta_0).
\end{eqnarray}
The allowed values of the dielectric function $\varepsilon_{lm}$ are then found to be:
\begin{eqnarray}\label{Ps-eps}
\varepsilon_{lm}	=	1- \frac{(l+m)!}	{(\zeta_0^2-1 )(l-m)!}
\left[ 	\frac{(-1)^{m}}	{P'^m_l(\zeta_0) Q^m_l(\zeta_0)	}	\right],
\end{eqnarray}
following similar steps for Eqs.~\eqref{pin-pout}--\eqref{5-wml}. In Fig.~\ref{5-eps-sp}, the different plots of dielectric function $\varepsilon_{lm}$ as a function of shape parameter $\zeta$ for different $l$ and $m$, where $m< l$ are obtained. 
\clearpage
\clearpage 
\subsection{Classical Energy \& Radiative Decay Rate}
We proceed to find the classical energy $E=T+V. $
The potential energy, taking the same steps as we did for all other cases, and using the orthogonality relation for spherical harmonics given in Eq.~\eqref{APs_Sh1},  can be calculated as:
\begin{equation}\label{PS-v2}
V=\frac{z_0}{8\pi}
\sum\limits_{l,m,p} 
\left[ {A}_{lmp}(t)\right] ^2
\frac{(-1)^{m+1}\, (l+m)!}{(l-m)!}
P^m_{l}(\zeta_0)	Q^m_{l}(\zeta_0).
\end{equation}
Similarly, kinetic energy is given by:
\begin{equation}\label{PS-K1}
T=\frac{z_0}{8\pi}
\sum\limits_{l,m,p} 
\frac{1}{\omega^2_{lm}}\, \left[	\dot{A}_{lmp}(t)\right] ^2\, \frac{(-1)^{m+1}\, (l+m)!}{(l-m)!}
P^m_{l}(\zeta_0) Q^m_{l}(\zeta_0).
\end{equation}
The classical energy $E$ could be obtained by adding kinetic and potential energies, given in Eqs.~\eqref{PS-K1} and \eqref{PS-v2}, respectively, as:
\begin{equation}\label{PS-E}
E=\frac{z_0}{8\pi}
\sum\limits_{l,m,p}
\frac{(-1)^{m+1}(l+m)!}{\omega^2_{lm}(l-m)!}
P^m_{l}(\zeta_0) 
Q^m_{l}(\zeta_0) 
\left\{ \left[ \dot{A}_{lmp}(t)\right] ^2+\omega^2_{lm} \left[{A}_{lmp}(t)\right] ^2\right\}.
\end{equation}
Real coefficients $A_{lmp}(t)$ obey the harmonic oscillators equation for motion, we can choose the form:
\begin{equation}
{A}_{lmp}(t)= \frac{ \alpha_{lmp}}{2\omega_{lm}} ~(a_{lmp}+a^*_{lmp}),
\end{equation}
for some complex time dependent functions $a_{lmp}$ which are proportional to $e^{-i\omega_{lm} t}$ and for some constant $\alpha_{lmp}$ which will be chosen later.
\begin{equation}\label{PS-E1}
E=\frac{z_0}{8\pi}
\sum\limits_{l,m,p}
\frac{(-1)^{m+1}(l+m)!}{\omega^2_{lm}(l-m)!}
P^m_{l}(\zeta_0) Q^m_{l}(\zeta_0) 
\frac{\alpha^2_{lmp}}{2}
\left[a^*_{lmp}a_{lmp}+a_{lmp}a^*_{lmp} \right].
\end{equation}
This gives $\alpha_{lmp}$ as:
\begin{equation}\label{PS-alpha}
\alpha^2_{lmp}=
\frac{8\pi\, \hbar}
{z_0}~
\frac{(-1)^{m+1} \, (l-m)!\, \omega_{lm}^3}{(l+m)!\, P^m_{l}(\zeta_0) \,Q^m_l(\zeta_0)}.
\end{equation}
Considering now the interaction Hamiltonian, followed by Eq.~\eqref{0-IntH2}, as:
\begin{eqnarray}\label{PS-H4}
{H}_{\text{int}} =   -\frac{n_0e}{c}
\int_0^{2\pi}~\int_{-1}^{1}
\left(	\dot{\Psi}\mathbf{A}	\cdot	\mathbf{\hat e}_r		\right)
h_\mu  	h_\varphi \,	d\mu	d\varphi,
\end{eqnarray}
where
\begin{eqnarray}\label{PS-psi}
\dot \Psi = -\frac{e}{m_0}
\sum\limits_{l,m,p}
Y^p_{ml}(\mu, \varphi) P^m_l(\zeta) 	Q^m_l(\zeta_0)
\frac{\alpha_{ml}}{ 2i \omega^2_{lm} }
(a^*_{ml}-a_{ml}) , 
\end{eqnarray}
and $\mathbf A$ is given by Eq.~\eqref{0-A3}, we could find:
\begin{multline}\label{Sp-H1}
{H}_{\text{int}} =  \frac{n_0e^2}{ m}	\sum\limits_\mathbf{s}
\sum\limits_{q=1,2}
\sqrt{\frac{\hbar }{\mathcal V	\omega_s}}
\sum\limits_{m,l,p}
\frac{\alpha_{ml}}{ 2i \omega^2_{lm} }
(a^*_{mlp}-a_{mlp})
P^m_l(\zeta_0) 	Q^m_l(\zeta_0)	 \\
\times 		
\int_{0}^{2\pi}
\int_{-1}^{1}
(\mathbf{\hat{e}}_{q}			\cdot		\mathbf{\hat{e}}_\zeta)
Y^p_{ml}(\mu, \varphi) \,  
\big ( 	c_{\mathbf{s}k}		e^{i\mathbf{s\cdot r}}		+
c^*_{\mathbf{s}k}		e^{-i\mathbf{s\cdot r}}\big )
h_\mu h_\varphi\,	d\mu d\varphi.			
\end{multline}
Emission matrix element in prolate spheroidal case following the same relations as in the paraboloidal and hyperboloidal cases, initiates by considering:
\begin{equation}\label{APS-M}
\mathcal{M}^{(j\mathbf s)}_{em} = 
\frac{n_0e^2}{m_0}
\sqrt{\frac{\hbar}{\mathcal V\omega_s}}
\frac{\alpha_{lmp}}{2i\omega^2_{lm}}
P^m_l(\zeta_0) Q^m_l(\zeta_0)  
\,\int	\int(\mathbf{\hat{e}}_\zeta	\cdot	\mathbf{\hat{e}}_j)
Y^p_{lm}(\mu,\varphi)
e^{-i\mathbf{s\cdot r}}
h_\mu h_\varphi  \,
d\mu d\varphi \Bigg|_{\zeta=\zeta_0}.
\end{equation}
We set:
\begin{equation}
\mathcal{I}_{lm} = \int	\int(\mathbf{\hat{e}}_\zeta	\cdot	\mathbf{\hat{e}}_j)
Y^p_{lm}(\mu,\varphi)
e^{-i\mathbf{s\cdot r}}
h_\mu h_\varphi  \,
d\mu d\varphi \Bigg|_{\zeta=\zeta_0}.
\end{equation}	
Assuming:
\begin{equation}
\mathbf{s}=(s_x\hat i, ~s_y\hat j, ~s_z\hat k)
\quad \quad \text{and}\quad\quad 	
\mathbf {\hat e_{j}}=(e_x\hat i,~ e_y\hat j, ~e_z\hat k),
\end{equation}
then
\begin{eqnarray}
(\mathbf{\hat{e}}_\zeta \cdot\mathbf{\hat{e}}_j)h_\mu h_\varphi   \Big|_{\zeta=\zeta_0} 
=z_0^2\left[-\zeta_0\sqrt{\mu^2-1}\,\cos\varphi\, e_x-\zeta_0\sqrt{\mu^2-1}\,\sin\varphi \,e_y+(1-\zeta_0^2)\, \mu\,e_z\right] ,
\end{eqnarray}	
and
\begin{eqnarray}
\mathbf{s}\cdot\mathbf r \Big|_{\zeta=\zeta_0}=
z_0\sqrt{(\mu^2-1)(1-\zeta_0^2)}\, \left(s_x\cos\varphi
+s_y\sin\varphi \right) +z_0\,s_z\mu\, \zeta_0. 
\end{eqnarray}	
We use the change of variable $
\mu=\cosh\xi, 
$
to write:
\begin{align}
\mathbf{s}\cdot\mathbf r &\Big|_{\zeta=\zeta_0}
=z_0\sqrt{1-\zeta_0^2}\left(s_x\sinh\xi \cos\varphi+s_y\sinh\xi\sin\varphi \right) 
+z_0\,s_z\,\zeta_0\cosh\xi, \\
(\mathbf{\hat{e}}_\zeta \cdot\mathbf{\hat{e}}_j)h_\varphi  h_\xi &\Big|_{\zeta=\zeta_0}
=z_0^2\left[-\zeta_0\sinh\xi e_x\cos\varphi-\zeta_0\sinh\xi e_y\sin\varphi 
+(1-\zeta_0^2)e_z\cosh\xi\right] .
\end{align}
We consider the following coordinate transformation as:
\begin{eqnarray}
&&s_x \quad\rightarrow \quad \frac{\zeta_0}{\sqrt{1-\zeta_0^2}}s'_x, \\
&&s_y \quad\rightarrow\quad \frac{\zeta_0}{\sqrt{1-\zeta_0^2}}s'_y, \\
&&s_z \quad\rightarrow\quad \frac{-(1-\zeta_0^2)}{\zeta_0}s'_z, 
\end{eqnarray}
using the fact that $\mu_0$ is fixed. We define the gradient operator $\vec \nabla'$ as
\begin{equation}
\vec \nabla^{'}=\frac{\partial}{\partial s'_x}
+\frac{\partial}{\partial s'_y}
+\frac{\partial}{\partial s'_z}.
\end{equation}
Then comparing with
\begin{align}
\mathbf{s}\cdot\mathbf r &\bigg|_{\zeta=\zeta_0}=z_0\left(s'_x\, \zeta_0\sinh\xi \cos\varphi+s_y\, \zeta_0\sinh\xi\sin\varphi \right) -z_0\, s_z\, (1-\zeta_0^2)\cosh\xi,\\
(\mathbf{\hat{e}}_\mu\cdot\mathbf{\hat{e}}_j)\, h_\varphi  h_\xi &\bigg|_{\zeta=\zeta_0}=z_0^2\, \left[-\zeta_0\,\sinh\xi\,  e_x\cos\varphi-\zeta_0\,\sinh\xi e_y\, \sin\varphi +(1-\zeta_0^2)\, e_z\,\cosh\xi\right] ,
\end{align}
we could write
\begin{equation}
(\mathbf{\hat{e}}_\zeta \cdot\mathbf{\hat{e}}_j)h_\varphi  h_\xi=
i\,z_0\, \mathbf{\hat{e}}_j\,\cdot\vec \nabla^{'}e^{-i\mathbf{s}\cdot\mathbf r}.
\end{equation}
We use the following integral identity  in the calculation of prolate spheroidal matrix element given by Eq.~(16.127)\cite{jackson}:

\begin{equation}\label{IN}
e^{i\mathbf{k}\cdot\mathbf x}=
4\pi\sum\limits_{l=0}^{\infty}i^l~j_l(\mathbf{k}r)
\sum\limits_{m=-l}^{l}
Y^*_{lm}(\theta, \varphi)
Y_{lm}(\theta', \varphi'),
\end{equation}
where $j_l(\mathbf{k}r)$ is the Spherical Bessel function of order $l$ whose relation with regular Bessel function $J_{l}( \cdot )$ is given by:
\begin{equation}\label{ASB}
j_{l}( x)=\sqrt{\frac{\pi}{2x}}\, J_{l+\frac12}(x ).
\end{equation}
We then obtain the matrix elements for photon emission:
\begin{equation}\label{PS-M1}
\mathcal M^{(j\mathbf s)}_{lm} = 
\frac{	z_0 \, \omega_{p}^2 \, (-i) ^l\, \alpha_{lmp}}{2\omega^2_{lm}}	
\sqrt{\frac{\hbar}{\mathcal V\omega_s}}
P^m_l(\zeta_0) \, 	Q^m_l(\zeta_0) 
\mathbf{\hat{e}}_j \cdot
\vec\nabla_{s '}	\bigg  [ j_{l}(z_0 \tilde s) \, Y^p_{lm} (\tilde \theta , \tilde{ \varphi}) \bigg ],
\end{equation}
using initial expression as in Eq.~\eqref{APS-M}, similar to the oblate spheroidal case in \cite{Little_Paper}, with $\mathbf{\hat e}_j$ being the unit polarization vector for the emitted light, $j_{l}( \cdot )$ spherical Bessel function of order $l$, (see Eq.~\eqref{ASB}) and $\vec\nabla_{s '}$ is the gradient of the wavevector given in spherical coordinate $(s', \theta', \varphi')$. 
Following the same quantization scheme as before, radiative decay rate of plasmons per unit solid angle $\Omega$, is then calculated to be:
\begin{equation}\label{Ps_RDR}
\frac	{d\gamma_{{lmp}}}	{d\Omega}=
\sum_{j=1,2}
\frac	{z_0 \omega^4_{lm}(1-\varepsilon_{lm}(\omega_{lm}))P^m_l{(\zeta_0)}}	{c^3(\zeta^2-1)P'^m_l(\zeta_0)}  
\left[ F^{(j)}_{lmp}	(s,\theta,\varphi)	\right] ^2,
\end{equation}
where
\begin{equation}\label{PS_Kp}
F^{(j)}_{lmp}(s,\theta,\varphi)=\frac{j_l(z_0   s'' )}{s}
\bigg [ \delta_{1j}\sqrt {\zeta_0^2-1}
\frac{\partial Y^p_{{lm}}( \theta'',\varphi'')}{\partial  \theta''}
+\delta_{2j}\frac{\zeta_0 }{ \sin\theta}
\frac{\partial Y^p_{{lm}}( \theta'',\varphi'')}{\partial \varphi''}\bigg ] ,
\end{equation}
for the two polarization states s and p \cite{Little_Paper}, with $( s'', \theta'',\varphi'')$ denoting the wavevector in spherical coordinates, for which the transformations are given  by taking the following transformations:
\begin{eqnarray}\label{PS-S}
\mathbf {s'}& =& \frac{s_x}{\zeta_0} \vec i + \frac{s_y}{\zeta_0} \vec j + \frac{\zeta_0 \, s_z}{\zeta_0^2-1 } \vec k, \\
\mathbf { s''} & =& \sqrt {\zeta_0^2-1} s_x \vec i +  \sqrt {\zeta_0^2-1} s_y \vec j + \zeta_0 \, s_z \vec k.
\end{eqnarray}

\begin{figure}
	\begin{center} 
		\includegraphics[width=6in]{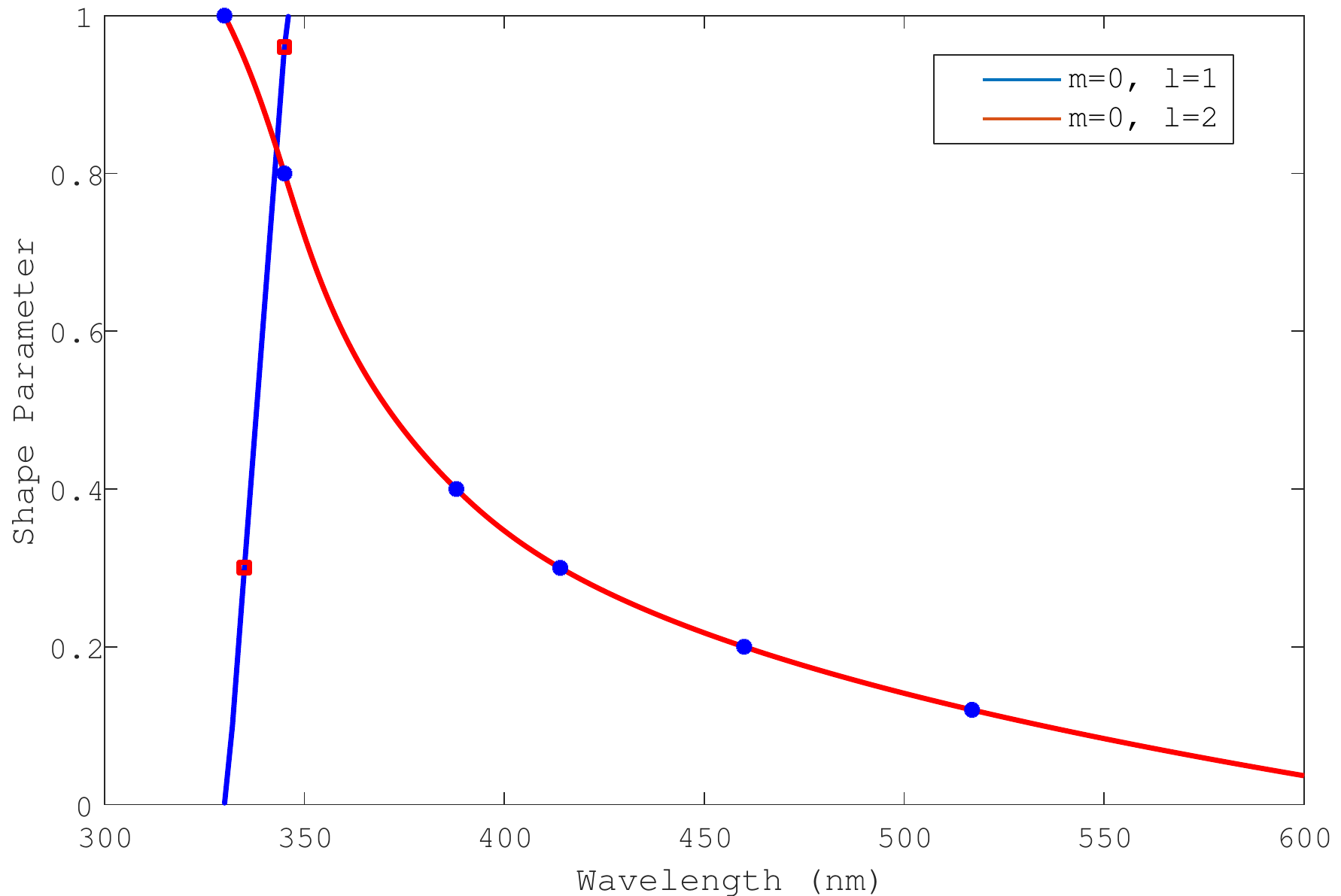}\\
		\caption[Here shows the calculations of the wavelength against ]{ Here shows the calculations of the wavelength against the shape parameter of a spheroid, here $\zeta$, which are obtained from the decay that occurs on surface plasmons for two modes: $(l,m)=(1,0)$ and $(2,0)$. The blue and red solid dots on the two curves are the exact match between the Hageman data and the frequency given in Eq.~\eqref{Ps-w1}. The rest is obtained by interpolating the values to the best it ws possible. 
		}
		\label{5-Hag}
	\end{center}
\end{figure}

\begin{figure}
	\begin{center} 
		\includegraphics[width=6.5in]{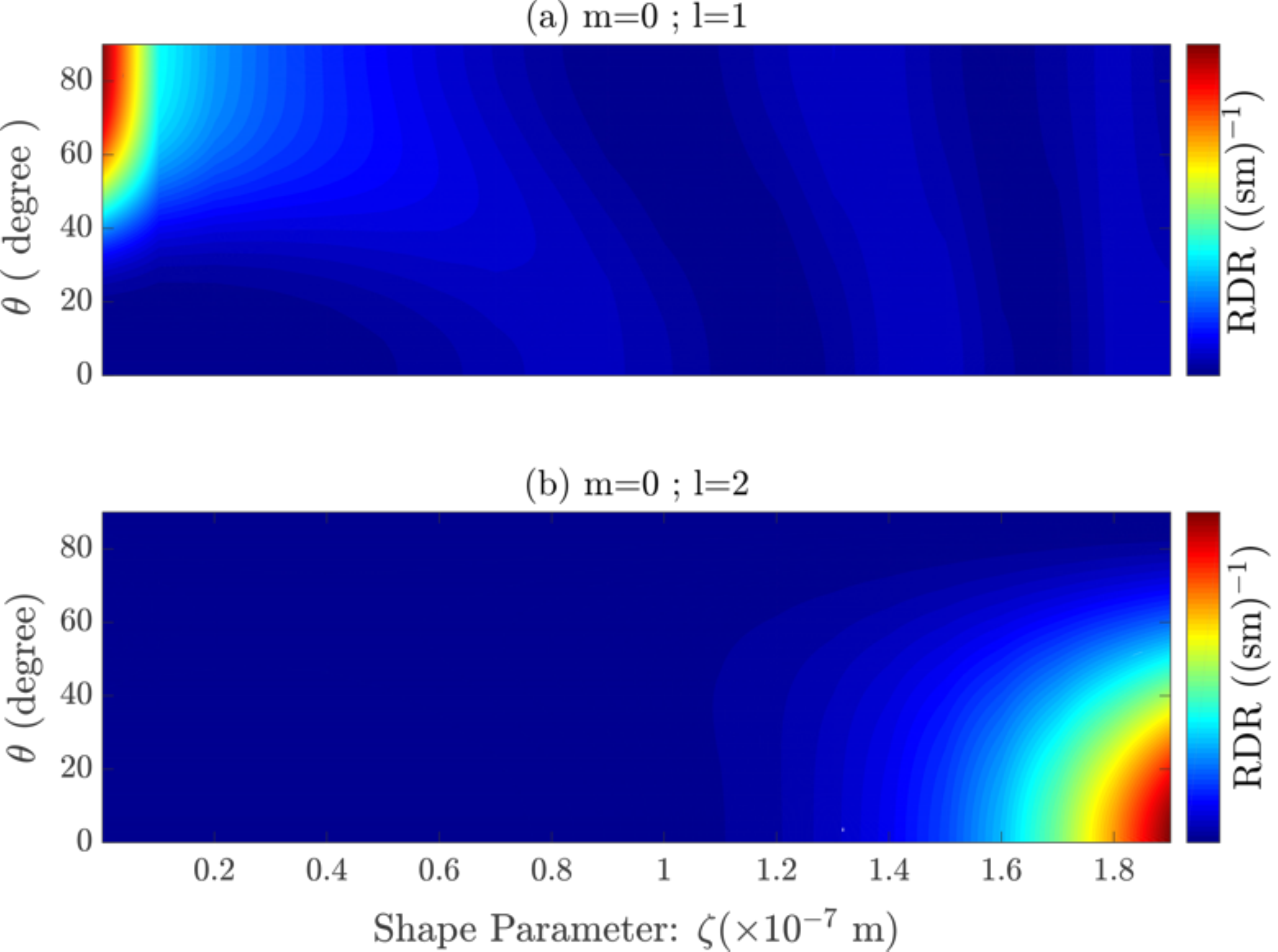}\\
		\caption[The contour plots are obtained by plotting decay rate given in]{ The contour plots are obtained by plotting decay rate given in in Eq.~\eqref{Ps_RDR} against the shape parameter, $\zeta$ and the emission angle $\theta$. We led the emission angle change between $0\le \theta \le \pi/2$ and shape parameter change between $0\le \zeta\le 2$. The intensity of radiative decay is defined by the color bar. Red color shows the highest intensity while dark blue shows the lowest. 
		}
		\label{5-Con}
	\end{center}
\end{figure}

\begin{figure}
	\begin{center} 
		\includegraphics[width=6in]{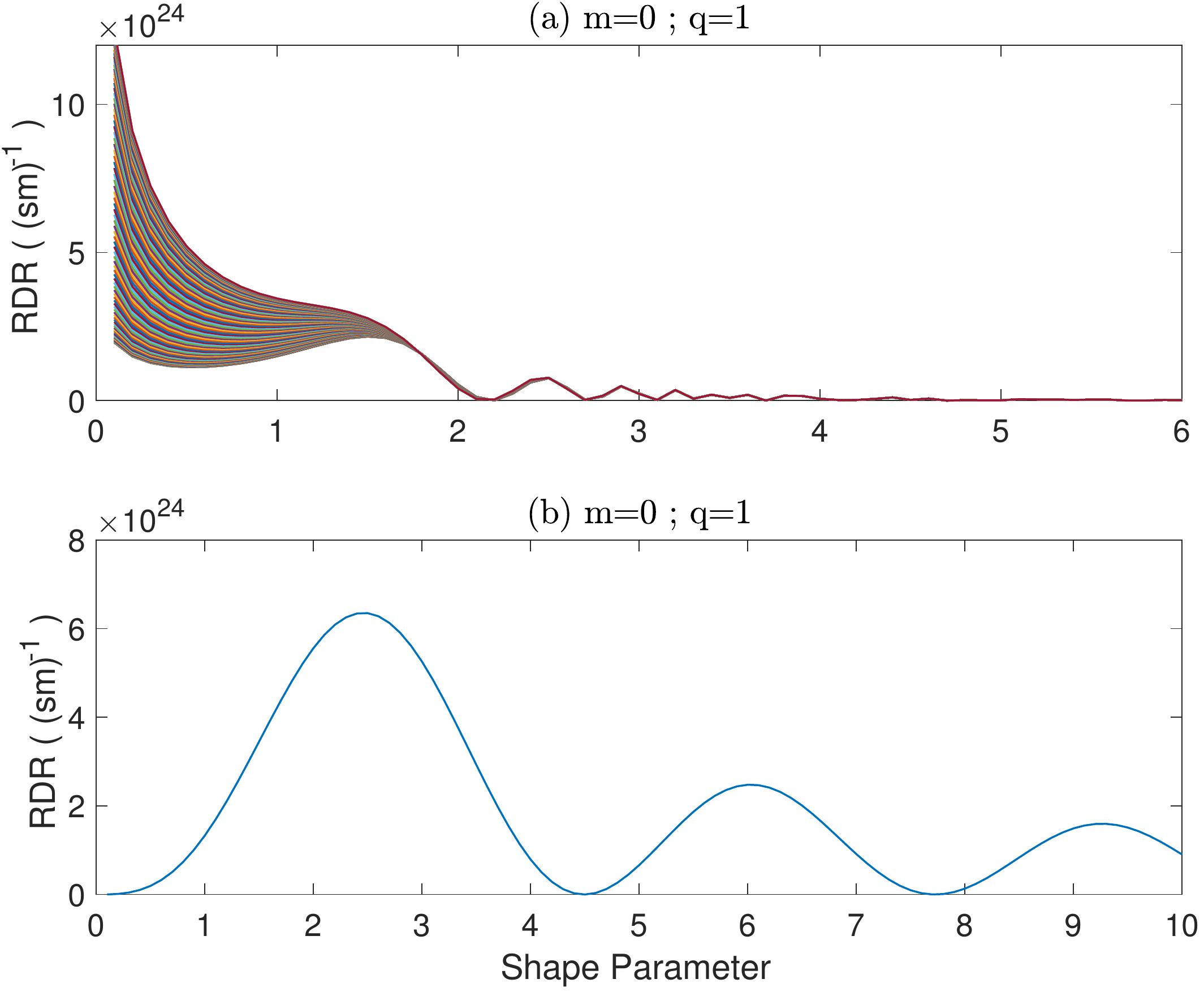}\\
		\caption[Here we consider a sphere as a limiting case for prolate spheroid. ]{ Here we consider a sphere as a limiting case for prolate spheroid.The contour plots are obtained by plotting decay rate given in in Eq.~\eqref{Ps_RDR} against the shape parameter, $\zeta$ and the emission angle $\theta$. We led the emission angle change between $0 \le \theta \le \pi/2$ and shape parameter change between $0 \le \zeta \le 2$. The intensity is defined by the color bar. The top figure is the contour plot for different $\theta$ which shows all the branches merge when shape parameter gets closer to $2$ and spheroid looks more like a sphere. It is noteworthy that a spheroid could never become a sphere for any shape parameter. 
		}
		\label{5-Lim}
	\end{center}
\end{figure}

Fig.~\ref{5-Hag} is obtained by using the experimental data for a silver spheroid provided by Hageman in \cite{Hag,oak}, in line with Fig.~2 in \cite{Little_Paper}. It shows, for the two specific modes $(l,m)=(1,0)$ and $(2,0)$, that the light emission reaches its maximal value at $\ang{80}$ and $\ang{20}$, respectively. Fig.~\ref{5-Con}, however, is a direct contour plot of Eq.~\eqref{Ps_RDR} which illustrates the same result as shown in Fig.~\ref{5-Hag}. Lastly, considering a sphere, as a limiting case of a spheroid by letting the shape parameter $\zeta \to 2$, Fig.~\ref{5-Lim} shows that the radiative decay rate becomes independent from emission angle $\theta$ as its shape approaches that of a sphere.  
\clearpage 
\clearpage 
\clearpage
\clearpage

\clearpage 

\section{Comparison}
\begin{figure}
	\begin{center}
		\includegraphics[width=6in]{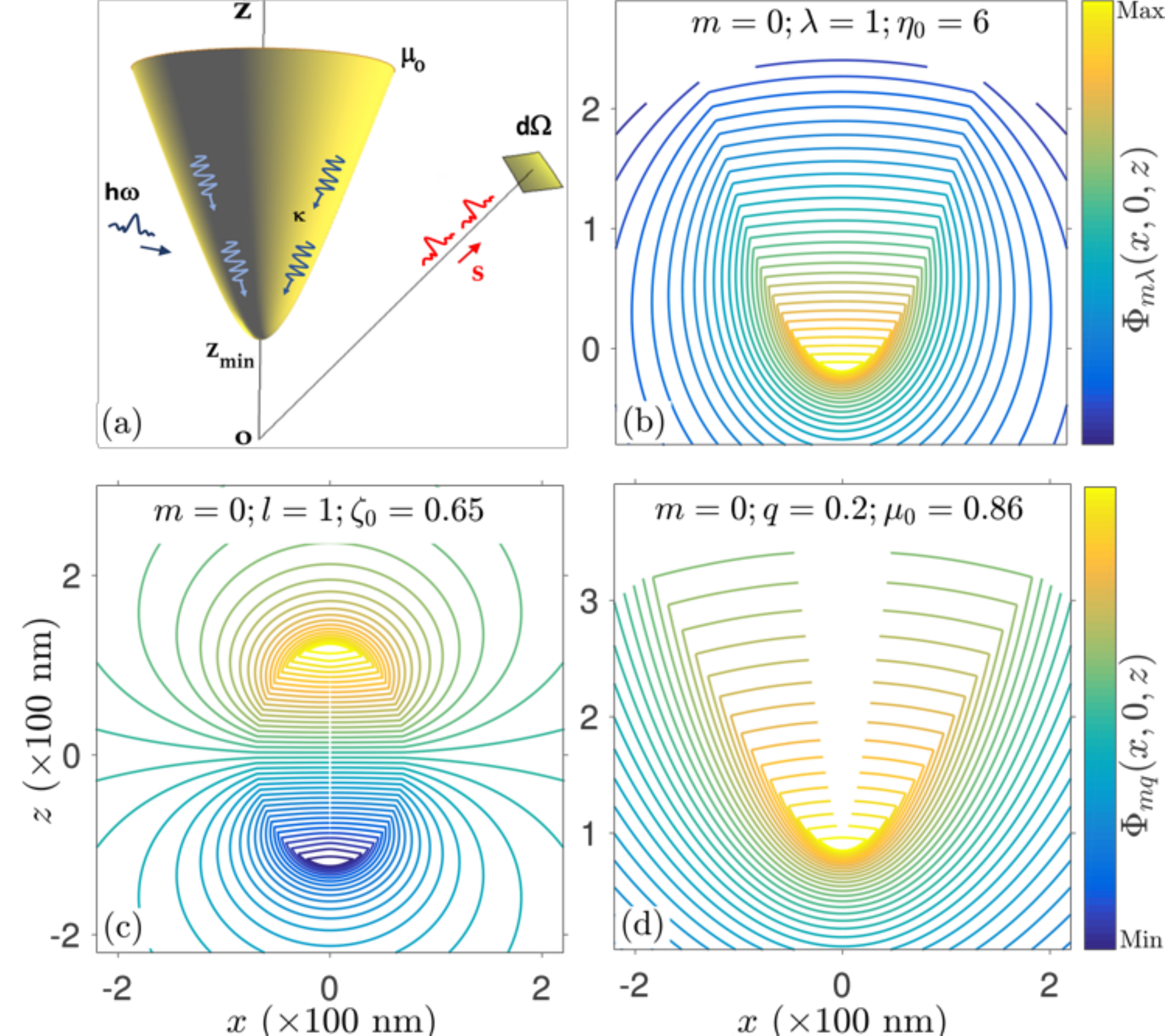}\\
		\caption[Modeling systems and their potential distributions ]{Modeling systems and their potential distributions. (a)One sheet of a two-sheeted hyperboloid  of revolution modeling a nanotip or a nanostructure with local curvature. Surface modes of momentum $\kappa$, e.g., excited by incoming photons $h\omega$, decay radiatively into a solid angle $d\Omega$. The curvature of the tip apex is set by the $\mu_0$ defining the hyperboloidal surface. Here, $\mu_0 = \cos\theta_0$, where $\theta_0$ is the angle between the $z$ axis and an asymptote to the hyperboloidal surface such that small $\theta_0$ yields a sharp probe while $\theta_0\to \pi/2$ corresponds to $xy$ plane. The apex point   $z_{\text{min}} = z_0 \mu_0,$  near the focal point of the hyperboloid, is set by the scale factor $z_0$, as in Eq.~\eqref{5-AHC1}. Figures (b), (c) and (d) show the spatial distribution of the lowest lying eigenmodes of the quasi-static electric potential for the three modeling domains investigated. For the same mode index $m$, optimizing the apex curvature overlap within the same spatial $zx$ domains, and analysing the potential distribution, leads to the determination of the corresponding continuous eigenvalues $\lambda$ of the paraboloid (b) and $q$ of the hyperboloid (d), respectively,  as well as the discrete eigenvalue  $l$ of the prolate spheroid (c). The geometric parameters $\eta_0$, $\mu_0$, and $\zeta_0$ determines the form of the considered domains. }
		\label{system_potential}
	\end{center}
\end{figure}

The harmonic functions, constituting the eigenmodes of the hyperboloidal and paraboloidal systems, represent the normal modes of the charge density. The infinite axial dimension of these domains results in the corresponding eigenvalues ($\lambda$ and $q$) to form continuous spectra as opposed to the discrete value spectrum ($l$) of finite domains such as spheroidal particles. While, this difference in the nature of the eigenvalues is less noticeable when visualizing the potentials, as seen in Fig.~\ref{system_potential},(b), (c) and (d), it becomes relevant when performing the quantization since now an integration over the corresponding eigenvalue spectrum enters the Hamiltonian, Eqs.~\eqref{5-H3} and \eqref{5-HH}. However, an inspection of the asymptotic properties of the integrands in the Hamiltonians reveal fast convergence, facilitating the calculation of the needed matrix elements. 

\begin{figure}
	\begin{center} 
		\includegraphics[width=3.6in]{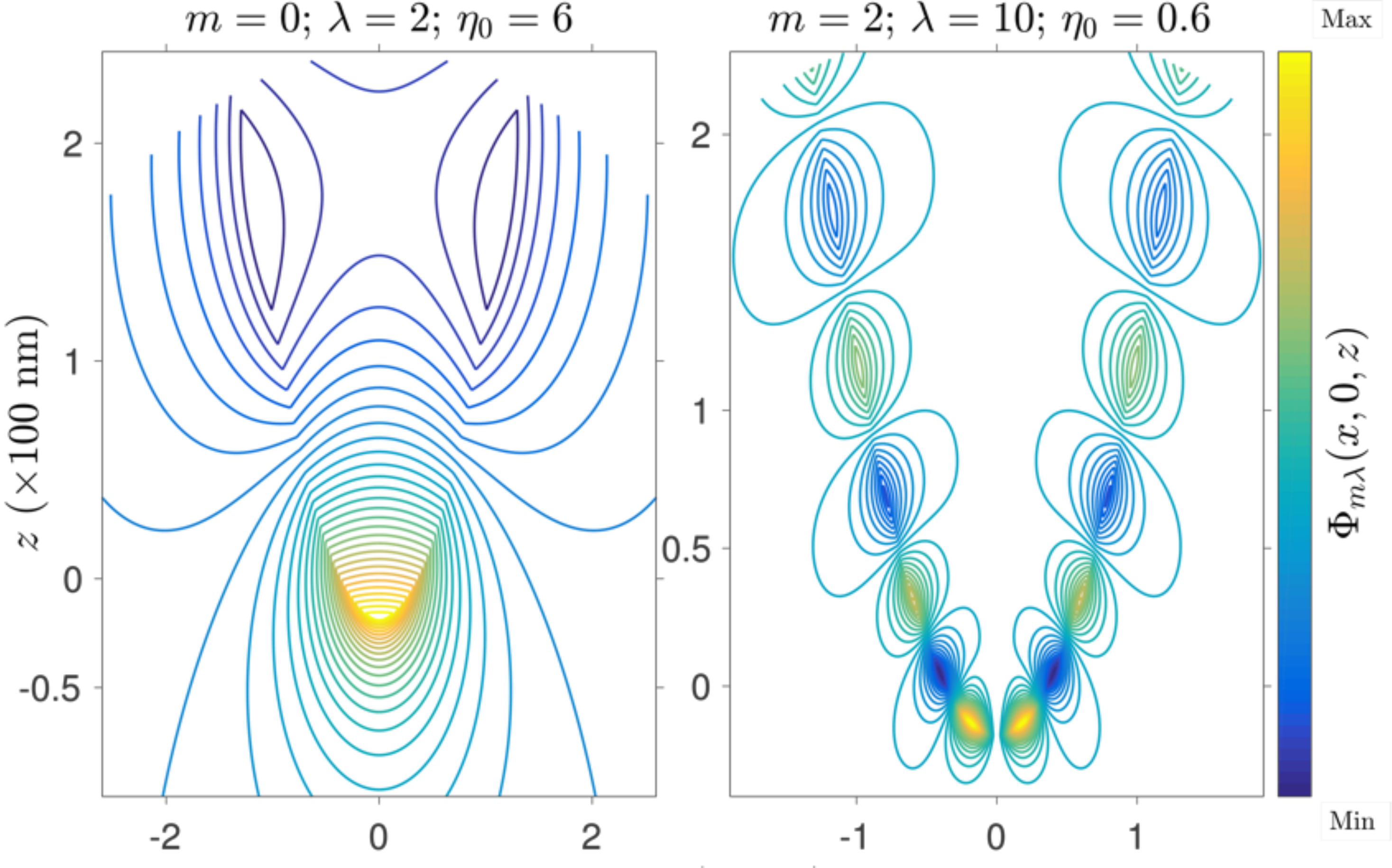} \\
		\includegraphics[width=3.6in]{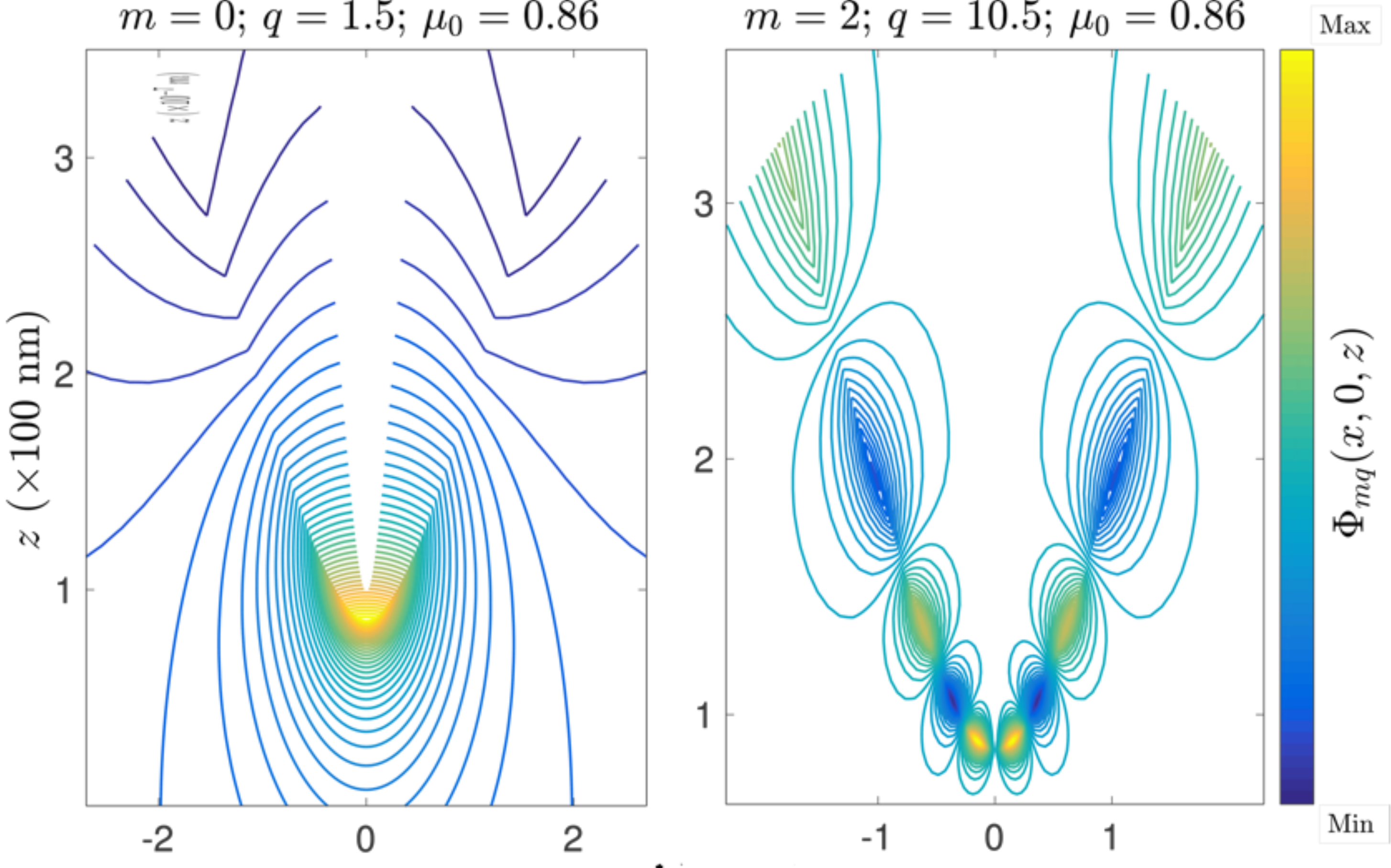}\\
		\includegraphics[width=3.6in]{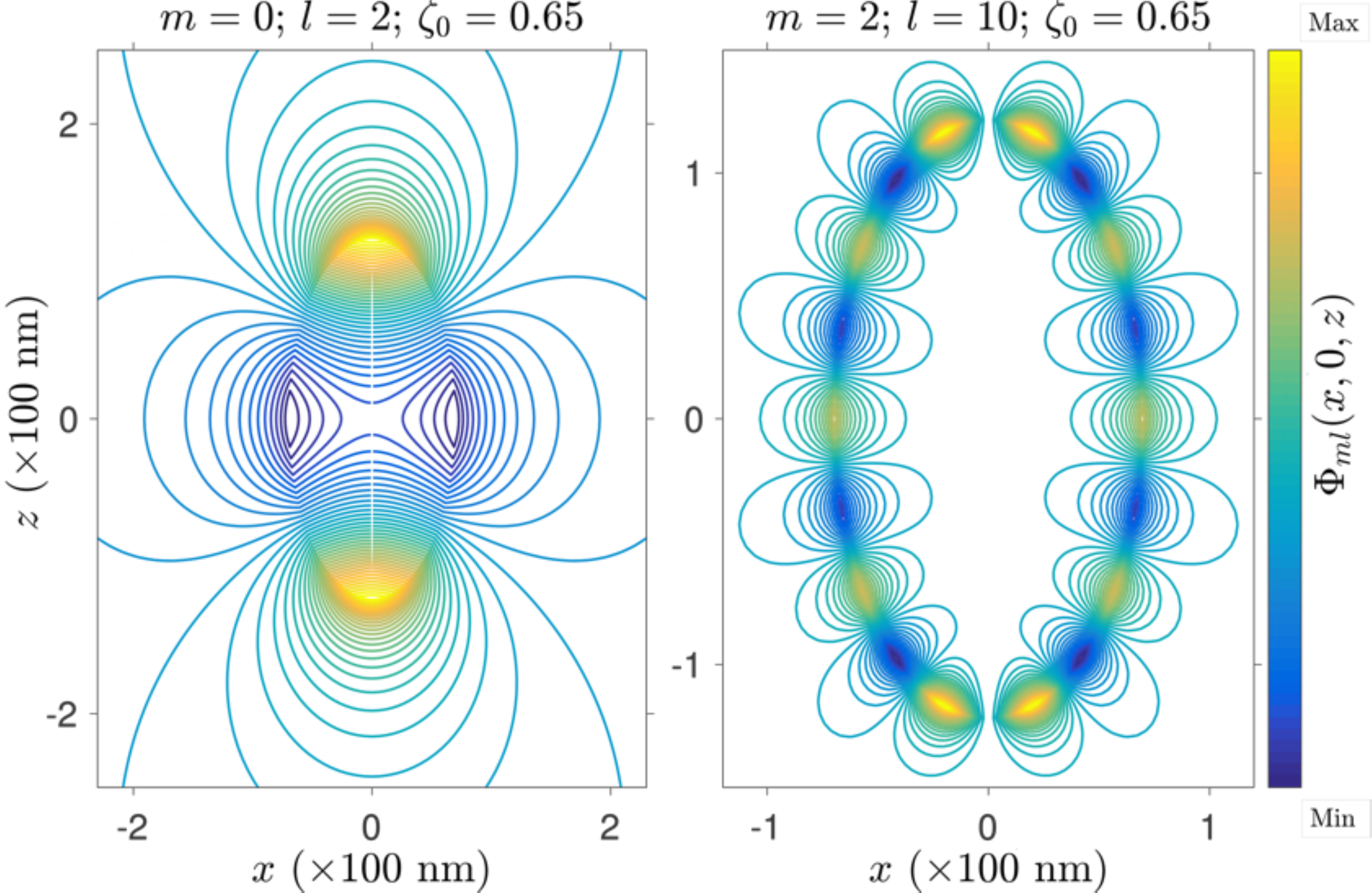}
		\caption[The spatial distribution of the one low and one high lying ]{
			The spatial distribution of the one low and one high lying eigenmodes of the quasistatic electric potential for the three modeling domains investigated. For the same mode index $m$, optimizing the apex curvature overlap within the same spatial $zx$ domains, and analyzing the potential distribution, leads to the determination of the corresponding continuous eigenvalues $\lambda$ of the paraboloid (top) and $q$ of the hyperboloid (middle), respectively,  as well as the discrete eigenvalue  $l$ of the prolate spheroid (bottom). The geometric parameters $\eta_0$, $\mu_0$, and $\zeta$ determines the form of the considered domains. For proper geometric and modal adjustments, the similarities in the potential distributions are clearly evident from the contour plots.  The discontinuity in the  contour lines near the symmetry axis of the hyperboloids is due to the singularity in the conical functions there. 
		}
		\label{Potential2}
	\end{center}
\end{figure} 
\begin{figure}
	\begin{center}
		\includegraphics[width=3.6in]{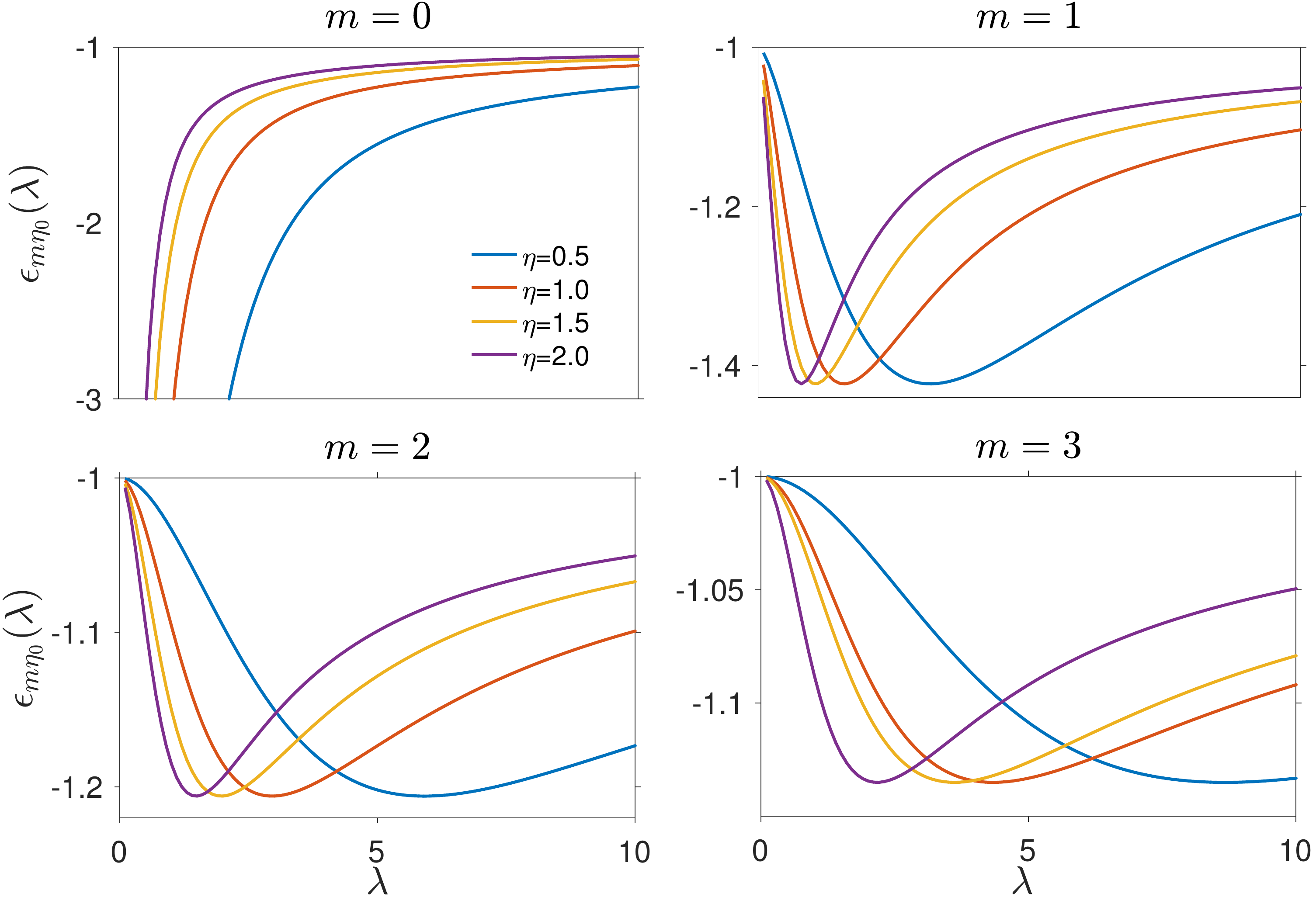}\\
		\includegraphics[width=3.6in]{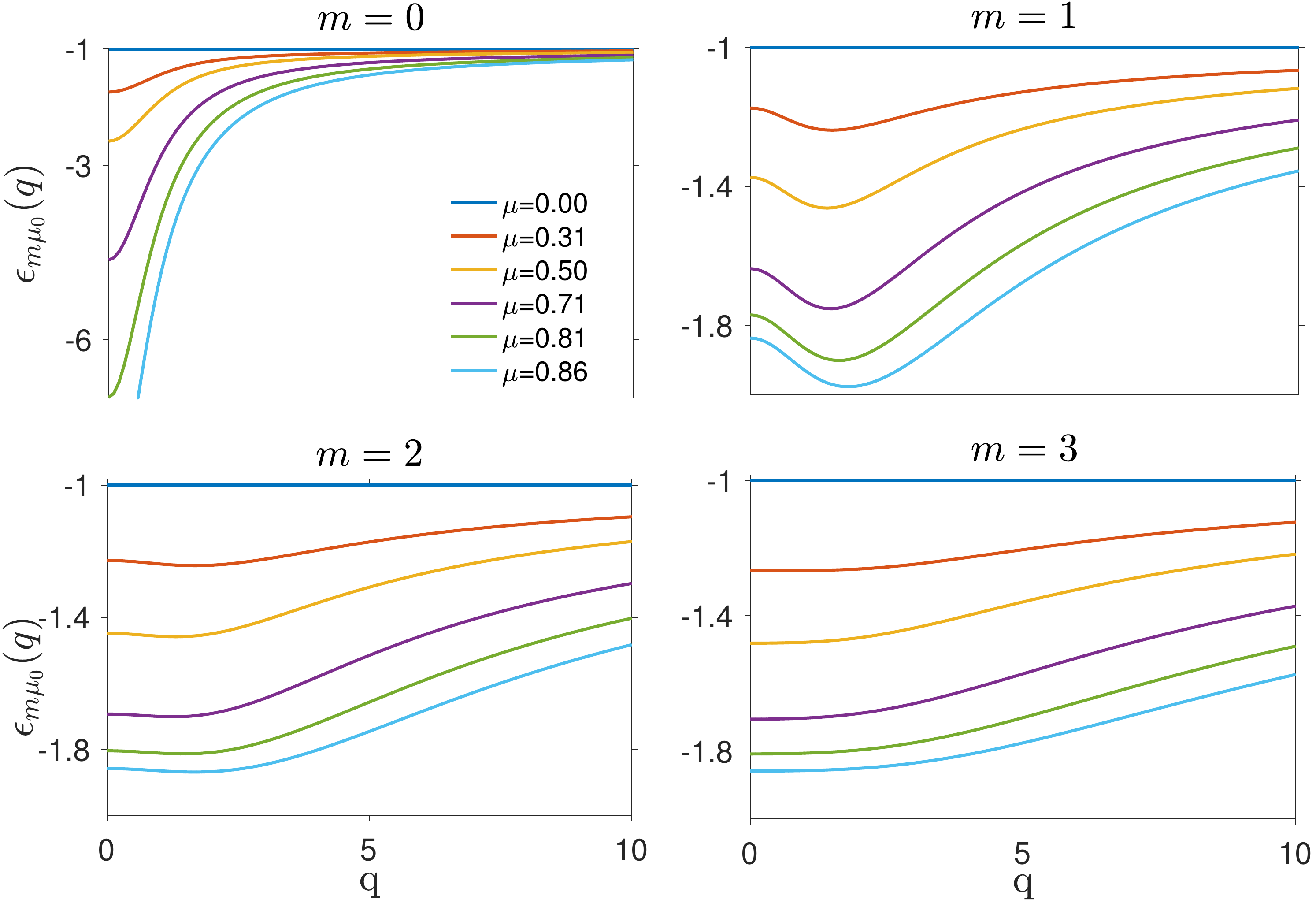}\\
		\includegraphics[width=3.6in]{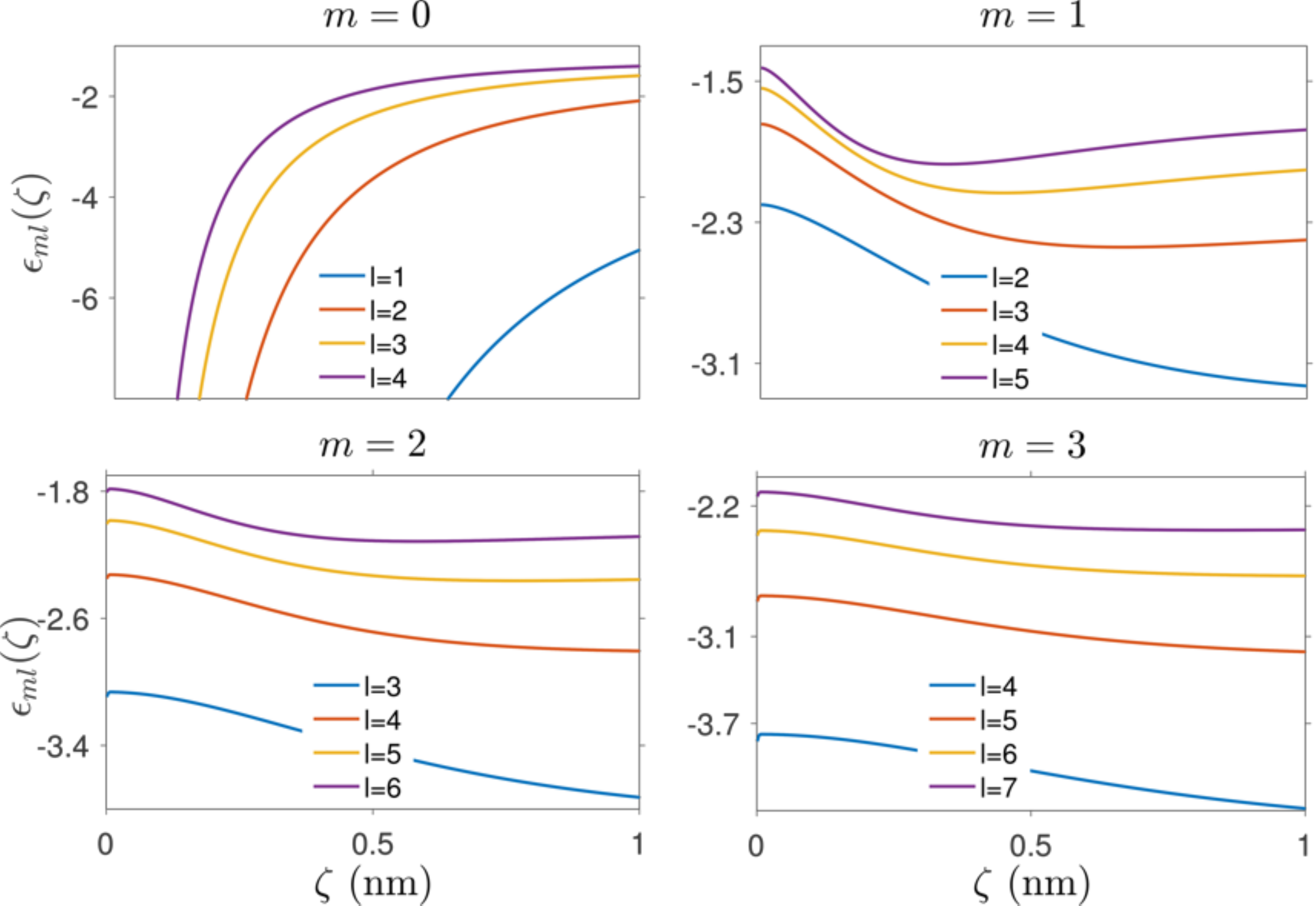}	
		\caption[Paraboloidal, hyperboloidal, and prolate spheroidal nonretarded  ]{
			Paraboloidal, hyperboloidal, and prolate spheroidal nonretarded surface plasmon  dispersion relations. The resonance values of the dielectric function $\varepsilon$ are shown for low lying modes as a function of  the continuous eigenvalue $\lambda$ for a paraboloid (top) and $q$ for a hyperboloid (middle), and  as a function of the shape parameter $\zeta$ for a prolate spheroid. The surfaces of the paraboloid and hyperboloids are set by the parameter $\eta_0$ and $\mu_0$, respectively, while $\zeta$ defines the form of the  spheroidal surface (bottom). The discrete modes are denoted by $m$ for the azimuthal oscillations and by $l$ in the spheroidal case.        
		}
		\label{Dispersion}
	\end{center}
\end{figure}

To proceed, we note that for a more reasonable comparison of the two systems, their physical dimensions must be made comparable. Therefore, we geometrically adjust the paraboloid and the hyperboloid  so as to control and match their apex curvature and further match it with that of a finite prolate spheroidal domain. The latter, due to its finite domain, makes a natural case to validate the findings for the two infinite domain cases investigated. 

For settings consistent with nanofabrication and photon scattering experiments involving gold probes, where strong plasmon excitation has been reported~\cite{sam_poster,muller}, we employ the following settings: $\eta_0=60$~nm  and $\mu_0=86$~nm for the systems in Fig.~\ref{system_potential}(b), (c) and (d).  
The close relationship of the calculated eigenvalues with the surface plasmon momenta are clearly observable from the ''wavelength'' of the charge density oscillations in Fig.~\ref{system_potential}(b), (c), (d) and Fig.~\ref{Dispersion}. For proper geometric and modal adjustments, the similarities in the potential distributions are clearly evident from the contour plots. Since for the argument values $\mu\approx 1, -1$,  the conical functions become singular, a numerical artifact in the form of a discontinuity in the  contour lines appear near the symmetry axis of the hyperboloids, where $\mu$ attains those values.
Comparison with the corresponding potential distribution in the spheroidal case can be facilitated by taking $\zeta_0=65$~nm 
matching its curvature with that of the apexes in paraboloidal and hyperboloidal domains.
With the dimensions adjusted, using Eqs.~\eqref{pin-pout}, \eqref{Hpin-Hpout} and \eqref{PS-pin-pout} to simulate the spatial distributions of the potentials for the three cases, one clearly observes the analogous role of $\lambda$ and $q$ to the discrete spectrum $l$. The lowest azimuthal mode $m=0$ for the three domains shown in Fig.~\ref{system_potential}(b), (c) and (d)
was simulated by taking the second indices as $\lambda=1$, $q=0.2$ and $l=1$, generating relatively same potential distributions. 
The potential distributions for higher modes are shown in Fig.~\ref{Potential2}, where two higher modes for each domain (shape parameters $\eta_0$, $\mu_0$ and $\zeta_0$) are shown. 
In doing so, the Legendre functions $P_{mq}(\mu)$ of imaginary order and their derivates have been calculated using the computational algorithm of Gil and Segura~\cite{Gil}, and the integral expansion of K{\" o}lbig~\cite{kolbig}.
Analogous to plasmon wavevector  in the case of excitations on a planar interface (or a Cartesian thin film), which can be emulated by $\mu_0 \to \pi/2$ in Fig.~\ref{system_potential}(a), the higher the $\lambda$, $q$, and $l$ for the same $m$, the higher the number of fluctuations for the same spatial domains, as shown in Fig.~\ref{Dispersion}. Furthermore, from the spheroidal nearfield distribution, one can readily observe the multipole order so that $(m,l)=(0,1)$ corresponds to a dipolar distributions, whereas $(m,l)=(0,2)$ leads to a quadrupolar behavior, etc. Similarly, for $(m,q)=(0,1.5)$ or $(m,\lambda)=(0,2)$ one obtains the corresponding multipolar nearfield distributions of the apex regions.   
Thus, guided by these charge density oscillations, controlled by the parameters $(m,\lambda,\eta_0),$ $(m,q,\mu_0)$, and $(m,l,\zeta_0)$ for the three cases and by a geometric matching of their apex curvatures, one may discuss the eigenmode dependent radiative decay rates. 
\clearpage
\begin{figure}
	\begin{center}
		\includegraphics[width=6.5in]{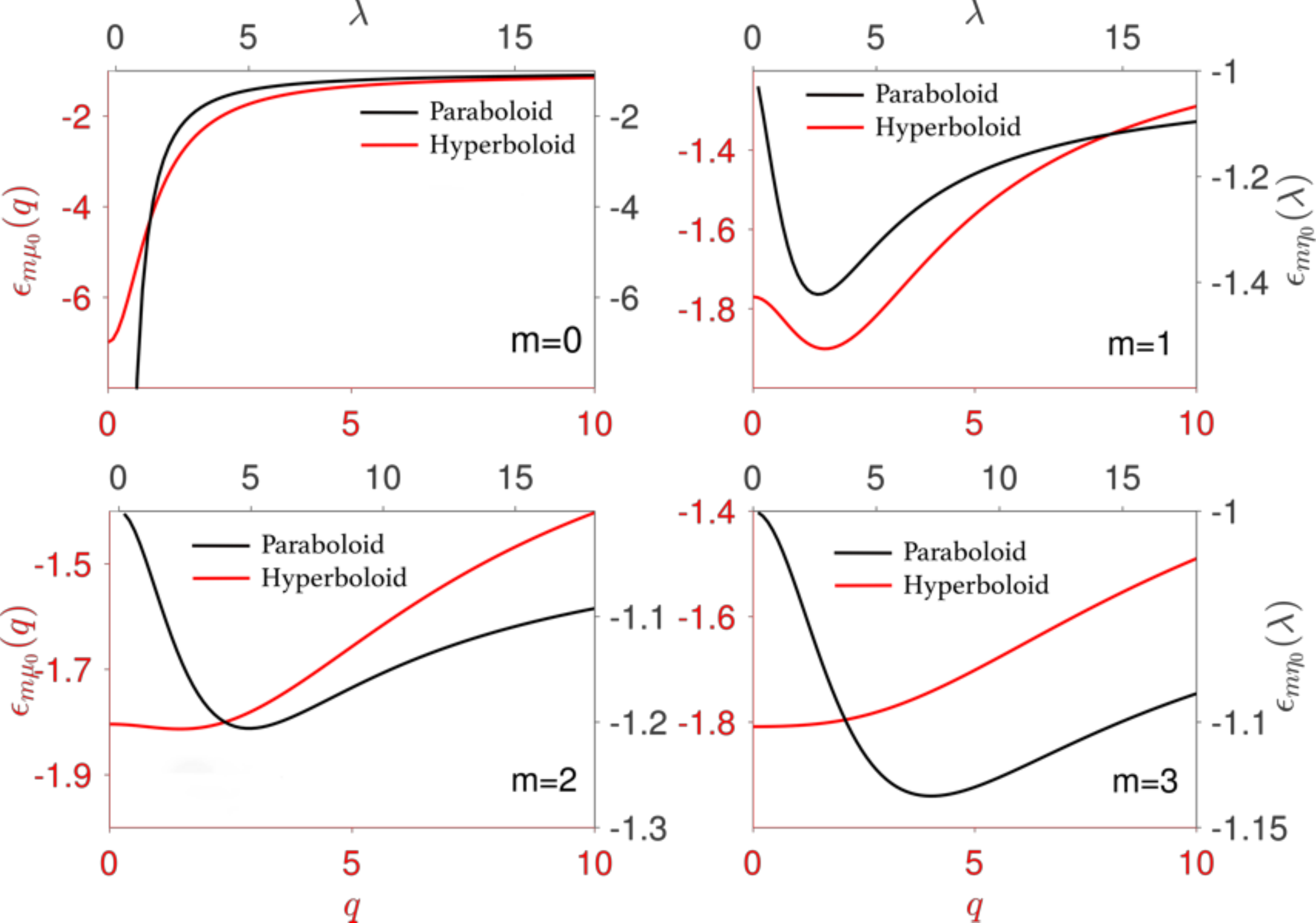}\\
		\caption[Paraboloidal and hyperboloidal nonretarded surface ]{
			Paraboloidal and hyperboloidal nonretarded surface plasmon dispersion relations. The resonance values of the dielectric function $\varepsilon$ are shown for low lying modes as a function of  the continuous eigenvalue $\lambda$ for a paraboloid (black) and $q$ for a hyperboloid (red). The surfaces of the paraboloid and hyperboloid are set by the parameter $\eta_0$ and $\mu_0$, respectively. The discrete modes are denoted by $m$ for the azimuthal oscillations.
		}
		\label{Dispersion_HP}
	\end{center}
\end{figure} 
\clearpage

In addition to the material-specific electronic and optical properties, in as far as the effect of the local curvature is concerned, we expect the considered cases to exhibit similarities in their resonance spectra. The plasmon dispersion relations are good indicators of this resemblance.
To study the relation between resonance modes of isolated solid paraboloids and hyperboloids in vacuum, we assume  a local dielectric function and calculate their eigenmode dependent allowed values $\varepsilon_{m\lambda}$ and $\varepsilon_{mq}$ using  Eqs.~\eqref{5-wml} and \eqref{Hwmq}. 
A comparison of the lowest lying plasmon modes ($m=0,1,2$ and $3$ of fixed probes $\eta_0$ and $\mu_0$) is given in Fig.~\ref{Dispersion_HP}. The modes may alternatively be displayed with reference to bulk plasma  frequency $\omega_{p}$.
Interestingly, the higher $m$ modes show a higher sensitivity to the morphological differences between the two systems, in particular for lower $\lambda$ and $q$ values, that is, in the long wavelength limit, which in analogy with the planar plasmons would be near the light line, where retardation effects are more pronounced~\cite{PPRB}.  
We also note that in the hyperboloidal case, in the limit $\mu_t\to 0$, we have $\varepsilon_{mq}\to -1$; that is, the modes asymptotically approach the surface plasmon resonance ($\omega \to \omega_p/\sqrt{2} $), as expected for a Cartesian half-space. This limit is also approached by large $m$ values as seen in Fig.~\ref{Dispersion_HP}.  In the short wavelength regime $\lambda, q \to \infty,$  the dielectric function reads: 
\begin{equation}\label{HepsAs}
\varepsilon_{mq}\sim -1-\frac{\cot (\theta_t)}{q},
\end{equation}
yielding the same surface plasmon  limit \cite{PPRB,z&k:table1}. Similar to  the paraboloidal case,  using the asymptotic behavior of modified Bessel functions $I_m(\lambda\eta)$ and $K_m(\lambda\eta)$ for large order $m$ and large arguments $\lambda$ and $\eta$ (see Eqs.~(9.7.8) and (9.7.9) in \cite{Abra}), we can write:
\begin{equation}\label{PepsAs}
\varepsilon_{m\lambda}\sim -1-\frac{1}{2\lambda\eta},
\end{equation}
which implies the same limit for large $\lambda$.
\clearpage
\begin{figure}
	\begin{center}
		\includegraphics[width=6.5in]{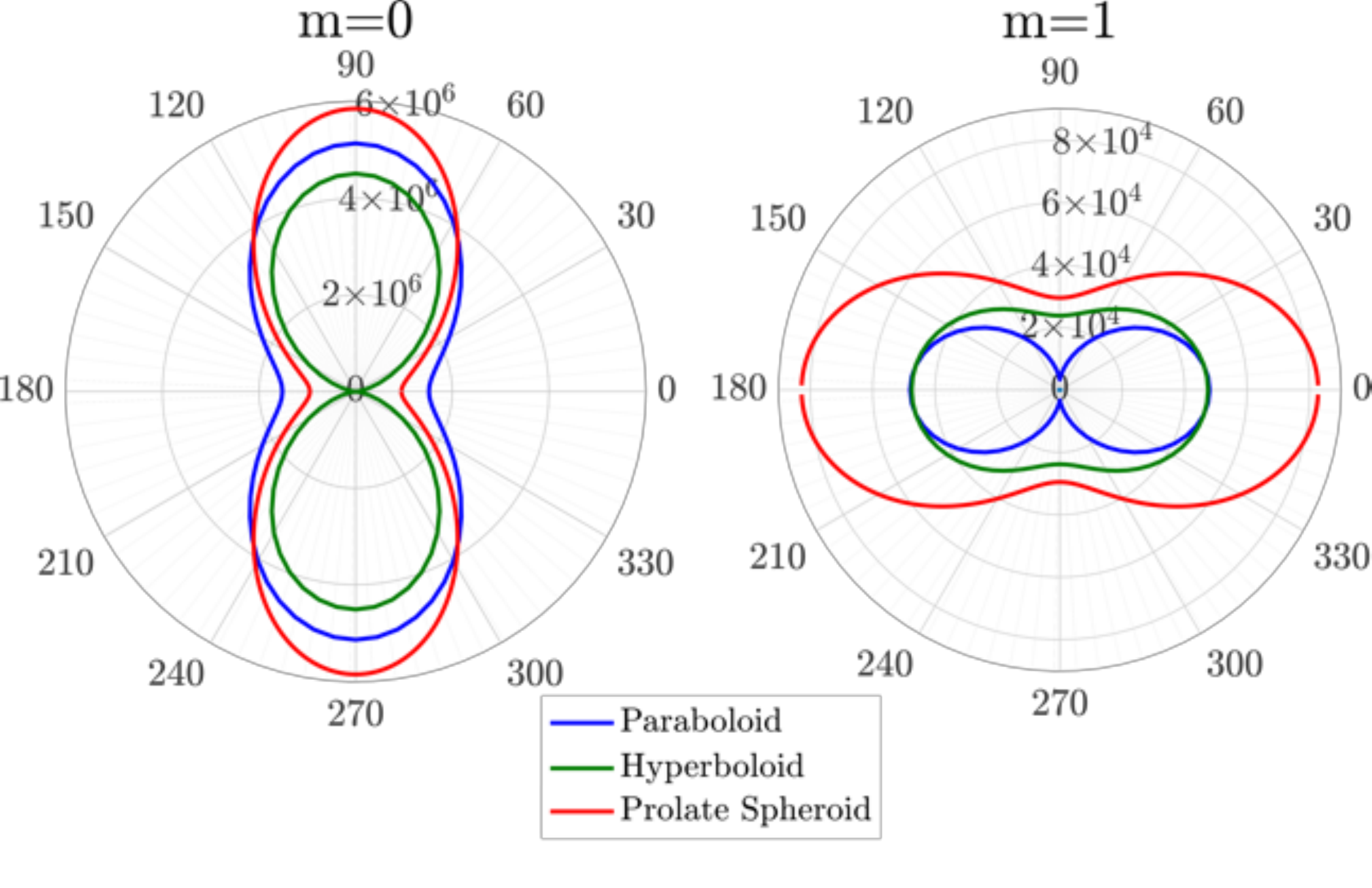}\\		
		\caption[The radiative decay rate of plasmons engendered ]{The radiative decay rate of plasmons engendered on paraboloidal, hyperboloidal, and prolate spheroidal surfaces.  The decay rate of paraboloidal plasmons (blue) is computed from Eq.~\eqref{5-R1} for the single eigenmode  $m=0$ (right), $m=1$ (left) and $\lambda=1$ when the shape parameter is $\eta_0=60$~nm.
			Similarly, the decay rate of hyperboloidal plasmons (green) computed from Eq.~\eqref{5-HR1} corresponds to the eigenmode $m=0$ and $q=0.2$ for a  shape parameter $\mu_0=86$~nm. For comparison, the radiative decay rate (red) for plasmons excited on a prolate spheroidal surface  is computed using Eq.~\eqref{Ps_RDR} for $\zeta_0=65$~nm and $m=0$, $l=1$, corresponding to the dipolar mode. To facilitate visual comparison within the same plot window, note that the spheroidal case for $m=0$ (right) has been multiplied by  $0.006$ and for the case $m=1$ (left) by $0.01$. The composition of the specific parameters for comparison was guided by the potential distribution in each case.}
		\label{RDR_low_modes}
	\end{center}
\end{figure} 
\clearpage
Following the field distribution and resonant dielectric values corresponding to the normal modes of the surface charge density oscillations,
we may assume that a plasmon has been created in a given eigenmode $(m,o,p)$ where $o=\lambda, q, l$ designate paraboloid, hyperboloid and spheroid, respectively. If the plasmons on the paraboloidal, hyperboloidal or spheroidal surface  are in the initial state  $\hat  a^\dagger_{(mop)_i} \ket{0},$ the probability amplitude for their emission into the final photon state 
$ \hat c^\dagger_{(\mathbf{s}j)_f} \ket{0}$ could be obtained using expressions \eqref{5-M1}, \eqref{5-HM} and \eqref{PS-M1}. The $\ket{0}$ indicates that the fields have been populated with 0 plasmons or photons 
(noting $ a_{mop} \ket{0} = c_{(\mathbf{s}j)} \ket{0}=0$) 
whereas a general photon-plasmon state is written as  
$\ket{\Psi} = \ket{n_{\mathbf{s}j}}\otimes \ket{n_{mop}},$ that is, a state with $n_{\mathbf{s}j}$ $j$-polarized photons of momentum $\mathbf{s},$ and $n_{mop}$ plasmons in the $(m,o,p)$ state. 

\begin{figure}
	\begin{center}
		\includegraphics[width=6in]{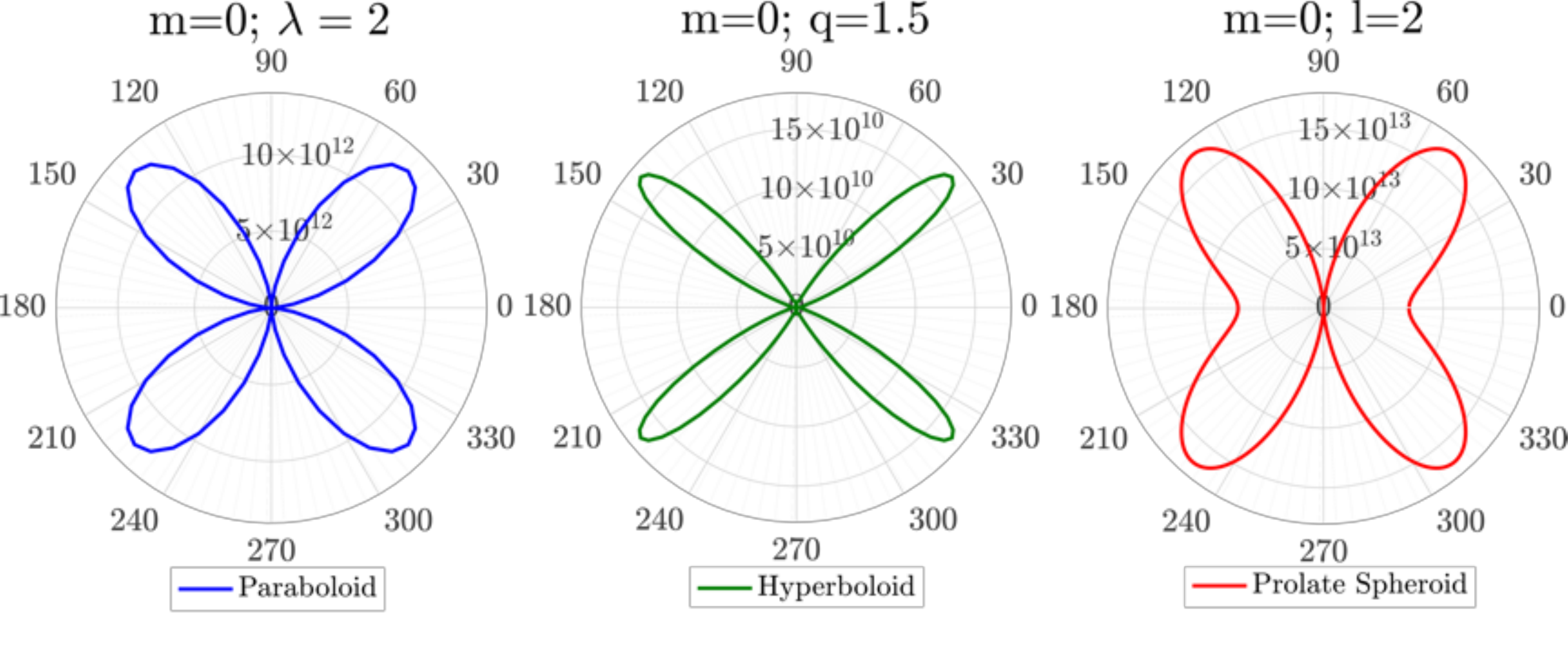}
		\caption[Comparison of the higher order modes' radiative  ]{
			Comparison of the higher order modes' radiative decay rates for the three different cases described in Fig.~\ref{RDR_low_modes}. In the case of the prolate spheroidal plasmons, the emitted radiation pattern corresponds to the quadropolar  charge density oscillations. 
		}
		\label{RDR_higher_modes}
	\end{center}
\end{figure}

Keeping the mode patterns, here, the $\varphi=0,\, \pi$-plane projection of the relative potential distributions in Fig.~\ref{system_potential}(b), (c) and (d) and dispersion relations in Fig.~\ref{Dispersion_HP} in mind, we now compare the radiative decay rate per solid angle of plasmons engendered on the three domains, using Eq.~\eqref{5-R1} (paraboloid with $\eta_0=60$~nm), Eq.~\eqref{5-HR1} (hyperboloid with $\mu_0=86$~nm) and Eq.~\eqref{Ps_RDR} (prolate spheroid with $\zeta_0=65$~nm) for two lowest azimuthal mods $m=0$ and $1$. 
To calculate the matrix elements given in Eqs.~\eqref{5-M1} and \eqref{5-HM}, the integrations $ \mathcal I^{(j)}_{m\lambda}$ and $ \mathcal I^{(j)}_{mq}$ (in case for paraboloid and hyperboloid, respectively) must be carried out numerically as they lack  analytical solutions in variable $\xi$ in the case for paraboloid and $\eta$ in the case for hyperboloid. The choice of polarization vectors $\mathbf{ \hat e}_j$ for $j=1,2$ corresponds to $\mathcal I^{(1)}_{m\lambda}$ given by Eq.~\eqref{5-I1} to represent the s-polarization, and $\mathcal I^{(2)}_{m\lambda}$ given by Eq.~\eqref{5-I2}, the  p-polarization. Inspecting the integrands for their convergence, we  compute the integrals  using an iterative numerical integration scheme (Runge-Kutta) due to lack of fast oscillations.
The result is shown in Fig.~\ref{RDR_low_modes}. Note that to facilitate visual comparison within the same plot window, the radiative decay rate for prolate spheroid for modes $m=0$  and $m=1$ have been multiplied by  $0.006$ and $0.01$, respectively.
The effect of higher index modes $\lambda$, $q$ and $l$ with the same azimuthal order $m$ on radiative decay rate per solid angle may also be studied, as  shown in Fig.~\ref{RDR_higher_modes}.
As one may expect from the nearfield patterns of higher $\lambda, $ and $q,$ analogous to $l=2$, a quadrupolar pattern appears for the emitted photons. Here, it is understood that an angular segment is occupied by the probe itself, as opposed to the $0-2\pi$ range for the finite volume spheroidal systems. It is further understood that for larger particles or probe apex size retardation effects may modify the higher order modes.

\begin{figure}
	\begin{center}
		\includegraphics[width=6.5in]{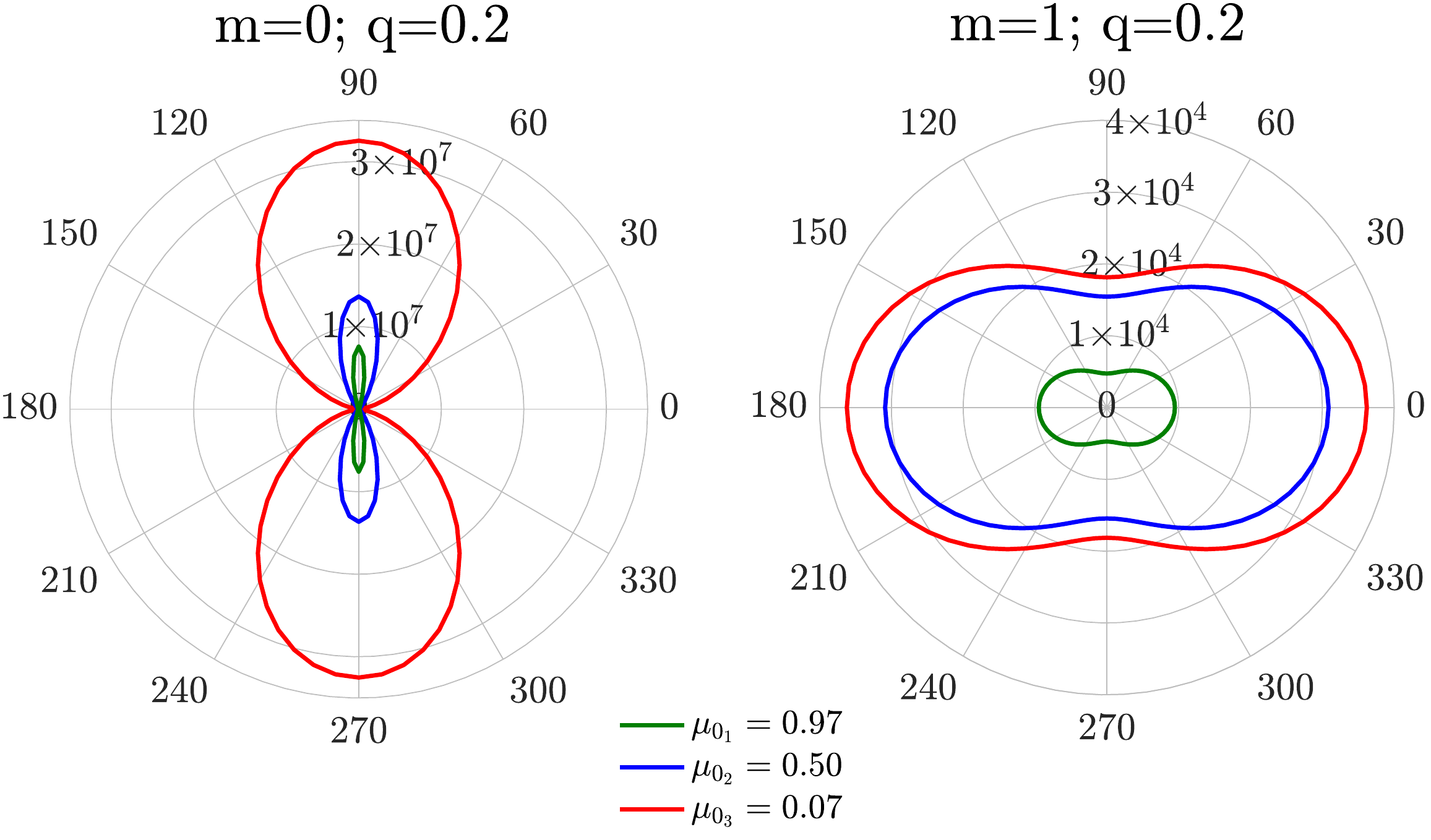}	
		\caption[Curvature induced shift in the radiation pattern  ]{Curvature induced shift in the radiation pattern associated with the decay of plasmons excited on the hyperboloidal surfaces for the modes $m=0$ (left) and $m=1$ (right) and $q=0.2$.
			To facilitate comparison within the same plot window, the case for $\mu_1$ has been multiplied by  $20$ (left) and by $50$ (right).   
		}
		\label{RDR_3hype}
	\end{center}
\end{figure}

The dependence of the radiative decay rate upon the parameter that sets the boundary, allows for control of the curvature and thus the photon emission patters. As can be seen from Fig.~\ref{RDR_3hype}, for the $m=0$ mode, the higher the curvature of the hyperboloidal apex, the lower the amplitude and the narrower the angular distribution of the emitted photons. Moreover, for $m=1$ the higher  curvature while resulting in a lower amplitude does not result in in a higher angular confinement. \\ \\
\clearpage
\clearpage

\setlength{\parindent}{0.5in}
\chapter[AN APPROXIMATION FOR RADIATIVE DECAY RATE ON T-\hspace{0.5in}\\
\null \hspace{0.25in} HE NON-SIMPLY CONNECTED SURFACE OF A TORUS]
{AN APPROXIMATION FOR RADIATIVE DECAY RATE ON THE NON-SIMPLY CONNECTED SURFACE OF A TORUS}\label{ring}

	\section{Introduction}\label{ch:6}

Photon scattering and excitation of surface plasmons in single rings, single composite rings and many-ring systems was recently shown to provide useful dispersion and field distributions \cite{garapati,garapati2} with specific applications in particle and molecular trapping \cite{alaee,salhi}. Following photon or electron excitation of metallic nanorings, both radiative and nonradiative decay channels are important in considering of the nanoparticle as a photon or phonon source, and in related applications.    
Here, we investigate the radiative decay channel for a vacuum bounded single solid nanoring by quantizing the fields associated with charge density oscillations on the nanoring surface. In quantizing the  fields, the  frequency spectrum of the charge density normal modes of  the nanoring is obtained and shown to agree with the exact quasi-static plasmon dispersion relations. We calculate the radiative decay rate per unit solid angle and show that both poloidal and toroidal modes contribute to the radiation field.

Particle scattering from material domains that take the form of nanoparticles, lead to useful excitations and surface modes \cite{B,amendola,pelaz}. 
Both electron and photon bombardment are known to lead to resonant collective electronic effects in thin films and particles, from which emitted light due to the damping of excited plasmons has been observed in many experiments \cite{mertens}. Following an excitation event of surface plasmon, the radiative decay rate of the associated modes on the bounding surfaces of the underlying material domains, and their radiation pattern can help understand and interpret experimental observations and emission measurements in the nearfield as well as in the far field.  In this chapter, we present a calculation of the differential scattering cross section for a nanoring, from quantizing the charge density oscillations. 
The presented formalism and results can be used to study the decay properties of other photon scattering and plasmon-related processes such as photoabsorption, photoemission, tip-enhanced, and surface-enhanced Raman scattering. By obtaining the angular, polarization and wavelength characteristics of the radiation from the excited surface-plasmon modes on various particles, experimental results may be interpreted. 

Our approach to calculate the radiative decay of the toroidal and poloidal plasmons closely follows previous works \cite{Little_Paper,BURMISTROVA,Leavitt} based on 
Bloch linearized hydrodynamic model \cite{Crowell}, leading to the Hamiltonian
\begin{eqnarray}
H=-\int dr
\left( \frac12 mn_0 \Psi \vec \nabla ^2\Psi	+
\frac12 e\varphi n-\frac{m\beta^2}{2n_0}n^2
\right). 
\end{eqnarray}
Specifically, the radiative decay of surface collective excitations in polarizable cylinders by \cite{BURMISTROVA}, calculation for a system of small metallic spheres embedded at random in a thin dielectric slab by \cite{Crowell}, investigation of a plane slab by \cite{Ritchie}, and oblate spheroids by \cite{Little_Paper} and its special case of a metallic 2D disc by \cite{Leavitt}. 
In these works, using nonretarded electrodynamics and bulk-metal optical properties, plasmons have been described on various particles with subwavelength dimensions, although the results may be made more accurate if surface-metal optical properties can be used. 

This chapter is organized as follows. We begin by modeling a nanoring as a solid ring toroid in toroidal coordinates. We then proceed to determine the surface plasmon field and dispersion relations assuming a local dielectric function and nonretarded electrodynamics. Relevant modes are discussed within the free electron gas model. Subsequently, the energy of the system is obtained and the field is quantized to obtain the surface-plasmon operators. Using the interaction Hamiltonian of Ritchie, and presenting the photon field via quantized vector potential, we determine the radiative decay rate of the plasmons per unit solid angle as a function of the geometric and excitation parameters.

\section{Nonretarded Potential \& Dispersion Relations}\label{r1}

The toroidal coordinates $(\mu,\eta,\varphi)$ are related to the Cartesian coordinates $(x,y,z)$ via:
\begin{eqnarray}\label{5R-C}
\begin{aligned}
x=a\frac{\sinh\mu\cos\varphi}{\cosh\mu-\cos\eta}~;~& 
y=a\frac{\sinh\mu\sin\varphi}{\cosh\mu-\cos\eta}~;&
z=a\frac{\sin\eta}{\cosh\mu-\cos\eta},
\end{aligned}
\end{eqnarray}
where $a>0$ is an overall scaling factor, $\mu \geq 0, $ and $\eta, \phi \in [0, 2\pi)$. The associated toroidal scale factors are given by
\begin{eqnarray}
\begin{aligned}
h_\mu=h_\eta=\frac{a}{\cosh\mu-\cos\eta};
\hspace{0.1in}
h_\varphi=\frac{a\sinh\mu}{\cosh\mu-\cos\eta}.
\end{aligned}
\end{eqnarray}
For  any fixed value $\mu_0 > 0,$ the equation  $\mu=\mu_0$ describes the surface of a torus whose inside and outside domains are given by  $\mu > \mu_0$ and 
$\mu  <\mu_0,$ respectively.  \cite{LSSFA}.\\
We consider a solid torus in the toroidal coordinate system $(\mu,\eta,\varphi)$
whose boundary surface is described by $\mu=\mu_0$ and its dielectric function is denoted by $\varepsilon.$ The surface modes of a single and a composite nanoring were recently reported in \cite{garapati2}. Specifically, following the quasi-static formulation of the toroidal  problem, the graphic presentation, and the plasmon dispersion relations of engendered normal modes in \cite{garapati}, we  write the scalar electric potential of a nanoring  as:
\begin{equation}\label{5Phi}
\Phi({\bf r},t)= 
\Theta(\mu-\mu_0) \,  
\Phi_{\text{i}}({\bf r},t)+ 
\Theta(\mu_0-\mu) \,  
\Phi_{\text{o}}({\bf r},t),
\end{equation}
where $\Phi_{\text{i}}$ and   $\Phi_{\text{o}}$ denote the inside- and the outside-potentials \cite{garapati2,KUYU}:
\begin{equation} \label{5R-Pin}
\Phi_{\text{i}}({\bf r},t)  =  f(\mu,\eta)
\sum\limits_{m,n} \, 
C_{mn}(t)\,
P_{n}^m	\, 
Q_n^m(\mu) 
e^{im\varphi}\, e^{in\eta},	
\qquad  \mu_0 \le  \mu
\end{equation}
\begin{equation} \label{5R-Pout}
\Phi_{\text{o}}({\bf r},t)  =  f(\mu,\eta)
\sum\limits_{m,n} \, 
C_{mn}(t)  \,
P_{n}^m(\mu)		\, 
Q_n^m\,	
e^{im\varphi} \, e^{in\eta}, 
\qquad \mu \le \mu_0, 
\end{equation}
where $m,n= 0, \pm 1, \pm 2, \cdots,$ $C_{mn}(t)$ are the complex-valued time dependent amplitudes,
\begin{equation}\label{5R-f}
f(\mu,\eta)=
\sqrt{\cosh\mu-\cos\eta} \, ,
\end{equation}
and $P_{n}^m(\mu),$ $Q_n^m(\mu),$ $P'^m_n(\mu),$ $Q;^m_n(\mu)$ stand for the associated Legendre functions of the first and second kind
$P^{m}_{n-\frac12}(\cosh\mu),$ $Q^{m}_{n-\frac12}(\cosh\mu),$ and their derivatives with respect to $\cosh\mu$. Their evaluations at the boundary $\mu=\mu_0$,
$P^{m}_{n-\frac12}(\cosh\mu_0),$ $Q^{m}_{n-\frac12}(\cosh\mu_0),$ are shown by $P_{n}^m$ and $Q_n^m,$ respectively. This convention will be used throughout the rest of the paper along with 
$f$ and $f_0$  to  denote $f(\mu,\eta)$ and $f(\mu_0, \eta)$, respectively. Moreover, the format adopted in Eqs.~\eqref{5R-Pin} and \eqref{5R-Pout} implies the continuity of the potential across the toroidal boundary; i.e., 
\begin{equation}
{\Phi_{\textrm{i}}}\Big|_{\mu=\mu_0}=
{\Phi_{\textrm{o}}}\Big|_{\mu=\mu_0}.
\end{equation}
Since $\vec \nabla ^2{\Phi}=0$ for the $\mu<\mu_0$ and $\mu>\mu_0,$ the Poisson equation takes the form 
\begin{equation}
\vec \nabla ^2{\Phi}=	 - \frac{4\pi}{h_\mu} \sigma
\delta(\mu-\mu_0),
\end{equation}
where $\sigma$ denotes the surface charge density on $\mu=\mu_0$ and $\delta(x)$ is the Dirac delta function.

In toroidal coordinate system (see \cite{morse}), the Laplacian of a function $F$ in the form $F=fg,$ where $f$ is given by Eq. \eqref{5R-f}, may be written as:
\begin{equation}\label{5F}
\vec \nabla ^2{F} = a^{-2} \, f^5  \mathcal{D} g,
\end{equation}
where the operator $D$ is given by:
\begin{equation}\label{5R-D}
\mathcal{D}= \coth\mu \frac{\partial}{\partial \mu} +
\frac{\partial^2}{\partial\mu^2 } 
+ \frac{\partial^2}{\partial\eta^2}+\frac{1}{\sinh^2\mu}
\frac{\partial^2}{\partial\varphi^2}+\frac14.
\end{equation}
In view of Eqs.~\eqref{5R-Pin} and \eqref{5R-Pout}, we may write the inside and outside potentials as $\Phi_{\text{i}}= f   \Psi_{\text{i}}$ and $ \Phi_{\text{o}}= f  \Psi_{\text{o}},$ where 
\begin{equation}\label{5R-PsiIn} 
\Psi_{\text{i}}({\bf r},t)  	= \sum\limits_{m,n} C_{mn}(t)
P_{n}^m\, Q_n^m(\mu)	e^{im\varphi}e^{in\eta},  
\end{equation}
\begin{equation}\label{5R-PsiOut} 
\Psi_{\text{o}}({\bf r},t)  	= \sum\limits_{m,n} C_{mn}(t)
P_{n}^m(\mu)\, Q_n^m\,	e^{im\varphi}e^{in\eta}. 
\end{equation}
Using Eq.~\eqref{5F}, the Laplacian of $\Phi$ given by 
Eq.~\eqref{5Phi} may  be calculated as:
\begin{equation}\label{5R-L2}
\vec \nabla ^2{\Phi} = 
a^{-2} f^5 
\mathcal{D} \Psi,
\end{equation}
where
\begin{equation}\label{5R-L3}
\Psi =	
\Theta(\mu-\mu_0) \Psi_{\text{i}} +
\Theta(\mu_0-\mu)  \Psi_{\text{o}}.
\end{equation}
To find $\mathcal{D} \Psi,$ noting that $\Theta$ only depends  on $\mu,$ it suffices to calculate the expressions for
$\frac{\partial \Psi}{\partial\mu}$ and $\frac{\partial^2 \Psi}{\partial\mu^2}.$ \\
Using the continuity of the potential $\Phi$ across the boundary surface,
it follows that $\Psi_{\text{i}}= \Psi_{\text{o}}$ at $\mu=\mu_0$ and thus the term  
$\delta(\mu-\mu_0)\Psi_{\text{i}}-\delta(\mu_0-\mu) \Psi_{\text{o}}$ vanishes identically yielding
\begin{equation}\label{5dPsi}
\frac{\partial\Psi}{\partial \mu}=\Theta(\mu-\mu_0) 
\frac{\partial\Psi_{\text{i}}}{\partial\mu}+
\Theta(\mu_0-\mu) \frac{\partial\Psi_{\text{o}}}{\partial\mu}.
\end{equation} 
Differentiating Eq.~\eqref{5dPsi} once again with respect to $\mu$ gives
\begin{equation}\label{5ddPsi}
\frac{\partial^2\Psi}{\partial \mu^2}=
\delta(\mu-\mu_0)\frac{\partial\Psi_{\text{i}}}{\partial\mu}-\delta(\mu_0-\mu)
\frac{\partial\Psi_{\text{o}}}  {\partial\mu}
+\Theta(\mu-\mu_0) \frac{\partial^2\Psi_{\text{i}}}{\partial\mu^2}+
\Theta(\mu_0-\mu) 		\frac{\partial^2\Psi_{\text{o}}}{\partial\mu^2}.
\end{equation}
Using the obtained expressions for $\frac{\partial \Psi}{\partial\mu}$ and $\frac{\partial^2 \Psi}{\partial\mu^2}$ given in Eq.~\eqref{5dPsi} and Eq.~\eqref{5ddPsi}, one may write  $\mathcal{D} \Psi$ as:
\begin{equation}\label{5R-D1}
\mathcal{D} \Psi
=\delta(\mu-\mu_0)
\left(\frac{\partial\Psi_{\text{i}}}{\partial\mu}-
\frac{\partial\Psi_{\text{o}}}{\partial\mu}\right)
+\Theta(\mu-\mu_0)\mathcal{D}\Psi_{\text{i}}
+\Theta(\mu_0-\mu)\mathcal{D}\Psi_{\text{o}}.
\end{equation}
Now, since $\vec \nabla ^2 \Phi_{\text{i}}=0$ for $\mu > \mu_0$ and 	$\vec \nabla ^2 \Phi_{\text{o}}=0$ for $\mu < \mu_0,$ it follows from Eq.~\eqref{5F} that
$\mathcal{D} \Psi_{\text{i}}=0$ for $\mu > \mu_0$ and $\mathcal{D} \Psi_{\text{o}}=0$ for $\mu < \mu_0,$ and thus the last two terms in Eq.~\eqref{5R-D1} both vanish identically. As a result, Eq.~\eqref{5R-D1} and Eq.~\eqref{5R-L2} imply:
\begin{equation}\label{5LPhi}
\vec \nabla ^2{\Phi} = 
a^{-2} f^5 \delta(\mu-\mu_0)
\left(\frac{\partial\Psi_{\text{i}}}{\partial\mu}-
\frac{\partial\Psi_{\text{o}}}{\partial\mu}\right),		
\end{equation}	
where from Eq.~\eqref{5R-PsiIn} and Eq.~\eqref{5R-PsiOut}
\begin{equation}\label{5s2}
\frac{\partial\Psi_{\text{i}}}{\partial\mu}-
\frac{\partial\Psi_{\text{o}}}{\partial\mu}=\sinh\mu
\sum_{m,n} C_{mn}(t)\, \mathcal{W}^m_n(\mu)\, e^{im\varphi}e^{in\eta},
\end{equation}
with $\mathcal{W}^m_n(\mu)$ denoting the Wronskian given as:
\begin{eqnarray}\label{5R-W}
\mathcal{W}^m_n(\mu)=P^m_n(\mu)Q'^m_n(\mu)-P'^m_n(\mu)Q^m_n(\mu) = -\frac{\Gamma(n+m+\frac{1}{2} )}{\Gamma(n-m+\frac{1}{2})\sinh\mu}~.
\end{eqnarray}
Finally, in view of the Poisson equation,  Eq.~\eqref{5LPhi} and  Eq.~\eqref{5s2}  imply the claimed expression for the surface charge density given as: 
\begin{equation}\label{5sigma}
\sigma=-\frac{\sinh\mu_0\, f_0^3}{4 \pi a}\sum_{m,n} C_{mn}(t)\mathcal{W}^m_n(\mu_0)e^{im\varphi}e^{in\eta}~.
\end{equation}
On the other hand, the surface charge density may be expressed using the polarization argument given in Eq.~\eqref{0-sddot1}.
To establish $\ddot{\sigma}$, we first observe that \cite{B}, 
\begin{equation}\label{5grad1}
\eval{\big (\mathbf{\hat{e}}_\mu\cdot\vec \nabla  \Phi_{\text{i}} \big )}_{\mu=\mu_0}=
\frac{1}{h_{\mu_0}}\eval{\frac{\partial \Phi_{\text{i}}}{\partial\mu}}_{\mu=\mu_0}.
\end{equation}
Letting  $\Phi_{\text{i}}= f   \Psi_{\text{i}}$  as above, together with the facts that $h_{\mu_0}=a f_0^{-2}$ and 
\begin{equation}
\frac{\partial}{\partial \mu}= \sinh \mu \frac{\partial}{\partial \cosh \mu},
\end{equation}
we may write the right-hand side of Eq.~\eqref{5grad1} as:
\begin{equation}\label{5grad2}
\frac{\sinh\mu_0\, 	f_0^3}{a}
\eval{\left( \frac{1}{2f^2}\Psi_{\text{i}}+
	\frac{\partial\Psi_{\text{i}}}{\partial\cosh\mu}\right)}_{\mu=\mu_0}.
\end{equation}
Utilizing the expression for $\Psi_{\text{i}}$ given in Eq.~\eqref{5R-PsiIn}, we arrive at Eq.~\eqref{5R-s3} in view of Eq.~\eqref{5grad2}.\\
A straightforward calculation shows that 
\begin{equation}\label{5R-s3}
\ddot{\sigma}= -\frac{\omega_p^2\sinh\mu_0\, f_0^3}{4\pi a} 
\sum\limits_{m,n} C_{mn}(t)
\bigg (\frac{1}{2f_0^2}P^m_n\,Q^m_n\,
+
P^m_n\,Q'^m_n\bigg )e^{im\varphi}e^{in\eta}.
\end{equation}

Differentiating Eq. \eqref{5sigma} twice with respect to time $t$ and comparing the result with Eq. \eqref{5R-s3} , it follows form the  orthogonality of the system 
$\{e^{im\varphi}\}$ that for each fixed $m=0, \pm 1, \dots,$ the amplitudes $C_{mn}$ satisfy
\begin{equation}\label{5R-HA}
\sum\limits_{n}\bigg\{ \ddot C_{mn}(t)\mathcal W(\mu_0)+\omega_p^2 C_{mn}(t)
\Big [ \frac{1}{2f_0^2}\, 	P^m_n\,Q^m_n
+P^m_n\,Q'^m_n\Big ]\bigg\}  e^{in\eta}=0.
\end{equation}
For $\mu_0 >0,$ function $\ \frac{1}{\cosh\mu_0 - \cos\eta}$
can be expanded in a uniformly convergent Fourier series on $[-\pi, \pi]$ as
\begin{equation}\label{5FS}
\frac{1}{f_0^2}= \sum_{l=-\infty}^\infty a_{l}
e^{il\eta}\ , 
\end{equation}
where the Fourier coefficients $a_l$ are given by:
\begin{equation}\label{5FSal}
a_l= \frac {1}{2\pi } \int_{-\pi}^{\pi} 
\frac{e^{-il\eta}}{\cosh\mu_0 - \cos\eta} \, d\eta.
\end{equation}
Using the substitution: $z=e^{i\eta},$ $\dfrac{1}{iz} \, dz=  d\eta,$ and $\cos\eta = \frac12 (z +z^{-1}),$ the integral in \eqref{5FSal} may be  transformed into a contour integral over the positively oriented unit circle $\partial\mathbb{D}$ in the complex plane as: 
\begin{eqnarray}
a_l=
\frac {1}{2\pi } \oint _{\partial\mathbb{D}} \frac{ z^{-l}}{\cosh\mu_0 - \frac12(z+\frac1z)}\, \frac{dz}{iz}		
=
-	\frac {1}{\pi i }  \oint _{\partial\mathbb{D}}   g(z) \, dz,
\end{eqnarray}
where
\begin{equation}\label{5g}
g(z)=  \frac{z^{-l}}{(z-e^{\mu_0})(z-e^{-\mu_0})}
\end{equation}
is a rational function having  two simple poles at $z= e^{\pm\mu_0}$  for all $l \in \mathbb{Z}$ with only $e^{-\mu_0}$  lying inside the unit circle and a pole of order $l$ at 
$z=0$ for $l \geq 1$. It follows from the Residue Theorem \cite{conway} that
\begin{equation}\label{5al}
a_l=
\begin{cases} 				
-2\text{Res\,}_{e^{-\mu_0}}(g)&  if  \quad l \leq 0 \\
-2\text{Res\,}_{e^{-\mu_0}} (g)  -2\text{Res\,}_0 (g) &  if \quad  l \geq 1 
\end{cases},
\end{equation} 
where
\begin{equation}\label{5Res1}
\text{Res\,}_{e^{-\mu_0}}(g)= \lim_{z\to e^{-\mu_0}} (z-e^{-\mu_0}) g(z)=
\frac{-e^{l\mu_0}}{2\sinh\mu_0}
\end{equation}
and,   for $l \geq1,$
\begin{equation}\label{5Res2}
\text{Res\,}_0(g)=\frac{1}{(l-1)!} \lim\limits_{z\to 0} \frac{d^{l-1}}{dz^{l-1}} \left (z^l g(z)\right).
\end{equation}
Noting that:
\begin{eqnarray}
z^l g(z)= \frac{1}{(z-e^{\mu_0})(z-e^{-\mu_0})}			
= \frac{1}{2\sinh\mu_0}
\left[\frac{1}{z-e^{\mu_0}}-\frac{1}{z-e^{-\mu_0}} \right],
\end{eqnarray}
together with the Leibniz's rule:
\begin{equation}
\frac{d^{l-1} }{dz^{l-1}}(z-e^{\pm \mu_0})^{-1}=(-1)^{l-1}\, (l-1)! (z-e^{\pm \mu_0})^{-l},
\end{equation}
it follows form Eq.~\eqref{5Res2} that:
\begin{eqnarray}\label{5Res3}
\text{Res\,}_0(g) =\frac{1}{2\sinh\mu_0} 
\bigg [ (-1)^{2l-1}\, e^{-l\mu_0}			
-(-1)^{2l-1}\, e^{l\mu_0}\bigg ] 	
=- \frac{e^{-l\mu_0}-e^{l\mu_0}}{2\sinh\mu_0} .
\end{eqnarray}
Finally, in view of Eqs.~\eqref{5al}--\eqref{5Res3}, we conclude that:
\begin{equation}\label{FSSal}
a_l= \frac{e^{-|l| \mu_o}}{\sinh\mu_0}, \quad l=0, \pm 1, \pm 2, \cdots.
\end{equation}
Now, replacing  $\dfrac{1}{2f_{0}^2}$ by its Fourier series expansion:
\begin{eqnarray}\label{5R-f4}
\frac{1}{2f_0^2}&=& \sum_{l=-\infty}^{\infty} \frac{e^{-|l|\mu_0}}{2\sinh\mu_0} e^{il\eta},
\end{eqnarray}
and utilizing the substitution $n+l \mapsto n,$ it  follows from the  orthogonality of the system $\{e^{in\eta}\}$ that for each fixed $n\in\mathbb{Z}$ in Eq.~\eqref{5R-HA}, we may write: 
\begin{equation}\label{5-HA}
\ddot C_{mn}(t)
+\sum_{ l \in \mathbb{Z}}	\mathcal V^m_{n,l}\,	
C_{m,n-l}(t)=0  , 
\end{equation}
where
\begin{eqnarray}\label{5R-U1}
\mathcal V^m_{n,0}= \frac{ \omega_p^2\, \left( \frac{1}{2\sinh\mu_0}\, P^m_n\,Q^m_n+P^m_n\,Q'^m_n \right)}{\mathcal{W}^m_n(\mu)}  , 
\end{eqnarray}
and
\begin{equation}\label{5R-V2}
\mathcal V^m_{n,l}= 	
\frac{ \omega_p^2\, 	e^{-|l|\mu_0} \, P^m_{n-l}\, Q^m_{n-l}}{2\sinh\mu_0\mathcal{W}^m_n(\mu)}, \qquad l\neq 0. 
\end{equation}

\noindent \textbf{ Cylindrical limit}

In order to use the well-developed theories of a cylindrical model, it is often useful to consider the cylindrical limit of the toroidal coordinate through a transformation from toroidal to cylindrical geometry by keeping the minor radius of a torus, $r$, fixed while the major radius $R$ increases indefinitely with $\mu=\mu_0$. Under these assumptions, the ring transforms into the column and one could utilize the theories of a cylindrical model. We refer the reader to \cite{zheng,Love_Plasma} for a detailed description of the cylindrical model. To this end, we define the ratio:
\begin{equation}\label{5-Cyl}
 k=\frac{m}{u}, 
 \end{equation}
 and we let $m$ and $\mu$ increase indefinitely such that $k$ remains fixed. Following Eq.~\eqref{5-Cyl}, we may write:
\begin{equation}
\frac{d}{dk}= \frac{d}{du}\frac{du}{dk}= \frac{-k^2}{m}\frac{d}{du}.
\end{equation}
Next, we consider the cylindrical limits of $\mathcal V^m_{n,l}$. Using Eq.~\eqref{5R-U1}, we can write:
\begin{eqnarray}\label{R-U}
\mathcal V^m_{n,0}& =& \frac{ \omega_p^2\, \left[ \frac{a_0}{2}\, P^m_n\,Q^m_n+P^m_n\,  Q'^m_n \right]}{ \mathcal{W}^m_n(\mu_0)} \notag\\
&=& \frac{ \omega_p^2\, P^m_n\, \left[ \frac{1}{2\sinh\mu_0}   Q_n^m  + Q'^m_n  \right]}{\mathcal{W}^m_n(\mu_0)}			\notag\\
&=& \frac{  \omega_p^2}{Q^m_n\,[\log\left| P^m_n/Q^m_n\right| ]'} \left[ \frac{Q_n^m}{2\sinh \mu_0  } +Q'^m_n \right] \notag \\
&=& \frac{\omega_p^2 }{\left(\log \left| P^m_n/Q^m_n\right| \right)'}
\left[\frac{ 1}{2\sinh \mu_0 }  +\left( \log \left|Q^m_n\right| \right)' \right], 
\end{eqnarray}
noting:
\begin{eqnarray}\label{5w}
\mathcal W^m_n(\mu)=P^m_n\, Q^m_n\, \left (\log\left| \frac{P^m_n}{Q^m_n}\right |\right )'. 
\end{eqnarray}	
Following Eq.~\eqref{5R-V2}, we obtain:
\begin{eqnarray}\label{5R-V}
\mathcal V^m_{n,l}&=&
\frac{ \omega_p^2\, 	a_l\, P^m_{n-l}\, Q^m_{n-l}}{2\mathcal{W}^m_n(\mu_0)}		\notag\\
&=&\frac{ \omega_p^2\, e^{-|l|\mu} \,P^m_{n-l}\, Q^m_{n-l}}{2\sinh \mu \,\mathcal{W}^m_n(\mu)}\notag \\
&=& \frac{\omega_p^2}{(\log\left| P^m_n/Q^m_n\right| )'} 
\times \left[ \frac{e^{-|l|\mu}}{2\sinh \mu_0}\, \frac{ P^m_{n-l}\, Q^m_{n-l}}{ P^m_n\, Q^m_n\,}\right]. 
\end{eqnarray}
The cylindrical limit of $P_n^m (\mu_0)$ and $Q_n^m (\mu_0)$ in terms of modified Bessel functions of  first and second kind are given by \cite{g&r:table}: 
\begin{eqnarray}\label{5pq}
P_n^m (\mu_0) &=& \frac{(-1)^{n}}{\pi^2}\, \mathcal{C}_m (\mu_0) \sqrt k\,
\left[  K_n \left( k\right) + \mathcal{O} \left( \frac{1}{m}\right) \right] ,\\
Q_n^m (\mu_0) &=&  \mathcal{C}_m (\mu_0) \sqrt k\,\left[ I_n \left( k\right) + \mathcal{O} \left( \frac{1}{m}\right) \right]~,
\end{eqnarray}
where 
\begin{equation}\label{5Cc}
\mathcal{C}_m (\mu_0) = (-1)^{m}\,  (m-1)!\,   \sqrt{\frac{\pi }{2}}.
\end{equation} 
Now, in view of  Eqs.~\eqref{5pq} --\eqref{5Cc}, 
\begin{eqnarray}\label{5log}
\left (\log\left |\frac{P^m_n}{Q^m_n}\right |\right )'&=& \frac{dr}{du}\bigg \{  \frac{K'_n(k)++ \mathcal{O} \left( \frac{1}{m}\right) }{\left[  K_n \left(k\right) + \mathcal{O} \left( \frac{1}{m}\right) \right] }
-\frac{I'_n(k)++ \mathcal{O} \left( \frac{1}{m}\right) }{\left[  I_n \left( k\right) + \mathcal{O} \left( \frac{1}{m}\right) \right] } \bigg \} \notag \\
&=& \frac{dk}{du}\bigg [  \frac{K'_n(k) I_n(k)-K_n(k) I'_n(k)}{  K_n \left( k\right) I_n \left( k\right)}\bigg ]
\left[ 1+ 
\mathcal{O} \left( \frac{1}{m}\right)\right]\notag \\
&=& \frac{-k^2}{m}\bigg [  \frac{1}{  k\, K_n \left( k\right) I_n \left( k\right)}\bigg ]\left[ 1+ 
\mathcal{O} \left( \frac{1}{m}\right)\right]\notag \\
&=&\frac{-k}{ m}\left[ \frac{1}{ K_n \left( k\right) I_n \left( k\right)}+ 
\mathcal{O} \left( \frac{1}{m}\right)\right] , 
\end{eqnarray}
and 
\begin{eqnarray}\label{5log2}
\left (\log |Q^m_n|\right )'		
&=& \frac{-k^2}{m}\frac{d}{du}\bigg \{  \log \left| \mathcal{C}_m (\mu_0) \right| + \log \left|  I_n \left(k\right) +\mathcal{O} \left( \frac{1}{m}\right)\right|\bigg \}'  \\
&=& \frac{-k}{m}\bigg [ \frac{1}{2k}+ \frac{I'_n(k)}{  I_n \left( k\right)}+\mathcal{O} \left( \frac{1}{m}\right)\bigg ],
\end{eqnarray}
where, in Eq.~\eqref{5log}, we have utilized the Wronskian identity for modified Bessel functions  \cite{LSSFA}: 
$$I(z)K'(z)-I'(z)K(z) = -\frac1z \quad \left( z\neq 0\right) . $$
Finally, noting that 
\begin{equation}
\sinh\mu= \sqrt{u^2-1}= \sqrt{\frac{m^2}{k^2}-1}= \frac mk \left[1+\mathcal O\left( \frac1m\right) \right], 
\end{equation}
and 
\begin{equation}
e^{-|l|\mu_0}=\left( \cosh\mu_0+ \sinh\mu_0\right) ^{-|l|}=  \left( \frac mk\right) ^{-|l|}\left[1+\mathcal O\left( \frac1m\right) \right], 
\end{equation}
we obtain:
\begin{eqnarray}
&&\mathcal{V}_{n,0}^m  =  \omega_{mn}^2 + \mathcal{O} \left( \frac{1}{m}\right) \ , \label{5U}\\
&&\mathcal{V}_{n,l}^m =  \mathcal{O} \left( \frac{1}{m^{|l|+1}} \right)~,\label{5V}
\end{eqnarray}
where
\begin{equation}\label{5R-nw}
\omega_{mn}^2  = \omega_p^2\, k\,  K_n \left( k\right) I'_n \left(k\right)~,
\end{equation}
which is the same as the cylindrical-limit frequency given in \cite{Love_Plasma}. \\

\section{Classical Energy}
Potential energy, following Eq.~\eqref{0-V}, and using 
Eq.~\eqref{5sigma}, may be expressed as:
\begin{multline}\label{5V2}
V=-\frac{a^2 \sinh^2\mu_0}{8\pi} 		
\sum\limits_{m,n} \sum\limits_{\hat{m},\hat{n}}
\overline {C}_{\hat{m}\hat{n}}(t){C}_{mn}(t) 
P^{\hat m}_{\hat n}\, 
Q^{\hat m}_{\hat n}\, 
\mathcal{W} (\mu_0)\\
\times \int_{0}^{2\pi} \int_{-\pi}^{\pi}
e^{i(n-\hat{n})\eta}\, 
e^{i(m-\hat{m})\phi}\, 
d{\eta}d{\varphi},
\end{multline}
where using orthogonality in $n$ and $\eta$ ($m$ and $\varphi$) as:
\begin{equation}
\int_0^{2\pi} e^{i(n-n')\theta}\, d\theta = 2\pi \delta_{nn'}, 
\end{equation}
and following the same procedure as was taken in Chapters 1 and 2, we may find the potential energy $V$ as: 
\begin{equation}\label{5R-V1}
V= \frac{\pi a \sinh^2\mu_0}{2} 
\sum\limits_{m,n} \big| C_{mn}(t)\big |^2P^m_n\,  Q^m_n\, \mathcal{W}^m_n(\mu_0). 
\end{equation}

Kinetic energy calculation's steps are different from the general procedure we introduced in Chapter~1 due to the coupling between the modes in case of nanoring. The fact that the geometric structure of a ring is not simply connected and that Laplace's equation is not separable in this coordinate system results in some changes. Eq.~\eqref{5-HA} shows that amplitudes in this case does not undergo pure harmonic oscillator equation of motion.

The kinetic energy is given by Eq.~\eqref{0-T}. The acceleration is the force per unit mass following Eq.~\eqref{0-udd1} gives:
\begin{equation}
\ddot{\vec{u}}=\frac{e}{m_e}
\nabla \, \Phi_{\text i}, 
\end{equation}
where 
\begin{equation}
\Phi_{\text i}= f(\mu,\eta)
\sum\limits_{m,n} 
C_{mn}(t) P^m_{n}
Q^m_{n}
e^{in\eta}e^{im\varphi}.
\end{equation}
Hence
\begin{equation}\label{udot}
\dot{\vec{u}}=\frac{e}{m_e}
\int^t \, \nabla \Phi_{\text i}\,  dt'
\end{equation}
and 
\begin{equation}\label{ring-uddot}
\dot{\vec{u}}\cdot \dot{\vec{u}} = \bigg( \frac{e}{m_e}\bigg) ^2
\int^t \int^t \, \nabla \Phi_{\text i}\cdot \nabla \Phi_{\text i}\,   dt'dt''. 
\end{equation}
Using the  identity:
\begin{equation}
\nabla{\dot\Phi_{\text i}}\cdot{\nabla{\dot\Phi_{\text i}}}=
\nabla\cdot\big{(}\dot\Phi_{\text i}{\nabla{\dot\Phi_{\text i}}}\big{)}.
\end{equation}
in Eq.~\eqref{ring-uddot}, we can write:
\begin{equation}
\dot{\vec{u}}\cdot \dot{\vec{u}} = \bigg( \frac{e}{m_e}\bigg) ^2
\int^t \int^{t}\, \nabla \cdot \left( \Phi_{\text i} \nabla \Phi_{\text i}\right) \,   dt' dt''. 
\end{equation}
Hence 

\begin{equation}\label{kinetic_0}
T= \frac{m_e n_0}{2} \bigg( \frac{e}{m_e}\bigg) ^2
\int_{ \textrm{volume}}
\int^t \int^{t}\, \nabla \cdot \left( \Phi_{\text i} \nabla \Phi_{\text i}\right) \,   dt' dt''\, d\mathcal V. 
\end{equation}
The Divergence theorem leads to:
\begin{eqnarray}
T= \frac{m_e n_0}{2} \bigg( \frac{e}{m_e}\bigg) ^2
\int_{ \textrm{surf}}
\int^t \int^{t}\,  \left( \Phi_{\text i} \nabla \Phi_{\text i}\right) \cdot{\hat{e}_{\mu}}  \,  dt' dt''\, d\mathcal A.
\end{eqnarray}
By means of the expression for inside potential given in Eqs.~\eqref{5R-Pin} and ~\eqref{5R-Pout}, we write: 
\begin{equation}
T= \frac{m_e n_0}{2} \bigg( \frac{e}{m_e}\bigg) ^2
\int^t \int^{t}\,dt dt' \times 
\int_{\eta}\int_{\varphi} \left( \overline{\Phi_{\text i}}\,  \frac{\partial \Phi_{\text i}}{\partial \mu}\right) \frac{1}{h_\mu}\, h_\eta h\varphi \, d\eta d\varphi, 
\end{equation}
where $\overline{\Phi_{\text i}}$ denotes the complex conjugate of inside potential given in Eq.~\eqref{5R-Pin} as well as using the fact:
\begin{equation}
\nabla \cdot e_\mu = \frac{\partial}{\partial \mu} \frac{1}{h_\mu}. 
\end{equation}
The chain rule gives:
\begin{equation}
\frac{\partial \Phi_{\text i}}{\partial \mu}= \frac{\sinh\mu}{f} \sum\limits_{m,n}C_{mn}(t) P^m_n\,  Q^m_n\,e^{in\eta}e^{im\varphi} + f\sum\limits_{m,n}C_{mn}(t) P^m_n\,  Q'^m_n\,e^{in\eta}e^{im\varphi}. 
\end{equation}
Hence: 
\begin{multline}
T= \frac{m_e n_0}{2} \bigg( \frac{e}{m_e}\bigg) ^2
\int^t \int^{t}\,dt' dt''  \\
\times\int\int \left[ \frac{\sinh\mu}{f_0} \sum\limits_{m,n}C_{mn}(t') P^m_n\,  Q^m_n\,e^{in\eta}e^{im\varphi} + f_0\sinh\mu\, \sum\limits_{m,n}C_{mn}(t') P^m_n\,  Q'^m_n\,e^{in\eta}e^{im\varphi}\right] \\
\times\left[ f_0\sum\limits_{\hat m,\hat n}\overline{C_{\hat m\hat n}(t'') } P^{\hat m}_{\hat n}\,  Q^{\hat m}_{\hat n}\,  e^{-i\hat n\eta}e^{-i\hat m\varphi} \right] 
\frac{1}{h_\mu}\, h_\eta h\varphi \, d\eta d\varphi. 
\end{multline}
Using scale factor relations:
\begin{multline}
T= \frac{m_e n_0}{2} \bigg( \frac{e}{m_e}\bigg) ^2
\int^t \int^{t}\,dt' dt''  \\
\times\int\int \left[ \frac{\sinh\mu}{f_0} \sum\limits_{m,n}C_{mn}(t') P^m_n\,  Q^m_n\,e^{in\eta}e^{im\varphi} + f_0\sinh\mu_0 \sum\limits_{m,n}C_{mn}(t') P^m_n\,  Q'^m_n\,e^{in\eta}e^{im\varphi}\right] \\
\times\left[ f_0\sum\limits_{\hat m,\hat n}\overline{C_{\hat m\hat n}(t'') } P^{\hat m}_{\hat n}\,  Q^{\hat m}_{\hat n}\,  e^{-i\hat n\eta}e^{-i\hat m\varphi} \right] 
\frac{\sinh\mu_0}{f_0^2} \, d\eta d\varphi. 
\end{multline}
and some simple calculations:
\begin{multline}
T=\frac{m_e n_0}{2} \bigg( \frac{e}{m_e}\bigg) ^2
\int^t \int^{t}\,dt' dt''  \\
\times\int\int \left[  \frac{1}{f_0^2}\sum\limits_{m,n}C_{mn}(t') P^m_n\,  Q^m_n\,e^{in\eta}e^{im\varphi} +\sum\limits_{m,n}C_{mn}(t') P^m_n\,  Q'^m_n\,e^{in\eta}e^{im\varphi}\right] \\
\times\left[ \sum\limits_{\hat m,\hat n}\overline{C_{\hat m\hat n}(t'') } P^{\hat m}_{\hat n}\,  Q^{\hat m}_{\hat n}\,  e^{-i\hat n\eta}e^{-i\hat m\varphi} \right] 
\sinh^2\mu_0 \, d\eta d\varphi. 
\end{multline}
Using expression for $\frac{1}{f_0^2}$, as given in Eq.~\eqref{5R-f4}, we get 
\begin{multline}
T=a^2\sinh^2\mu_0\, \frac{m_e n_0}{2} \bigg( \frac{e}{m_e}\bigg) ^2
\int^t \int^{t}\,dt' dt'' \\
\times\int\int \left[ \sum\limits_{m,n}\, \sum\limits_l a_l C_{mn}(t') P^m_n\,  Q^m_n\,\, e^{i(l+n)\eta}\,e^{im\varphi} +\sum\limits_{m,n}C_{mn}(t') P^m_n\,  Q'^m_n\,e^{in\eta}e^{im\varphi}\right] \\
\times\left[ \sum\limits_{\hat m,\hat n}\overline{C_{\hat m\hat n}(t'') } P^{\hat m}_{\hat n}\,  Q^{\hat m}_{\hat n}\,  e^{-i\hat n\eta}e^{-i\hat m\varphi} \right] 
\, d\eta d\varphi. 
\end{multline}
Letting $n+l\rightarrow n$ and isolating the term $l=0$, the orthogonality in $\eta$ and $\varphi$ implies:
\begin{multline}
T= 4\pi^2\,a^2\, \sinh^2\mu_0\, \frac{m_e n_0}{2} \bigg( \frac{e}{m_e}\bigg) ^2\, 	\int^t \int^{t}\,dt' dt''\\
\times \sum\limits_{mn} \bigg[ \overline{C_{mn}(t')}\, C_{mn}(t')\,\left( P^m_n\right) ^2\, Q^m_n\, Q'^m_n\\
+ \overline{C_{mn}(t')}P^m_n\, Q^m_n \sum\limits _l a_l\, C_{m, n-l}(t'')P^m_{n-l}\, Q^m_{n-l}\bigg], 
\end{multline}
with the bulk frequency identity given in Eq.~\eqref{1-29}. 
If we multiply and divide each term in $m,n$ by the Wronskian, and use the relations given in Eqs.~\eqref{5R-U1} and \eqref{5R-V2}, we can write:
\begin{multline}\label{5-Kin}
T= \pi\, \sinh^2\mu_0\,	a^2\,\int^t \int^{t}\,dt' dt''\\
\times \sum\limits_{mn}  \overline{C_{mn}(t')}\, P^m_n\, Q^m_n\, \mathcal W^m_n(\mu_0)\, \bigg[\mathcal V^m_{n,0} C_{mn}(t'') 
+  \sum\limits _{l\neq 0} \mathcal V^m_{n,l}\, C_{m, n-l}(t'')\bigg].
\end{multline}
From Eq.~\eqref{5-HA}, it follows that: 
\begin{equation}\label{5-NewC}
\ddot C_{mn}(t'')
=-\omega_p^2 \sum_{ l \in \mathbb{Z}}	\mathcal V^m_{n,l} \,	
C_{m,n-l}(t''). 
\end{equation}
Using Eq.~\eqref{5-NewC} in Eq.~\eqref{5-Kin} and integrating with respect to time $t$, we may write:
\begin{equation}\label{Tf}
T=-\frac{a^2 \pi\, \sinh^2\mu_0}{2}\,	
\sum\limits_{mn}   P^m_n\, Q^m_n\, \mathcal W^m_n(\mu_0)\, \dot C_{mn}(t)\, \int^t \overline{C_{mn}(t')}\, dt'.
\end{equation}
Inspired by cylindrical limit, we may assume amplitudes undergo harmonic oscillator equation of motion (see Eqs.~\eqref{5w}--\eqref{5R-nw} and \cite{Love_Plasma}). We therefore set:
\begin{equation}\label{5r_A}
{C}_{mn}(t)= -\frac{\ddot C_{mn}(t)}{\omega^2_{mn}}, 
\end{equation}
in Eq.~\eqref{Tf} to obtain:
\begin{equation}\label{Tf1}
T=\frac{a^2 \pi\, \sinh^2\mu_0}{2}\,	
\sum\limits_{mn}   P^m_n\, Q^m_n\, \mathcal W^m_n(\mu_0)\, \big|\dot C_{mn}(t)\big|^2. 
\end{equation}
One can easily find the total classic energy $E=T+V$, using Eqs.~\eqref{5R-V1}, \eqref{Tf1} as:
\begin{equation}\label{5R-E}
E= \frac{ a^2\pi\sinh^2\mu_0}{2}
\sum\limits_{m,n}\frac{1}{a\omega_{mn}^2} P^m_n\, Q^m_n\, \mathcal{W}^m_n(\mu_0)\, 
\left[|\dot{C}_{mn}(t)|^2+\omega_{mn}^2 |C_{mn}(t)|^2\right].						
\end{equation}
Complex coefficients $C_{mn}(t)$ could be written in the form \cite{Folland}:
\begin{equation}\label{5R-C1}
{C}_{mn}(t)= \frac{\gamma_{mn}}{2\omega_{mn}}c_{mn},
\end{equation}
for some complex time dependent functions $c_{mn}$ which are proportional to $e^{-i\omega_{mn} t}$ and for some constants $\gamma_{mn}$ which will be chosen later. The derivative with respect to time gives us:
\begin{equation}
\dot{C}_{mn}(t)= -i \frac{ \gamma_{mn}}{2} {c}_{mn}.				
\end{equation}
Using this, total energy could be written as:
\begin{equation}\label{5R-E1}
E= \frac{ a\pi\sinh^2\mu_0}{2}
\sum\limits_{m,n}\frac{P^m_n\, Q^m_n\, \mathcal{W}(\mu_0)}{\omega_{mn}^2}
\bigg [\frac{ \gamma^2_{mn}}{2}\big (\bar{c}_{mn}c_{mn}
+c_{mn}\bar{c}_{mn}\big )\bigg ], 			
\end{equation}
where $\bar c_{mn}$ denotes the complex conjugate of $c_{mn}$. \\

\section{Interaction Hamiltonian \& Radiative Decay Rate}
The full classic Hamiltonian could be written as: 
\begin{eqnarray}\label{5R-H4}
{H} = \sum\limits_{m,n}\frac{\hbar\omega_{mn}}{2}(\hat{c}^*_{mn}\hat{c}_{mn}+\hat{c}_{mn}\hat{c}^*_{mn}).
\end{eqnarray}
This is understood to act on bosonic Fock space \cite{Folland}. States of which we will denote using the usual Dirac ket notation. Using Wronskian in terms of Gamma function \cite{Abra} given in Eq.~\eqref{5R-W}, we can choose:
\begin{eqnarray}\label{alpha}
\gamma^2_{mn}= 
\frac{4 \hbar(-1)^{m}   \omega^3_{mn}}{\pi a\, P^m_n\, Q^m_n}
\frac{{\Gamma(n-m+\frac12)}}{ {\Gamma(n+m+\frac12)}}, 				
\end{eqnarray}
in Eq.~\eqref{5R-E1}. 

In analogy with other cases, the interaction of these two fields is described via the minimal coupling:
\begin{multline}\label{5Hem}
 H_{em} = \frac{n_0e^2}{2m_0c}
\sum\limits_{m,n} \sum\limits_\mathbf{s}\sum\limits_{q=1,2}
\int_{0}^{2\pi}\, \int_{0}^\infty
\frac{(c_{mn}-c^*_{mn})P^m_n\,	
	Q^m_n}{\omega^2_{mn}} \\
\times \sqrt{\frac{\hbar c^2} {\mathcal V \, \omega_s}} 			
\Big [ a_{\mathbf{s}q}e^{i\mathbf{s\cdot r}}
+a^*_{\mathbf{s}q}e^{-i\mathbf{s\cdot r}}\Big ]
(\mathbf{\hat{e}}_\mu\cdot\mathbf{\hat{e}}_q)\,
e^{im\varphi}e^{in\eta}h_\eta h_\varphi d\eta d\varphi. 
\end{multline}
then the matrix element for emission reduces to
\begin{equation}\label{5Mem1}
\mathcal M^{(q)}_{em}=\frac{n_0e^2}{m_0}\sqrt{\frac{\hbar}{\mathcal V\, \omega_s}}
\frac{\gamma_{mn}}{2i\omega^2_{mn}}
P^m_n \, Q^m_n\, I_{mn}^{(q)},
\end{equation}
with 
\begin{equation}
I_{mn}^{(1)} =\pi \int_{-\pi}^{\pi}\frac{1-\cosh\mu\cos\eta}{(\cosh\mu-\cos\eta)^{5/2}} 
\tilde E(\eta, \varphi)
\bigg [ J_{m+1}(K(\eta))
-J_{m-1}(K(\eta))\bigg ]d\eta,
\end{equation}
and
\begin{multline}
I_{mn}^{(2)} =\pi \int_{-\pi}^{\pi}\frac{\sin\psi(1-\cosh\mu\cos\eta)}{(\cosh\mu-\cos\eta)^{5/2}}
\tilde E(\eta, \varphi) 
\bigg [J_{m+1}(K(\eta))
+J_{m-1}(K(\eta))\bigg ]d\eta \\
- 2\pi \int\frac{\cos\psi\sinh\mu\sin\eta}{(\cosh\mu-\cos\eta)^{5/2}}J_{m}(K(\eta))\tilde E(\eta, \varphi) d\eta,
\end{multline}
where 
\begin{equation}
K(\eta)=\frac{a\omega_s \cos\psi\sinh\mu}{\cosh\mu-\cos\eta},
\end{equation}       
and
\begin{equation}
\tilde E(\eta, \varphi)=\exp\left\lbrace i(n\eta-\frac{a\omega_s\sin\psi\sin\eta}{\cosh\mu-\cos\eta})\right\rbrace.
\end{equation}
Lastly,  the decay rate per solid angle $\Omega$ is given by: 
\begin{equation}\label{5RDR-reduced}
\frac{d\gamma_{mn}}{d \Omega}  =\sum\limits_{q=1,2}
\frac{(n_0\, e^2)^2\left| P^m_n\,  Q^m_n\, \right| }{4\pi^3\, a \, m_0^2 \, c^3}
\left| \frac{ \Gamma(n-m+\frac12)}{\Gamma(n+m+\frac12)}\right|
\left|I_{mn}^{(q)}\right|^2.
\end{equation}

\setlength{\parindent}{0.5in}
\chapter{CONCLUSION}
\footnotetext[4]{Portions of this chapter were reprinted from: Bagherian, M. and Kouchekian, S. and Rothstein, I. and Passian, A., Quantization of surface charge density on hyperboloidal and paraboloidal domains with application to plasmon decay rate on nanoprobes,  125413, vol. 98, Sep 2018, with permission from \emph{Phys. Rev. B} \\
	Permission is included in Appendix C.}
On the basis of the simulated results, the presented quantization scheme emerges as a reasonable approach to obtaining the qualitative behavior of extended electronic systems, for which many body calculations may be demanding or unfeasible, analytically or computationally. 
  The obtained analytical results  
for the charge density oscillations on the surface of geometric entities with local curvature but with an extended dimension constitute first time results with potential for modeling quantum effects in plasmonics. 
The obtained analytical expressions for the operators associated with surface plasmon quantities,  help calculating 
interactions with other quantized fields, e.g., the interaction of the probe with a nearby quantum emitter, or with the radiation field of a quantum dipole. 
Owing to their apex symmetry and curvature, the comparison of the calculated quantities show that hyperboloidal and paraboloidal plasmons qualitatively exhibit similar  dispersion relations and radiative decay rates. 
We conclude that, the quantitative differences observed in the allowed resonance values of the dielectric function and the emitted radiation patterns are therefore primarily attributed to the difference in curvature  in the asymptotic region, away from the apex. From a comparison with the modes excited on a prolate spheroidal surface, for which experimentally typically only low energy dipolar and quadrupolar eigenmodes or their mixtures have been observed to contribute to far field radiation in the visible spectrum, we expect that also only low lying modes will contribute to the emission spectra of the probe. 
Unlike the all discrete spectrum of the quantum numbers associated with  the spheroidal modes, the eigenvalues characterizing the hyperboloidal or paraboloidal plasmons along their infinite dimensions exhibit a continuous spectrum, making a direct comparison of plasmon wavevectors unclear. However, visualization of the eigenmode field patterns, that is, how the fields fluctuate for given continuous eigenvalues in the case of hyperboloids and paraboloids, can facilitate the comparison  with the discrete eigenvalues for the spheroidal case.  For a given set of eigenmodes, photon emission from the higher apex curvature tips occurs with a more localized radiation pattern. Following the presented results for the allowed values of the dielectric functions $\varepsilon_{ml}(\zeta_0)$, $\varepsilon_{m\lambda}(\eta_0)$ and $\varepsilon_{mq}(\mu_0)$, the corresponding dispersion relations and radiative decay rates can be obtained for real materials from a comparison with the experimentally determined optical properties of solids (such the compilation by Hageman~\cite{Hag}, or by Palik~\cite{palik1997}).
In summary, the presented results can aid the design and fabrication of tips with specific photonic and plasmonic characteristics.
For example for a gold or silver tip, such as those used in electron emission experiments, using the presented results one may determine both the fabrication design parameters and the excitation laser wavelength and polarization. In such applications, a comparison of Fig.~\ref{Dispersion_HP} and \ref{Dispersion} with the optical properties of, for example silver~\cite{Hag}, indicates the availability of several resonance modes in the visible. 
The results can help the analysis of the radiation emitted  from the nanotips  as a result of electron or photon scattering, which are of importance to plasmonics in experiments such as EELS (electron energy loss spectroscopy) and SPM. For a specific material typically employed in plasmonics such as gold and silver, the results  offer estimates of the polarization, angular,  and spectral properties of the emitted radiation. In such instances, the emitted photons may be analyzed for the specific eigenmode of charge density oscillation $(m,\lambda)$ that created them. In light of the obtained multi-parameter dependence, any agreement with the theory would require fabrication control of the geometric features of the nanostructure. 
Undoubtedly, many features remain to be explored from a mathematical as well as physical point of view.  For example, the geometric origin of the mode coupling in the toroidal solutions and its significance within a bigger topological theory or morphological picture warrant further exploration. What are the class of systems for which similar couplings occur? What are the implication of  the geometric characteristics  for the dynamics of the electronic system within the many-body theory? What are some of the implications of the specifics of the quantization volumes for the unbounded material systems? Devising appropriate truncation criteria or introducing suitable perturbation theoretical approaches to calculating the Hamiltonian of the system are other examples of topics needing further investigation. The connection between the underlying physical processes and the corresponding mathematical formulation, such as presented here, offers a rich spectrum of heretofore unknown mathematics and physics. Examples of the productivity of such connections are many including the discovery of a new orthogonality relation for the MacDonald's functions \cite{PASSIAN2009380}
and new integral expansions \cite{PassianIndex} . 
\cleardoublepage
\addcontentsline{toc}{chapter}
{\normalsize\textnormal{{\MakeUppercase{References}}}}
\titlespacing{\chapter}{0pt}{0.65in}{10ex}
\titleformat{\chapter}[block]
{\normalfont\Large\filcenter\bf}
{ \vspace{3pc}%
  \Large\MakeUppercase{\chaptertitlename}
  \thechapter  
}{0pt}
{	\Large
	\vspace*{-0.5in}
}
\setstretch{1}

\newpage
\setlength{\parindent}{0.5in}
\begin{center}
	\vspace*{0.75in}
	\addcontentsline{toc}{chapter}{\appendixname
		\hspace*{2pt} A : THE CASES FOR SPHERE \& CYLINDER \dotfill}
	\appendix{\bf APPENDIX A : THE CASES FOR SPHERE \& CYLINDER}
\end{center}

\label{App:AppendixA}
\renewcommand\thesection{A.\arabic{section}}
\renewcommand{\theequation}{A-\arabic{equation}}
\setcounter{equation}{0}

Here, we present the calculation of radiatve decay rate for the two spherical and cylindrical geometries. In section \ref{1-sphy}, we consider particles with spherical cross sections and particles with cylindrical cross sections  are discussed in section \ref{1-cyli}. The spherical case is an example of a finite geometry while the cylindrical case could be considered as an example of both finite and infinite geometry, depending on how the problem is set up  and aimed for. The results for these two cases have been provided in certain references, mainly \cite{B, Crowell, BURMISTROVA}. However, the detailed calculations  presented here, to the best of our knowledge, were not found in the literature. 
\section{Radiative Decay on Spherical Surfaces } \label{1-sphy}

\textit {Spherical} coordinates are given by \cite{LSSFA}:
	\begin{equation}\label{1-2}
			x=a_0\,	r  \sin\theta \cos\varphi, \quad  
			y=a_0\,	r \sin\theta \sin\varphi, \quad 
			z= a_0 \, r \cos\theta,
	\end{equation}
where $a_0$ is the overall scale factor and:
	\begin{equation}\label{1-3}
			0\le r <\infty, \quad 
			0 \le \theta	\le \pi,\quad 
			0 \le \varphi	\le 2\pi.
	\end{equation}
The scale factors for a parametrization in each component are given by \cite{LSSFA}:
	\begin{equation}\label{1-4}
			h_r = a_0, \qquad 
			h_\theta= a_0\,  r,\qquad 
			h_\varphi =a_0\, r\,  \sin\theta.
	\end{equation}
Laplacian  in spherical coordinate is given by \cite{LSUWPAM}:
	\begin{equation}\label{1-5}
			\vec \nabla^2\Phi=\frac{1}{h_r\, h_\theta\, h_\varphi}
			\left[ \frac{\partial}{\partial r}
			\left( 	\frac{h_\theta\, h_\varphi}{h_r} \frac{\partial \Phi}{\partial r}	\right )+
			\frac{\partial}{\partial \theta}
			\left( 	\frac{h_r\, h_\varphi}{h_\theta} \frac{\partial \Phi}{\partial \theta}	\right )+
			\frac{\partial}{\partial \varphi}
			\left( 	\frac{h_r\, h_\theta}{h_\varphi} \frac{\partial \Phi}{\partial \varphi}	\right )
			\right]=0. 
	\end{equation}
	Using scaling factors given in Eq.~\eqref{1-4}, we can write:
	\begin{equation}\label{1-6}
			\vec \nabla^2\Phi=\frac{1}{a_0^2}
			\left[
			\left( 	\frac{\partial^2}{\partial r^2}+\frac{2}{r}
			\frac{\partial}{\partial r}\right) +
			\frac{1}{r^2 \sin\theta}\frac{\partial}{\partial \theta }	
			\left(	\sin\theta 	\frac{\partial}{\partial\theta }	\right)+
			\frac{1}	{r^2 \sin^2\theta }
			\frac{\partial^2}	{\partial \varphi^2}
			\right] \Phi=0. 
	\end{equation}
Assuming ansatz, using separation of variable:
	\begin{equation}\label{1-7}
			\Phi(r, \theta, \varphi) = R(r)\,  \Theta (\theta)\, \Psi(\varphi), 
	\end{equation}
where $R(r)$, $\Theta(\theta)$ and $\Psi(\varphi)$ are functions of variables $r, \theta$ and $\varphi$, respectively. Therefore, 
	\begin{equation}\label{1-8}
			\frac{\partial \Phi(r, \theta, \varphi)}{\partial r}= \frac{d R(r)}{d r}\,  \Theta (\theta),
	\end{equation}
and,
	\begin{equation}\label{1-9}
			\frac{\partial \Phi(r, \theta, \varphi)}{\partial \theta}= R(r)\,  \frac{d \Theta (\theta)}{d \theta}.
	\end{equation}
Substituting Eqs.~\eqref{1-8} and \eqref{1-9} into Eq.~\eqref{1-6}, we get:
	\begin{equation}\label{1-10}
			\vec \nabla^2\Phi=\frac{1}{a_0^2}
			\left\{
			\frac{\Theta (\theta)}{r^2} \frac{\partial }{\partial r}\left[ r^2\, R'(r)\right] +
			\frac{R(r)}{r^2 \, \sin\theta }
			\frac{\partial}{\partial \theta} 
			\left[	\sin\theta 	\Theta'(\theta)	\right]\right\} =0. 
	\end{equation}
Having separated Laplace's equation into two ordinary differential equations, the general solution given by the suitable ansatz for the potential $\Phi(\mathbf r,t)$ as:
	\begin{equation}\label{1-12}
			\Phi(\mathbf r,t)=
			\sum\limits_{l=0}^{\infty} 
			\sum\limits_{m=-l}^{l}
			\sum\limits_{p=0}^{1}
			\left( A_l(t) \, r^l + B_l(t) \, r^{-(l+1)}\right)  Y^p_{lm}(\theta, \varphi), 
	\end{equation}
for some real time-dependent amplitudes $A_l(t)$ and $B_l(t)$ and real \textit{spherical harmonics} $Y^p_{lm}(\theta, \varphi)$ given by: \cite{LSSFA} :
	\begin{equation}\label{1-13}
			Y^p_{lm}(\theta, \varphi)= \sqrt{\frac{(2- \delta_{0m}) (2l+1)(l-m)!}{4\pi (l+m)!}} 
			P^m_l(\cos\theta) (\delta_{0p} \cos m\varphi+\delta_{1p}\sin m\varphi),
	\end{equation}
where $\delta_{kp}$ is the {Kroneker delta} symbol and $P^m_l(\cos\theta )$ denotes real \textit{associated Legendre functions} of first kind \cite{g&r:table,Bell}.
Inside and outside potential of a sphere can be constructed by imposing proper boundary conditions over the solution of Laplace's equation given in Eq.~\eqref{1-12}. Inside and outside potential, denoted by $\Phi_{\text{i}}$ and $\Phi_{\text{o}}$ respectively, should satisfy the following conditions along the boundary of a solid sphere defined by $r=r_0$ using Eqs.~\eqref{BD1} and \eqref{BD2}, we may write:
	\begin{eqnarray}\label{1-14}
			\Phi_{\text{i}}	(\mathbf r,t)\, \Big|_{r=r_0}&=&\Phi_{\text{o}}	(\mathbf r,t)\, \Big|_{r=r_0},\\
			\varepsilon \frac{\partial \Phi_{\text{i}}	(\mathbf r,t)}{\partial r}\Big|_{r=r_0} &=&	\frac{\partial \Phi_{\text{o}}	(\mathbf r,t)}{\partial r}\Big|_{r=r_0} .
	\end{eqnarray}
Therefore, we could write inside and outside potentials as:
		\begin{eqnarray}
				\Phi_{\text{i}}	(\mathbf r,t)&=&
				\sum\limits_{l=0}^{\infty} 
				\sum\limits_{m=-l}^{l}
				\sum\limits_{p=0}^{1}
				A_l(t) \, r_0^{-(l+1)}\, r^l \,  Y^p_{lm}(\theta, \varphi), 
				\qquad	 \text{for} \quad r		\le r_0, 	\label{Sp-in}\\
				\Phi_{\text{o}}	(\mathbf r,t)&=&
				\sum\limits_{l=0}^{\infty} 
				\sum\limits_{m=-l}^{l}
				\sum\limits_{p=0}^{1}
				A_l(t) \, r^{-(l+1)}\, r_0^l  \, Y^p_{lm}(\theta, \varphi), 
				\qquad	\text{for} \quad r_0		\le r .	\label{Sp-out}
		\end{eqnarray}
Utilizing the \textit{Heaviside} function $\Theta(x)$:
	\begin{equation}\label{1-15}
			\Theta(x)= \frac12 \left( \frac{|x|}{x}+1\right), 
	\end{equation}
whose derivative gives:
	\begin{equation}\label{1-15d}
			\frac{d\Theta(x)}{dx}= \delta(x), 
	\end{equation}
where $\delta$ denotes the {Dirac Delta} function, we define potential function implicitly in terms of inside and outside potentail $\Phi_{\text{i}}$ and $\Phi_{\text{o}}$ given by Eqs.~\eqref{Sp-in} and \eqref{Sp-out} as:
	\begin{equation}\label{1-16}
			\Phi(\mathbf r,t)=
			\Theta(r_0-r)
			\Phi_{\text{i}}(\mathbf r,t)+
			\Theta(r-r_0)
			\Phi_{\text{o}}(\mathbf r,t).
	\end{equation}
Its derivative with respect to variable $r$ gives:
	\begin{multline}\label{1-17}
			\frac{\partial \Phi(\mathbf r,t)}{\partial r}= 
			\Theta(r_0-r)
			\frac{\partial \Phi_{\text{i}}(\mathbf r,t)}{\partial r}+
			\Theta(r-r_0)
			\frac{\partial \Phi_{\text{o}}(\mathbf r,t)}{\partial r}\\-
			\delta(r_0-r)
			\Phi_{\text{i}}(\mathbf r,t)+
			\delta(r-r_0)
			\Phi_{\text{o}}(\mathbf r,t).
	\end{multline}
Using that fact that on the boundary $r=r_0$, 
	\begin{equation}
			-\delta(r_0-r)
			\Phi_{\text{i}}(\mathbf r,t)+
			\delta(r-r_0)
			\Phi_{\text{o}}(\mathbf r,t)= \delta(r-r_0) \left[\Phi_{\text{o}}(\mathbf r,t)-\Phi_{\text{i}}(\mathbf r,t)\right] \equiv 0, 
	\end{equation}
we can write:
	\begin{equation}\label{1-18}
			\frac{\partial \Phi(\mathbf r,t)}{\partial r}= 
			\Theta(r_0-r)
			\frac{\partial \Phi_{\text{i}}(\mathbf r,t)}{\partial r}+
			\Theta(r-r_0)
			\frac{\partial \Phi_{\text{o}}(\mathbf r,t)}{\partial r}.
	\end{equation}
Similarly, 
	\begin{equation}\label{1-19}
			\frac{\partial \Phi(\mathbf r,t)}{\partial \theta}= 
			\Theta(r_0-r)
			\frac{\partial \Phi_{\text{i}}(\mathbf r,t)}{\partial \theta}+
			\Theta(r-r_0)
			\frac{\partial \Phi_{\text{o}}(\mathbf r,t)}{\partial \theta},
	\end{equation}
and, 
	\begin{equation}\label{1-20}
			\frac{\partial \Phi(\mathbf r,t)}{\partial \varphi}= 
			\Theta(r_0-r)
			\frac{\partial \Phi_{\text{i}}(\mathbf r,t)}{\partial \varphi}+
			\Theta(r-r_0)
			\frac{\partial \Phi_{\text{o}}(\mathbf r,t)}{\partial \varphi}.
	\end{equation}
Second derivatives can be calculated as:
		\begin{multline}\label{1-21}
				\frac{\partial^2 \Phi(\mathbf r,t)}{\partial r^2}= 
				\Theta(r_0-r)
				\frac{\partial^2 \Phi_{\text{i}}(\mathbf r,t)}{\partial r^2}+
				\Theta(r-r_0)
				\frac{\partial^2 \Phi_{\text{o}}(\mathbf r,t)}{\partial r^2}\\-
				\delta(r_0-r)
				\frac{\partial \Phi_{\text{i}}(\mathbf r,t)}{\partial r}+
				\delta(r-r_0)
				\frac{\partial \Phi_{\text{o}}(\mathbf r,t)}{\partial r},
		\end{multline}
and,
	\begin{equation}\label{1-22}
			\frac{\partial^2 \Phi(\mathbf r,t)}{\partial \theta^2}= 
			\Theta(r_0-r)
			\frac{\partial^2 \Phi_{\text{i}}(\mathbf r,t)}{\partial \theta^2}+
			\Theta(r-r_0)
			\frac{\partial^2 \Phi_{\text{o}}(\mathbf r,t)}{\partial \theta^2},
	\end{equation}
and,
	\begin{equation}\label{1-23}
			\frac{\partial^2 \Phi(\mathbf r,t)}{\partial \varphi^2}= 
			\Theta(r_0-r)
			\frac{\partial^2 \Phi_{\text{i}}(\mathbf r,t)}{\partial \varphi^2}+
			\Theta(r-r_0)
			\frac{\partial^2 \Phi_{\text{o}}(\mathbf r,t)}{\partial \varphi^2}.
	\end{equation}
Using Eqs.~\eqref{1-21}--\eqref{1-23} in Eq.~\eqref{1-10}, we have:
	\begin{multline}\label{1Sp-L1}
			\vec	\nabla^2	\Phi(\mathbf r,t)=
			\frac	{\delta(r-r_0)}	{a_0^2}
			\left(	\frac{\partial	\Phi_{\text{o}}	}	{\partial r}-
			\frac{\partial	\Phi_{\text{i}}	}	{\partial r}		\right)+
			\Theta	(r_0-r)
			\vec \nabla^2	\Phi_{\text{i}}	(\mathbf r,t)+
			\Theta	(r-r_0)
			\vec \nabla^2	\Phi_{\text{o}}	(\mathbf r,t).
	\end{multline}
Since Laplacian vanishes inside and outside but on the surface of the sphere, then we can write:
	\begin{equation}\label{1Sp-L2}
			\vec \nabla^2	\Phi(\mathbf r,t)=
			\frac	{\delta(r-r_0)}	{a_0^2}
			\left( 	\frac{\partial	\Phi_{\text{o}}	}	{\partial r}-
			\frac{\partial	\Phi_{\text{i}}	}	{\partial r}		\right), 
	\end{equation}
where using \eqref{Sp-in} and \eqref{Sp-out}:
	\begin{equation}\label{Sp-L3}
			\frac{	\partial\Phi_{\text{o}}	}	{\partial r}-
			\frac{	\partial\Phi_{\text{i}}	}	{\partial r}=
			\sum\limits_{m,l,p} 
			A_l(t) \left[ -(l+1)\,  r_0^l\, r^{-(l+2)}- l\,  r_0^{-(l+l)} r^{l-1} \,  \right] Y^p_{lm}(\theta, \varphi).
	\end{equation}
Using this in Eq.~\eqref{1Sp-L2}, we can write:
	\begin{equation}\label{1-27}
			\vec \nabla^2	\Phi(\mathbf r,t)=
			\frac	{\delta(r-r_0)}	{a_0^2}
			\sum\limits_{m,l,p} 
			A_l(t) \left[ -(l+1)\,  r_0^l\, r^{-(l+2)}- l\,  r_0^{-(l+l)} r^{l-1} \,  \right] Y^p_{lm}(\theta, \varphi).
	\end{equation}
In order to find surface charge density $\sigma$, we use Eq.~\eqref{1-27} in Eq.~\eqref{0-26} to write: 
	\begin{eqnarray}\label{Sp-s}
			\sigma=-\frac{1}	{4\pi a_0	}
			\sum\limits_{m,l,p}
			A_l(t) \left[ -(l+1)\,  r_0^l\, r^{-(l+2)}- l\,  r_0^{-(l+l)} r^{l-1} \,  \right] Y^p_{lm}(\theta, \varphi).
	\end{eqnarray}
The only time-dependent component in the latter equation are amplitudes $A_l(t)$. Hence, the second time-derivative of Eq.~\eqref{Sp-s} gives:
	\begin{eqnarray}\label{Sp-s1}
			\ddot \sigma \big|_{r=r_0}= -
			\frac{1}	{4\pi a_0	}
			\sum\limits_{m,l,p}
			\ddot A_l(t) \left[ -(2l+1) r_0^{-2} \right] Y^p_{lm}(\theta, \varphi).
	\end{eqnarray}
\noindent Amplitudes $A_l(t)$ undergo harmonic oscillations at continuous frequencies $\omega_l$, that is:
	\begin{equation}\label{1-32}
			\ddot A_l(t)  + 
			\omega^2_l\, A_l(t)=0. 
	\end{equation}	
We utilize the relation given in  Eq.~\eqref{0-sddot1} to write:
	\begin{eqnarray}\label{Sp-s2}
			\ddot \sigma\big |_{r=r_0}=-\frac{\omega_p^2}	{4\pi a_0	}
			\sum\limits_{m,l,p}
			A_l(t) \left( l r_0^{-2} \right) Y^p_{lm}(\theta, \varphi).
	\end{eqnarray}
Using the equality of \eqref{Sp-s1} and \eqref{Sp-s2}, we find:
	\begin{eqnarray}\label{Sp-o}
			\ddot A_l(t)+ 	\omega^2_{l}	A_{l}(t)=0,
	\end{eqnarray}
where,
	\begin{eqnarray}\label{Sp-w}
			\omega^2_{l}=  \omega_p^2\, \frac{  l}{2l+1},
	\end{eqnarray}
which is known as the surface plasmon eigen-frequency and is given by \cite{B}, p.~129.	It is noteworthy that the frequency in case of a sphere is independent of index $m$. One can use the latter expression for frequency in Drude model and finds dielectric function of a sphere as:
	\begin{equation}\label{1-34}
			\varepsilon (\omega_l) = \frac{1-l}{l }.
	\end{equation}
The potential energy, using Eq.~\eqref{Sp-in} and \eqref{Sp-s}, could be written:
	\begin{multline}\label{Sp-V1}
			V = \frac{a_0 }{8\pi }   	
			\int_{0}	^{2\pi} \int_{0}  ^{\pi}  \left[ \sum\limits_{m,l,p}
			A_l(t)  (2l+1) r_0^{-2}  Y^p_{lm}(\theta, \varphi)\right] \\
			\times \left[	\sum\limits_{m',l',p'}
			\overline{A_{l'}}(t) \, r_0^{-1}\, \overline {Y^{p'}_{m'l'}}(\theta, \varphi)\right] r_0^2 \sin\theta \, d\theta d\varphi , 	
	\end{multline}
as we replaced real-valued potential with its complex conjugate as they are equal. This allows us to utilize the \textit{orthogonality relations} for the orthogonal system $\{Y^p_{lm}\}_{l,m,p}$ known as spherical harmonics system. This relation is given by \cite{LSSFA,g&r:table}: 
	\begin{equation}\label{Sp-o1}
			\int_{0}^{2\pi}  \int_{0}^{\pi} 
			Y^p_{lm}(\theta, \varphi)	\overline {Y^{m'}_{l'}}(\theta, \varphi)
			\sin\theta\,  d\theta d\varphi=\delta_{mm'}\delta_{ll'}. 		
	\end{equation}
Applying Eq.~\eqref{Sp-o1} to Eq.~\eqref{Sp-V1}, we may write:
	\begin{eqnarray}\label{Sp-V2}
			V=\frac{ a_0}{8\pi r_0 }   
			\sum\limits_{l=0}^{\infty} 	
			\sum\limits_{m=-l}^{l}
			\left| A_l(t)\right| ^2 \left(2l+1 \right), 
	\end{eqnarray}
using Eq.~\eqref{Sp-w}, the potential energy could be written as:
	\begin{eqnarray}\label{1-V}
			V=\frac{ a_0}{8\pi r_0 }   
			\sum\limits_{l} 	
			\frac{ \omega^2_p}{\omega^2_{l}}\,l\,  \left| A_l(t)\right| ^2.
	\end{eqnarray}
 Similar calculations, using Eq.~\eqref{Sp-o1}, give the kinetic energy as:
	\begin{eqnarray}\label{1-T}
			T=\frac{a_0 }{8\pi r_0}   
			\sum\limits_{l} 	
			\frac{\omega_p^2}{\omega^4_{l}}\, l \, \left| \dot A_l(t)\right| ^2.
	\end{eqnarray}
By means of Eqs.~\eqref{1-V} and \eqref{1-T}, total energy $E$ could be found as:
	\begin{eqnarray}\label{Sp-E}
			E=\frac{a_0 }{8\pi r_0}  
			\sum\limits_{l} 	
			\frac{2l+1}{\omega^2_{l}}
			\bigg [  
			\big|\dot A_l(t)\big|^2+
			{\omega^2_{l}}\big |A_l(t)\big|^2
			\bigg ].
	\end{eqnarray}
As described in Chapter \ref{1}, in order to convert $E$  into the Hamiltonian operator for a scalar boson field (plasmons are spinless quasi-particles), we write:
	\begin{equation}\label{Sp-a}
			A_l(t)=\frac{\alpha_{lmp}}{2\omega_{l}} (a_{lmp}+a^*_{lmp}),
	\end{equation}
where $a_{lmp}$ are some complex time-dependent functions, proportional to $e^{-i\omega_{l}t}$, $a^*_{lmp}$ its complex conjugate and coefficients $\alpha_{lmp}$ would be determined later. Its time derivative gives:
	\begin{equation}\label{Sp-a1}
			\dot A_l(t)= 
			- i\frac{\alpha_{lmp}}{2}~
			(a^*_{lmp}-a_{lmp}).
	\end{equation}
Using Eqs.~\eqref{Sp-a} and \eqref{Sp-a1}  in Eq.~\eqref{Sp-E}, we have:
	\begin{equation}\label{1-E}
			E=\frac{a_0 }{8\pi r_0}  
			\sum\limits_{l} 	
			\left( \frac{2l+1}{\omega^2_l}\right) \, 
			\frac{\alpha^2_{lmp}}{2} \, a_{lmp}  a^*_{lmp}.
	\end{equation}
Performing field quantization~\cite{B,SIRQFT}, we replace the amplitudes $a_{lmp}(t)$ and $a^*_{lmp}(t)$ with operator valued distributions $\hat{a}_{lmp}$ and its conjugate $\hat{a}^\dagger_{lm p},$ and note the commutation relations:
	\begin{equation}\label{1-56}
			[\hat{a}_{lm p},\hat{a}^\dagger_{l'm'p'}]=\delta_{ll'}\delta_{mm'}\delta_{pp'}.
	\end{equation}
Taking the normal ordered expansion of the noncommuting boson creation $\hat{a}^\dagger_{lm p}$ and annihilation  $\hat{a}_{lmp} $ operators, a  comparison  with the normal ordered expression of the Hamiltonian operator for a scalar boson field~\cite{B}, yields the Hamiltonian:
	\begin{eqnarray}\label{1-E2}
			\mathcal{H}_{sp} = \sum\limits_{l,m,p}
			\hbar\, \omega_{m\lambda}\, 
			\hat{a}^{\dagger}_{lm p}\hat{a}_{lm p}.
	\end{eqnarray}
Comparing Eq.~\eqref{1-E} with Eq.~\eqref{1-E2}, one finds:
	\begin{eqnarray}\label{Sp-alpha}
			\alpha^2_{lmp}=
			\frac{8\pi \,  r_0\, \hbar}{a_0\, (2l+1)	}
			\omega^3_{l}.
	\end{eqnarray}
Next, in order to calculate the current $\mathbf J$ as given in Eq.~\eqref{0-J}, following Eq.~\eqref{0-psidot}, we set:
	\begin{equation}
			\dot \Psi =-\frac{e}{m} \sum\limits_{m,l} \frac{1 }{\omega^2_{l}}\, \dot \Phi^{lm}_{\text{i}}.
	\end{equation}
Using the inside potential expression given in Eq.~\eqref{Sp-in}, one can write: 
	\begin{equation}\label{Sp-Psi}
			\dot \Psi	=	-\frac{e}{m_0}
			\sum\limits_{m,l,p} 
			r_0^{-(l+1)}\, r^l \frac{\alpha_{lmp}}{ 2i \omega^2_{l} }\, Y^p_{lm}(\theta, \varphi) 
			(a^*_{lmp}-a_{lmp}), 				
	\end{equation}
upon making the replacement $\dot{A}_{l}(t)\to- i\frac{\alpha_{lmp}}{2}~(a^*_{lmp}-a_{lmp})$ given by Eq.~\eqref{Sp-a1} in the expression for $\dot{\Psi}$. Given interaction Hamiltonian by Eq.~\eqref{0-IntH}, using Eqs.\eqref{A} and \eqref{Sp-Psi}, we can write:
	\begin{multline}\label{Sp-H2}
		H_{em} =   -\frac{n_0e}{c}
		\int_0^{2\pi}~\int_0^{\pi}
		\left[	-\frac{e}{m_0}
		\sum\limits_{m,l,p} 
		r_0^{-(l+1)}\, r^l \frac{\alpha_{lmp}}{ 2i \omega^2_{l} }\, Y^p_{lm}(\theta, \varphi) 
		(a^*_{ml}-a_{ml}) \right]\\
		\times
		\left[\sum\limits_{q=1,2} 
		\sum\limits_{\mathbf s} 
		\sqrt{\frac{\hbar c^2}{\mathcal V\omega_s}}\, 
		\mathbf{\hat e}_q 
		\, \big ( 	a_{\mathbf{s}q}		e^{i\mathbf{s\cdot r}}		+	
		a^*_{\mathbf{s}q}	e^{-i\mathbf{s\cdot r}}\big )\, 
		\right]\cdot	\mathbf{\hat e}_r	\, 
		h_\theta  	h_\varphi \,	d\theta	d\varphi	,															
	\end{multline}
which, after using scaling factors given in Eq.~\eqref{1-4}, simplifies to:
	\begin{multline}\label{2-Sp-H1}
			H_{em} =  \frac{n_0e^2}{ m}	\sum\limits_\mathbf{s}
			\sum\limits_{q=1,2}
			\sqrt{\frac{\hbar }{\mathcal V	\omega_s}}
			(\mathbf{\hat{e}}_{q}			\cdot		\mathbf{\hat{e}}_r)
			\big ( 	a_{\mathbf{s}q}		e^{i\mathbf{s\cdot r}}		+		a^*_{\mathbf{s}q}		e^{-i\mathbf{s\cdot r}}\big )\\
			\times 
			\sum\limits_{l,m,p}
			\int_{0}^{2\pi}
			\int_{0}^{\pi}
			r_0^{-1}\frac{\alpha_{lmp}}{ 2i \omega^2_{l} }\, Y^p_{lm}(\theta, \varphi) 
			(a^*_{lmp}-a_{lmp})		h_\theta h_\varphi\,	d\theta	d\varphi.			
	\end{multline}
The relevant transition matrix element for photon emission, given by Eq.~\eqref{Mem}, using Eq.~\eqref{2-Sp-H1}, could be obtained as:
	\begin{eqnarray}\label{Sp-M}
			\mathcal {M}^{(q\mathbf{s})}_{em} =
			\frac{n_0e^2 }{m\, r_0} 
			\sqrt{\frac{\hbar}{\mathcal V \omega_s}}
			\frac{\alpha_{lmp}}{2i\omega^2_{l}}
			\int_{0}^{2\pi}
			\int_{0}^{\pi} 	
			Y^p_{lm}(\theta, \varphi)
			(\mathbf{\hat{e}}_r		\cdot		\mathbf{\hat{e}}_q	)
			e^{-i\mathbf{s\cdot r}}	
			h_\theta h_\varphi\, d\theta d\varphi.
	\end{eqnarray}
Using the identities: 
	\begin{eqnarray}\label{Sp-r}
			&&{\bf r} = a_0\, r (\sin\theta \cos\varphi, \sin\theta\sin\varphi, \cos\theta),\\
			&&{\bf \hat e}_q =q_x \vec i + q_y \vec j + q_z \vec k,\\
			&&{\bf s} = s_x \vec i+ s_y \vec j+ s_z \vec k,\\
			&&{\bf s}\cdot {\bf r}\big |_{r=r_0} = a_0 r_0 (s_x \sin\theta \cos\varphi, s_y\sin\theta\sin\varphi, s_z\cos\theta),\\
			&&\mathbf{\hat{e}}_{r_0} = \frac{1}{h_{r_0}} \frac{\partial \bf r}{\partial r}= (\sin\theta \cos\varphi, \sin\theta\sin\varphi, \cos\theta),
	\end{eqnarray}
we can write:
	\begin{multline}\label{Sp-r1}
			h_\theta h_\varphi(\mathbf{\hat{e}}_r		\cdot		\mathbf{\hat{e}}_q	) e^{-i\mathbf s \cdot \mathbf r} = a_0^2 r_0^2 \sin\theta 	\, \mathbf{\hat{e}}_q \cdot (\sin\theta \cos\varphi \vec i+ \sin\theta\sin\varphi \vec j+ \cos\theta \vec k)\\
			\times  \exp\{{-ia_0r_0 (s_x\sin\theta \cos\varphi \vec i+ s_y \sin\theta\sin\varphi \vec j+ s_z \cos\theta \vec k)}\}, 
	\end{multline}
hence,
	\begin{eqnarray}\label{Sp-r2}
			h_\theta h_\varphi(\mathbf{\hat{e}}_r		\cdot	
			\mathbf{\hat{e}}_q	) e^{-i\mathbf s \cdot \mathbf r} 
			= i a _0r_0 \sin\theta 	\, \mathbf{\hat{e}}_q \cdot \vec \nabla_{\bf s}\, e^{-i\bf s \cdot \bf r}.
	\end{eqnarray}
Let:
	\begin{eqnarray}\label{Sp-I}
			\mathcal I_{lmp} =
			\int_{0}^{2\pi}
			\int_{0}^{\pi} 	
			Y^p_{lm}(\theta, \varphi)
			(\mathbf{\hat{e}}_r		\cdot		\mathbf{\hat{e}}_q	)
			e^{-i\mathbf{s\cdot r}}	
			h_\theta h_\varphi  \, d\theta d\varphi,
	\end{eqnarray}
using Eq.~\eqref{Sp-r2}, we write:
	\begin{eqnarray}\label{Sp-I1}
			\mathcal I_{lmp} = i a_0 r_0 \mathbf{\hat{e}}_q	\cdot \vec  \nabla_{\bf s}
			\int_{0}^{2\pi}
			\int_{0}^{\pi} 	
			Y^p_{lm}(\theta, \varphi)
			e^{-i\mathbf{s\cdot r}}	
			\sin\theta  d\theta d\varphi.
	\end{eqnarray}
Using the identity (see \cite{jackson}, p. 767):
	\begin{eqnarray}\label{Sp-I2}
			e^{i\mathbf{s\cdot r}}	=4\pi \sum\limits_{l=0}^{\infty}
			i^{l} j_l({\bf s} r) \sum\limits_{m=-l}^{l}	
			Y^{*p}_{lm}(\theta, \varphi) Y^p_{lm}(\theta', \varphi')
	\end{eqnarray}
where $\bf s$ is a wave vector in spherical coordinate $(r, \theta',\varphi' )$. Replacing $i$ with $-i$, we have:
	\begin{eqnarray}\label{Sp-I3}
			\mathcal I_{lmp}= 4\pi \, a_0\,  r_0  
			\, i \, (-i)^{l} \, 
			j_l({\bf s} r)\, 
			\mathbf{\hat{e}}_q 
			\cdot 
			\vec \nabla_{\bf s} Y^p_{lm}(\theta, \varphi).
	\end{eqnarray}
Hence the emission matrix element becomes:
	\begin{eqnarray}\label{Sp-M1}
			\mathcal {M}^{(q\mathbf{s})}_{em} =
			\frac{2\pi a n_0e^2 r}{m}  (-i)^{l} 
			\sqrt{\frac{\hbar}{\omega_s}}
			\frac{\alpha_{lmp}}{\omega^2_{l}}
			\frac{j_l({\bf s} r)}{r}\, 
			\mathbf{\hat{e}}_q 
			\cdot 	\vec 	\nabla_{\bf s} Y^{p}_{lm}(\theta, \varphi).
	\end{eqnarray}
Using Eq.~\eqref{Sp-alpha}, 
we write:
	\begin{eqnarray}\label{Sp-M2}
			\mathcal{M}^{(q\mathbf{s})}_{ml} &=&
			(-i)^{l}\, 	\frac{2\pi a_0 n_0e^2 r}{m_0}  \,
			\sqrt{\frac{\hbar}{\omega_s}}
			\frac{\alpha_{lmp}}{\omega^2_{l}}\,
			\frac{j_l({\bf s} r)}{r}\, 
			\mathbf{\hat{e}}_q \,
			\cdot 	\vec 	\nabla_{\bf s} Y^{p}_{lm}(\theta, \varphi)\\
			&=&(-i)^{l}\, \frac{4\pi a_0 n_0e^2 r}{2m_0}   
			\sqrt{\frac{\hbar}{\omega_s}}
			\frac{\alpha_{lmp}}{\omega^2_{l}}\,
			\frac{j_l({\bf s} r)}{r}\, 
			\mathbf{\hat{e}}_q 
			\cdot 	\vec 	\nabla_{\bf s} Y^{p}_{lm}(\theta, \varphi)\\
			&=&(-i)^{l}\, \frac{a_0 \omega^2_{\mathcal P} r}{2} \, 
			\sqrt{\frac{\hbar}{\omega_s}}
			\frac{\alpha_{lmp}}{\omega^2_{l}}\,
			\frac{j_l({\bf s} r)}{r}\, 
			\mathbf{\hat{e}}_q 
			\cdot 	\vec 	\nabla_{\bf s} Y^{p}_{lm}(\theta, \varphi)\\
			&=&(-i)^{l}\, \sqrt{\frac{a_0 (2\pi\hbar)^2\omega^4_{\mathcal P} r^3}{\omega_s(2l+1)\omega_{l}}} \, 
			\frac{j_l({\bf s} r)}{r}\, 
			\mathbf{\hat{e}}_q 
			\cdot 	\vec 	\nabla_{\bf s} Y^{p}_{lm}(\theta, \varphi), 
	\end{eqnarray}
which for scale factor $a_0=1$ is the same as matrix element given in  by Eq.~(51) in \cite{B}, p.~129, as:
	\begin{eqnarray}\label{Sp-M3}
			\mathcal{M}^{(q\mathbf{s})}_{ml} =	(-i)^{l} 
			\sqrt{\frac{(2\pi\hbar)^2  \omega_{\mathcal P}^4 r^3}{ \omega_s\,(2l+1)\,\omega_{l}}}\, \times			
			\frac{j_l({\bf s} r)}{r}\, 
			\mathbf{\hat{e}}_q 
			\cdot 	\vec 	\nabla_{\bf s} Y^{p}_{lm}(\theta, \varphi).
	\end{eqnarray}
On the other hand, considering $\bf s$ being a vector in spherical coordinate, we have:  
	\begin{equation}\label{Sp-e1}
			\mathbf{\hat{e}}_q \cdot 	\vec 	\nabla_{\bf s} =\frac{\mathbf{\hat{e}}_q}{s}  \cdot
			\left(	\mathbf{\hat{e}}_r  \frac{\partial }{\partial r }+
			\mathbf{\hat{e}}_\theta  \frac{1}{r } \frac{\partial }{\partial \theta }+
			\mathbf{\hat{e}}_\varphi \frac{1}{r \sin\theta } \frac{\partial }{\partial \varphi }\right).
	\end{equation}
Since the polarization vector must be perpendicular to the direction of travel, thus  $(\mathbf{\hat{e}}_q \cdot \mathbf{\hat{e}}_r )=0$, hence:
	\begin{eqnarray}\label{Sp-e2}
		\mathbf{\hat{e}}_q 
		\cdot 		\vec \nabla_{\bf s} \, Y^p_{lm}(\theta, \varphi)&=& 	\frac{1}{s}\left( \mathbf{\hat{e}}_q 
		\cdot \mathbf{\hat{e}}_\theta \right)  \frac{1}{r } \frac{\partial Y^p_{lm}(\theta, \varphi)}{\partial \theta }+
		\frac{1}{s}\left( \mathbf{\hat{e}}_q \cdot \mathbf{\hat{e}}_\varphi \right)  \frac{1}{r \sin\theta }
		\frac{\partial Y^p_{lm}(\theta, \varphi)}{\partial \varphi }, 					.
	\end{eqnarray}
where:
	\begin{eqnarray}\label{Sp-Y}
			\frac{\partial Y^p_{lm}(\theta, \varphi)}{\partial \theta }=
			\sqrt{\frac{(2- \delta_{0m}) (2l+1)(l-m)!}{4\pi (l+m)!}} 
			P'^m_l(\cos\theta) (\delta_{0p} \cos m\varphi+\delta_{1p}\sin m\varphi),
	\end{eqnarray}	
and,
	\begin{eqnarray}\label{Sp-Y1}
			\frac{\partial Y^p_{lm}(\theta, \varphi)}{\partial \varphi }=
			m\,	\sqrt{\frac{(2- \delta_{0m}) (2l+1)(l-m)!}{4\pi (l+m)!}} 
			P^m_l(\cos\theta) (\delta_{1p} \cos m\varphi-\delta_{0p}\sin m\varphi).
	\end{eqnarray}		
Taking the unit vectors $\mathbf{\hat{e}}_\theta$ and $\mathbf{\hat{e}}_\varphi$ as: 
	\begin{eqnarray}\label{Sp-1e}
			\mathbf{\hat{e}}_\theta &=& \cos\theta \, \cos\varphi \vec i +  \cos\theta \, \sin\varphi \vec j-\sin\theta \vec k ,\\							
			\mathbf{\hat{e}}_\varphi &=& -\sin\varphi \vec i + \cos\varphi \vec j. 
	\end{eqnarray}	
Taking $(	\mathbf{\hat{e}}_{q_1} =	\mathbf{\hat{e}}_\theta )$ and $(	\mathbf{\hat{e}}_{q_2} =	\mathbf{\hat{e}}_\varphi )$ for s and p-polarization vectors, respectively, then: 
	\begin{align}
		(	\mathbf{\hat{e}}_{q_1} \cdot	\mathbf{\hat{e}}_\theta )=1\qquad \text{and} \qquad
		(	\mathbf{\hat{e}}_{q_1} \cdot	\mathbf{\hat{e}}_\varphi )=0, \\
		(	\mathbf{\hat{e}}_{q_2} \cdot	\mathbf{\hat{e}}_\varphi)=1\qquad \text{and} \qquad
		(	\mathbf{\hat{e}}_{q_2} \cdot	\mathbf{\hat{e}}_\theta )=0.
	\end{align}
Letting:
	\begin{equation}\label{Sp-k}
			K^{q_1}_{ml}(r, \theta, \varphi)=
			\sqrt{\frac{(2- \delta_{0m}) (2l+1)(l-m)!}{4\pi (l+m)!}} 
			P'^m_l(\cos\theta) (\delta_{0p} \cos m\varphi+\delta_{1p}\sin m\varphi) ,
	\end{equation}
	\begin{multline}\label{Sp-k1}
			K^{q_2}_{ml}(r, \theta, \varphi) =
			\frac{m}{ \sin\theta }
			\,	\sqrt{\frac{(2- \delta_{0m}) (2l+1)(l-m)!}{4\pi (l+m)!}} \\
			\times P^m_l(\cos\theta) (\delta_{1p} \cos m\varphi-\delta_{0p}\sin m\varphi),
	\end{multline}
Eq.~\eqref{Sp-M3}, by taking $s=\omega_s/c$, could be rewritten as:
	\begin{equation}\label{Sp-mf}
			\mathcal {M}^{(q\mathbf{s})}_{ml}= (-i)^{l} 
			\sqrt{\frac{(2\pi\hbar)^2  \omega_{\mathcal P}^4 r^3}{ \omega_s\,(2l+1)\,\omega_{l}}}\, \times			
			\frac{j_l(r\, \omega_s/c)}{r \, \omega_s/c}\, \,K^{q}_{ml}(r, \theta, \varphi).
	\end{equation}
Therefore, radiative decay rate, using Eq.~\eqref{Sp-mf} in Eq.~\eqref{0-g1}, could be found as:
	\begin{equation}\label{Sp-rdr1}
			\frac{d\gamma_{l}}{d\Omega} =
			\sum\limits_{q=1,2} \frac{\omega^4_{p} r^3}{\omega_l (2l+1) \omega_l} 
			\left[ \frac{j_l(r\, \omega_s/c)}{r\, \omega_s/c}\right] ^2 
			\left|K^{q}_{ml}(r, \theta, \varphi) \right|^2   \frac{\omega_{l}^2}{c^3}. 
	\end{equation}
which leads to:
	\begin{equation}\label{Sp-rdr3}
			\frac{d\gamma_{l}}{d\Omega} =
			\sum\limits_{q=1,2} \frac{\omega^4_{p} r^3}{ c^3(2l+1)} 
			\left[ \frac{j_l(r\, \omega_s/c)}{r\, \omega_s/c}\right] ^2 
			\left|K^{q}_{ml}(r, \theta, \varphi) \right|^2  .
	\end{equation}
Using frequency given in Eq.~\eqref{Sp-w}, we can write:
	\begin{eqnarray}\label{Sp-rdr2}
			\frac{d\gamma}{d\Omega} =
			\sum\limits_{q=1,2} \frac{\omega^4_{l} (2l+1 )r^3}{ c^3\, l^2} 
			\left[ \frac{j_l(r\, \omega_s/c)}{r\, \omega_s/c}\right] ^2 
			\left|K^{qs}_{ml}(r, \theta, \varphi) \right|^2  ,
	\end{eqnarray}
where:
	\begin{multline}\label{Sp-k12}
			\left|K^{q_1}_{ml}(r, \theta, \varphi) \right|^2 + 	\left|K^{q_2}_{ml}(r, \theta, \varphi) \right|^2= 
			\frac{(2-\delta_{0m}) (2l+1)(l-m)!}{4\pi(l+m)!}\\
			\times \left\{  \Big [ P'^m_l(\cos\theta)\Big ]^2 + \frac{m^2}{\sin^2\theta }
			\Big [P^m_l(\cos\theta)\Big ]^2\right\}.
	\end{multline}
Knowing:
	\begin{equation}\label{Sp-solid}
		d\Omega=  \sin\theta \, d\theta d\varphi,
	\end{equation}
and,
	\begin{equation}\label{Sp-rdr4}
			\gamma_l=  \int \frac{d\gamma_{l}}{d\Omega} d\Omega=\int  \int \frac{d\gamma_{l}}{d\Omega} \sin\theta \, d\theta d\varphi,
	\end{equation}
hence:
	\begin{multline}\label{Sp-rdr5}
			\gamma_{l}= \frac{\omega^4_{l} (2l+1 )r^3}{ c^3\, l^2} 
			\left[ \frac{j_l(r\, \omega_s/c)}{r\, \omega_s/c}\right] ^2 \\
			\times  \int_0^{2\pi}   \int_0^\pi
			\left[ 	\left|K^{q_1}_{ml}(r, \theta, \varphi) \right|^2 + 	\left|K^{q_2}_{ml}(r, \theta, \varphi) \right|^2\right]  
			\sin\theta \, d\theta d\varphi.
	\end{multline}
It simplifies to:
	\begin{multline}\label{Sp-rdr6}
			\gamma_l= 2\pi \times \frac{\omega^4_{l} (2l+1 )r^3}{ c^3\, l^2} 
			\left[ \frac{j_l(r\, \omega_s/c)}{r\, \omega_s/c}\right] ^2 \\
			\times  \int _0^\pi  
			\left[ 	\left|K^{q_1}_{ml}(r, \theta, \varphi) \right|^2 + 	\left|K^{q_2}_{ml}(r, \theta, \varphi) \right|^2\right]  
			\sin\theta \, d\theta.
	\end{multline}
Let:
	\begin{equation}\label{Sp-solid2}
			\mathcal I_{lmp}=\int _0^\pi  
			\left[ 	\left|K^{q_1}_{ml}(r, \theta, \varphi) \right|^2 + 	\left|K^{q_2}_{ml}(r, \theta, \varphi) \right|^2\right]  
			\sin\theta \, d\theta, 
	\end{equation}
therefore:
	\begin{multline}\label{Sp-solid4}
			\mathcal I_{lmp}=
			\frac{(2-\delta_{0m}) (2l+1)(l-m)!}{4\pi(l+m)!}\\
			\times  \int _0^\pi  
			\left\{  \left [ P'^m_l(\cos\theta)\right ] ^2 + \frac{m^2}{\sin^2\theta }
			\left[ P^m_l(\cos\theta)\right ] ^2\right\}\, \sin\theta \, d\theta. 
	\end{multline}
Looking at the fixed modes $m=0$ and $l=1$, one could see: 
	\begin{equation}\label{Sp-solid3}
			\mathcal I=	\frac{ 3}{4\pi}\int 
			\left[  P'^0_1(\cos\theta)\right] ^2 \sin\theta d\theta= \frac{ 3}{4\pi}\int 
			\left[ -\sin\theta\right] ^2 \sin\theta d\theta= \frac1\pi, 
	\end{equation}
which $l=1$ gives the same result as the one:
	\begin{equation}\label{Sp-rdr7}
			\gamma_{l}= \frac{6\, \omega^4_{1} \, r^3}{ c^3} 
			\left[ \frac{j_1(r\, \omega_1/c)}{r\, \omega_1/c}\right] ^2,
	\end{equation}
given in \cite[p.~52]{B}. Next, using the asymptotic behavior for small argument in spherical Bessel functions:
	\begin{equation}\label{Sp-sp1}
			\lim\limits_{x\to 0} j_l(x)= \frac{2^l \, l!} {(2l+1)!}\, x^l, 
	\end{equation}
hence:
	\begin{equation}\label{Sp-sp2}
		\lim\limits_{x\to 0}\frac{ j_l(x)}{x^l}= \frac{2^l \, l!} {(2l+1)!},
	\end{equation}
and,
	\begin{equation}\label{Sp-rdr8}
			\gamma_l= \frac{6\, \omega^4_{1} \, r^3}{ c^3} 
			\left[  \frac26\right] ^2 =\frac 23\frac{\omega^4_{1} \, r^3}{ c^3}.
	\end{equation}
which is the same as Eq.~(14) in \cite{Crowell}. 
\section{Radiative Decay on Cylindrical Surfaces}\label{1-cyli}
The interaction of surface collective excitation, like surface plasmons, in the vicinity of a cylindrical boundary with external probes has been studied in \cite{BURMISTROVA} following the general case argued in \cite{B}. Authors in \cite{BURMISTROVA} studied one specific case of an extremely long and thin cylinder for which the $z$-axis coincides with the axis of the cylinder. The case investigated in \cite{B} yet is more general and is considered the first infinite geometry being quantized. Throughout this section, we take a general cylinder and we follow the exact steps we have taken in Chapter I, to drive the results in \cite{B}. \\
Cylindrical coordinate is given by:
	\begin{equation}
			x=b_0\,\rho\cos\varphi, \quad 
			y=b_0\,\rho\sin\varphi, \quad 
				z=b_0\, z,
	\end{equation}
where:
	\begin{equation}
			0\le \rho<\infty, \quad 
			-\pi<\varphi\le \pi,\quad 
			-\infty<z<\infty,
	\end{equation}
with scale factors:
	\begin{equation}
			h_\rho=h_z=b_0, \quad 
			h_\varphi=b_0\rho.
	\end{equation}
Laplace equation in Cylindrical coordinate \cite{LSSFA} $(\rho, \varphi,z)$ is:
	\begin{equation}\label{1-LCy}
			\vec \nabla^2\Phi=\frac{1}{b_0^2}
			\left\{ 
			\frac{1}{\rho}
			\frac{\partial}{\partial \rho}
			\left(	\rho	\frac{\partial}{\partial \rho}	\right)+
			\frac{1}	{\rho^2}
			\frac{\partial^2}	{\partial \varphi^2}+
			\frac{\partial^2}	{\partial z^2}
			\right\} \Phi, 
	\end{equation}
which has infinitely many solutions of the form:
	\begin{equation}\label{1-Cy}
			\Phi(\rho,\varphi,z)=R( \rho)\Theta(\varphi)Z(z). 
	\end{equation}
Applying separation of variable techniques, one obtains the equations:  
	\begin{align}
			&	\frac	{d^2Z}	{dz^2}	-	k^2	Z=0,\\ 
			&	\frac	{d^2\Theta}	{d\varphi^2}	+	m^2	\Theta=0,\\ 
			&	\frac	{d^2R}	{d \rho^2}	+	
			\frac	{1}	{ \rho}	
			\frac{dR}	{d\rho}+	\left( k^2-
			\frac{m^2}	{ \rho^2}\right) R=0,
	\end{align}
replacing $k$ with $ik$ gives: 
	\begin{align}
			&\frac{d^2Z}	{dz^2}+	k^2		Z=0,\\ 
			&\frac{d^2\Theta}	{d\varphi^2}+	m^2		\Theta=0,\\ 
			&\frac{d^2R}	{d\rho^2}+
			\frac{1}	{\rho}
			\frac{dR}{d\rho}-\left( k^2+
			\frac{m^2}{ \rho^2}\right) R=0.
	\end{align}
Scalar inside and outside potentials of an infinite cylinder with base radius $\rho_0$ could be written as:
	\begin{eqnarray}
			\Phi_{\text{i}}	(\mathbf r,t)=
			\sum\limits_{m=-\infty}^{\infty} 
			e^{im\varphi}
			\int_{-\infty}^{\infty}
			B_{mk}(t)
			I_m(|k| \rho)	K_m(|k|\rho_0)
			e^{ikz} \, dk,
			\qquad	 \rho		\le \rho_0, 	\label{2-Phi}\\
			\Phi_{\text{o}}	(\mathbf r,t)=
			\sum\limits_{m=-\infty}^{\infty} 
			e^{im\varphi}
			\int_{-\infty}^{\infty}
			B_{mk}(t)
			I_m(|k| \rho_0)
			K_m(|k| \rho)
			e^{ikz} \, dk,
			\qquad	 \rho_0		\le \rho .		\label{2-Pho}
	\end{eqnarray}
Utilizing Heaviside function $\Theta$, we define:
	\begin{equation}
			\Phi(\mathbf r,t)=
			\Theta(\rho_0-\rho)
			\Phi_{\text{i}}(\mathbf r,t)+
			\Theta(\rho-\rho_0)
			\Phi_{\text{o}}(\mathbf r,t),
	\end{equation}
hence Laplacian becomes:
	\begin{equation}
			\vec \nabla^2	\Phi(\mathbf r,t)=
			\frac	{\delta(\rho-\rho_0)}	{b_0^2}
			\left(	\frac{\partial	\Phi_{\text{o}}	}	{\partial \rho}-
			\frac{\partial	\Phi_{\text{i}}	}	{\partial \rho}		\right)+
			\Theta	(\rho_0-\rho)
			\vec \nabla^2	\Phi_{\text{i}}	(\mathbf r,t)+
			\Theta	(\rho-\rho_0)
			\vec \nabla^2	\Phi_{\text{o}}	(\mathbf r,t).
	\end{equation}
Since we are only interested on the surface and since Laplacian vanishes inside and outside, then:
	\begin{equation}\label{2-C_Laplace}
			\vec \nabla^2	\Phi(\mathbf r,t)=
			\frac	{\delta(\rho-\rho_0)}	{b_0^2}
			\left( 	\frac{\partial	\Phi_{\text{o}}	}	{\partial \rho}-
			\frac{\partial	\Phi_{\text{i}}	}	{\partial \rho}		\right).
	\end{equation}
Using \eqref{2-Phi} and \eqref{2-Pho}, we have:
	\begin{eqnarray}
			\frac{	\partial\Phi_{\text{o}}	}	{\partial \rho}-
			\frac{	\partial\Phi_{\text{i}}	}	{\partial \rho}&=&
			\sum\limits_{m=-\infty}^{\infty} 
			e^{im\varphi}
			\int_{-\infty}^{\infty}
			|k|~B_{mk}(t)
			\mathcal W\{I_m(|k| \rho),K_m(|k| \rho)\}
			e^{ikz} \, dk \notag \\
			&=&-
			\sum\limits_{m=-\infty}^{\infty} 
			e^{im\varphi}
			\int_{-\infty}^{\infty}
			\rho^{-1}~B_{mk}(t)
			e^{ikz} \, dk
	\end{eqnarray}
using Wronskian identity \cite{LSSFA}:
	\begin{equation}
			\mathcal W\{	I_m(|k| \rho),	K_m(|k| \rho)	\}=	- \frac{1}{|k|\rho}.
	\end{equation}
On the one hand, using the relation
$ h_\rho \vec \nabla^2 \Phi = -4\pi \delta(\rho-\rho_0) \sigma, $ we get:
	\begin{equation}\label{2-S}
		\sigma=\frac{1}	{4\pi b_0	\rho_0}
		\sum\limits_{m=-\infty}^{\infty}
		e^{im\varphi}
		\int_{-\infty}^{\infty}
		B_{mk}(t)
		e^{ikz} \, dk,
	\end{equation}
therefore:
	\begin{equation}\label{2-Sd1}
		\ddot \sigma=\frac{1}	{4\pi b_0	\rho_0}
		\sum\limits_{m=-\infty}^{\infty}
		e^{im\varphi}
		\int_{-\infty}^{\infty}
		\ddot B_{mk}(t)
		e^{ikz} \, dk. 
	\end{equation}
On the other hand,
	\begin{equation}\label{2-Sd2}
		\ddot \sigma=-\frac{\omega_{p}^2}	{4\pi b_0	}
		\sum\limits_{m=-\infty}^{\infty}
		e^{im\varphi}
		\int_{-\infty}^{\infty}
		|k|~B_{mk}(t) 
		I'_m(|k| \rho_0)	K_m(|k| \rho_0)							 
		e^{ikz} \, dk.
	\end{equation}
Putting \eqref{2-Sd1} and \eqref{2-Sd2} equal and using the orthogonality in $m$ and $\varphi$ and also $q$ and $z$, we get:
	\begin{equation}
		\ddot B_{mk}(t)+ 	\omega^2_{mk}	B_{mk}(t)=0,
	\end{equation}
where:
	\begin{equation}\label{2-w}
			\omega^2_{mk}= \omega_{p}^2 \,		|k|	\,	\rho_0	 \, 	I'_m(|k| \rho_0)	\,		K_m(|k| \rho_0), 
	\end{equation}
is known as the frequency in cylindrical domain. \\
Next, we proceed by the intention of calculating classical energy $E$. for which we need to find potential and kinetic energy. We shall follow the steps provided in Chapter 1, through Eqs.~\eqref{0-39}--\eqref{Sp-K}. Rewriting Eq.~\eqref{0-V} for the specific cylinder, potential energy is given by:
	\begin{equation}
			V =\frac12 \int_{-\pi}	^{\pi} \int_{-\infty}  ^{\infty}  
			\sigma  {\Phi_{\text{i}}}\bigg|_{\rho=\rho_0}  
			h_{z}h_{\varphi} \, dz d\varphi. 
	\end{equation}
Using Eq.~\eqref{2-Phi} for inside potential $\Phi_{\text i}$ and Eq.~\eqref{2-S} for surface charge density $\sigma$, one finds:
		\begin{multline}\label{2-V1}
				V=\frac{b \rho_0 \omega_{\mathcal B}^2}{8\pi}   	
				\sum\limits_{m,m'}^{\infty} 
				\int_{-\pi}	^{\pi}  e^{i(m-m')\varphi}  \, d\varphi  \\
				\times\int_{-\infty}  ^{\infty} 
				\int_{-\infty}  ^{\infty}  \int_{-\infty}  ^{\infty}    
				|k|	\,		\frac{B_{mk}(t)}{\omega^2_{mk}}~	\overline{B_{m'k'}}(t)
				I'_m(|k|	\rho_0) I_{m'}	(|k'|\rho_0)	\\
				\times K_m(|k|\rho_0)  K_{m'}	(|k'|\rho_0)			 
				e^{i(k-k')z}\, dk	\,	dk'	\,dz,	
		\end{multline}
as we replaced real-valued potential with its complex conjugate. Using the orthogonality relation:
	\begin{equation}\label{2-O1}
			\int_{-\pi}^{\pi}  
			e^{i(l-l')x }dx = 	2\pi ~\delta_{ll}, 
	\end{equation}
It follows that \eqref{2-V1} can be written as:
	\begin{equation}\label{2-V2}
			V=\frac{\pi b  \rho_0 \omega_{\mathcal B}^2}{2}   	
			\sum\limits_{m=-\infty}^\infty
			\int_{-\infty}  ^{\infty}    
			\frac{|k|}	{\omega^2_{mk}}	~	
			\big| B_{mk}(t)\big|^2
			I_m 	(|k|\rho_0)	I'_m	(|k|\rho_0) 
			\big [ K_m(|k|\rho_0) \big ]^2
			\, dk	.
	\end{equation}
We use frequency relation we have obtained in Eq.~ \eqref{2-w} to write: 
	\begin{equation}\label{2-V3}
			V=\frac{\pi b  }{2}   	
			\sum\limits_{m=-\infty}^\infty
			\int_{-\infty}  ^{\infty}    
			\big| B_{mk}(t)\big|^2
			I_m 	(|k|\rho_0)	 K_m(|k|\rho_0) 	\, dk	.
	\end{equation}
Similar calculations gives kinetic energy as:
		\begin{equation}\label{2-K}
				T=\frac{\pi b  \rho_0 \omega_{\mathcal B}^2}{2}   
				\sum\limits_{m=-\infty}^\infty
				\int_{-\infty}  ^{\infty}    
				\frac	{|k|}	{\omega^4_{km}}
				\big|\dot B_{mk}(t)	\big|^2
				I_m		(|k|	\rho_0)		I'_m	(|k|\rho_0)
				\big [	K_m	(|k|	\rho_0)	\big ]^2
				\,dk.
		\end{equation}
Therefore, total classic energy is found by adding Eqs.~\eqref{2-V3} and \eqref{2-K} as:
	\begin{equation}\label{2-E}
			E=\frac{b \pi  }{2}  
			\sum\limits_{m=-\infty}^\infty
			\int_{-\infty}  ^{\infty} 					
			\frac{I_ m(|k|\rho_0) \, 	K_m(|k|\rho_0)}{\omega^2_{mk}}
			\bigg [  
			\big|\dot B_{mk}(t)\big|^2+
			{\omega^2_{mk}}\big |B_{mk}(t)\big|^2
			\bigg ]	 \,dk.
	\end{equation}
Before moving forward,we note the following symmetry relations for modified Bessel functions given by Eq. ~(5.7.10) (see \cite{LSSFA}, p.~110) as:
	\begin{align}\label{2-Sym}
		\begin{cases}
				I_{m}(z) = I_{-m}(z); \qquad \text{for} \quad m\in \mathbb{Z},\\
				k_{\nu}(z) = k_{-\nu}(z);\qquad \text{for} \quad \nu\in \mathbb{R},
		\end{cases}
	\end{align}
where for integers $\nu=n$, $k_{n}(z) = \lim\limits_{\nu\to n}k_{\nu}(z)$, then:
	\begin{eqnarray}
			k_{n}(z) = \lim\limits_{\nu\to n}k_{\nu}(z)=
			\lim\limits_{-\nu\to n}k_{-\nu}(z)\overset{\eqref{2-Sym}}{=} \lim\limits_{-\nu\to n}k_{\nu}(z)=
			\lim\limits_{\nu\to -n}k_{\nu}(z)=k_{-n}(z),
	\end{eqnarray}
hence $	k_{n}(z) = k_{-n}(z)$ for $n\in \mathbb{Z}$. Considering inside potential given in Eq.~\eqref{2-Phi}, the conjugate of potential is hence given by:
	\begin{equation}
			\overline{\Phi_{\text{i}}}	(\mathbf r,t)=
			\sum\limits_{m=-\infty}^{\infty} 
			e^{-im\varphi}
			\int_{-\infty}^{\infty}
			\overline{B_{mk}}(t)
			I_m(|k| \rho)	K_m(|k|\rho_0)
			e^{-ikz} \, dk,	
	\end{equation}
letting $m\to -m$ and $k\to -k$, we have:
	\begin{eqnarray}
			\overline{\Phi_{\text{i}}}	(\mathbf r,t)=
			\sum\limits_{m=-\infty}^{\infty} 
			e^{im\varphi}
			\int_{-\infty}^{\infty}
			\overline{B_{-m-k}}(t)
			I_{-m}(|k| \rho)	K_{-m}(|k|\rho_0)
			e^{ikz} \, dk,	
	\end{eqnarray}
using symmetry relations for modified Bessel functions given in \eqref{2-Sym} \cite{LSSFA}, we may write:
	\begin{eqnarray}
			\overline{\Phi_{\text{i}}}	(\mathbf r,t)=
			\sum\limits_{m=-\infty}^{\infty} 
			e^{im\varphi}
			\int_{-\infty}^{\infty}
			\overline{B_{-m-k}}(t)
			I_m(|k| \rho)	K_m(|k|\rho_0)
			e^{ikz} \, dk,	
	\end{eqnarray}
since potential is real-valued, then $\Phi_{\text{i}}	(\mathbf r,t)=\overline{\Phi_{\text{i}}	}(\mathbf r,t)$, implies:
	\begin{equation}\label{2-B}
			B_{mk}(t)=	\overline{B_{-m-k}}(t),
	\end{equation}
for all $m$ and $k$. The complex coefficients $B_{mk}(t)$ could be written as:
	\begin{equation}\label{2-Bmk}
			B_{mk}(t)=\frac{\beta_{mk}}{\omega_{mk}} b_{mk}, 
	\end{equation}
where $b_{mk}$ are some complex function (of time) proportional to $e^{-i\omega_{mk}t}$ in which symmetric (in $m$ and $k$) coefficients $\beta_{mk}$ would be determined later. We could follow the identities:
	\begin{eqnarray}\label{2-C1}
			B_{mk}(t)&=&\frac12	\left[ 	B_{mk}(t)	+	B_{mk}(t)	\right] \\
			\overset{\eqref{2-B}}{=}&&
			\frac12	\left[ 	B_{mk}(t)	+	\overline{B_{-m-k}}(t)	\right] \\
			&=&\frac	{\beta_{mk}}	{2\omega_{mk}}	
			\left[ 	b_{mk}+\overline{b_{-m-k}}	\right] \\
			&=&\frac	{\beta_{mk}}	{2\omega_{mk}}	
			\left[ 	b_{mk}+b^*_{-m-k}	\right] ,
	\end{eqnarray}
noting that according to Eq.~\eqref{2-B}, since $\omega_{mk}=\omega_{-m-k}$, we could replace one with another. The time derivative of coefficients given in Eq.~\eqref{2-Bmk} gives:
	\begin{equation}\label{2-Bdot}
			\dot B_{mk}(t)= 
			- i\beta_{mk}~
			b^*_{mk},
	\end{equation}
then we follow similar steps to write:
	\begin{eqnarray}\label{2-B3}
			\dot B_{mk}(t)&=&\frac12	\left[ 	\dot B_{mk}(t)	+	\dot B_{mk}(t)	\right] \\
			&=&\frac12	\left[ 	\dot B_{mk}(t)	+	\overline{\dot C_{-m-k}}(t)	\right] \\
			&\overset{\eqref{2-Bdot}}{=}&
			\frac{\beta_{mk}}{2}	\left[ -i	b^*_{mk}+i b_{-m-k}\right] \\
			&=&	\frac{i\beta_{mk}}{2}	\left[ b_{mk}- b^*_{-m-k}\right]. 
	\end{eqnarray}
Therefore:
	\begin{equation}
			\big|\dot B_{mk}(t)\big|^2+
			\omega^2_{mk}\big |B_{mk}(t)\big|^2= \frac	{\beta^2_{mk}}	{2}	
			\left(	b_{mk}b^*_{-m-k} +	b^*_{mk}b_{-m-k}	\right), 
	\end{equation}
and,
	\begin{eqnarray}\label{2-E1}
			E=\frac{b \pi  }{2}  
			\sum\limits_{m=-\infty}^\infty
			\int_{-\infty}  ^{\infty}    
			I_{ m}(|k|\rho_0)
			K_m(|k|\rho_0)						
			\frac{ \beta^2_{mk}}{2\omega^2_{mk}}
			\left(
			b_{mk}	b^*_{-m-k}+	b^*_{mk}	b_{-m-k}						
			\right).
	\end{eqnarray}
Comparing Eq.~\eqref{2-E1} with the quantized Hamiltonian in terms of operators as:
	\begin{eqnarray}
			\hat{\mathbf H} =
			\sum\limits_{m}\int_{\Omega}
			\frac{\hbar\omega_{mk}}{2}
			\left( 
			\hat{b}^{\dagger}_{mk}	\hat{b}_{mk}+
			\hat{b}_{mk}	\hat{b}^{\dagger}_{mk}
			\right)	\, dk, 
	\end{eqnarray}
where $\hat{b}_{mk}$ and $\hat{b}^{\dagger}_{mk}$ are annihilation and creation operators respectively. We find $\beta_{mk}$ as:
	\begin{eqnarray}\label{2-beta}
			\beta^2_{mk}=
			\frac{2\hbar}{b_0\pi	}
			\frac{ \omega^3_{mk}}		{I_{ m}	(|k|\rho_0)	K_m	(|k|\rho_0)}	.
	\end{eqnarray}	
Using Eq.~\eqref{2-beta}, Eq.~\eqref{2-Bmk} could be rewritten as:
	\begin{eqnarray}\label{2-Bnew}
			B_{mk}(t)&=& \frac{1}{2\omega_{mk}}  	\sqrt{\frac{2\hbar}{b_0\pi	}
				\frac{ \omega^3_{mk}}		{I_{ m}	(|k|\rho_0)	K_m	(|k|\rho_0)}} ~			(b_{mk}+b^*_{-m-k})\\
			&=&   	\sqrt{
				\frac{ \hbar\omega_{mk}}		{2b_0\pi I_{ m}	(|k|\rho_0)	K_m	(|k|\rho_0)}} ~			(b_{mk}+b^*_{-m-k}).
	\end{eqnarray}
Replacing Eq.~\eqref{2-Bnew} in Eqs.~ \eqref{2-Phi} and \eqref{2-Pho}, we can write the quantized inside and outside potentials as:
	\begin{multline}
			\Phi	(\mathbf r,t)=
			\sum\limits_{m=-\infty}^{\infty} 		
			\int_{-\infty}^{\infty}
			\mathcal B_{mk} 	(t)
			\bigg [ \Theta(\rho_0-\rho) I_m(|k| \rho)	K_m(|k|\rho_0) \\
			+
			\Theta(\rho-\rho_0) I_m(|k| \rho_0)	K_m(|k|\rho)						
			\bigg ] 
			(b_{mk}+b^*_{-m-k}) 
			e^{i(kz+m\varphi)}	 \, dk,
	\end{multline}
where:
	\begin{equation}\label{2-mathB}
			\mathcal B_{mk}=\sqrt{
				\frac{ \hbar\omega_{mk}}		{2b_0\pi I_{ m}	(|k|\rho_0)	K_m	(|k|\rho_0)}}, 
	\end{equation}
also see \cite{BURMISTROVA}.	Carrying from integral to sum, over a finite box $\mathcal B$ with side $L$, we have:
	\begin{multline}
			\Phi	(\mathbf r,t)=
			\sum\limits_{m=-\infty}^{\infty} 		
			\sum\limits_{k} 	
			\mathcal B^*_{mk}
			\bigg [\Theta(\rho_0-\rho) I_m(|k| \rho)	K_m(|k|\rho_0) \\+
			\Theta(\rho-\rho_0) I_m(|k| \rho_0)	K_m(|k|\rho)						
			\bigg ] \\
			\times (b_{mk}+b^*_{-m-k}) 
			e^{i(kz+m\varphi)}	,
	\end{multline}
with one-dimensional wave vector $k$ and: 
	\begin{eqnarray}\label{2-mathB*}
			\mathcal B^*_{mk}=\sqrt{
				\frac{ \hbar \omega_{mk}}		{b_0 L\,  I_{ m}	(|k|\rho_0)	K_m	(|k|\rho_0)}}.
	\end{eqnarray}
Next, we continue by calculating the current given by $\mathbf J=-n_0\, e \,\dot{\vec u}$. For the charge accerelation vector $\dot{\vec u}$, we use the relation given in Eq.~\eqref{0-udot} to write: 
	\begin{equation}
			\dot{\vec{u}}=
			-\frac{e}{m_e}
			\vec \nabla
			\sum_{m} e^{im\varphi}
			\int_{-\infty}^{\infty}
			\frac{\dot B_{mk}(t)}{\omega_{mk}^2} 
			I_{m}(|k|\rho)
			K_{m}(|k|\rho_0)
			e^{ikz} \, dq,
	\end{equation}
using:
	\begin{equation}
			\dot B_{mk}(t)= 
			\frac{i\beta_{mk}}{2} 
			(b_{mk}-b^*_{-m-k})=	\frac{i}{2}\sqrt{\frac{2\hbar}{b_0\pi	}
				\frac{ \omega_{mk}}		{I_{ m}	(|k|\rho_0)	K_m	(|k|\rho_0)}}	(b_{mk}-b^*_{-m-k})	,		
	\end{equation}
we find the current as:
	\begin{multline}
			J=-\frac{i}{2}\frac{n_0 e^2}{m_e}
			\vec \nabla
			\sum_{m} 
			\int_{-\infty}^{\infty}
			\frac{1}{\omega_{mk}^2} 
			\sqrt{\frac{2\hbar}{b_0\pi	}
				\frac{ \omega^3_{mk}}		{I_ m	(|k|\rho_0)	K_m	(|k|\rho_0)}}\\
			\times (b_{mk}-b^*_{-m-k})
			I_m	(|k|\rho)
			K_m (|k|\rho_0)
			e^{ikz} e^{im\varphi}\, dk. 
	\end{multline}
We may use the bulk frequency relation, given in Eq.~\eqref{1-29} to write:
	\begin{equation}\label{2-Jdot}
			J=	-\frac{i\omega_p^2}{8\pi }
			\sum_{m} 
			\int_{-\infty}^{\infty}
			\sqrt{
				\frac{2 \hbar}{b_0 \pi}	\frac{  K_{m}(|k|\rho_0)}		{ \omega_{mk}	 I_m(|k|\rho_0)	}}
			(b_{mk}-b^*_{-m-k})
			\vec \nabla \left[ I_{m}(|k|\rho)
			e^{ikz} e^{im\varphi}\right] \, dk.
	\end{equation}
Here if we follow \cite{B,JRTPE,zeid,maggiore}, every formula remains correct if we carry the substitutions: 
	\begin{eqnarray}
&&	\sum\limits_s \rightarrow	\sqrt{\frac{V}{(2\pi)^3}}	\int d^3 s;\\
&& \sum\limits_k \rightarrow	\sqrt{\frac{L}{2\pi}} \int \, dk.
\end{eqnarray}
using the normalization factor $\frac{1}{\sqrt {\mathcal V}}$. 
Using this in Eq.~\eqref{2-Jdot}, gives:
	\begin{eqnarray}
			J=	-\frac{i\omega_{\mathcal B}^2}{8\pi }\sqrt{\frac{2\pi}{L}}
			\sum\limits_{m} 
			\sum\limits_k
			\sqrt{
				\frac{2 \hbar}{b _0\pi}	\frac{  K_{m}(|k|\rho_0)}		{ \omega_{mk} I_{ m}	(|k|\rho_0)	}}
			(b_{mk}-b^*_{-m-k})
			\vec \nabla \left[ I_{m}(|k|\rho)
			e^{ikz} e^{im\varphi}\right]. 
	\end{eqnarray}
For $b_0=1$, it gives the  Eq. ~(70) on \cite{B}, p.~134, as:
	\begin{eqnarray}
			J	=	-\frac{i\omega_{p}^2}{4\pi\sqrt{L}}
			\sum\limits_{m} 
			\sum\limits_k
			\sqrt{
				\frac{ \hbar}{\pi}	\frac{  K_{m}(|k|\rho_0)}		{ \omega_{mk} I_{ m}	(|k|\rho_0)	}}
			(b_{mk}-b^*_{-m-k})
			\vec \nabla \left[ I_{m}(|k|\rho)
			e^{ikz} e^{im\varphi}\right].								
	\end{eqnarray}
Considering Eq~\eqref{0-IntH}  in case for a cylinder:
	\begin{eqnarray}\label{2-H2}
			{H}_{em} =   -\frac{n_0e}{c}
			\int_0^{2\pi}~\int_0^\infty
			\left(	\dot{\Psi}\mathbf{A}	\cdot	\mathbf{\hat e}_{\eta} 			\right)
			h_z	h_\varphi \,	dz 	d\varphi, 
	\end{eqnarray}
we could find the interaction Hamiltonian as:
		\begin{multline}\label{CH}
				{H}_{em}  = \frac{n_0e^2}{ m_0\, c}	\sum\limits_\mathbf{s}
				\sum\limits_{k=1,2}
				\sqrt{\frac{\hbar }{	\mathcal V	\omega_s}}
				(\mathbf{\hat{e}}_\rho			\cdot		\mathbf{\hat{e}}_k)
				\big(	a_{\mathbf{s}k}		e^{i\mathbf{s\cdot r}}		+		a^*_{\mathbf{s}k}		e^{-i\mathbf{s\cdot r}}\big) \\
				\times 		\sum\limits_{m}
				\int_{0}^{2\pi}
				\int_{-L}^{L} \Bigg\{
				\int_0^{\infty}
				I_{m}(|k|\rho_0)	K_{m}(|k|\rho_0)	  e^{im\varphi}		e^{ikz}
				\\
				\times
				\frac{	\beta_{mk}	}{	2i \omega^2_{mk}	}(	b_{mk}	-	b^*_{-m-k})
				dk \Bigg\}  h_\varphi h_z \, d\varphi		 dz ,
		\end{multline}
where:
	\begin{equation}
			\mathcal V=\pi \rho^2_0 L, 
	\end{equation}
is the volume of the finite cylindrical box with length $L$. The emission matrix element can be calculate as:
	\begin{eqnarray}\label{2-CM}
			\mathcal {M}^{(q\mathbf{s})}_{mk} =
			\frac{n_0e^2}{m_0}
			\sqrt{\frac{\hbar}{\mathcal V\omega_s}}
			\frac{\beta_{mk}}{2i\omega^2_{mk}}
			I_{m}(|k|\rho_0)			K_{m}(|k|\rho_0)	
			\mathcal I^q_{mk},
	\end{eqnarray}
where the integrals $	\mathcal I_{mk}$ are given as:
	\begin{align}\label{2-CI}
			\mathcal I^q_{mk}=	\int_{0}^{2\pi}
			\int_{-L}^{L} 	(\mathbf{\hat{e}}_\rho		\cdot		\mathbf{\hat{e}}_k	)
			e^{-i\mathbf{s\cdot r}}	e^{im\varphi}		e^{ikz}
			h_\varphi h_z \, d\varphi		 dz .
	\end{align}
Taking:
	\begin{equation}
			\mathbf{s}=\omega_s	
			(\cos\psi,	0	,\sin\psi),
	\end{equation}
allows us to take $\hat{\mathbf{e}}_{q}$, for $q=1,2$ as:
	\begin{equation}
			\mathbf {\hat e}_{1}=(0,1,0)
			\qquad 	\text{and} 	\qquad
			\mathbf {\hat e}_{2}=	(\sin\psi,	0	,-\cos\psi).	\notag 
	\end{equation}
Leaving us with two different integrals:
	\begin{eqnarray}\label{2-CI1}
			\mathcal I^{(1)}_{mk}=b_0\rho_0	
			\int_{0}^{2\pi} 	\int_{-L}^{L}					\sin\varphi 
			e^{im\varphi}		e^{ikz}		E_{mk}(\varphi,z)
			\, d\varphi		 dz .
	\end{eqnarray}
and,
	\begin{eqnarray}\label{2-CI2}
			\mathcal I^{(2)}_{mk}=b_0\rho_0	
			\int_{0}^{2\pi} 	\int_{-L}^{L}					\sin\psi 		\cos\varphi 
			e^{im\varphi}		e^{ikz}		E_{mk}(\varphi,z)
			\, d\varphi		 dz .
	\end{eqnarray}
where:
	\begin{eqnarray}\label{CIE}
			E_{mk}(\varphi,z)	=
			e^{-i\omega_s \big ( \rho_0	\cos\psi	\cos\varphi 	+	\sin\psi	z\big )}.
	\end{eqnarray}
Radiative decay rate per solid angel $\Omega$, after using Eqs.~\eqref{2-beta} and \eqref{2-CM} in Eq.~\eqref{0-dr}, takes the form:
	\begin{equation}\label{2-CRDR}
			\frac{	d\beta_{mk}	}{	d \Omega}	=	
			\frac{1}{8\pi b_0 c^3}	\left[ 	\frac{n_0e^2}{m_e}	\right] ^2
			I_{m}(|k|\rho_0)			K_{m}(|k|\rho_0) 
			\left[		\left(  \mathcal I^{(1)}_{mq}	\right)^2	+	
			\left(  \mathcal I^{(2)}_{mq}\right)^2	\right].
	\end{equation}
The analytic solution for $\mathcal I^{(1)}_{mk}$ and $\mathcal I^{(2)}_{mk}$ are given as following:
	\begin{eqnarray}\label{2-CI1S}
			\mathcal I^{(1)}_{mk}=\frac{-2ib_0\rho_0}{\mathcal C\, \mathcal D}		\sin(L\mathcal C)\left[ (-1)^m	e^{-i\mathcal D}	-
			\frac{im}{2\pi}		J_m(\mathcal D)\right], 
	\end{eqnarray}
and,
	\begin{eqnarray}\label{2-CI2S}
			\mathcal I^{(2)}_{mk}=\frac{2b_0\rho_0	\sin\psi}{4\pi 	\mathcal C}		\sin(L\mathcal C)\Big [ J_{m+1}(\mathcal D)	+
			J_{m-1}(\mathcal D)\Big ].
	\end{eqnarray}
where:
		\begin{align}\label{2-CAB}
				&\mathcal C=k-\omega_s \sin\psi;\\
				&	\mathcal D=\rho_0	\omega_s	\cos\psi. 
		\end{align}

\clearpage
\newpage
\setlength{\parindent}{0.5in}
\begin{center}
\vspace*{0.75in}
\addcontentsline{toc}{chapter}{\appendixname
\hspace*{2pt} B : COLLECTED FACTS\dotfill}
\appendix{\bf APPENDIX B : COLLECTED FACTS}
\end{center}

\label{App:AppendixB}
\renewcommand\thesection{B.\arabic{section}}
\renewcommand{\theequation}{B-\arabic{equation}}
\setcounter{equation}{0} 
\section{Collected Facts about Coordinate Systems}

\subsection{An Arbitrary Coordinate System}\label{cor}
Let $(\hat i, \hat j, \hat k )$ denote the Cartesian unit vectors and $(\alpha, \beta, \gamma)$ be an arbitrary coordinate system. Let $\vec r$ be the radius vector in $(\alpha, \beta, \gamma)$-coordinate. Lastly, let $h_\alpha, h_\beta$ and $h_\gamma$ be the scale factors given by:
\begin{equation}
h_\alpha= \left| \frac{\partial  {\vec r}}{\partial \alpha}\right |, \quad
h_\beta= \left| \frac{\partial  {\vec r}}{\partial \beta}\right |, \quad
h_\gamma= \left| \frac{\partial { \vec r}}{\partial \gamma}\right |. 
\end{equation}
with $ {\vec r} = x \hat i + y \hat j + z\hat k$ with $(\hat i, \hat j, \hat k)$ the Cartesian unit  vectors. Let us assume the surface of revolution is defined by fixing $\zeta=\zeta_0$, 
then unit vectors for this coordinate system are defined by \cite{AWMMP} :
\begin{align}
&\hat e _\alpha=\frac{1}{h_\alpha} \, \frac{\partial \vec r}{\partial \alpha}, \\
&\hat e _\beta=\frac{1}{h_\beta} \, \frac{\partial \vec r}{\partial \beta}, \\
&\hat e _\gamma=\frac{1}{h_\gamma} \, \frac{\partial \vec r}{\partial \gamma}. 
\end{align}

Laplace equation in an arbitrary coordinate system is given by:
\begin{equation}
\vec \nabla^2 \Phi= \frac{1}{h_\alpha\, h_\beta\, h_\gamma}\, \Bigg\{\frac{\partial}{\partial \alpha} \left[ \frac{h_\beta\, h_\gamma}{h_\alpha}\, \frac{\partial \Phi}{\partial \alpha}\right] + \frac{\partial}{\partial \beta} \left[ \frac{h_\alpha\,  h_\gamma}{h_\beta}\, \frac{\partial \Phi}{\partial \beta}\right]+\frac{\partial}{\partial \gamma} \left[ \frac{h_\alpha\, h_\beta}{h_\gamma}\, \frac{\partial \Phi}{\partial \gamma}\right]
\Bigg \}.
\end{equation}

Here are some collected facts about the specific coordinate systems we have considered throughout this manuscript.

\subsection{Spherical Coordinate} 
\begin{eqnarray}\label{b1-2}
x&=&r \, \sin\theta \cos\varphi, \\
y&=&	r \, \sin\theta \sin\varphi, \\
z&=&  r\, \cos\theta,
\end{eqnarray}
with 
\begin{align}\label{b1-3}
0\le r <\infty, \qquad 
0 \le \theta	\le \pi,\qquad 
0 \le \varphi	\le 2\pi.
\end{align}
The scale factors for a parametrization in each component are given by \cite{LSSFA}:
\begin{align}\label{b1-4}
h_r& =  1, \\
h_\theta &=  r \\
h_\varphi & =r\,  \sin\theta.
\end{align}

\subsection{Cylindrical Coordinate }
\begin{align}\label{B-cyl}
&x=\rho\cos\varphi, \\
&y=\rho\sin\varphi, \\
&	z= z,
\end{align}
where
\begin{align}
0\le \rho<\infty, \qquad 
-\pi<\varphi\le \pi,\qquad
-\infty<z<\infty,
\end{align}
with scale factors
\begin{align}\label{B-cyl1}
&h_\rho=h_z=1, \\
&h_\varphi=\rho.
\end{align}

\subsection{Parabaloidal Coordinate}
are related to the rectangular coordinates by
\begin{align}\label{b5-Pc}
x&=\xi~\eta\cos\varphi, \\
y&=\xi~\eta\sin\varphi, \\
z&=\frac{1}{2}~(\xi^2-\eta^2),
\end{align}
with the corresponding  scale factors
\begin{align}\label{b5-sf}
&h_\xi=h_\eta=\sqrt{\xi^2 + \eta^2},\\
&h_\varphi=\xi  \,\eta,
\end{align}
where $0\leq\varphi < 2\pi$ denotes the usual azimuthal angle and the two coordinates $\eta,\xi\geq 0$ are such that the  surfaces of constant $\eta>0$ and  $\xi>0$ describe upward and downward paraboloids of revolution about the $z$-axis.

\subsection{Prolate Spheroidal Coordinate}
\begin{align}
\begin{split}
x(\zeta,\theta,\varphi)&= 	\sinh{\zeta}	\sin{\theta}	\cos{\varphi},\\
y(\zeta,\theta,\varphi)&=\sinh{\zeta}	\sin{\theta}	\sin{\varphi},\\
z(\zeta,\theta,\varphi)&=	\cosh{\zeta}	\cos{\theta},
\end{split}
\end{align}
in the domain defined by :
\begin{equation}
0	\le	\zeta<\infty	,\qquad 	0\le	\theta	\le	\pi, 	\qquad 0\le\varphi\le2\pi,
\end{equation}
or alternatively, 
\begin{align}\label{b5-AHC1}
x&=	\sqrt{({\eta}^2-1)(1-\mu^2)}	\cos{\varphi},\notag\\
y&= 	\sqrt{({\eta}^2-1)(1-\mu^2)}	\sin{\varphi},\\
z&=	{\eta}\, 	{\mu},\notag
\end{align}
with the corresponding scale factors:
\begin{align}\label{b5-AH_SF}
&h_{\eta}=	\sqrt{	\frac{\eta^2-\mu^2}	{\eta^2-1}},\\
&h_{\mu}=	\sqrt{	\frac{\eta^2-\mu^2}{1-\mu^2}},\\
&h_{\varphi}= 	\sqrt{	(1-\mu^2)(\eta^2-1)}.
\end{align}

\subsection{Toroidal Coordinate} 

\begin{align}\label{b5R-C}
&x=\sinh\mu\cos\varphi / (\cosh\mu-\cos\eta),\\
&y=\sinh\mu\sin\varphi / (\cosh\mu-\cos\eta),\\
&z=\sin\eta / (\cosh\mu-\cos\eta),
\end{align}
where 
\begin{equation}
\mu \geq 0, \qquad  0 \le  \eta, \varphi \le  2\pi.
\end{equation}
The associated toroidal scale factors are given by
\begin{align}\label{bR-Ch}
&h_\mu=h_\eta= 1/ (\cosh\mu-\cos\eta),\\
&h_\varphi= \sinh\mu / (\cosh\mu-\cos\eta).
\end{align}

\section{Collected Facts about Special Functions}
\subsection{Bessel Functions}\label{Bess}

\noindent \textbf {Asymptotic behaviour for large arguments and fixed order $m$:}

For large arguments, the series expansions are Eqs. (5.11.6), (5.11.8) and (5.11.9) \cite{LSSFA}, p.122, 123.
\begin{multline}\label{7-B-Asym} 
	J_m(z)=(\frac{2}{\pi z})^{1/2}
	\cos (z-\frac12m\pi-\frac14\pi)	
	\left[ \sum_{k=0}^n			(-1)^k(m,2k)(2z)^{-2k}+	\mathcal O(|z|^{-2n-2})\right] \\
	-	(\frac{2}{\pi z})^{1/2}			\sin (z-\frac12		m\pi-	\frac14\pi) 
	\left[ \sum_{k=0}^n 	(-1)^k 	(m,2k+1)		(2z)^{-2k-1}+	\mathcal O(|z|^{-2n-3})\right],
\end{multline}
\begin{multline}\label{7-B-Asym1} 
	I_m(z)=e^z(2\pi z)^{-1/2}
	\left[ \sum_{k=0}^n
	(-1)^k(m,k)(2z)^{-k}+
	\mathcal O(|z|^{-n-1})\right]\\
	+e^{-z\pm \pi i(m+1/2)}(2\pi z)^{-1/2}
		\left[ \sum_{k=0}^n
	(m,k)(2z)^{-k}+
	\mathcal O(|z|^{-n-1})\right],
\end{multline}
	\begin{equation}\label{7-B-Asym2} 
	K_m(z)=(\frac{\pi}{2z})^{1/2}
	e^{-z}	\left[ \sum_{k=0}^n
	(m,k)(2z)^{-k}+
	\mathcal O(|z|^{-n-1})\right], 
\end{equation}
where $|arg z|\le \pi - \delta$ and 
\begin{eqnarray}
	(m,k)=\frac{(4m^2-1)(4m^2-3^2)...(4m^2-(2k-1)^2)}{2^{2k}k!}, \quad \text{and} \quad (m,0)=1.
\end{eqnarray}
In general \cite{LSSFA,BIBF}:
\begin{align}\label{B-Bessel}
	&	J_m(z)\to\infty\mbox{ as }z\to 0, \\
	&	J_m(z)\to 0\mbox{ as }z\to 0,\qquad m\neq0,\\
	&	I_\nu(z)=e^{-i\nu\pi/2}		J_\nu(ze^{i\pi/2}),  {\small -\pi<\arg(z)<\frac{\pi}{2}},\\
	& 	K_\nu(z)=\frac{i\pi}{2}e^{i\nu\pi/2}H_\nu(ze^{i\pi/2}).
\end{align}

\noindent The Bessel functions of the first kind:
\begin{eqnarray}
	&J_m(z)\to\infty\mbox{ as }z\to\infty \\
	&J_m(z)\to 0\mbox{ as }z\to 0
\end{eqnarray}
The Bessel functions of the second kind:
\begin{eqnarray}
	a(z)	\approx	\frac{2}{\pi}
	\log\frac{z}{2}		\mbox{ as }z\to 0&& \\
	Y_m(z)\approx-	\frac{(n-1)!}{\pi}
	\left(	\frac{z}{2}	\right)^{-n}	\mbox{ as }z\to 0&& \mbox{ for }m=1,2,\dots
\end{eqnarray}

\noindent \textbf {Bessel functions of imaginary argument}

Also called the modified Bessel functions of first, $I_\nu(z)$ and second, $K_\nu(z)$, kind, respectively. Their relation with Bessel functions of first and second kind are given as \cite{LSSFA,BIBF}:
\begin{align}
	&I_\nu(z)=e^{-i\nu\pi/2}J_\nu(ze^{i\pi/2}),\quad \text{for}\quad -\pi<\arg(z)<\frac{\pi}{2}, \\
	&K_\nu(z)=\frac{i\pi}{2}\, e^{i\nu\pi/2}\, H_\nu(ze^{i\pi/2}).
\end{align}
The asymtotic behaviour for modified Bessel functions for a fixed argument $z$ and large complex order $\nu$ when $\nu \to \infty$ are given by \cite{Abra}:
\begin{equation}\label{9BAsy1}
	I'_\nu(\nu \,z)\sim
	\frac{(1+z^2)^{1/4}e^{\nu z}}{\sqrt{2\pi\nu}\,z}
	\sum\limits_{k=0}^\infty .	
\end{equation}	
\subsection{Associated Legendre Functions}\label{leg}
Here are some collected facts and indentities about Associated Legendre Functions \cite{LSSFA, Abra,g&r:table}. 
The associated Legendre functions are symmetric in $n$: 
	\begin{align}\label{b-1}
		P^{m}_{n-\frac12}(\cosh\mu)&= P^{m}_{-n-\frac12}(\cosh\mu),  \\
		Q^{m}_{n-\frac12}(\cosh\mu)&= Q^{m}_{-n-\frac12}(\cosh\mu). 
	\end{align}	
Following relations hold upon making $m\to -m$: 
		\begin{align}\label{b-2}
			P^{-m}_{n-\frac12}(\cosh\mu)&= \frac{\Gamma(n+\frac12-m)}{\Gamma(n+\frac12-m)} P^M_{n-\frac12}(\cosh\mu), \\
			Q^{-m}_{n-\frac12}(\cosh\mu)&= \frac{\Gamma(n+\frac12-m)}{\Gamma(n+\frac12-m)} P^M_{n-\frac12}(\cosh\mu). 
		\end{align}	


\subsection{Conical Functions}\label{con}
\noindent \cite{nist,PK} Integral representation (Mehler integrals) of conical functions:
	\begin{align}\label{b-3}
P^{m}_{-\frac12+iq}(\cos\theta)&= (-1)^m\, \frac{2^{m+1/2}}{\pi(2m-1)!!}\, \frac{Z^m_q}{\sin^m\theta}\, \int_0^\theta \frac{\cosh qx}{(\cos x-\cos\theta)^{1/2-m}}\, dx, \\
P^{m}_{-\frac12+iq}(\cosh\xi)&= \frac{2^{m+1/2}}{\pi(2m-1)!!}\, \frac{Z^m_q}{\sinh^m\xi}\, \int_0^\xi \frac{\cos qx}{(\cosh \xi-\cosh x)^{1/2-m}}\, dx,   
\end{align}	
with $Z^m_q$ given in Eq. ~\eqref{AHZ}. Moreover, 
	\begin{align}\label{b-4}
		P^{m}_{-\frac12+iq}(z)=P^{m}_{-\frac12-iq}(z).
	\end{align}	
Successive relations for Conical functions: 
	\begin{align}\label{b-5}
P^{m}_{-\frac12+iq}(\mu)&= (-1)^m\, (1-\mu^2)^{m/2}\, \frac{d^m P^0_{-\frac12+iq}(\mu)}{d\mu^m}, \quad \mu=\cos\theta, \\
P^{m}_{-\frac12+iq}(\eta)&= (\eta^2-1)^{m/2}\, \frac{d^m P^0_{-\frac12+iq}(\eta)}{d\eta^m}, \qquad \eta=\cosh\xi.  
\end{align}	

Asymptotic representations for fixed $0<\xi\le \alpha<\infty$ and $0<\theta\pi$ can be derived from the asymptotic form of Lagendre functions as \cite{robin}:
	\begin{align}\label{b-6}
		P^{m}_{-\frac12+iq}(\cos\theta)= \frac{q^{m-1/2}}{\sqrt{2\pi\, \sin\theta}}\,e^{\theta q}\bigg [1-\left( \frac{m^2}{2q} -\frac{1}{8q}\right) \cot\theta+ \mathcal O(q^{-2})\bigg ],
	\end{align}	
	\begin{multline}\label{b-7}
P^{m}_{-\frac12+iq}(\cosh\xi)= \sqrt{\frac{2}{\pi\sinh\xi}} \, q^{m-1/2}\, \\
\times \bigg\{ \cos\big[    q\xi + \frac{(2m-1)\pi }{4}\big ]+ 
\left( \frac{m^2}{2q} -\frac{1}{8q}\right)\, \cos \big[    q\xi + \frac{(2m+1)\pi }{4}\big ]\coth\xi+
\mathcal O(q^{-2})
\bigg\}. 
\end{multline}	
\section{Sturm-Liouville Problem}

A {Sturm-Liouville system} is defined by the self-adjoint operator $\mathcal L$ as:
\begin{equation}\label{0-SL}
	\mathcal L u = \frac{d}{dx} \left[ p(x)\frac{du}{dx} \right] + q(x)\, u(x) , 
\end{equation}
where $p>0$ and $q$ are two continuous functions on an interval $[a,b]$, (see \cite{zettl,al2008sturm}). The \emph{regular} Sturm-Liouville \emph{eigenvalue problem} has the form:
\begin{equation}\label{0-SL5}
	-\mathcal L u(x)=\lambda \rho(x) u(x), \quad x\in(a,b),  
\end{equation}
 subject to the separated homogeneous boundary conditions :
\begin{align}\label{0-SL2}
	\begin{cases}
		\alpha_1 u(a)+\alpha_2 u'(a)&=0, \quad |\alpha_1|+ |\alpha_2|>0, \quad \alpha_{1,2}\in \mathbb R, \\
		\beta_1 u(b)+\beta_2 u'(b)&=0, \quad |\beta_1|+ |\beta_2|>0, \quad \beta_{1,2}\in \mathbb R, 
	\end{cases}
\end{align}
where such values of $\lambda$, if any, are called the eigenvalues of the system and $\rho(x)$ is a weight function. 
Since $\mathcal L$ is a self-adjoint operator under the boundary conditions given in Eqs.~\eqref{0-SL2}, eigenvalues of Eq.~\eqref{0-SL5} are real and can be ordered in an ascending manner. To each eigenvalue $\lambda_n$, there corresponds a unique eigenfunction $u_n(x)$ which is considered as the $n$-th fundamental solution to Eq.~\eqref{0-SL5}. The set of all these fundamental solutions $\{u_n(x)\}_{n=0}^\infty$, up to a normalization constant, form an orthonormal basis in $L^2([a,b], \rho(x)dx)$ as they satisfy the following orthogonality relation:
\begin{equation}\label{0-SL7}
\int_a^b u_n(x)\, u_m(x)\, \rho(x)dx=\delta_{nm}.   
\end{equation}
The spectrum of an operator, e.g. $\mathcal L$, is a set consists of all $\lambda$ (in $\mathbb C$) such that the operator (eigenvalue problem) $\lambda I-\mathcal L$ fails to be an invertible element of the respective algebra. The spectrum can be partitioned to many disjoint parts one of which is called the \emph{point spectrum} and consists of all eigenvalues $\lambda_n$. 
Eigenvalue problems over finite intervals lead to discrete spectrum whereas infinite intervals result in a continuous spectrum. Among the geometries investigated here, sphere, prolate spheroid and torus are examples in which the spectrum space is discrete and paraboloid, hyperboloid and infinite cylinder are examples in which a continuous spectrum is assumed \cite{conway,AWMMP,spectral,sturmL}.

\clearpage
\newpage
\setlength{\parindent}{0.5in}
\begin{center}
\vspace*{0.75in}
\addcontentsline{toc}{chapter}{\appendixname
\hspace*{2pt}C : COPYRIGHT CLEARANCE FORM \dotfill}
\appendix{\bf APPENDIX C : COPYRIGHT CLEARANCE FORM}
\end{center}

\label{App:AppendixC}

\begin{doublespace}
Below is the permission for Chapters 2 and 5.
\end{doublespace}
\begin{figure}[!htb]
  \centering
  \includegraphics[scale=0.48]{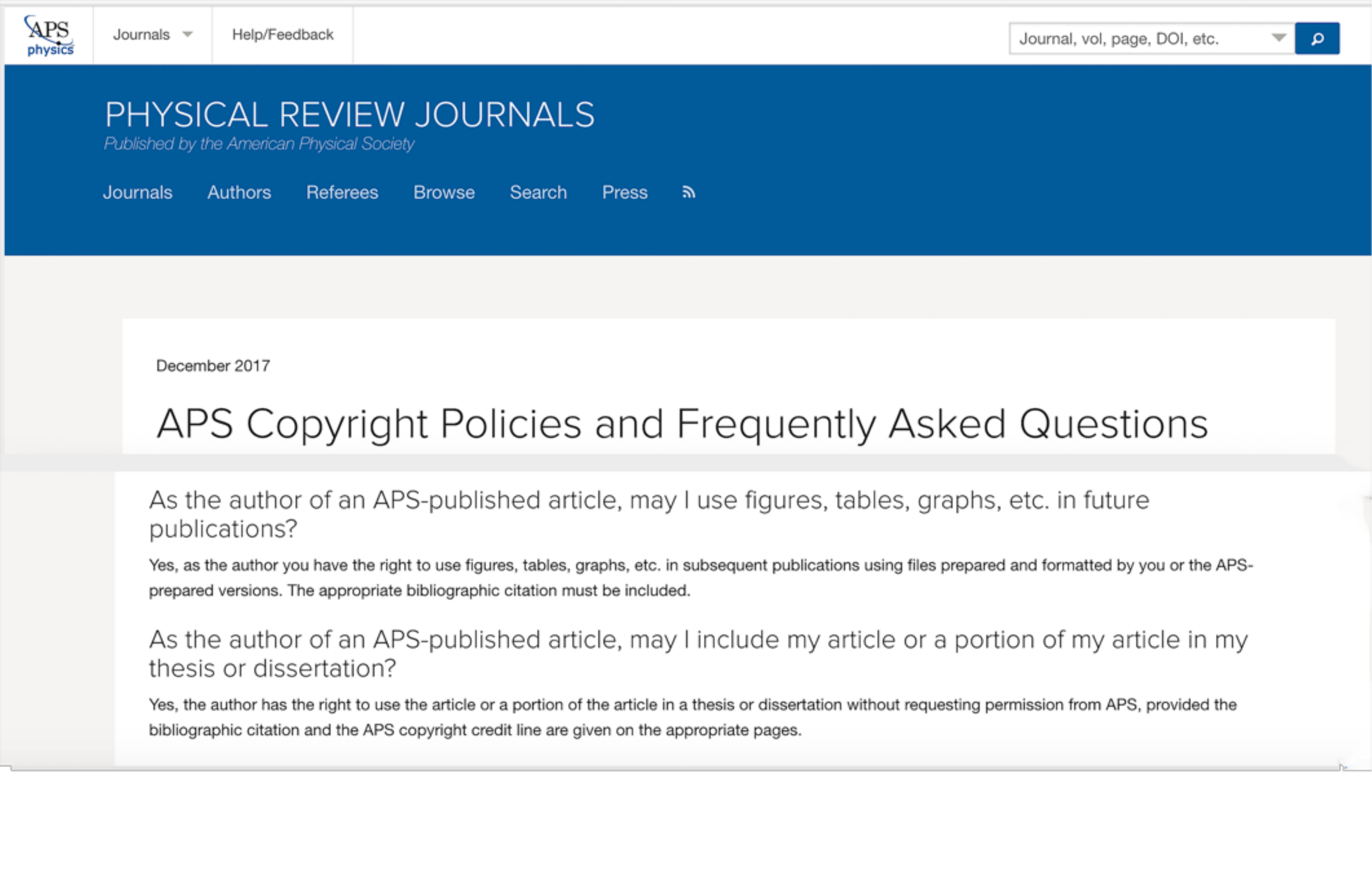}
   \label{fig:copyright_1}
\end{figure}

\clearpage
\setlength{\parindent}{0.5in}

\pagestyle{fancy}
\renewcommand{\headrulewidth}{0pt}
\cfoot{}

\begin{center}
\vspace*{0.75in}
\addtocontents{toc}{\protect\contentsline{chapter}{\normalsize\normalfont ABOUT THE AUTHOR}{\hspace*{-9ex}END PAGE}}
\appendix{\bf ABOUT THE AUTHOR}
\end{center}

\vspace*{0.1in}
\renewcommand\baselinestretch{2.0}\selectfont
Maryam Bagherian received her Bachelor of Science in pure mathematics from the Shahid Beheshti University in 2006.  She started to pursue the Doctor of Philosophy in  Applied Mathematics at the University of South Florida in August 2014. In between, she earned a Master of Arts in applied mathematics with mathematical physics concentration in 2016.  Her research interests include nanoparticles, modeling and simulation and signal/image processing. Some of her non-academic pursuits are yoga, classical music, photography and traveling. 

\end{document}